# Midinfrared Semiconductor Photonics – A Roadmap

J. R. Meyer, I. Vurgaftman, S.-Q. Yu, R. Q. Yang, A. M. Andrews, G. Strasser, B. Schwarz, M. Razeghi, L. Shterengas, G. Kipshidze, G. Belenky, L. Sterczewski, W. Zhou, S. Lee, M. Pan, R. Szedlak, N. Schäfer, J. Koeth, R. Weih, A. Rogalski, A. Piotrowski, J. Sobieski, P. Leszcz, J. Piotrowski, M. R. Mirzaei, R. Kim, J. H. Park, D. Z. Ting, M. B. Santos, V. Trinite, S. Pes, J.-L. Reverchon, N. Gajowski, S. Krishna, W. Du, R. Soref, E. Tournié, J.-B. Rodriguez, L. Cerutti, A. Spott, S. Jung, N. Nookala, A. Vasanelli, B. Chomet, C. Sirtori, N. P. Li, M. A. Zondlo, S. Jain, J. Midkiff, M. Hlaing, K.-C. Fan, R. T. Chen, F. Grillot, S. Zaminga, P. T. Camp, P.-Y. Hsiao, G. Daligou, S. Molesky, and O. Moutanabbir

## Abstract

Semiconductor photonic devices operating in the midwave infrared (mid-IR, which we roughly define here as wavelengths spanning 3 to 14 µm) uniquely address a wide range of current practical needs. These include chemical sensing, environmental monitoring, industrial process control, medical diagnostics, thermal imaging, LIDAR, free space optical communication, and security monitoring. However, mid-IR device technologies are currently still works in progress that are generally much less mature than their near infrared and visible counterparts. Not only are most of the relevant materials more difficult to grow and process, but attainment of the desired optical device performance is often fundamentally more challenging. This Roadmap will review the leading applications for mid-IR optoelectronics, summarize the status and deficiencies of current device technologies, and then suggest possible roadmaps for improving and maturing the performance, manufacturability, and cost of each device type so the critical needs that are uniquely addressed by mid-IR photonics can be satisfied.

## Introduction

J. R. Meyer, I. Vurgaftman, S.-Q. Yu, R. Q. Yang, A. M. Andrews, G. Strasser, and B. Schwarz

Semiconductor photonic devices operating at midwave (MWIR, λ = 3-5 µm) and longwave (LWIR, λ = 8-14 µm) infrared wavelengths, which for convenience in this article will often be referred to collectively as the mid-IR, have enormous potential to uniquely address a wide range of practical needs. For example, most greenhouse gases and other chemicals have unique fingerprint absorption features that provide definitive identification of the given substance, and are orders of magnitude stronger than the overtone lines at shorter wavelengths. Night vision is provided by thermal signatures that are efficiently transmitted with low loss through the MWIR and LWIR atmospheric spectral windows. These and other unique capabilities may be exploited in applications including chemical sensing, environmental monitoring, industrial process control, medical diagnostics, thermal power conversion, thermal imaging, free space optical communication, surveillance, quantum optics, remote sensing, LIDAR, and security monitoring, as well numerous defense applications.

However, at this stage many mid-IR device technologies are still works in progress, being far less mature than their near-IR and visible counterparts. Not only are the required constituent materials generally more difficult to grow and process, but attainment of the desired optical device performance often faces additional fundamental challenges. Furthermore, at present the marketplace for mid-IR systems is significantly constrained by the lower manufacturability and higher cost per device that accompany the immature development status.

Our objective in this Roadmap article is to address the pronounced maturity gap for semiconductor mid-IR optoelectronics by comprehensively reviewing the leading applications, summarizing the status and deficiencies of current components and systems, and then identifying practical pathways to improvement and maturation of the performance, manufacturability, and cost of each device type. The overarching goal is to broaden the scope and impact of mid-IR photonics technology, so it can more readily satisfy the critical needs it uniquely addresses within the foreseeable future.

The Roadmap that follows consists of 25 sub-topic articles that review the status and challenges of each major device type, platform, and application comprising semiconductor mid-IR optoelectronics, and then provide individual roadmaps for guiding the future development and maturation of each technology. The first group of sub-topics treats light emitters, which include quantum cascade lasers (QCLs, Section 1), type-II interband cascade lasers (ICLs, Section 2), type-I ICL and diode lasers (Section 3), frequency combs (Section 4), photonic crystal surface-emitting lasers (PCSELs, Section 5), ring lasers (Section 6), and incoherent light-emitting devices (LEDs, Section 7). The second group comprises light-detecting devices, including focal plane arrays (Section 8), HgCdTe (Section 9), lead salt (Section 10), and III-V structures based on type-II nBn (Section 11), interband and quantum cascade detectors (Sections 12 and 13), resonant cavity detectors (Section 14), and avalanche photodiodes (Section 15). The third group reviews certain specialized platform configurations: silicon-based lasers and detectors (Section 16), growth of III-V devices on silicon substrates (Section 17), and photonic integrated circuits fabricated on both silicon and III-V platforms (Sections 18 and 19), as well as the introduction of metasurfaces for enhancing the photonic properties (Section 20). Finally, after the status and future prospects of these various MWIR device technologies have been established, leading applications most likely to see profound impact will be reviewed. In particular, we will discuss low-cost spectroscopic chemical sensing (Section 21), sensing systems that can be integrated on an ultra-compact semiconductor chip (Section 22), free space optical communication (Section 23), light detection and ranging (LIDAR, Section 24)), and mid-IR thermophotovoltaic devices for energy harvesting (Section 25).

Nonlinear optics (NLO) was not assigned a separate topic, since semiconductors are not the primary mid-IR nonlinear materials. However, several of the articles mention the importance of NLO in various semiconductor technologies (e.g., Sections 1, 4, 18, 19, 25).

Although mid-IR topological materials were also not reviewed, preliminary investigations have recently received a degree of attention [1,2]. Mid-IR photonics for defense is not discussed here to avoid potential concerns over security and export control.

We also mention that B. Hinkov, J. Kunsch, W. Mäntele, and L Sterczewski are Guest-Editing a Roadmap article on the more specialized complementary topic "Infrared Photonics for Healthcare: A Roadmap for Proactive and Predictive Health Management" [3], which will appear in a Focus Issue on Photonics in Medical Diagnostics and Monitoring in the *Journal of Physics: Photonics*.

# 1. Quantum Cascade Lasers

**MANIJEH RAZEGHI** [1,*]

**[1]Center for Quantum Devices, Department of Electrical Engineering and Computer Science, Northwestern University, Evanston, Illinois 60208, USA**

*\*razeghi@northwestern.edu*

## Overview

Mid-wave infrared (IR) quantum cascade lasers (QCLs) offer high output power, excellent efficiency, broad wavelength tunability, and elevated operating temperatures, especially when operating in the 3–12 μm wavelength range. These characteristics make them highly promising for a wide range of applications, including high-resolution molecular spectroscopy, ultra-low-loss optical fiber communications using fluoride-based glasses (with attenuation below $2.5 \times 10^{-4}$ dB/km), trace gas detection, air pollution monitoring (as many molecules, particularly hydrocarbons, exhibiting strong absorption lines in this spectral region), and medical diagnostics. This article presents a comprehensive overview of the development of QCLs, highlighting key milestones, the current state of the technology, and future directions, framed within the broader context of the Semiconductor Mid-Infrared Photonics Roadmap.

## Current status

The development of QCLs operating in the mid-IR range (λ~3–12 μm) is the result of decades of foundational discoveries and technological advancements in semiconductor physics, quantum engineering, and epitaxial growth techniques. A series of key milestones has progressively shaped the state-of-the-art in QCL technology within this spectral range, as illustrated in Fig. 1.

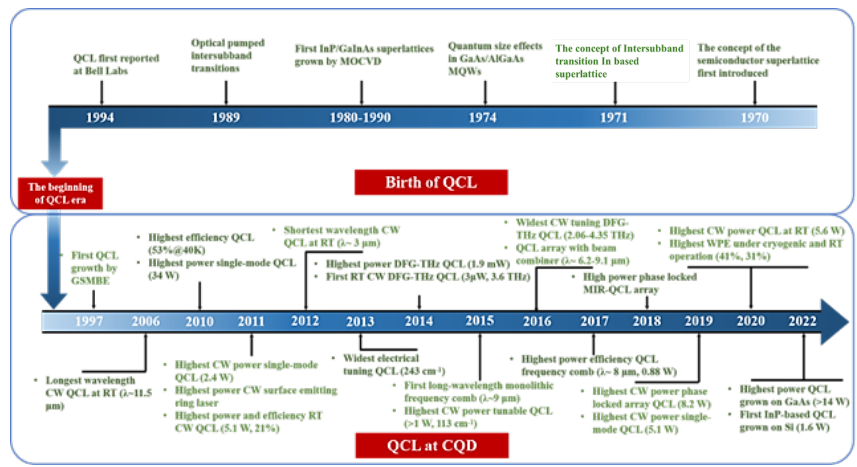

Fig. 1 Birth of QCL technology (upper panel) and the evolution of QCL and its timeline of records set at QCL by CQD at Northwestern University (lower panel).

The concept of the semiconductor superlattice was introduced by Leo Esaki and Raphael Tsu in 1970 [1], establishing the idea of artificial periodic structures to control electronic band structures and carrier transport. This innovation laid the groundwork for modern bandgap and quantum engineering. The subsequent year, Kazarinov and Suris proposed the concept of a laser based on intersubband transitions in a biased superlattice [2]. Although they predicted optical gain from transitions between quantized subbands, they recognized the challenge of electrical instability under strong bias, which precluded the realization of an electrically pumped laser at that time. In 1974, quantum size effects observed experimentally in GaAs/AlGaAs multi-quantum wells by Dingle et al. [3] verified the existence of discrete energy levels in quantum-confined systems and validated quantum well theory. During the 1980s and 1990s, a major leap forward came with the pioneering work of [4] that played a transformative role in advancing III–V compound semiconductor technologies, particularly through the development of metal-organic chemical vapor deposition (MOCVD). At a time when molecular beam epitaxy (MBE) dominated quantum device research, MOCVD developed with competitive merits and not only matched, but in many cases surpassed MBE in terms of material quality, scalability, and reproducibility - crucial factors for industrial applications. Specifically, superlattice material systems including GaAs/GaInP/GaInAsP [4] InP/GaInAs and InP/GaInAsP [5] were successfully grown by MOCVD, a critical development for long-wavelength telecommunication lasers operating at 1.3 and 1.55 μm. In 1989, Helm et al. reported the first observation of optically pumped intersubband emission in quantum wells [6], providing experimental confirmation that subband transitions could be harnessed for mid-IR light emission. Then in 1994, the first electrically pumped intersubband laser, or quantum cascade laser, was demonstrated by Faist et al. in Bell Labs [7]. In 1994, Yang, proposed the interband cascade laser [8, 9]. Intersubband lasers leverage engineered subband transitions in multiple quantum wells and utilize a cascading mechanism to produce multiple photons per injected electron, enabling high power and spectral tailorability.

By tracing the evolution from telecommunication laser development to state-of-the-art intersubband QCLs, we aim to provide not only a comprehensive overview of the field but also an in-depth understanding of how fundamental innovations [10] have shaped modern intersubband optoelectronics.

This paper presents a comprehensive overview of both experimental and theoretical advances that have led to high-power, high wall-plug efficiency (WPE), and exceptional long-term reliability of QCL devices [10]. The scientific and technological basis for these results originates from extensive and pioneering work in telecommunication laser structures. The methods and insights from this background directly underpin many of the experimental results and modeling approaches summarized in this review [11]. With this context, our aim is not only to summarize the state-of-the-art performance of QCLs, but also to clarify how these unmatched levels of power, efficiency, and reliability were achieved, offering a detailed references that may guide future device design.

## 1. Foundational Breakthroughs in QCL Technology

The journey of QCL development began with the realization that traditional diode lasers could not operate in the mid-IR (MWIR) and long-wave infrared (LWIR) spectral ranges. The concept of unipolar QCLs, where electrons transit between the subbands within quantum wells to emit photons, enables lasing in these previously inaccessible wavelengths. This innovation laid the groundwork for the development of QCLs [7], which have since become essential tools in various applications, including chemical sensing, environmental monitoring, and defense technologies.

## 2. Enhancing Performance and Efficiency

The unique capabilities of QCLs have led to their adoption in diverse application areas. In the field of environmental monitoring, QCLs are employed in portable spectrometers for detecting trace gases and pollutants. QCLs have also had important applications in the fields of long-range hazardous chemical and explosives detection, biomedicine, infrared countermeasures and long-range free-space optical communication. In defense and security, they are used in standoff chemical detection systems. Since the sensitivity and range are directly proportional to the laser output power, it is crucial to further improve the power and electrical-to-optical power conversion efficiency of the devices.

Historically in 1997 a QCL at $\lambda \sim 8.5$ μm was produced by a single-step epitaxial growth using customized gas-source molecular beam epitaxy [9]. The first continuous wave (CW) operation of a QCL at room temperature was reported in 2002 [12] . In 2003, the first room-temperature (RT) operation of QCLs emitting at $\lambda \sim 6$ μm with a maximum temperature of 308 K was attained by using improved thermal packaging of a 3-μm electroplated Au contact layer and epi-down bonding on a copper heat sink. The CW optical output power was 132 mW at 293 K and 21 mW at 308 K [13]. In 2010, QCLs emitting at $\lambda \sim 4.85$ μm achieved wall-plug efficiencies up to 53% at cryogenic temperatures, marking the first ever  intersubband laser emitting more light than heat [14]. The P-I-V and WPE vs current curves of a one-well-injector QCL under pulsed mode operation at 40 K is shown in Fig. 2(c). The injection region was designed as a single-well structure, which significantly reduced the thermal backfill effect of carriers and voltage defect, and improved the injection efficiency of carriers at the same time. This achievement was pivotal in reducing thermal management requirements and expanding the operational scope of QCLs [14]. Liu et al. similarly observed 50% WPE for CW operation at 9 K, and 47% at 80 K [15].

For QCLs to reach shorter wavelengths requires higher strain and higher barriers to provide sufficient confinement of the electrons [16] , which poses a significant challenge to the epitaxy. Extremely high strain may lead to defects, dislocations, and even material relaxations, which could deteriorate the material quality. In addition, the carrier leakage into indirect valleys becomes more pronounced when operating at short wavelengths, leading to a gradual degradation of laser performance with increased scattering between the valleys. Therefore, increasing the band offset through strain balancing and minimizing the intervalley scattering by engineering the wave function in the indirect valley states led to the first demonstration of a QCL at $\lambda \sim 3.0$-$3.2$ $\mu$m, which operated at RT with CW output power of 20 mW [17]. Fig 2 illustrates the device structural design, material epitaxial growth technology, and device preparation process. These modifications optimized key performance parameters such as output power, WPE, beam quality, and threshold current density.

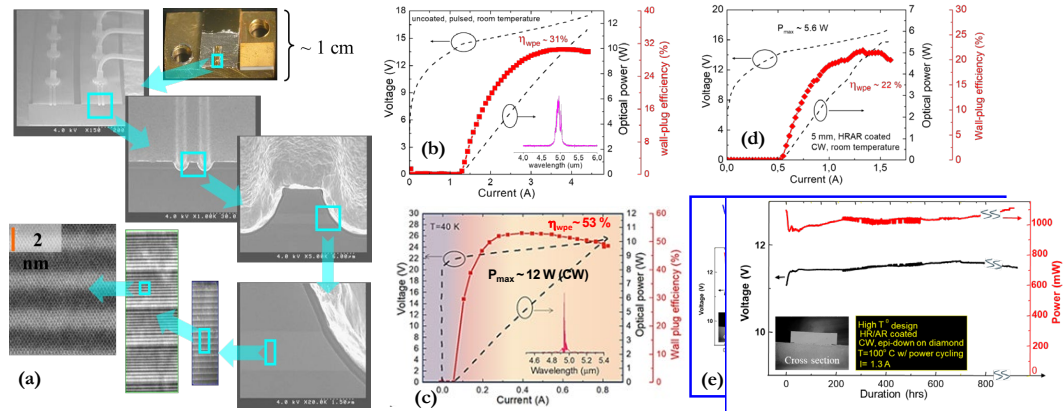

Fig. 2(a) Images of a QCL at different magnification. The top image is a packaged QCL with the tiny laser bar. As each box is zoomed in successively with SEM and TEM, we see gold wires, the double channel structure, the ridge defined by these two channels, the brighter layer core region, the stages and 20 layers in each stage and an image with atomic resolution of the individual layers. (b) PIV curve and WPE of the buried laser as a function of current. Inset: Spectrum of the laser emission at 1.6 A. This test was conducted in pulsed mode with a pulse width of 500 ns and duty cycle of 2% [18]. (c) P-I-V and WPE vs current curves of a one-well-injector intersubband laser under pulse mode operation at 40 K. The inset shows the emission spectrum at 4.9 $\mu$m [14]. (d) P-I-V and WPE vs current curves of a shallow-well QCL with 50 active stages, bonded epi-down under CW mode operation at room temperature [19]. (e) The high temperature life testing of a QCL with power cycling.

A monochromatic spectrum is highly desired in many intersubband laser applications. The first distributed feedback (DFB) QCL was reported in 1997 [20]. In [21], a first-order metallic surface DFB grating achieved the highest single-facet CW output power for a QCL, with side-mode suppression ratio (SMSR) at $\lambda$~4.8 μm of 30 dB. Following further optimization of the DFB coupling coefficient and facet coating to enhance the modal loss difference, the maximum single-facet CW output power was 5.1 W at RT, with a WPE of 16.6% [22]. To realize a high-power mid-IR broad-area intersubband laser with single spatial and spectral modes, a spectrally pure and almost diffraction-limited QCL at $\lambda$~4.36 μm was achieved with high peak power in pulsed mode of 34 W by integrating a first-order photonic-crystal distributed feedback (PCDFB) structure in a cavity length of 3 mm and ridge width of 400 μm [23]. The structure was designed with different coupling coefficients in the lateral and longitudinal directions to compensate for the under coupling and over coupling in the two directions, so that single mode operation in both spectral and spatial realms was achieved [24]. The PCDFB design was then evolved into a β-DFB intersubband laser design within an angled cavity design. A diffraction-limited beam output at $\lambda$~10 μm was demonstrated [25]. The angled cavity design eventually led to the brightest intersubband laser at $\lambda$~4.8 μm, with a record-high pulsed peak output power of 203 W and brightness 156 MW cm$^{-2}$sr$^{-1}$, which are still the records to date [26].

CW wall-plug efficiency and output power at RT are widely deemed the most important metrics for the wide applications of QCL technology. With respect to the quantum design of power efficiency in the active region, a $\lambda$~4.9 μm InP-based QCL displayed a RT CW output power of 5.1 W and wall-plug efficiency of 21% [27]. Its active region was based on a shallow-well (In$_{0.53}$Ga$_{0.47}$As/In$_{0.52}$Al$_{0.48}$As) high-barrier (In$_{0.36}$Al$_{0.64}$As/AlAs) design, which can greatly reduce the carrier leakage and enhance the carrier injection efficiency. By optimizing the device waveguide and fabrication process, a QCL with peak power of 23 W and WPE of 31% at room temperature was reported [18]. Meanwhile, the device surface was flattened by the buried ridge regrowth processing and post-polishing technique. The flattened device geometry improved the thermal conduction and reliability, and most importantly, increased the power and efficiency during CW operation. Figure 2(d) shows the P-I-V and WPE vs. current curves of a high-power/efficiency QCL that produced an even higher output power of 5.6 W, with WPE and 22% under continuous operation at room temperature [28].

At the long wavelength of λ~8 μm, a pocket-injector active region with high barriers was designed to cool the thermalized carriers and reduce their leakage into the continuum. This led to a high WPE of 20.5% under pulsed operation [29]. When an on-chip integrated optical phased array was proposed to address the heat dissipation issue, a RT CW power of 8.2 W was achieved, the highest for any on-chip intersubband laser operating at any wavelength was achieved [29].

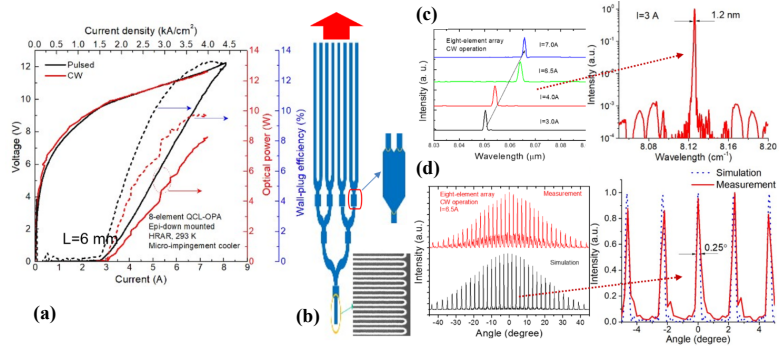

Fig 3 - Eight-Element LWIR Phase-Locked 6-mm-Long Buried-Ridge QCL Array: (a) P-I-V and WPE characteristics for the phase-locked QCL array operating in pulse and CW mode; (b) Schematic structure of the eight-element QCL array that is combined by a tree-array multimode interferometer (MMI). The two sides of the MMI are tapered to suppress total reflection of the out-of-phase mode; (c) Emission spectrum for CW operation, which demonstrates the single mode operation at 8.1 μm and 1.2 nm spectral linewidth; and (d) Comparison of measured and simulated far field distributions of the 8-element array operating in pulsed and CW mode [29].

A QCL with band structure based on a highly localized diagonal laser transition strategy was designed as in Fig 4(a), and with out-coupling by an electrically-isolated taper structure as shown in Fig 4(b). The device emitted at wavelengths spanning λ ≈ 8 – 10 μm. The maximum CW LWIR brightness from a single QCL facet was extended considerably to 2.0 MW cm⁻² sr⁻¹ for λ ≈ 10 μm, 2.2 MW cm⁻² sr⁻¹ for λ ≈ 9 μm, and 5.0 MW cm⁻² sr⁻¹ for λ ≈ 8 μm from 5-mm-long, single-element devices [30].

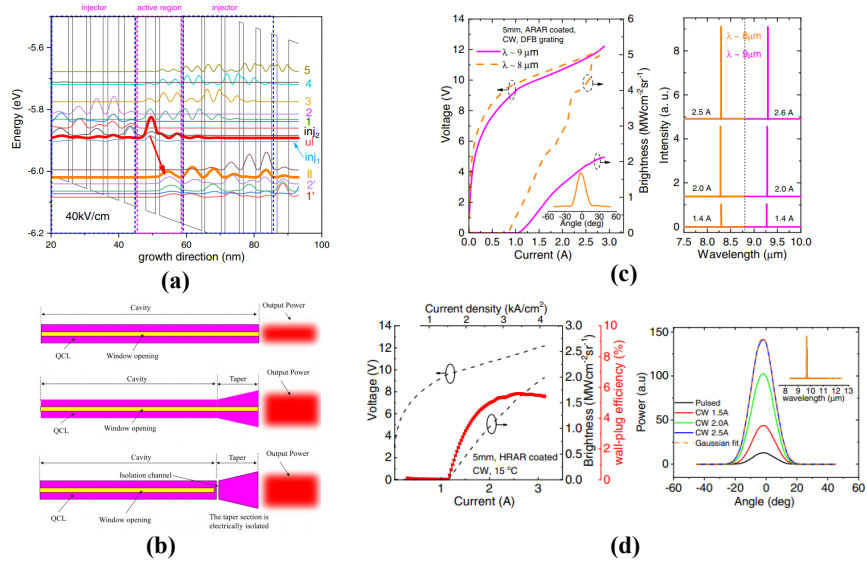

Fig.4. (a) Engineered band diagram of a QCL emitting at wavelength λ ≈ 10 μm. (b) QCL with a normal FP cavity (top), a tapered output structure (middle), and an electrically isolated tapered output structure (bottom). (c) LIV curves of the tapered QCLs with DFB gratings targeted at λ≈ 8 and 9 μm, respectively. Inset: Far field of the λ ≈ 8 μm QCL at an injection current of 1.8 A (left), and spectra of the two lasers as a function of current from 1.4 to 2.6 A (right). (d) LIV-WPE curves of the QCL with a tapered structure (left), and far field of the QCL at different current injections in CW and pulsed operation (right). Inset: Spectrum of the laser emission at 3 A [30].

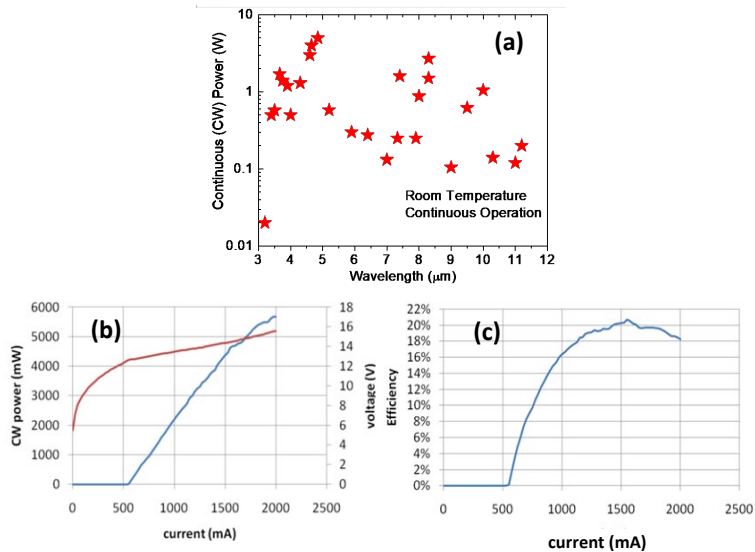

Fig. 5 (a) Maximum CW output powers for  QCLs at room temperature *vs*. wavelength [29, 19]; and (b) L-I-V and (c) WPE for the QCL that Daylight Solutions fabricated and packaged at CQD Northwestern in 2011.

## *Challenges and opportunities*

Manufacturing scalability and cost reduction have presented persistent challenges limiting broader adoption of the QCL in technology.  These issues can be improved through such innovative approaches as optimizing the MOCVD process for uniform growth on larger wafers, which would enhance the yield and reduce the cost. Collaborations with industry partners have also focused on integrating QCLs with complementary metal-oxide-semiconductor (CMOS) platforms, facilitating high-volume production and integration with electronic systems.

In order to be compatible with photonic integrated circuits (PICs) [31, 32, 33], effort has recently been devoted to structures grown on heterogeneous substrates (discussed in a separate sub-topic article of this Roadmap). For example, an InP metastable buffer layer was grown directly on a GaAs substrate to better match the lattice constants for high-quality core growth. The processed laser grown on a GaAs substrate emitted a peak power >14 W at $\lambda$~5.3 $\mu$m at room temperature. This success of intersubband lasers grown on GaAs shed light on the technology for intersubband laser growth on Si substrates (see Fig. 6). An InP-on-Si template consisting of layers of GaAs and $Ga_{0.47}In_{0.53}As$/InP superlattice (SL) was used as a dislocation filter for growth of the InP-based intersubband laser structure. QCLs grown on Si were demonstrated at $\lambda$~4.82 and 10.8 $\mu$m, with peak output powers up to 1.6 W and 4 W at room temperature, respectively [32, 33]. Further optimization of the growth condition and device fabrication resulted in significant improvements of the threshold current density and power efficiency. The RT CW operation of intersubband lasers on Si at $\lambda$~8.35 $\mu$m was reported for the first time in [34]. The maximum output power was >0.7 W at 343 K, and the WPE reached 11.1% at room temperature. While this breakthrough provides an appealing approach to integrating QCLs on Si for PICs, further technological challenges that include the integration of mid-IR detectors, modulators, etc. must also be addressed to fully realize PICs operating in the mid-IR.

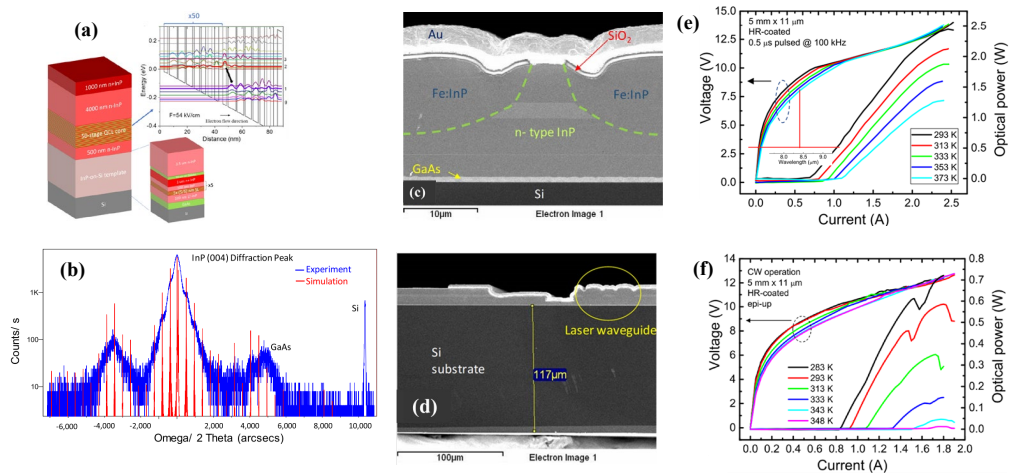

Fig. 6(a) Schematic of the layer sequence for an LWIR QCL grown on Si. The band structure and wave functions of relevant energy levels for one cascade stage of the strain-balanced laser core are also shown. Photons are emitted during electron transitions from upper laser level 2 to lower laser level 1; (b) X-ray diffraction and simulation of a laser wafer after growth of the laser core; (c) Cross-sectional image of the QCL waveguide, showing the planarized Fe:InP regrowth; (d) Cross sectional scanning electron microscopy (SEM) image of an entire device, including the lateral contact scheme, Si substrate, and lapping; (e) Pulsed mode voltage and output power as a function of current for different heat sink temperatures. Inset shows the emission spectrum at 1.5 A. (f) Continuous wave voltage and output power as a function of current for different heat sink temperatures [34].

The spectral range for QCLs has expanded substantially over the past three decades. For example, the group at University of Montpelier demonstrated low threshold and very long-wavelength InAs/AlSb QCLs emitting near 15 μm operating CW at room temperature [35]. In particular, intersubband lasers have become an enabling technology for the terahertz (THz) band (1~6 THz) [36]. The first THz QCL was demonstrated by R. Köhler [37]. A subsequent GaAs/AlGaAs-based QCL emitting at 5.7 THz operated in a pulsed mode at 68 K [38]. The device demonstrated broadband gain and lasing modes ranging from 4.76 to 6.03 THz, with the highest reported frequency for a THz QCL in pulse mode. A QCL emitting at ≈ 6.5 THz recently operated in pulse mode up to -12°C [39]. The improved performance compared to previous designs indicated progress toward room-temperature operation. THz QCLs grown on InP substrates have also been reported recently. GaInP/AlGaInP-based QCLs operating at 6.8 THz achieved modal gain > 90 cm⁻¹ and a maximum operating temperature of 104 K [40]. The study highlighted the potential of phosphide-based materials for compact semiconductor sources in the 5.5–7 THz range.

By harnessing the giant nonlinear susceptibility that is characteristic of mid-IR QCLs, intracavity difference-frequency generation (DFG) has been used to generate THz radiation at RT and beyond [41, 42]. Unlike QCLs on GaAs substrates, which employ intersubband transitions to generate THz radiation, DFG-based THz devices inherit the advantages of mid-IR intersubband lasers [43, 44]. By optimizing the active region nonlinear design and phase-matched packaging scheme of the mid-IR pump lasers, powers of 2.4 mW for pulsed operation and 14 μW for CW at 3.4-3.5 THz were achieved at room temperature [45, 46]. A sampled-grating DFB design was integrated into the DFG THz QCL to achieve broad wavelength tuning [47] over a frequency range of 2.06-4.35 THz, with output power up to 4.2 μW at room temperature [45].

A high-performance top-emitting RT monolithic THz QCL based on DFG and a buried modulated distributed feedback grating was reported in [48]. The laser exhibited single-mode operation d at 3.37 THz, with a maximum output power of 100 μW at room temperature.

Widely tunable intersubband lasers have important applications in chemistry, astronomy, physics and biology. In particular, by using tunable diode laser absorption spectra (TDLAS) in conjunction with broadly tunable single-mode QCLs, multiple molecular absorption features in the mid-IR spectrum can be detected simultaneously, enabling high-resolution spectroscopic analysis of a wide range of trace gases. A wide range covering 5.2~11.8 μm was realized by a strain-balanced $Ga_{0.35}In_{0.65}As/Ga_{0.47}In_{0.53}As$ homogeneous heterostructure [49]. The widest tuning range to date has been achieved by on-chip integration of a digital concatenated grating-sampled grating DFB for on-chip integrated intersubband lasers, with a value of 236 $cm^{-1}$ [50].

Frequency combs, which emit phase-coherent equidistant spectral lines, have revolutionized time and frequency metrology [51]. The recently developed QCL comb exhibits great potential, with high power and broad spectral bandwidth [52]. A wide flat-top gain with near-zero group-velocity dispersion was engineered using a dual-core active-region structure. This facilitates the locking of dispersed FP modes into equally-spaced frequency lines by four-wave mixing. A wide spectral coverage of 65 $cm^{-1}$ and  high output power of 180 mW was realized at 9 µm [53]. Subsequently, a 2.2-3.3 THz room-temperature THz harmonic frequency comb was demonstrated based on difference-frequency generation in a mid-IR QCL [54].

Recent QCL frequency comb results include:

- Combs utilizing multi-section waveguides to manage dispersion. The devices exhibited continuous comb operation over a wide current range (>300 mA), with output power exceeding 380 mW and optical bandwidth > 55 cm$^{-1}$ [55].

- An 8.1 µm QCL was phase-locked to a commercial mid-IR frequency comb, resulting in linewidth narrowing to sub-kHz levels at 1 ms observation time. This advance is pivotal for precision measurements and frequency metrology in the mid-IR [56].

- Demonstration of on-chip mid-IR supercontinuum generation covering 3-13 µm wavelengths, paving the way to broadband mid-IR sources on a single chip [57].

- Monolithic integration of mid-IR QCLs with low-loss passive waveguides via butt-coupling, demonstrating CW frequency comb operation at room temperature and advancing toward on-chip mid-IR dual-comb sensors [58].

- Monolithic integration of a mid-IR QCL frequency comb at 8.6 µm, achieving a maximum CW output power of 141 mW at 20°C. The device demonstrates spectral broadening from 10 to 34 cm$^{-1}$ under microwave injection, indicating the potential for high-precision spectroscopy applications. [59]

- Development of a tightly-locked dual QCL-comb system that achieved residual phase noise below 600 mrad across all comb lines. This system demonstrated coherent averaging with figure-of-merit $7 \times 10^5$ Hz$^{-1/2}$, enhancing precision in mid-IR spectroscopy [60].

- A novel waveguide design incorporating two highly doped InP plasmon layers, which enabled dispersion compensation in short-wavelength QCLs. The device operating at 5.3 µm produced up to 350 mW output power and maintained coherent comb operation up to 50°C, demonstrating robustness for practical applications [61].

- Optical injection locking of a QCL frequency comb's repetition frequency, using intensity-modulated near-infrared light at 1.55 µm. The technique achieved < 1 mrad residual phase noise at 1 s integration time, offering a more stable and coherent source for mid-IR spectroscopy and metrology applications [62].

Other QCL research breakthroughs include external-cavity QCLs [63], QCLs containing two substacks that were optimized to emit simultaneously at 5.2 μm and 8.0 μm wavelengths [64], high output powers in the LWIR [65], ring-cavity surface-emitting lasers based on quantum cascade structures [66], QCL master-oscillator power-amplifiers [67], QCLs with tapered buried heterostructure geometry [68], coherent combining of five QCLs emitting at 4.65 μm [69], chaotic light at mid-IR wavelengths [70], etc. The versatility of QCLs continues to drive innovation in these and other sectors. Some of the key milestones illustrated in Figure 1, highlighting global contributions to QCLs, are summarized in References [71, 72, 73, 74, 75, 76].

### *Future developments to address challenges*

Looking ahead, several promising avenues for advancing QCL technology include:

• Wavelength Extension: Developing QCLs that operate in the far infrared and terahertz regions at room temperature to broaden the spectrum of detectable substances.

• Room-Temperature Operation: Achieving continuous-wave operation at room temperature under extended conditions to simplify system integration and reduce cooling requirements.

• Integration with Photonic Circuits: Integrating QCLs with photonic integrated circuits to enable compact, multifunctional devices.

• Cost Reduction: Implementing advanced manufacturing techniques to further reduce production costs, making QCLs more accessible for a wider range of applications.

### *Concluding Remarks*

Quantum Cascade Lasers have achieved remarkable progress over the past three decades. Innovative research and development around the world has addressed key challenges in performance, manufacturability, and cost, establishing QCLs as critical components in mid-IR photonics. As the field progresses, ongoing efforts promise to unlock new capabilities and applications that will continue to drive advancements in semiconductor mid-IR photonics.

Acknowledgments
The author would like to acknowledge the support of the Walter P. Murphy Chair Professorship at the McCormick School of Engineering, Northwestern University. The author declares no conflicts of interest.

## 2. Interband Cascade Lasers

**I**GOR **V**URGAFTMAN* AND **J**ERRY **R. M**EYER

**Naval Research Laboratory, Washington, DC 20375, USA**
*\*MWIR_laser@nrl.navy.mil*

*Overview*

Since the first proposal over thirty years ago [1], interband cascade lasers (ICLs) have evolved into low-operating-power, highly compact sources of coherent mid-IR radiation [2,3]. While the best performance levels are currently achieved in the 3-4 μm spectral range, ICL operation has been extended to wavelengths as long as 14 μm [4]. On the short-wavelength side, cascaded lasers with type-I quantum wells have operated to wavelengths as short as 2 μm, as covered in a separate sub-topic article of this Roadmap. In spite of this undoubted success and ensuing rapid commercial development, understanding of the physics governing ICLs remains incomplete. More specifically, optimization of the type-II active region for suppressing intervalence absorption and Auger recombination losses, as well as the electron and hole injector designs and doping levels at different emission wavelengths, are still being addressed. Filling these and other gaps in the understanding could lead to significant further performance improvements, particularly for operation at wavelengths much longer than 4 μm.

**Current status**

Interband cascade lasers were proposed by Rui Yang in 1994 [1], soon after the initiation of intensive work on quantum cascade lasers (QCLs) [5]. Numerous investigations of ICL designs for the 3-4 μm spectral window were carried out in subsequent years, resulting in major improvements and modifications. In contrast to QCLs, the ICL active transitions are between subbands in the conduction and valence bands (rather than in the conduction band only), most often using type-II InAs/GaInSb/InAs quantum wells with a "W"-like alignment of the band profiles [6]. Similar to the QCL, the ICL features a cascaded geometry with $M$ stages that trades lower threshold current density $J_{th}$ (here varying slowly with $M$) for higher threshold voltage $V_{th}$ (approximately scaling with $M$) [2,7,8]. While its type-II active medium is characterized by electron and hole wavefunctions peaking in different, neighboring layers, the available gain is proportional to the square of the electron-hole wavefunction overlap [8]. Only electrons are injected and extracted at the contacts, which requires the internal generation of holes at the semimetallic interface separating the InAs/AlSb superlattice electron injector and the GaSb/AlSb hole injector. The carriers are produced by the application of an external electric field that drops approximately one photon energy per stage at threshold, as illustrated in Fig. 1(a).

The electron injector is typically much thicker than the hole injector, particularly at shorter wavelengths, where the ICL design is considerably more mature. This design choice favors significantly more carrier occupation in the electron injector, which creates an imbalance between the electron and hole populations residing in the active wells at the lasing threshold. Therefore, to minimize $J_{th}$ the carrier populations are rebalanced via substantial doping of the electron injector to a sheet density of $\sim 5 \times 10^{12}$ cm$^{-2}$ [9], much higher than the threshold electron concentration $n_{th} \sim 6\text{-}9 \times 10^{11}$ cm$^{-2}$ in the active wells. Note that $n_{th}$ in the ICL is considerably lower than for quantum-well lasers emitting in the near-IR [8,10], owing to the reduced density of states near the band edges that are characteristic of heavily-strained narrow-gap materials. This leads to a relatively low $J_{th}$ for mid-IR ICLs despite dominance of the nonradiative Auger recombination process, in which electron-hole pairs recombine while promoting other electrons and holes to higher positions in their respective bands [8].

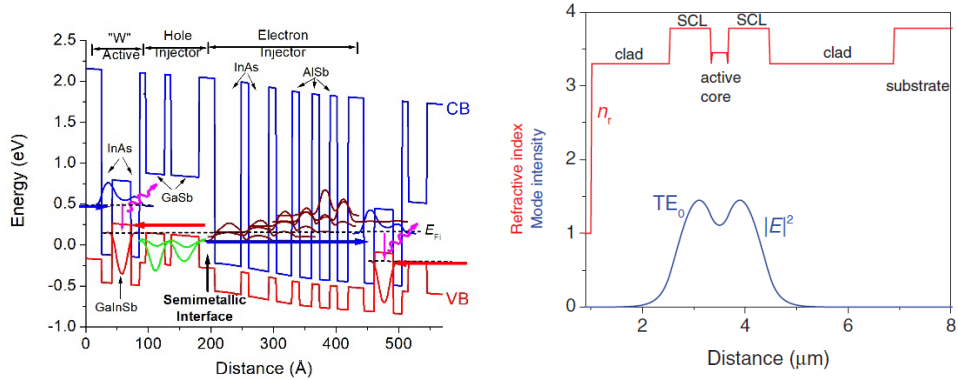

Band diagram for one and a half stages in the ICL active core. (b) Spatial profiles for the optical mode (blue) and refractive index (red) in the ICL waveguide. Reproduced with permission from Ref. 2.

The highly beneficial effect of carrier rebalancing was verified experimentally via measurements on a series of ICLs with different doping levels in the electron injector [9]. It is attributed theoretically to comparable values of the multi-electron ($n^2p$) and multi-hole ($p^2n$) Auger coefficients in the type-II wells. This can be understood intuitively by considering the case where the $n^2p$ Auger process is dominant. A decrease in the electron density at the expense of the hole density would then attain the same gain with a lower recombination rate, insofar as the recombination scales faster with $n$ [8]. However, the lowest values of $J_{th}$ were attained when the electron-injector doping was neither too light nor too heavy, which motivates the conclusion stated above [9]. We note in passing that interband cascade light emitting devices (ICLEDs) and interband cascade infrared photodetectors (ICIPs) employ very similar or identical active stages to the ICL, as discussed in separate sub-topic articles of this Roadmap.

Most of the initial work on designing high-performance ICLs was carried out in the wavelength range between 3 and 4 μm, where it becomes quite difficult for a QCL to realize low threshold and high output power owing to the limited height of the conduction band offset, in particular those available in the technologically-mature InP-based material system [11]. An ICL typically includes only $M$ = 5-7 stages, as compared to 20-40 in state-of-the-art MWIR QCLs [12]. This reduction is enabled by the much higher differential gain per unit current density for the ICL interband transition, which can be expressed as: $\frac{dg}{dJ}\big|_{th} = \Gamma \frac{dg}{dn}\big|_{th} \frac{dn}{dJ}\big|_{th} = \Gamma \frac{dg}{dn}\big|_{th} \frac{\eta_i \tau_{ul}}{q n_{th}}$. Here $\Gamma$ is the optical confinement factor, which is approximately proportional to the number of stages when $M$ is small, and $\eta_i$ is the internal efficiency [7]. Most of the difference is due to the QCL's much shorter upper-level lifetime that is limited to $\tau_{ul} \approx 1$ ps by phonon and interface-roughness scattering to the lower subband [12]. By contrast, the carrier lifetime in an ICL can be as long as 1 ns at room temperature (RT), with more typical values of several hundred ps resulting from the sensitive dependence of lifetime on the threshold carrier density ($\propto n_{th}^{-1} p_{th}^{-1}$) [**Error! Bookmark not defined.**]. The large lifetime ratio is only partially mitigated in the QCL by its higher differential gain per unit carrier density $\frac{dg}{dn}\big|_{th}$, increased optical confinement $\Gamma$ resulting from the larger number of stages, and lower waveguide loss.

Since the ICL's active core with M = 5-7 is much thinner than the lasing wavelength, it cannot sustain a guided optical mode on its own. To enable waveguiding, it is surrounded by high-index ($n_r$ = 3.76-3.78) GaSb separate-confinement layers (SCLs), which are in turn enclosed by low-index ($n_r$ = 3.3-3.4) InAs/AlSb superlattice claddings. The SCLs are doped very lightly $n$ type, while the claddings are divided into low-doped inner and high-doped outer sections, with both choices designed to minimize the loss due to free carrier absorption [13]. Figure 1(b) illustrates a typical ICL waveguide designed for emission in the 3-4 µm spectral window.

The design developments described above resulted in a series of advances in the ICL performance. These includes $J_{th}$ below 100 A/cm² in pulsed mode at $T$ = 20ºC [14], and a maximum continuous-wave (cw) operating temperature of 107ºC [9]. In cw mode at room temperature (RT): $P_{th} = J_{th}V_{th}$ as low as 300-350 W/cm² [3,15], external differential quantum efficiency (EDQE) per stage as high as 56% [3], threshold drive power as low as 29 mW [9], output power as high as 500 mW with beam quality less than 2.5× the diffraction limit [16], and wallplug efficiency (WPE) as high as 18% for short anti-reflection-coated ICL cavities [**Error! Bookmark not defined.**], which is only 4% below the record for QCLs [17]. Other demonstrations include distributed-feedback (DFB) ICLs operating cw at RT in a single spectral mode over a wide spectral range from 3 to 6 µm [18,19], vertical cavity surface-emitting ICLs operating in pulsed mode at RT (λ = 3.6 µm) [20], and ICL frequency combs [21-23] (discussed in a separate sub-topic article of the Roadmap). In spite of these achievements, several unexpected effects were observed, including apparent internal efficiency $\eta_i$ well below 100%, a pronounced decrease in the slope efficiency in pulsed mode at high currents (a factor of 3 between threshold and $J$ = 2.8 kA/cm²), lower than the projected differential gain $\frac{dg}{dJ}\big|_{th}$ [13], and unexpected increases of the threshold current density and internal loss at wavelengths shorter than ≈ 3.4 µm [3].

The design guidelines presented above were developed based on experimental work that was largely confined to the 3.2-3.9 µm region. When the ICL operation is extended outside these bounds, some degradations of the observed threshold and efficiency are observed on both the short and long-wavelength sides. In particular, the threshold current density at RT increases to 300-400 A/cm² for ICLs emitting at λ = 2.8-3.0 µm [3]. At longer wavelengths it grows much more gradually [3,24], but seemingly monotonically, to values in the 500-900 A/cm² range at λ = 6.0 µm for the GaSb-based ICL, while lower values of 330-400 A/cm² have been reported for InAs-based ICLs [25], perhaps as a result of additional optimization of the waveguide structure. Assuming a constant internal efficiency $\eta_i$ and threshold gain estimated from the slope efficiency near threshold, and adopting the threshold carrier density from band structure calculations, these findings appear to indicate a steady increase of the Auger coefficient with wavelength [3,26]. The corresponding internal loss also increases, from ≈ 4 cm⁻¹ at λ ≈ 3.0–3.8 µm to > 20 cm⁻¹ at λ = 6.0 µm [3]. Nevertheless, $J_{th}$ and $P_{th}$ in ICLs remain well below those in QCLs, which display typical state-of-art values of $J_{th}$ ~ 1 kA/cm² and $P_{th}$ ~ 10 kW/cm². This remains true even at λ = 4-6 µm, where the performance of the QCLs have been extensively explored and optimized [3]. Room-temperature operation of ICLs is still out of reach for wavelengths significantly exceeding 6 µm, as we discuss next.

### Challenges and opportunities

As described above, experimental results to date indicate that the Auger coefficient and internal loss increase at emission wavelengths both shorter and longer than the optimal range of λ ≈ 3.2-3.8 µm. While this may be expected at longer wavelengths, origins of the degradation at shorter wavelengths remain poorly understood. And even at longer wavelengths, stronger free-carrier absorption in the GaSb SCLs and superlattice claddings appear to account only for a small fraction of the overall increase. For example, most of the mode in a state-of-the-art ICL resides in the $n$⁻-GaSb SCLs, where the loss is expected to increase from ≈ 0.9 to only 2.3 cm⁻¹ as the wavelength increases from 3.7 to 6 μm. Even accounting for uncertainties in this estimate and the cladding loss, the implication is that the active core dominates the increase.

The loss in the active core is expected to arise primarily from intervalence transitions in the GaInSb hole well and GaSb hole injector wells. Even though the total electron sheet density in the active core greatly exceeds the hole density, the electron free-carrier absorption cross section is generally smaller by at least an order of magnitude. The intervalence absorption coefficient can be calculated from the band structure and wavefunctions for all the relevant states in the valence band, with typical results shown in Fig. 2 [27,28]. Near the zone center, the interaction between the first light-hole subband (LH1) and the second heavy-hole subband (HH2) leads to an electron-like dispersion. Since the band admixture increases with wavevector [8], Fig. 2 shows that this dispersion leads to a local maximum in the intervalence absorption between 4 and 5 μm, typically followed by a dip in the 5-6 μm range and a broad background at longer wavelengths. Somewhat counterintuitively, the peaks and dips in the loss spectra for wider GaInSb hole wells occur at slightly longer wavelengths owing to reduced mixing between the hole states. While these theoretical results are somewhat sensitive to the band parameters and envelope-function approximation used in the calculations, the trends with thickness and composition should be robust. One direct experimental measurement showed relatively low loss at λ = 3.8 μm and much higher loss at λ = 4.6 and 5.4 μm, with no clear difference between the latter wavelengths [29].

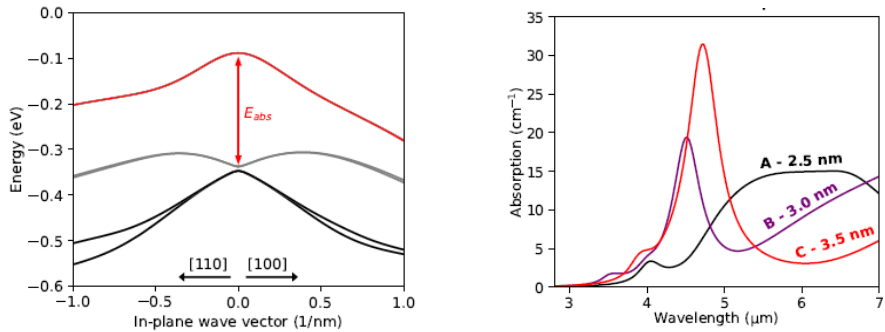

Fig. 2. (a) Typical valence subband dispersions in the plane of the type-II "W" quantum well employed in the active core of an ICL. (b) Calculated intervalence absorption loss for several different thicknesses of the GaInSb hole well in this structure. Reproduced with permission from Ref. 27.

Using the trends discussed above, improved ICLs with thinner active hole wells have been demonstrated to have lower $J_{th}$ (247 vs. 310 A/cm²), higher characteristic temperature $T_0$, and higher slope efficiency (176 vs. 127 mW/A, equivalent to 21 vs. 15% EDQE) at λ = 4.4 μm [27]. However, we note that similarly low thresholds were obtained without valence-band engineering [3], possibly because of other, uncontrolled differences in the waveguide characteristics and material properties. The strategy of optimizing the hole well thickness was later extended to demonstrate broad-area and single-mode DFB ICLs grown on GaSb with emission wavelengths just beyond 6 μm and operating cw at RT [28,30]. A broad-area $J_{th}$ as low as 500 A/cm$^2$ was achieved by using 3.5 nm thick GaInSb wells, as well as an electron injector with only three wells and lower injector doping in the nine active stages [28]. The reduction in size of the electron reservoir likely resulted in a higher fraction of electrons per donor dopant transferring to the active wells, which could in turn allow lower doping for a lower lasing threshold.

Modeling also indicated that the efficiency droop at higher injection currents could be caused by increased occupation and higher intervalence absorption in the hole injector wells [31]. Furthermore, the apparent internal efficiency of well below 100% that is often inferred from cavity-length studies was reported to be an artifact of the internal loss's relatively strong dependence on cavity length. When corrected for these effects, almost all of the reduction in EDQE can be assigned to various types of internal optical loss, namely intervalence absorption in the active region and hole injector, as well as free-carrier absorption in the SCLs and the inner part of the claddings.

Since the optical cladding thickness scales with wavelength (neglecting the weak spectral dependence of the refractive index), the conventional approach of employing InAs/AlSb superlattice layers can lead to prohibitively long epitaxial growths when extended to longer wavelengths. For example, the required bottom cladding thickness increases from 2.5-3.5 μm at λ = 3 μm to > 5 μm at λ = 6 μm. The SCL thicknesses should also increase to maintain the same confinement factor at the active core. Even beyond the impractical thickness of the superlattice clads, the high index of the GaSb substrate makes the waveguide modes leaky. This leakage into the substrate causes the loss to scale exponentially with inverse clad thickness [32,33].

These issues with the optical confinement at longer wavelengths can be ameliorated by (partially) replacing the superlattice claddings with heavily doped $n^{++}$-InAs(Sb) layers [34-36]. The electron plasma in these layers reduces the refractive index as the plasma wavelength blue shifts toward the operating wavelength. However, the improvement in optical confinement comes at the expense of some increase in the free-carrier absorption loss. Initial devices grown on InAs substrates employed plasma-assisted claddings that completely replaced the InAs/AlSb clads [37]. However, it quickly became clear that this architecture produces too much loss, so the optimal strategy is to employ lightly-doped inner superlattice claddings (in the regions with highest mode intensity next to the GaSb SCLs), in combination with $n^{++}$-InAs(Sb) outer claddings to suppress the residual mode leakage [34]. This concept has been applied successfully to ICLs grown on both InAs and GaSb substrates [38-40].

Long-wavelength ICLs are more commonly grown on InAs substrates. Since the higher InAs fraction in each active stage makes it more difficult to strain balance the growth on a GaSb substrate, the spectral coverage of InAs-based ICLs is considerably wider on the long-wavelength side [41-43]. In particular, devices emitting at wavelengths up to 14 μm have been reported [4], although their performance remains far short of what can be achieved in the 3-6 μm window. For example, the maximum operating temperature in pulsed mode was 212 K at λ = 12 μm and 150 K at λ = 14 μm. The faster degradation with temperature most likely reflects the combined influences of stronger Auger recombination and increased intervalence absorption in the active core, as well as other types of free-carrier absorption. Lower electron-hole wavefunction overlap caused by the wider InAs electron wells is another limitation of long-wavelength ICLs. One approach to increasing the overlap and hence the differential gain is to use InAs$_{0.5}$P$_{0.5}$ barriers that allow the InAs wells to be thinner [4,42]. The downside of this approach is that phosphide molecular beam epitaxy is usually unavailable in the machines dedicated to growing antimonides, and significant improvements due to these design change remain to be observed experimentally.

Another research thrust explored whether the GaSb and InAs substrates could be replaced with alternatives that allow for greater photonic integration such as Si and GaAs (discussed in other sub-topic articles of the Roadmap). This can be accomplished by either direct epitaxial growth on these substrates [44-46], or by wafer bonding the material grown on a more conventional substrate [47], followed by removal of the GaSb or InAs substrate. Successful demonstrations of both approaches have been reported in the literature. It was found that the growth on lattice-mismatched substrates does not significantly degrade the ICL threshold current, presumably because the additional Shockley-Read-Hall nonradiative recombination due to the induced dislocations remains far below the Auger rate. Nevertheless, the overall wall-plug efficiency was reduced from 14.3% for the devices grown on GaSb to 9.6% on GaAs and 8.4% on Si [45].

Even at this late date, the effects of active-region design on the ICL performance remain incompletely understood. When normalized to bulk units, the Auger coefficients are considerably suppressed compared to bulk alloys (such as HgCdTe and InAsSb) with the same gap, and steadily increase with wavelength. Figure 3 illustrates that nearly all of the points derived from lasing thresholds remain within a factor of 2 of the trendline [3,26]. It has been suggested in the literature that a variety of Auger mechanisms are present, which could explain the apparent insensitivity of the estimated recombination coefficient to design variations [48]. However, other than those derived from the laser thresholds, relatively few Auger measurements are available at this point. Independent cross-checking would provide greater confidence in the numerical extractions of the Auger coefficient as a function of energy gap.

The optical cavity structures for ICLs are also less developed than for many other semiconductor lasers. DFB ICLs in cw mode with top gratings have operated up to 80 °C, and emitted 55 mW at RT [49,50]. Nevertheless, loss associated with the reduced top cladding thickness significantly limits the output power. For this reason, gratings etched into the ridge sidewalls and lateral metal gratings have been investigated with some success, as mentioned above [18,19,46]. Higher-order gratings can often reduce the fabrication tolerances by increasing the grating pitch. The basic trade-off is between low loss realized in a more standard ICL waveguide against lower coupling coefficient due to the smaller overlap between the optical mode and the index grating. Alternative strategies for achieving single-spectral-mode operation have also been investigated, including V-coupled cavities [51,52] and slotted waveguides [53], as well as vertical-cavity [**Error! Bookmark not defined.**] and ring-cavity [54,55] surface-emitting ICLs. Because the lateral size of a vertical-cavity surface-emitting laser (VCSEL) can be much smaller than that of an edge emitter, it should ultimately provide much lower threshold and operating power. That outcome would be quite useful for battery-operated and other power-limited applications. However, the requirement for careful optimization of the lateral current injection and diffraction loss in such devices has so far limited their operation to pulsed mode [20].

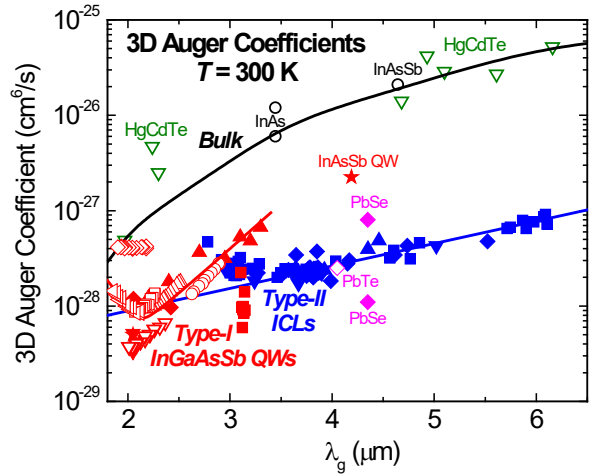

Fig. 3. Auger coefficients in 3D units for type-I and type-II quantum wells as a function of wavelength corresponding to the energy gap, as obtained from reported laser thresholds. For comparison, the values for bulk materials measured and estimated by various means are also shown. Reproduced from Ref. 26.

### Future developments to address challenges

In spite of considerable successes in the development of high-performance ICLs, the following major challenges require significant future development: (1) The ICL output power into a good beam is more limited than for QCLs, even in the sweet spot near λ = 3.5 μm; (2) High output power in a single spectral mode remains to be demonstrated in a robust and practical manner; (3) The ICL performance still drops off rather rapidly for λ > 4 μm, which prevents cw operation at RT in the long-wave IR (LWIR) region ; (4) Further reduction in the drive power would be beneficial for portable and battery-operated systems; (5) Spectroscopic and high-frequency applications will also benefit from reduced spectral and relative intensity noise (RIN) [56,57].

While the output power from a single-lateral-mode ridge is limited because the ridge cannot be more than ≈ 5 μm wide at λ = 3-4 μm without emitting into multiple lateral modes, various approaches have been proposed for scaling the power by increasing the emitting area without degrading the beam quality. One approach is the tapered laser or amplifier geometry, in which the lateral width is gradually (e.g. linearly) opened up [58,59]. This structure works best if feedback from the output mirror is quite small. Another approach is the photonic crystal geometry, usually of the surface-emitting variety (PCSEL) [60] (discussed in another sub-topic article of this Roadmap), which is capable of oscillating in a single mode over a large area owing to two-dimensional distributed feedback. A further possibility recently proposed is a double-ridge architecture that realizes weak index guiding by heterogeneous bonding to a germanium-on-silicon waveguide [61]. The full potential of these technologies remains to be realized, in part because of the relative immaturity of the GaSb-based material system. For example, the epitaxial regrowth on patterned surfaces, which is commonly used to fabricate near-IR PCSELs with strong index modulation, is not well developed for antimonide materials in spite of the efforts of workers at Stony Brook University [62]. Furthermore, current injection and heat dissipation for large-area GaSb-based and InAs-based ICLs are quite challenging. The initial work on scaling the output power was done near the sweet spot of type-II and type-I ICLs, and it is not yet clear whether it can be profitably extended farther into the MWIR or beyond into the LWIR spectral range.

The tapered amplifier approach to power scaling discussed above may also provide a simple route tow high-power single-mode operation. Because the maximum output power in a semiconductor optical amplifier is of the same order of magnitude as the saturation power, which is in turn determined by the differential gain and carrier lifetime in the active medium in conjunction with the lateral dimension of the structure, it is not significantly affected by the input power in a correctly designed amplifier. Therefore, the input signal can be provided by a low-power single-mode master oscillator, which is already well known from work at shorter emission wavelengths [63].

While improvements have been made in ICLs emitting at longer wavelengths, their performance outside the MWIR window still falls well below that of the better-developed QCLs. This applies not only to the output power and wall-plug efficiency, but also to RT cw operation. As discussed above, it appears thus far that most design changes have little impact on the Auger recombination coefficient, while the internal loss is only somewhat easier to manipulate by adjusting the layer thicknesses, compositions, and doping levels. Nevertheless, considerable room for optimization remains, and devices operating at cryogenic temperatures with sufficiently low operating powers may be acceptable for some applications.

More generally, an ICL's drive power can be reduced by optimizing the design, which has largely been accomplished in the 3-4 μm spectral window and will be soon for the 4-5 μm range, as well as by scaling down the device size. Obviously, the cavity length of a Fabry-Perot ICLs can only be reduced while maintaining the same threshold by increasing the mirror reflectivity. This generally requires multilayer facet coatings (e.g. Bragg stacks), particularly if both reflection and transmission are desired, which complicates the device fabrication significantly. For ICLs based on ring cavities with surface emission [54,55], the cavity length is constrained by the need to minimize bend loss by maintaining a sufficiently large radius of curvature compared to the wavelength. Although the VCSEL geometry already has high-reflectivity mirrors with transmission built into the basic design, much work remains to combine efficient current injection with minimum diffraction and scattering loss as required for the very short (wavelength-size) cavity. In particular, a deposited dielectric Bragg stack requires lateral injection, with the current-spreading layer placed precisely at a node of the cavity mode. Nevertheless, it appears that the limitations in this area are largely technological rather than fundamental. The possibilities of exploring photonic-crystal or even plasmonic cavities appear quite remote at this point, because thus far the technology for producing an undercut membrane does not exist in the antimonide material system. And as described above, even a small additional loss from the carrier plasma makes the ICL unattractive at longer emission wavelengths.

## Concluding Remarks

Ever since its original proposal 30 years ago, the ICL has improved by leaps and bounds. It has demonstrated RT operation at low input powers in cw mode, single-mode lasing, and emission of moderately high powers into a high-quality beam. The advantages of single-mode output from a compact device requiring modest drive power has increasingly made ICLs the go-to source for many chemical sensing systems based on MWIR spectroscopy. However, this impact has the potential for substantial further expansion if: the drive power can be minimized further; the output in a single spectral mode can be increased to >> 100 mW; the output in a single spatial mode can be further enhanced; the favorable spectral range can be extended to longer wavelengths; and/or the ICL source can be integrated onto an ultra-compact chemical sensing chip that incorporates source, detector, and sensing waveguide within a photonic integrated circuit (PIC). The integration of an ICL source with a sorbent-coated sensing waveguide for detecting dimethyl methylphosphonate (DMMP) was recently demonstrated [64], along with the potential for integrating an interband cascade detector on the same chip as well.

The spectral region providing the best performance to date can be extended once the physical limitations constraining longer-wavelength operation are fully understood, so that effective mitigations can be incorporated into improved active-core and optical-waveguide designs. This work is in progress, and continues to produce qualitative advances. ICL operation in the LWIR may run into what was originally considered to be the Achilles' heel of long-wavelength lasers: higher thresholds and lower efficiencies driven by strong Auger recombination and free-carrier absorption. However, the current longer-wavelength ICLs are most likely still far from these fundamental barriers, while the focus for MWIR ICLs has in the meantime shifted mostly to novel device concepts such as frequency combs and coherent emitters for specific applications.

Funding

This work was supported by the Office of Naval Research through the NRL Base Program.

Acknowledgments

We thank Drs. W. W Bewley, J. R. Lindle, C. D. Merritt, C. S. Kim, M. Kim, C. L. Canedy, and J. A. Massengale for their many contributions to the ICL work at the Naval Research Laboratory.

Disclosures

The authors declare no conflicts of interest.

3. GaSb-based type-I quantum well diode and cascade diode lasers

LEON SHTERENGAS,* GELA KIPSHIDZE, AND GREGORY BELENKY

**Stony Brook University, Stony Brook, New York 11794, USA**

*leon.shterengas@stonybrook.edu

### Introduction

At the early stages of semiconductor laser development, it was believed that diode lasers intended for operation in infrared spectral region would exhibit excessively high threshold current densities and, hence, very low if any output power levels in continuous wave (CW) regime at room temperature (RT). This perspective was primarily grounded in the fundamental increase of free carrier absorption rate with wavelength and increase of nonradiative Auger recombination rate in narrow bandgap semiconductors. Continuous advances in understanding the true limiting factors of specific designs and breakthroughs in related epitaxial and fabrication technologies successfully pushed CW RT operation of interband lasers to mid-infrared region of spectrum. In the case of III-V-Sb type-I quantum well (QW) diode laser heterostructures grown by molecular beam epitaxy on Te-doped GaSb substrates it required: (a) optimization of the epitaxial growth conditions and the use of broadened waveguide designs resulting in low threshold device operation near 2 μm [1,2], and (b) development of the active region designs with improved hole confinement to enhance lasers operating at wavelengths up to 3 μm and above [3-5]. The latter implied recognition of the critical role of the GaInAsSb QW compressive strain above 1% [6,7] and utilization of the quinternary AlGaInAsSb material as the QW barrier [8,9]. Several focused reviews have been published which provide extended reference lists. [10,11].

The fundamental reason behind successful experimental demonstrations of the CW RT operation of GaSb-based type-I QW diode lasers in the spectral region initially above 2 and later above 3 μm has been well recognized. Namely, reduction of the electron density of states at the conduction band edge with bangap in III V alloys [12] and reduced in-plane hole effective mass in heavily compressively strained QWs [13] facilitate separation of the quasi-Fermi levels in narrow bandgap active QWs under injection. Hence, it is expected that mid-infrared diode lasers can have much lower transparency and, possibly, lower threshold carrier concentrations as compared to their near-infrared counterparts. The Auger recombination rate depends superlinearly on carrier concentration; hence, the threshold current density of mid-infrared diode lasers can be rather low despite the anticipated increase of the probability of the individual Auger events in narrow bandgap materials. The related decrease of the maximum achievable optical gain in QWs with low values of reduced density of states can be addressed by employing multiple-QW active regions. Potential injection nonuniformity in the MQW stack in antimonide-based lasers was eliminated by the cascade pumping scheme proposed in [14] and perfected later resulting in a new class of interband cascade lasers [15].

## Current status

In its most classic form, the GaSb-based type-I QW diode laser heterostructures utilize GaIn(As)Sb QWs with bandgap adjustable by varying the indium and arsenic compositions and well thickness. The barriers are made of either AlGaAsSb quaternary or AlGaInAsSb quinary alloys depending on operating wavelength. The quinary barriers can confine the holes in QWs with high In and As contents - a decisive factor in the development of high-power diode lasers emitting near and above 3 μm. Figure 1 below illustrates possible designs of the active QWs providing optical gain for diode lasers operating at room temperature near 2 μm (left) and near 3 μm (right). Of critical importance is the adequate confinement of electrons and holes, which implies QW compressive strain of ~1.5% and transition from quaternary to quinternary barrier alloys as the device operating wavelength goes above ~2.5 μm. It should be noted that utilization of the quinternary alloy complicates carrier transport and/or promotes nonradiative defect recombination as well as free carrier absorption in the waveguide core layer. Hence, it might be difficult to benefit from broadened waveguide (BW) laser design [16] for diodes designed to operate near and above 3 μm [17], while for devices operating at shorter wavelengths and based on quaternary waveguide core and barrier materials, the BW approach helps to maximize output power.

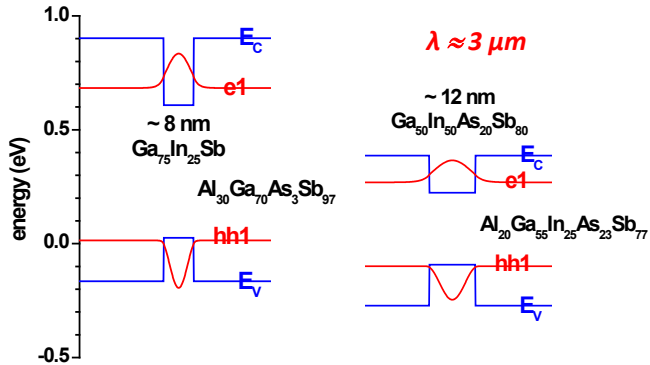

Fig. 1. Calculated band diagrams and electron and hole envelope wavefunctions of III-V-Sb compressively strained QWs designed to operate near 2 μm (left) and near 3 μm (right).

Currently, coated diode lasers with 2-3 mm cavity length and emitting in the range of 1.9 – 2.5 μm generate above 1 W of CW output power from 100-μm-wide apertures and demonstrate threshold current densities near 100 A/cm². High power coated diode lasers operating near 3 μm generate ~350 mW of CW output power and have RT thresholds of 220 A/cm². The diode lasers operating at longer wavelength near and above 3.3 μm demonstrate higher threshold current densities and lower CW output power levels approaching ~500 A/cm² and below 100 mW, respectively. Single frequency distributed feedback (DFB) lasers have been designed and developed for both 2 μm (2.05 μm lasers targeting $CO_2$ absorption band [18]) and 3 μm (3.27 μm lasers targeting the methane absorption band) [19] spectral regions. Linear arrays of wide stripe 2.2 - 2.3 μm emitting lasers generated above 10 W of CW power from 1 cm wide bars [20,21]. Tapered laser designs have been explored to improve the laser and laser array brightness, though with somewhat limited success [22]. The apparent tendency to form filaments in the near field have been confirmed experimentally in GaSb-based wide stripe diode lasers [23].

The cascade pumping scheme has been integrated with GaSb-based type-I QW diode laser heterostructures to further improve device performance parameters [24]. Broken gap band alignment at the heterointerface between InAs and GaSb can be utilized in the design of GaSb-based cascade bipolar devices [14]. Efficient band-to-band tunneling at the GaSb/InAs heterointerface can be realized without excessive doping and, hence, without severe free carrier absorption penalty. This unique feature can be viewed as a critical advantage of the GaSb-based material system for development of high-power interband cascade lasers emitting in the spectral region from below 2 to over 3 μm.

Figure 2 illustrates the design of a cascade diode laser operating near 3 μm at room temperature. In that particular example, the n- and p-cladding layers were made of $Al_{0.8}Ga_{0.2}As_{0.07}Sb_{0.93}$ doped with Te and Be, respectively. The barrier and waveguide core layers were nominally undoped $Al_{0.2}Ga_{0.55}In_{0.25}As_{0.23}Sb_{0.77}$ [25]. Compressively strained (~1.5%) ~12 nm-wide GaInAsSb layers, with ~50% Indium content, comprised the QWs responsible for the room temperature optical gain near 3 μm. The graded layer was nominally undoped AlGaAsSb with Al composition changing from 50% down to ~5% over a thickness of 100-nm. The graded layer was followed by the tunnel junction and electron injector comprised of nominally-undoped 10-nm-thick GaSb, 2.5-nm-thick AlSb and a Te-doped 6-period InAs/AlSb chirped superlattice (SL). The arrangement of the tunnel junction and SL electron injectors transfers electrons from the valence band of the preceding active QW to the conduction band of the subsequent QW, implementing a cascade pumping scheme that links the active QWs in series instead of in parallel.

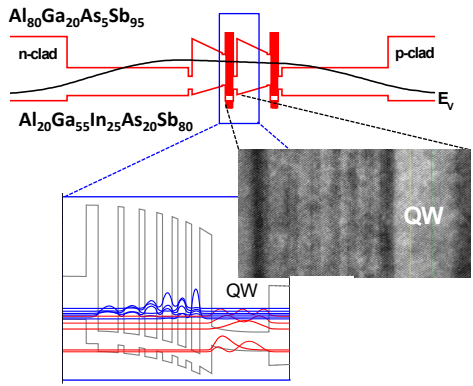

Fig. 2. The calculated band diagram of a three-stage cascade laser heterostructure, designed to operate at 3 μm at room temperature, under flat band conditions. This band diagram is superimposed with the calculated intensity distribution of the fundamental vertical lasing mode. The insets display a transmission electron microscope image of the InAs/AlSb superlattice injector and the active quantum well (top), along with a simulated band diagram of the same section under bias, highlighting the relevant electron and hole envelope wavefunctions.

Connecting the active quantum wells in series can greatly lower the device's threshold current and enhance laser slope efficiency due to the electron recycling effect. Carrier recycling between two QWs emitting near 3 μm led to reduction of the threshold current density from ~200 down to ~100 A/cm$^2$ and increase of the maximum CW power level from ~350 up to ~650 mW compared to non-cascade diodes. It also nearly doubled the device peak power conversion efficiency, reaching above 16%. Narrow shallow ridge lasers fabricated from the same material demonstrated record 100 mW of CW power in a nearly-diffraction-limited beam. Two-stage cascade diode lasers operating near 2 μm demonstrated threshold current densities ~80 A/cm$^2$ and CW output power ~2 W per 100-μm-wide stripe, and ~ 250 mW from ~ 5-μm-wide narrow ridge apertures [26]. Increasing the number of cascades from two to three led to further improvement of the CW output power level, reaching 960 mW, 500 mW and 350 mW in multimode beams for operation near 3, 3.15 and 3.25 μm, respectively.

A fundamental advantage of the cascade pumping scheme is its ability to provide uniform injection over a large number of stages, which can enable efficient lasing despite limited optical gain in each stage. This benefit was realized in various designs of the intra- and interband cascade lasers discussed at length in other sections of this review. Optimization of the number of cascades in cascade diode laser heterostructures has yet to be performed to unlock its full potential. Besides facilitating the carrier transport through multiple QW active regions, the cascade pumping scheme offers additional benefits. It helps to dilute an adverse effect of the excessive voltage drop across laser heterostructure elements outside the active QWs, leading to improved conversion efficiency at high currents. In the context of GaSb-based 3 μm range lasers, the cascade pumping model offers an avenue to eliminate a need for growing quintenary alloys, since carriers in the active QWs can be confined by InAs/AlSb SL and graded layers. Consequently, type-I quantum well lasers capable of efficiently emitting light at wavelengths of 3 μm and higher have been created without employing quinary barriers, utilizing both cascade diode [27] and interband cascade [28] designs.

The advancements in design and epitaxial growth encouraged efforts focused on developing more sophisticated laser architectures within the III-V-Sb material system. It should be noted that a majority of the reported GaSb-based diode and cascade diode DFB lasers are based on the so-called laterally coupled (LC-DFB) architecture [29]. The LC-DFB architecture does not require regrowth over an etched grating, as the periodic structure is defined (etched or deposited) in the ridge sidewalls and/or nearby in the etched section outside of ridge. For the 3 μm range, the LC-DFB lasers based on cascade diode laser heterostructures demonstrated improved performance parameters compared to non-cascade diodes [30]. In addition to general enhancement of the device operating parameters, the cascade pumping scheme allowed variation of the bandgaps of QWs in the stack. This resulted in a widened optical gain spectrum. The corresponding structures have been successfully utilized as gain elements in external cavity configurations to demonstrate a record-wide tuning range extending from 2.8 to 3.25 μm [27]. Antimonide-based diode laser heterostructures that are compatible with buried grating DFB technology have been reported [31], but they have not yet achieved widespread dominance. This is presumably due to challenges associated with the regrowth epitaxy. It was demonstrated that the emission spectrum of GaSb-based diode lasers can be stabilized by replacing one of the facets of the ridge laser cavity with a high-order Bragg Reflector (BR) [32]. Stable high-power narrow-linewidth operation of 2-μm-range GaSb-based diode lasers was achieved by utilizing 6th-order surface-etched BR mirrors. The BR multimode devices with 100 μm wide ridge waveguides generated ~850 mW in the CW regime at room temperature. A comparable top-down etched BR technology was utilized to attain a CW power level of 100 mW for dual wavelength operation in Y-branch narrow ridge GaSb-based devices. [33].

Epitaxial regrowth within the antimonide material system was successfully applied to the development of tunnel-junction-confined vertical cavity surface emitting lasers [34,35]. Temperature stable operation was targeted in VCSELs with optimized detuning [36]. The available power of the VCSELs remains in a low mW level but is adequate for some sensing applications. For applications requiring high power levels, GaSb-based optically pumped disks can be used in the vertical external cavity laser (VECSEL) configuration [37]. VECSELs emitting multi-watt power levels in narrow spectrum and diffraction limited beams have been demonstrated [38]. These optically pumped disk laser gain elements can be placed in specially designed cavity arrangements containing saturable absorbers to achieve mode-lock operation [39]. Ongoing research is focused also on the development of compact, passively mode-locked, edge-emitting GaSb-based type-I quantum well lasers that are electrically pumped. Several reports were published on the diode lasers emitting near 2 μm and utilizing split-contact architecture in which a section of the ridge was forward biased to provide gain and a separate section of the same ridge was reverse biased to act as a saturable absorber [40-42]. Split-contact passively mode-locked GaSb-based type-I QW cascade diode lasers operating near 2.7 and 3.2 μm have been demonstrated [43,44]. The lasers generated trains of 10 - 15 ps chirped pulses at repetition frequencies about 13 GHz and average power up to 10 mW. The intermodal beat notes of the free running devices, which demonstrated a

Lorentzian linewidth of about 30 kHz. The laser emission spectral bandwidth was up to 20 nm. Reduction of the intermodal beat note linewidth was demonstrated in devices stabilized by external feedback, and multiheterodyne beat notes were observed. Compensation of the cavity dispersion is required for the type-I QW passively-mode locked cascade diode lasers to generate wide bandwidth optical frequency combs and transform limited pulses. Figure 3 illustrates the geometry and presents some parameters of passively mode-locked cascade diode lasers emitting near 2.7 and 3.2 μm [43,44] that were fabricated at Stony Brook University. Figure 3b shows the comb-like laser emission spectra measured at bias conditions corresponding to the presence of strong intermodal beat notes in the RF spectrum (Figure 3c) and autocorrelation traces (Figure 3d) that indicate the formation of pulse trains. The observed spectral bandwidth is sufficient to support the generation of sub-picosecond pulses if transform-limited device operation could be achieved. However, uncompensated cavity dispersion resulted in the generation of chirped pulses with temporal widths exceeding 10 picoseconds (Figure 3d).

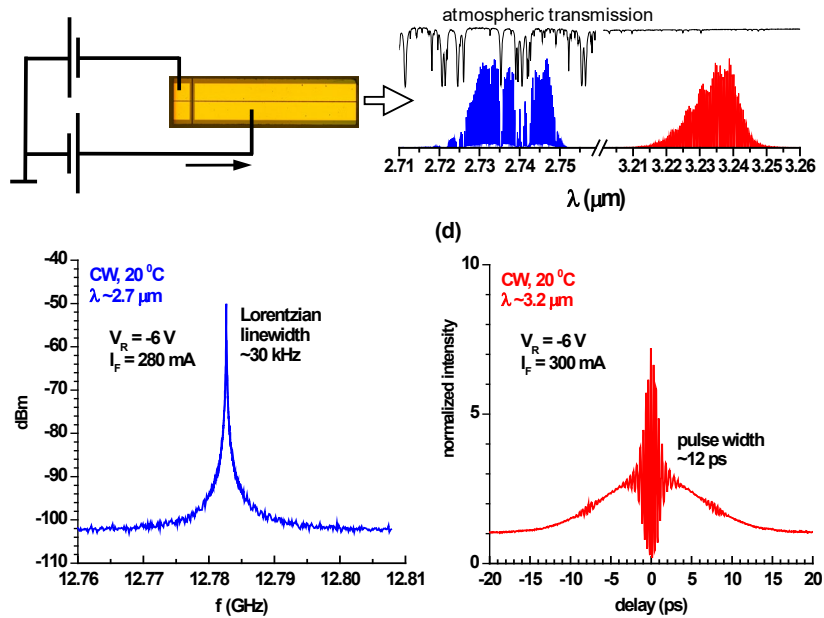

Fig. 3. (a) Top view of the split-contact narrow ridge waveguide passively mode-locked cascade diode laser, and schematic illustration of the reverse bias applied to the absorber and forward bias applied to the gain section; (b) Emission spectra of 2.7 and 3.2 μm passively mode-locked cascade diode lasers overlapped with the atmospheric transmission spectrum; (c) Intermodal beat note for the 2.7 μm laser measured with 280 mA of gain section current at -6 V bias of the absorber section;

(d) Second order interferometric autocorrelation trace measured for the 3.2 μm device with 300 mA of gain section current at -6 V bias of the absorber section.

### *Challenges and opportunities*

A major challenge that, once addressed, could lead to a substantial increase in the output power from GaSb-based diode lasers is the excessive voltage drop occurring across the device's heterostructure. However, it will not eliminate the fundamental problem of edge emitting devices associated with limited beam quality. The exploration of different approaches to creating surface-emitting device architectures that can produce high-quality beams with significant output power is crucial. The capabilities of optically pumped disk lasers, stabilized by external cavity components, have already been successfully demonstrated.[37]. Achieving comparable performance in a more compact, monolithic, and electrically pumped design would be highly advantageous. A potential solution is the Photonic Crystal Surface Emitting Laser (PCSEL) architecture [45]. PCSELs based on GaSb-based type-I QW active regions are actively being developed at the time of this publication [46-48].

The ability to achieve coherent emission from the extensive surface area in a PCSEL allows for the production of narrow-spectrum, ultra-low divergent beams with beam quality that is significantly enhanced compared to those of traditional edge-emitting semiconductor lasers. Furthermore, the PCSEL fabrication process does not necessitate facet cleaving or coating, and the ultra-low divergence of the emitted beam reduces or even eliminates the need for external optics. In scenarios where high wattage output is not essential, the post-growth fabrication and packaging of PCSELs closely resemble that of cost-effective light-emitting diodes. Monolithic PCSELs employing nitride, arsenide, and phosphide diode laser technologies have been successfully demonstrated, functioning within the visible and near-infrared telecom regions [49-51]. Additionally, the adaptation of PCSEL technology to longer wavelengths exceeding 4 μm has been achieved through quantum cascade laser designs [52]. Stony Brook University research team reported the first CW operation at room temperature of monolithic epitaxially-regrown III-V-Sb PCSELs operating near 2 μm [53,54]. We have introduced air-pocket-retaining epitaxial regrowth within cascade diode heterostructures aimed at enabling PCSEL operation close to 3 μm; however, device functionality has only been demonstrated at lower temperatures thus far.

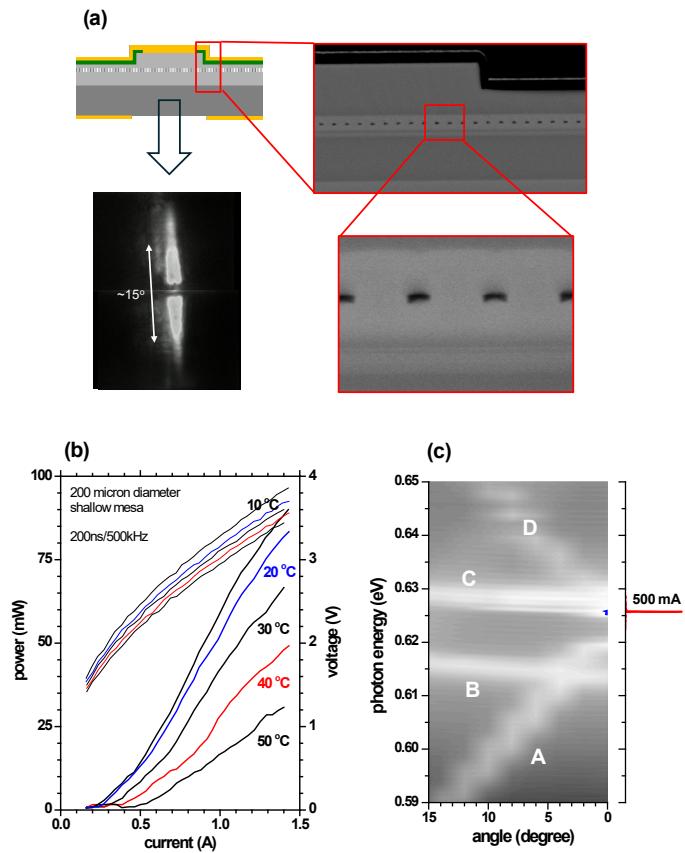

Fig. 4. (a) This diagram illustrating the cross-section of a shallow-mesa PCSEL includes insets featuring a scanning electron microscope image of the cleaved mesa device that emphasizes the hidden array of voids. Another inset presents a camera image capturing the far-field pattern of the beam emitted by this device; (b) Temperature dependence of the light-current-voltage characteristics of the 200 μm diameter shallow mesa PCSEL; (c) An angle-resolved electroluminescence intensity map recorded at room temperature along the X symmetry direction. On the right is the emission spectrum of the PCSEL, obtained in the surface normal direction at a current of 500 mA and a temperature of 20°C.

The PCSEL fabrication involves two epitaxial stages. The initial growth produces an incomplete laser heterostructure. Once this incomplete device heterostructure is taken out of the high vacuum environment, it undergoes a nanopatterning process. This step acts as a foundation for creating the integrated high index contrast 2D photonic crystal layer. The procedure concludes with a second epitaxial regrowth, which encapsulates the photonic crystal layer beneath the top cladding and additional auxiliary device heterostructure layers. Figure 4a illustrates the generic structure of the substrate-outcoupled PCSEL, with insets showing scanning electron microscope images of the device cross-section. The buried voids are clearly visible. The corresponding device power-current-voltage characteristics are plotted in Figure 4b. These particular PCSELs emit tens of mW of optical power from the window in the substrate contact. An angle-resolved electroluminescence map and laser emission spectrum are shown in Figure 4c. The electroluminescence map clearly shows four subbands (marked A, B, C and D) that are expected in the vicinity of the $\Gamma_2$ symmetry points of the buried square photonic crystal. The lowest threshold was achieved for the band edge corresponding to subband C.

GaSb-based PCSELs operating in the 2 – 3 μm range that emit high power in a single lobe with ultra-low divergence have not yet been achieved. Currently, the brightness is constrained by filamentation and a propensity for multimode operation. Furthermore, higher order spatial modes linked to subband C, which are emitted at angles other than the surface normal, are expected to exhibit low radiative losses that may potentially rival the fundamental mode emitted at the surface normal. To enhance the stability of the spatial modal structure and facilitate the generation of an ultra-low divergence beam at elevated injection currents, precise adjustments to the arrangement of buried voids are necessary. In situ generation of the vector-vortex beam [55] can be achieved in PCSELs by specially designed photonic band edge states. This functionality can enhance resilience of the sensor and communication systems.

### *Future developments to address challenges*

The GaSb-based diode and cascade diode laser technology currently faces the challenge of limited power conversion efficiency (below 30% in peak), which restricts the achievable CW power levels. Significant improvements in the device performance will be realized by reducing the voltage drop across the laser heterostructure. Elimination of the parasitic voltage drop across auxiliary layers and reduction of the quantum defect in each stage of the cascaded structures are required. This optimization will allow for an increased number of stages in the cascade diode structure, further enhancing both power and efficiency. To advance the next generation of Distributed Feedback (DFB), Distributed Bragg Reflector (DBR) lasers, and Photonic Crystal Surface Emitting Lasers (PCSELs), a crucial technological breakthrough in versatile epitaxial regrowth within the III-V-Sb material system is essential. This development will also necessitate optimizations in the PCSEL heterostructure and buried photonic crystal designs. To fully leverage the significant potential of passively mode-locked GaSb-based cascade diode laser technology for creating compact ultra-fast optical frequency comb sources, it is vital to develop device architectures that control cavity dispersion. Additionally, there exists a vast and largely untapped potential for the hybrid integration of GaSb-based emitters with advanced photonic integrated circuit platforms. Recently, a hybrid integration of multiple GaSb-based gain elements and photodetectors onto a specially designed silicon photonics integrated circuit was achieved [56]. This spectroscopic sensor architecture offers an impressive net tuning span by combining the ranges of several gain elements on a single chip. Dramatic phase noise reduction in passively mode-locked 2 μm lasers (resulting into sub-kHz RF linewidth) was achieved through its hybrid integration of GaSb-based diode laser with a silicon photonic circuit

[57]. Further progress in the hybrid integration of antimonide-based lasers and other active photonic components is poised to transform the field of compact sensor technology.

*Concluding Remarks*

GaSb-based cascade diode lasers have proven to be the most efficient semiconductor coherent emitters in the spectral range of 2 to 3 μm. These devices can effectively compete with other technologies in the wavelength ranges just below 2 μm and extending to 3.5 μm. The significant reduction in the reduced density of states associated with a smaller bandgap, along with established band engineering techniques, allows these devices to operate with low thresholds and high efficiency. Recent advancements in epitaxial and hybrid integration methods are set to pave the way for a new generation of compact and efficient devices suitable for a wide array of applications, including eye-safe lidars, remote sensing, free-space communication, compact lab-on-chip module development, and various medical diagnostic and treatment devices.

Acknowledgments

The GaSb-based diode and cascade diode PCSEL development at Stony Brook University was supported by US Army Research Office, grant W911NF2210068. The authors acknowledge Drs. A. Stein and D. Zakharov (Center of Functional Nanomaterials at Brookhaven National Laboratory) for development of the e-beam lithography and TEM studies, supported by the U.S. Department of Energy, Office of Basic Energy Sciences, through the Center for Functional Nanomaterials, Brookhaven National Laboratory, under Contract DE-SC0012704.

Disclosures

The authors declare no conflicts of interest.

4. Semiconductor Laser Frequency Combs
in the Mid-Wave- and Long-Wave Infrared

## Lukasz A. Sterczewski[1]

**[1]Faculty of Electronics, Photonics and Microsystems, Wroclaw University of Science and Technology, Wybrzeże Wyspiańskiego 27, 50-370 Wrocław, Poland**
*\*lukasz.sterczewski@pwr.edu.pl*

**Current status**

Semiconductor lasers offer the remarkable convenience of coherent light emission from an electrically pumped heterostructure. Some of their unique strengths include high optical power, native operation in relevant spectral regions, and nearly-diffraction-limited beam quality. Historically, access to the mid- and longwave-infrared (MWIR/LWIR, 3–5 μm / 5–12 μm) has relied on problematic lead salt lasers [1,2], but this has changed with the conceptions of the quantum cascade (QCL) [3] and interband cascade laser (ICL) [4] (both are discussed in separate sub-topic articles of this Roadmap). The wavelength flexibility offered by quantum engineering of the heterostructure triggered the first revolution in compact MWIR/LWIR light sources and more widespread exploitation of this region. Although early efforts focused on ensuring single longitudinal mode operation with high spectral purity [5–7], a second revolution came with the discovery of frequency comb generation by a QCL in 2012 [8]. Such a source emits an array of phase-locked equidistant lines [9], which use nonlinearity to overcome the native intracavity group velocity [10,11]. Very soon, other sources, including ICLs [12–15], quantum-well diode lasers [16], and quantum-well cascade diode lasers [17,18] expanded the portfolio of electrically-pumped optical frequency combs (OFCs) with varying levels of absolute coherence (optical linewidth). The importance of the OFC mode of operation stems from its definition of the optical frequency for each emitted mode by only two frequencies: the offset frequency ($f_0$) and the repetition rate ($f_{rep}$, or line spacing) [19].

The traditional picture of a frequency comb is a train of optical pulses obtained via passive mode-locking of the longitudinal modes in the cavity; however, it rarely applies to single-section Fabry-Perot devices [20]. From a mathematical perspective, the only requirement for a comb is to have a (perfectly) periodic waveform in the time domain [21]. So in principle, there are no constraints on the temporal profile of the electric field emitted by an OFC. This inevitably sparks a heated debate in the ultrafast optics community, as the textbook picture of an OFC does not include so-called frequency-modulated (FM) combs [10]. Regardless of the convention and nomenclature, however, FM-locked QCLs, ICLs, and quantum well diode lasers have proven compatibility with typical experimental techniques applied to OFCs. In particular, they display rigorous modal equidistance [8,22,23] and phase reproducibility between power cycling [22], proving the non-random nature of the emission.

Several review and perspective papers have discussed in detail the progress of chip-based MWIR to THz optical frequency combs [24,25], microcombs [26], and spectroscopic techniques [27]. Some have focused on a particular platform like QCLs [24] or ICLs [28,29], or cross-platform FM combs [21]. Therefore, the goal of this roadmap is not to review the state-of-the-art of physics of the devices in great detail, but rather to identify potential knowledge gaps and roadblocks that currently hamper the more widespread adoption of these sources. We will first briefly introduce the concepts required to understand the topic.

**MWIR/LWIR comb sources**

### Prerequisites

Several requirements must be fulfilled for a chip-scale semiconductor laser to operate as a frequency comb. For edge-emitting single-section devices, which promote FM operation, the group velocity dispersion (GVD) and Kerr nonlinearity should be carefully balanced. A highly simplistic model derived from Ref. [30] assumes that multimode operation is first enabled by standing wave effects and spatial hole burning (SHB), with the mode spacing governed by dispersion. Next, the Kerr nonlinearity originating from the third-order susceptibility $\chi^{(3)}$ promotes four-wave mixing (FWM) gain (rather than the bulk material gain), which equalizes the mode spacing and turns a multimode laser into an FM comb. This mode of operation seems to be a universal phenomenon for both fast (QCL) and slower (ICL, diode lasers) gain media, which are all proven to have a fast gain component [31] regardless of whether they exploit intraband or interband transitions. Naturally, the GVD should be moderate, or at least not extreme (but not zero), to facilitate broadband OFC formation. In simple terms, FWM will struggle to compensate modes that deviate too much from a uniform grid of lasing frequencies. Note that in contrast to microresonators [32], whose cavities have extremely high (>$10^6$) quality factors, the Kerr effect in a Fabry-Pérot (FP) semiconductor laser with orders of magnitude lower $Q$ factor is rather insufficient to produce new optical modes [30].

This picture changes in the case of ring devices [33] (discussed in a separate sub-topic of this Roadmap), which natively promote single longitudinal emission since standing wave effects and hence SHB are greatly suppressed. However, in real life, defects in the waveguide can induce phase turbulence and also lead to comb operation in the form of bright or dark optical solitons [34]. This platform is, however, very new and still subject to intensive research.

Another option is strong microwave modulation, which can turn an FM comb into an amplitude-modulated (AM) one that generates pulses [35–37], or control its emission properties. This implies the need for microwave engineering to ensure low insertion losses, enabling efficient modulation or stabilization of the comb's repetition rate. Some devices utilize dedicated modulator sections [35,37] with geometrically small areas to lower the capacitance and hence reduce the RC time constant. RF engineering is also important for ring devices with the era of quantum walk combs (QWC) – sources whose optical spectra are very predictable and almost deterministically governed by the microwave injection signal frequency and power [38]. A QWC is the semiconductor laser implementation of a quantum walk in a synthetic frequency lattice, which has a Hermite–Gaussian intensity distribution. A prerequisite is support for single-mode operation without RF modulation in a ring cavity, along with unidirectional (either clockwise or counter-clockwise) operation via symmetry breaking.

Passive mode locking [39] induced by an on-chip saturable absorber [18,30] can also attain OFC operation with an AM character. A short biasable section of the cavity acts as an on-chip intensity-dependent modulator that opens a brief temporal transmission window. The most spectacular results have been obtained with type-I cascaded diode lasers operating at 3.27 µm [17], albeit this mode of operation should be possible using type-II ICLs as well. QCLs do not natively support passive mode locking directly from the cavity because their picosecond dynamics preclude energy storage between consecutive roundtrips.

The final requirement for an OFC device is operation on the lower-order spatial mode. Broad-area multimode devices often develop too many groups of modes with different spacings to promote the harmony of comb operation. Nevertheless, with the help of strong microwave injection it is possible to control the transverse mode and hence make other modes lossy [40].

We next briefly discuss the most popular device families for MWIR/LWIR OFCs. For convenience, Fig.1 plots key milestones in the field over the past 3 decades.

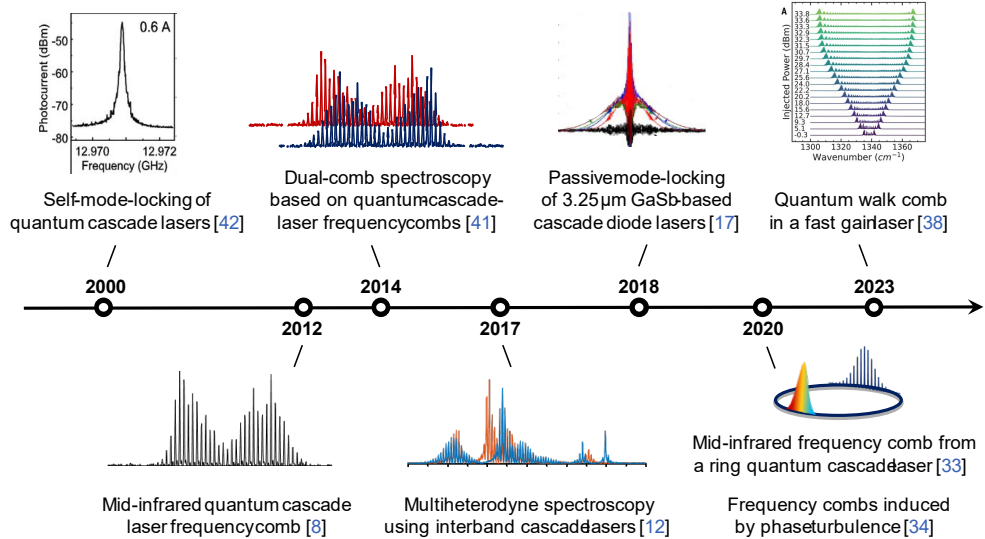

Fig. 1. Key milestones in MWIR/LWIR in semiconductor laser frequency combs. Images reprinted with permission from the American Association for the Advancement of Science and American Chemical Society.

### Quantum cascade lasers

Arguably, QCLs are the most popular OFC source at wavelengths spanning 4.5 μm to the THz (150 μm). Since their first demonstration in 1994, it took 18 years to prove that the phenomenon of FM self-mode locking natively supported by these devices has a comb nature [8]. Two years later, in 2014, Villares et al. demonstrated dual-comb spectroscopy (DCS) of water vapor and a polished GaAs etalon wafer at ~7.1 μm, with interleaving to surpass the device's coarse sampling grid of ~7.5 GHz (dictated by the cavity length) [41]. The multi-heterodyne experiment unambiguously proved the repetitive nature of the electric field waveform emitted by the QCLs, and marked a new chapter in broadband molecular spectroscopy at high speeds. It should be noted that the first observation of so-called "self-mode locking" in QCLs was observed by Paiella in 2000 [42], but at the time no suitable characterization techniques were available to prove the comb characteristics and coherence.

The wavelength portfolio of QCL OFCs has expanded substantially [24]. Most of the initial devices operated in the optimal (from a QCL performance perspective) LWIR region of 7–8 μm. This was partially dictated by the negative dispersion of InP (used as a material) , which compensated the waveguide's positive dispersion. The resulting group velocity dispersion was low enough for the FWM to equalize the line spacing.

QCLs can operate as combs out to the THz range [43], with coverage reaching 1 THz and center frequencies between 2–4 THz [25,44]. Native OFC emission is typically attainable only at cryogenic temperatures (20–40 K), although low-power room temperature THz comb emission has been obtained via difference frequency generation from an MWIR QCL [45] comb that is designed to emit two widely-spaced groups of modes, one of which is single while the second has an OFC character [46].

The popularity of the QCL OFC platform has driven extensive research on cavity dispersion engineering [47] and dynamics [48]. To ensure reproducibility and high device yield from a single batch, several schemes to control the GVD profile have been proposed. Because its value at MWIR/LWIR wavelengths is not as extreme as in the THz, there is no need to corrugate the waveguide [43], which is a microwave/optical chip-based analogue of chirped fiber gratings. Instead, various techniques involve layered reflective structures known as Gires-Turnois interferometers (GTI) [49,50], passive waveguides [51,52], dual-core active region designs [53], or micromechanical (MEMS) mirrors that induce an external, dispersion-compensating passive cavity [54]. Control over the GVD has unlocked operation at both shorter wavelengths extending toward the MWIR (5 μm) [50] and longer ones (9–10 μm) [53,55]. Even shorter wavelengths (4.5 μm) are possible, although with a harmonic / mode skipping emission spectrum [56].

The problems of comb reproducibility and spectral match between the two sources required for dual-comb spectroscopy can now be solved conveniently using a pair of QWCs [38]. Since their emission spectra are almost entirely governed by the microwave signal, it is easy to ensure excellent spectral overlap, in contrast to conventional QCL OFCs for which the overlap is rather partial.

Despite the supposed inability of QCL OFCs to emit short pulses, there is ongoing research to compress the *E*-field chirp of FM combs from FP cavities by using external prisms [57] or dispersive waveguides [58]. Whereas the combination of spectral filtering, grating compression, and microwave injection locking allowed femtosecond pulse generation at 8 μm [57], a recent work from ETH and Université Paris-Saclay showed chip-based ~1.4 ps pulse generation. Further power scaling should eventually allow spectral broadening using nonlinear waveguides. However, at present the platform still requires an optical isolator, which is difficult to realize using on-chip technology.

A remarkable feature of QCL combs is their high output power. Whereas one typically obtains hundreds of mW from the laser facet, some devices offer 1 W of optical output with excellent intermodal coherence properties [59]. This high power can sometimes be problematic, as the device is also sensitive to residual amounts of radiation backreflected to the cavity. On the other hand, a combination of high power and microwave injection has enabled researchers from TU Wien to demonstrate an on-chip dual-comb spectrometer in the LWIR [60]. Light from one comb was injected into a second QCL comb, and both were subject to microwave injection. Although the coupling efficiency was rather low, a clear self-detected DCS spectrum was observed.

### Interband cascade lasers

While the wavelength coverage of ICLs reaches ~2.7 µm to 14.4 µm (up to 6 µm at room temperature in continuous wave mode) [61], comb operation has typically been obtained in the MWIR 3–5 µm range [28] with ~1% fractional bandwidth. The main limitation stems from the  material contribution to the GVD near the energy gap of the active core, particularly at shorter (3–4 µm) wavelengths. Historically, the first signature of ICL comb operation was observed by Princeton University researchers Sterczewski & Westberg et al. in 2017 using multiheterodyne beating [12] (20 years after the first experimental demonstration of a working ICL [62]). A pair of 2-mm-long FP ICLs operating at ~3.2 µm with repetition frequencies of 19.2 GHz was optically beaten on a fast MWIR photodetector to observe a GHz-spanning microwave down-converted comb with MHz-wide lines. The optical coverage in this pioneering demonstration was only 8 cm$^{-1}$ (240 GHz), but later much broader combs were obtained. The high modal coherence in this work was proven using a computational (digital) phase correction technique, which used information from two neighboring microwave lines to globally compensate for the phase noise of all the comb lines. The interferogram (electrical signal measured in DCS) representing a cross-correlation of the electric fields lacked bursts or pronounced pulses, thus signifying a highly chirped, FM-like mode of operation analogous to already established QCL OFCs. It was rather unexpected to observe such an emission pattern, considering the radically-different nature and dynamics of interband and intraband transitions. Unfortunately, at the time the intermode beat note was not experimentally accessible to fully assess the comb properties using established characterization techniques [8,63].

In a follow-up work from 2018 [13], researchers from NASA Jet Propulsion Laboratory (JPL) and Naval Research Laboratory (NRL) demonstrated a two-section cavity design for ICLs lasing at 3.6 µm, wherein the gain section (~4 mm long) was independently biased from a short-section (50–150 µm) saturable absorber (SA) section. To promote passive mode-locking, the SA section was ion-bombarded to facilitate the formation of optical pulses at high negative biases. The devices emitted broadband OFCs (~35 cm$^{-1}$ / 1 THz) with ~10 mW of power per facet with sub-kHz intermode beat note linewidths at ~9.7 GHz. However, the observed interferometric autocorrelation characterizing the emission waveform traces lacked the 8:1 peak-to-background ratio expected for passive mode-locking [64]. Instead, the waveform featured a profile with the ~8:3 ratio observed earlier for FM QCLs [8]. A major advantage of the absorber, whose role is still under investigation, was an extension of the ICL comb coverage up to ~50 cm$^{-1}$ (1.5 THz) for some devices, while preserving a sub-kHz intermode beat note linewidth [15]. This feature was essential in ICL DCS experiments.

Later, in 2019, Schwarz et al. of TU Wien unequivocally proved the FM character of ICL combs [14]. Their devices, operating at ~3.8 μm and spanning approximately 25 cm$^{-1}$ (750 GHz), were characterized using the microwave interferometric technique SWIFTS [63]. This enabled reconstruction of the instantaneous intensity and frequency profiles. As expected, the devices emitted an almost constant intensity over time with a positive frequency sweep (chirp), thus proving the speculated similarity to QCL combs. The gain-modulator cavity design (but not absorber) enabled the researchers to actively stabilize the repetition rate and demonstrate an on-chip microwave link with GHz bandwidths. This was made possible by the ICL's bifunctionality: the same laser structure could function as both a source and an efficient detector. Using the same laser structure, but subject to strong microwave injection (>1 W, 30 dBm), Hillbrand et al. (also from TU Wien) showed switching from the FM to AM regime [35]. The intermodal phases synchronized in the presence of microwaves rather than being splayed over the $-\pi$ to $\pi$ range, which ultimately yielded ps pulses at MWIR wavelengths.

Despite these impressive results, the lack of a native sub-kHz intermode beat note, reaching a few kHz width at optimal injection currents and temperatures, has prevented the realization of mode-resolved multiheterodyne spectra by the same group, even with microwave injection. Empirically, it is now expected that MWIR/LWIR ICL devices should possess sub-kHz-wide intermode beat notes that are suitable for such experiments. Since then, broadband single-section ICL combs have also been developed by researchers at Sorbonne Université [65]. Using an InAs-rich design, the devices featured intermode beat notes of width ~40 kHz.

To facilitate future referencing to optical frequency combs operating at convenient near-infrared telecommunication wavelengths, ICL combs offer intracavity frequency doubling. In 2019, Sterczewski et al. characterized the intracavity sum-frequency and second harmonic generation of an ICL operating at ~3.6 μm, which produced NIR radiation at 1.8 μm. To prove that the emission was not an artifact, a spectroscopic experiment was conducted. Selective absorption of water vapor at NIR but not MWIR wavelengths proved the real character of the second harmonic, which could otherwise arise from the nonlinear response of the photodetector.

Diode lasers

In addition to ICLs, GaSb-based type-I quantum well diode lasers that implement a simple or cascaded design also offer MWIR OFCs. In 2018, researchers from Stony Brook University passively mode-locked a type-I ICL operating at 3.25 μm [17]. A two-section geometry with an on-chip absorber generated 10-ps long pulses with ~180 kHz intermode beat note linewidth and mW-level output optical power. However, the phase-noisy operation was incompatible with mode-resolved DCS without resorting to optical stabilization using feedback [18].

Much more stable native MWIR combs have been offered by ordinary (non-cascaded) type-I quantum well diode lasers, which supported FM operation by a single-section cavity geometry. Initially demonstrated in the NIR at 2 μm [66], MWIR diodes with a modified quantum well design supported OFCs centered at 3.05 μm that spanned 1 THz with ~10 GHz repetition rate [16], as demonstrated by Sterczewski et al. in 2023 [16]. Although the free-running intermode beat note had a multi-kHz linewidth, the high anti-correlation of the offset frequency and repetition rate frequency noise permitted the observation of a mode-resolved multiheterodyne spectrum. A remarkable feature of this platform is its almost flat-top emission spectrum accompanied by monotonic tuning without comb lines developing in one spectral lobe. Another practical advantage relates to the low compliance voltage (<2.5 V) and resulting low power consumption (<1 W). When biased directly by an AA battery without any temperature control, a pair of devices provided ~0.5 THz of spectral coverage at 3 μm. This promises that efficient OFC dual-comb generators will become available in the future. The devices have already proven useful in broadband, high-resolution molecular spectroscopy [67], yet their limited optical bandwidth calls for dispersion engineering strategies similar to ICLs.

**Spectroscopic techniques**

The full potential of MWIR/LWIR OFCs can be unlocked by access to the individual comb lines. This section will discuss some popular experimental techniques employed for spectroscopy in this challenging spectral region. They are schematically depicted in Fig. 2.

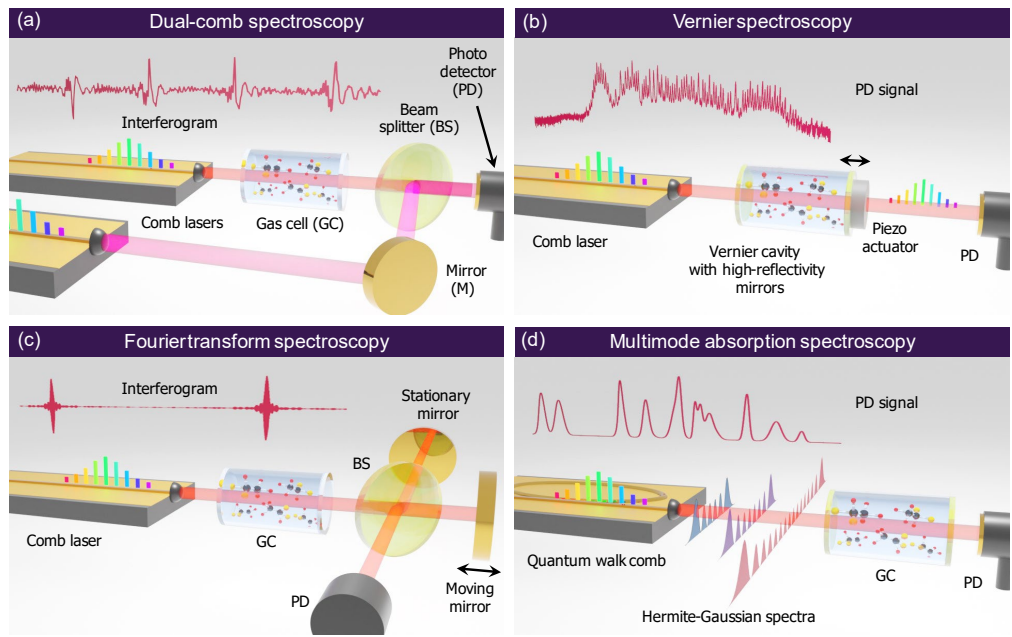

Fig. 2. Spectroscopic techniques compatible with chip-based MWIR/LWIR frequency combs. The experimental setups are simplified to illustrate the main idea. In real-life implementation, additional optical isolators or lenses may be needed. (a) Dual-comb spectroscopy; (b) Mode-resolved Vernier spectroscopy; (c). Fourier transform spectroscopy; (d). Multimode absorption spectroscopy (MUMAS). In principle, all these techniques offer tooth-limited optical resolution.

**Dual-comb spectroscopy**

Dual-comb spectroscopy is a moving-parts-free down-conversion technique that maps the MWIR optical spectrum to microwave frequencies via multi-heterodyne beating of two sources with mismatched repetition rates and spectral overlap [68]. Details of the technique are thoroughly discussed in a multi-platform review paper by Coddington et al. [69], while a recent work by Hayden et al. lists progress in QCL-based DCS [70]. Therefore, here we only briefly overview the state-of-the-art. Since 2014, following the pioneering demonstration of QCL DCS by Villares et al. [41], all of the OFC platforms discussed above have proven compatibility with DCS. In addition to coarse single-shot measurements with lines spaced by GHz-rate $f_{\text{rep}}$, [12,15,71,72], some experiments have employed gapless tuning supported either natively by the comb [12,73] or with help from optical feedback [74] or microwave injection.

Most DCS experiments probe only the absorption spectrum, although some have probed the dispersion spectrum [12,73] as well since DCS also measures the optical phase. Other modalities employed in DCS include wavelength modulation spectroscopy [75], swept dispersion spectroscopy with inter-locked combs [76], and Faraday rotation [77].

While DCS is still a challenging experimental technique, some of the roadblocks have been removed. For example, self-detected DCS systems lift the requirement for an external photodetector [60,72,78]. Problems with mutual coherence between the sources can be addressed by computational phase correction algorithms for free-running systems [79–82] rather than frequency- [41] or phase-locked loops [71], or optical injection [83]. The technological maturity of DCS is reflected in a recent demonstration from Princeton U. by Westberg et al. [55], which showed the first field deployment of a QCL-based dual-comb spectrometer for remotely measuring urban air over 230 m at ~10 μm wavelength. This promises an era of real-time air quality monitoring in the molecular fingerprint region on a wider scale.

**Vernier spectroscopy**

Whereas DCS is arguably the most popular technique for MWIR/LWIR molecular spectroscopy using chip-based OFCs, it suffers from the problematic requirement of a fast (near-GHz bandwidth) photodetector that ideally operates at room temperature. The second challenge relates to the high data rates and computational power required to process DCS interferograms. The billions of samples per second (GS/s) needed to capture DCS spectra spanning hundreds of MHz typically requires high electrical power consumption and demanding computational resources.

Vernier spectroscopy provides an interesting remedy to this issue, which also enables high-rate and high-resolution measurements. This experimental technique uses an intentionally-mismatched optical cavity to selectively filter one comb line at a time while simultaneously serving as an optical-path-enhancing mechanism. Although the dissimilarity of the frequency axes makes Vernier spectroscopy somewhat similar to DCS, here the key concept is coincidence – only one comb line coincides with the cavity mode.

In a proof-of-concept demonstration, a Vernier cavity resolved the full bandwidth (30 cm$^{-1}$ / 0.9 THz) of an ICL comb operating at ~3.6 μm [84]. A ~3 cm long cavity provided an equivalent optical path of ~30 m, enabling the researchers to detect ppm levels of a broadband hydrocarbon over milliseconds. In a follow-up work, researchers from Caltech (Chen et al.) employed an ICL comb at ~3.3 μm to probe toluene using a combination of Vernier spectroscopy with ringdown measurements [85]. Rather than measuring modal intensities, the technique probed the cavity ringdown time, which depends on the loss but is virtually insensitive to laser noise. This modality is expected to find application in physical chemistry for multiplexed measurements of multiple species at low concentration. Further improvements in mode-resolved Vernier spectroscopy with MWIR/LWIR combs are expected using techniques that probe the cavity mode frequency shift rather than light intensity, like mode dispersion spectroscopy [86], although this requires a stabilized rather than free-running frequency comb.

**Fourier transform spectroscopy**

Fourier transform spectroscopy (FTS) is a class of experimental techniques that measure an optical signal (interferogram) whose numerical Fourier transform provides the optical spectrum. While DCS belongs to the family of FTS techniques, more typical implementation involves a mechanical Michelson or Mach-Zehnder interferometer in tandem with one comb source rather than two. Such advantages of frequency combs in mechanical FTS as brightness and power per spectral element have been quickly recognized by researchers (Mandon et al.) [87]. However, the notion that mechanical FTS provides poor resolution and spectral artifacts is still widespread in the community, and is often used to favor DCS over mechanical FTS. It stems from the fact that mechanically-scanned interferometers have a finite delay range that truncates the interferogram. In the frequency domain, it convolutes the true spectrum with a sinc function, which introduces ringing artifacts and loss of resolution.

It is fortunate, however, that the discrete structure of the comb spectrum allows one to employ FTS techniques with sub-nominal resolution. The original idea conceived by Maslowski et al. [88] allows for kHz-resolution spectroscopic investigations with equivalent km-long interferometer path differences. It relies on the fact that the convolution disappears if one properly samples the interferogram from a comb source, because neighboring comb lines coincide with zero-crossings of the sinc function that resulted from the interferogram truncation. Therefore, they are not at all affected by ringing artifacts or loss of intensity. Although the original implementation required knowledge of the comb parameters ($f_0$ and $f_{rep}$), adapted versions of the algorithm lift the requirement for $f_0$. This has practical implications, as $f_{rep}$ is easily accessible from the laser cavity using a microwave spectrum analyzer, while $f_0$ is not. Using an offset-extraction algorithm (directly from the measured interferogram) tailored to chip-based combs, Sterczewski et al. [67] obtained MHz-resolution interleaved measurements of methane and acetylene using MWIR ICL and quantum well diode combs, respectively. The interferometer with a native resolution of 10 GHz required merely 15 mm of mechanical scanning (3 cm optical path difference), which already provided resolution enhancement to 30 MHz. The ultimate resolution and spectral coverage (~1 THz) were then limited by the source, not the technique. An impressive demonstration of a rotary delay line (also with interleaved, sub-nominal resolution) was presented by Markman et al. of ETH, Zürich [89]. The researchers utilized an LWIR QCL comb (at 8.6 μm) to probe methane at lowered pressures with millisecond response rates.

Both demonstrations clearly prove that mechanical FTS should not be treated as a sub-optimal choice for resolving lines from a chip-based OFC. A practical advantage is that any laser is more coherent with itself than with another one. Therefore, phase noise does not corrupt the spectroscopic measurement quality as much as in DCS. The frequency calibration can also be very accurate, as the wavelength accuracy of the reference laser (convenient telecom or HeNe) is naturally translated to the comb's lasing region without requiring help from auxiliary lasers operating at more challenging MWIR/LWIR wavelengths.

**Multimode absorption spectroscopy**

Before ICLs were used for OFC spectroscopy, their multimode nature had been exploited for simultaneous multi-wavelength sensing using a single-pixel detector[90]. The technique termed MUMAS – multimode absorption spectroscopy [91] – relies on knowledge and reproducibility of the emission spectrum at a given injection current and temperature. Using numerical reconstruction, it is possible to retrieve the analyte's absorption spectrum from a scan of the injection current. A drawback is the semiconductor laser emission spectrum's tendency to change with time, particularly during the burn-in period (up to ~1000 h of operation). However, a potential mitigation is the emerging quantum-walk comb platform [38], which is already known to perform well for absorption spectroscopy with a single-pixel detector [92]. By adjusting the RF power, researchers have gradually broadened the QCL spectrum to measure the analyte's response. Using dual-tone injection with a variable phase shift, it was possible to control the spectral asymmetry and spectral center of mass to locate higher intensity at arbitrary positions rather than symmetrically around the center. With more control over the laser emission properties, this technique may well be suited for on-chip integration.

## Challenges, opportunities, and future directions

MWIR/LWIR combs have matured into practical broadband light sources for demanding spectroscopic and optical communication applications. However, their more widespread adoption is still hampered by numerous factors, which are schematically listed in Fig. 3. In this section, we will discuss them to identify potential knowledge gaps and routes for improvement.

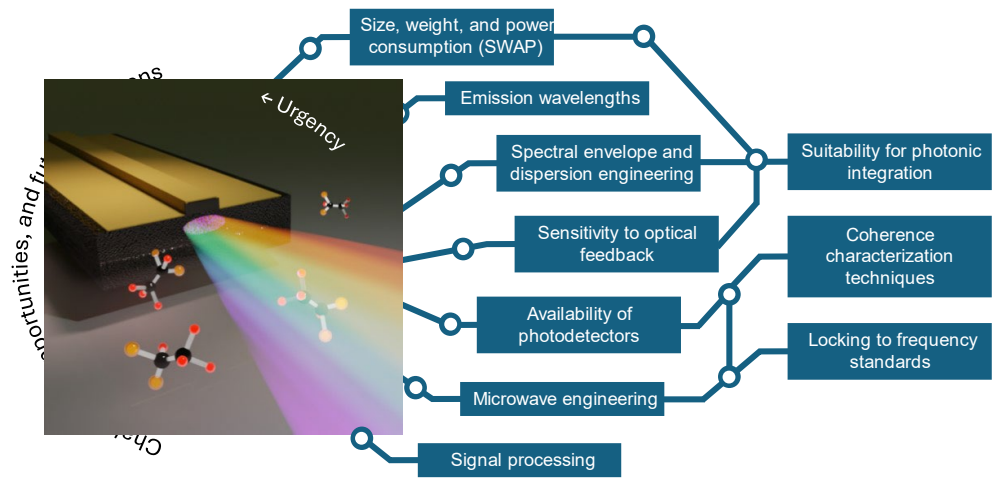

Fig. 3. Challenges, opportunities, and future directions for MWIR/LWIR semiconductor laser frequency combs shown as a network of mutual dependencies. Items are listed in decreasing urgency order from top to bottom. Some advancements require progress in multiple fields. For instance, suitability for photonic integration relies on size, weight, and power consumption (SWAP) optimization and minimization of sensitivity to optical feedback, which is here visualized as electrically connected tracks.

## SWAP – size, weight, and power consumption

One of the biggest challenges facing chip-based MWIR/LWIR combs is their power consumption and cooling requirements. Whereas the laser cavity is no larger than the millimeter scale, typically 4–6 mm long and several μm wide for FP devices, the submount, radiator with a thermoelectric cooler, housing, and collimation optics make such devices bulky, particularly when operating below the dew point. While MWIR devices often consume less than 1 W of electrical power, significant advances are needed in the LWIR to facilitate the battery-powered operation of broadband combs at elevated (room) temperature. A potential remedy may come from advances in LWIR ICLs, whose performance is continuously improving at wavelengths reaching 14 μm [61].

**Emission wavelengths**

While the wavelength portfolio of MWIR and LWIR combs spans 3–10.5 μm, a significant gap remains between 10.5 μm and THz wavelengths (60–150 μm). Unfortunately, it may not be trivial to push the emission wavelength further toward the far-infrared, as many semiconductor materials have phonon overtones accompanied by higher group velocity dispersion. Sophisticated waveguide and dispersion engineering may be needed to ensure favorable conditions for comb formation.

**Spectral envelope and dispersion engineering**

The spectral profile of an ideal OFC spectroscopic source should be flat-top, or at least continuous. This is not always possible, however, as some devices (particularly ICLs) emit light in mutually coherent spectral clusters spaced many wavenumbers apart [93], due to modal leakage to the high-index substrate [94]. This implies the need for further gain and dispersion engineering, as the two are connected via the Kramers-Kronig relation and have a profound effect on the emission profile of an FM comb [95]. Quantum well diode lasers [16] are a prime example of the expected spectral profile, lacking only a broader bandwidth of multiple THz. Dispersion engineering is also important from a quantum walk comb perspective, as it limits the spectral coverage at high microwave injection powers [96]. Optimization of the microwave signal, e.g., dual-tone (harmonic) injection with a variable phase shift, can also help tailor the spectral properties.

Notable improvement of the spectral properties has been obtained recently in ICLs by employing hybrid bottom clad layers (superlattice followed by n$^{++}$-InAsSb) with a low refractive index and significant loss at the cladding-substrate interface [29,97,98]. In combination with surface roughening of the bottom contact [29], this has virtually eliminated modal leakage and is expected to improve the bandwidth of future ICL combs due to lower GVD. Smooth spectral profiles are also expected for passively mode-locked ICL combs (operating AM) once sufficiently fast carrier extraction becomes available [29,99].

**Microwave engineering**

The ability to rapidly modulate an OFC has paramount importance for spectroscopy (stabilization of $f_{rep}$) and free-space optical communication. However, chip-based OFCs are still sub-optimal from a microwave standpoint due to poor impedance matching and strong insertion losses. Microwave engineering is also important in the context of microwave-light interaction because an interplay between microwaves and light triggers various spatiotemporal phenomena [100]. For instance, at THz wavelengths it was possible to obtain unprecedented flexibility of the repetition rate (induce a tunable group velocity) when the microwave injection efficiency was optimized [101].

**Sensitivity to optical feedback**

Semiconductor lasers are known to be sensitive to optical feedback. In many cases this feature can be used to narrow the optical linewidth from a high-finesse optical cavity. Unfortunately, however, in real-life scenarios it is uncontrolled and triggers noisy or even chaotic operation. A potential remedy used in many OFC experiments is to incorporate an optical isolator. Such devices utilize a permanent magnet, a special crystal with a strong Faraday effect, and two polarizers. A wider availability of such optical components, particularly at more exotic wavelengths, is key for MWIR/LWIR combs. Nevertheless, the holy grail will be to make chip-based combs virtually insensitive to optical feedback (as in ring devices) or to realize an on-chip optical isolator.

**Suitability for photonic integration**

Rather than relying on bulk optical components, it would be beneficial to couple the light from chip-based OFCs into passive waveguides for improved optical sensing using tight modal confinement and local field enhancement. First results have proved the QCL comb's suitability for this purpose [102]. Very recently, wavelength multiplexing on InP enabled devices emitting at different wavelengths (5.2 μm and 8 μm) to be coupled into the same waveguide [103]. Although the coupled devices were not explicit combs, this demonstration marks a significant breakthrough towards multi-wavelength sensing with the potential for future extension of the number of sources.

Another important aspect of future on-chip sensing is optical waveguiding. To date, slot-type [104] and suspended waveguides [105] with varying losses and confinement factor have been used for molecular spectroscopy in the MWIR region. Future optimization of the waveguide's geometry and losses will be needed to ensure the optimal ratio of modal confinement to losses, which translates into the attainable equivalent optical path length for sensing.

**Availability of photodetectors**

A fast (GHz-bandwidth) photodetector is needed to fully unlock the potential of MWIR/LWIR OFC sources. The optical response up to ~2.6 μm of typical commercially-available extended-SWIR InGaAs photodetectors is unsuitable for use at longer wavelengths. This niche is addressed by more sophisticated quantum well infrared photodetectors (QWIPs) [106], interband cascade photodetectors (ICP) [107,108], or resonant cavity infrared detectors [109]. Whereas a semiconductor OFC source (laser chip) can be used as a working GHz-rate photodetector, tailored designs (also commercial [110]) offer better responsivities and detection speeds. However, considering the high optical power of semiconductor OFCs, a sub-optimal design may still be practical. For instance, bi-functional ICLs [14,72] and QCLs [60,78] have been used to detect GHz-rate multiheterodyne spectra.

Photodetector arrays are also desired for MWIR/LWIR comb spectroscopy (used, for instance, with a grating in direct frequency comb spectroscopy schemes). Finally, the wider availability of longer-wavelength balanced detectors will greatly improve nearly all the spectroscopic techniques. This is because single-ended detection does not allow one to suppress many environmentally-induced noise sources, which can be canceled in a balanced detection setup.

**Locking to frequency standards**

To fully unlock the metrology potential of chip-based MWIR/LWIR comb devices, it will be desirable to reference the emitted frequencies to frequency standards [111]. This implies the need for control loops and often auxiliary lasers, which will be non-trivial to implement at longer wavelengths.

Another major challenge of MWIR/LWIR devices is detection and stabilization of the offset frequency ($f_0$). Whereas the gold standard for MHz-rate femtosecond pulses emitted by fiber- or solid-state lasers employs the $f$-to-$2f$ self-referencing technique [112], it is virtually impossible or at least impractical to apply this concept to chip-based electrically-pumped combs. This is because a prerequisite for offset detection is octave-spanning coverage via spectral broadening in a nonlinear medium, together with second harmonic generation of the red part of the broadened spectrum. For lasers with multi-GHz rates and sub-ps or ps pulses, the available peak powers in the watt range (pulse energy in the single pJ range) [57] are weakly compatible with nonlinear spectral broadening schemes, which typically require nJ-energy pulses with sub-kW peak power. It may be expected that further power scaling or emerging MWIR materials will promote low-threshold nonlinear broadening with pJ-energy pulses with fs durations [113].

As of today, one solution for detecting and stabilizing the offset frequency without resorting to the $f$-to-$2f$ technique is delayed heterodyne detection of the waveform emitted by the laser [114]. With the emergence of new integrated photonic platforms, it may be feasible to realize this concept on-chip, as it merely requires an acousto-optic modulator and a thermally-stabilized delay line (here, potentially a waveguide with a heater).

Fortunately, it is much more straightforward to stabilize the repetition rate ($f_{rep}$). One of the most widely used approaches is to inject a strong microwave signal into one end of the cavity. This promotes operation with an externally-defined line spacing, which sometimes enables significant spectral broadening [115]. The locking range of a QCL or ICL is usually a small fraction of the repetition frequency, with typical values in the 1–5% range.

In principle, the repetition rate can be controlled by simply changing the injection current. However, an anti-correlation between $f_0$ and $f_{rep}$ [116] exists, which sometimes allows one to obtain narrower optical linewidths than expected from $f_{rep}$ only (cumulative broadening due to the order number of the lasing line). This effect is particularly strong in quantum well diode lasers, where the fixed point [117] is located within the lasing spectrum. This anti-correlation exists even in the presence of residual optical feedback from a moving mirror [67]. Unfortunately, the anti-correlation between the two degrees of freedom of a semiconductor laser OFC means that improving the stability of one typically has a destructive effect on the other. Therefore, one should either stabilize both frequencies or leave the laser free-running.

**Coherence characterization techniques**

The development of new comb architectures and devices requires non-trivial characterization techniques. Arguably, the most popular employs a modality of Fourier transform spectroscopy, which detects a microwave signal resulting from collective intermode beating at $f_{rep}$. When demodulated using a lock-in amplifier, it provides information about the intermodal phases and coherence. The technique known as SWIFTS – shifted wave interference Fourier transform spectroscopy – was proposed by Burghoff et al. in 2015 [63]. Its main advantage is that, unlike popular non-linear interferometry techniques developed for ultrafast lasers, SWIFTS employs a linear photodetector that is much easier to obtain at MWIR/LWIR wavelengths. Non-linear interferometry is used sporadically to probe the intensity autocorrelation, but is not very informative. Due to its experimental complexity, the SWIFTS technique will require, in the near future, dedicated instruments and photodetectors, particularly in the self-referenced arrangement.

Criticism of the SWIFTS approach led to conception of the FACE technique – Fourier-transform analysis of comb emission [23], which relies on direct beating of a mode-locked MWIR/LWIR OFC with a laser under test. The authors have proven agreement between the characterizations obtained using SWIFTS and FACE. Another useful variant of the asynchronous interaction between a mode-locked laser and the comb under test is the asynchronous upconversion sampling (ASUPS) technique [55,112], which generates a sum frequency in a nonlinear crystal when a NIR pulse interacts with light from a QCL. Again, ASUPS data fully agreed with the SWIFTS reconstruction.

Clearly the development of MWIR/LWIR comb sources requires experimentally-challenging characterization techniques. The ease of their implementation will strongly depend on the progress and availability of high-$f_{rep}$ mode-locked lasers (also long-wavelength), and fast photodetectors. Notably, fast oscilloscopes with ~50–100 GHz electrical bandwidth may enable direct observation of the emitted waveform without the need for waveform reconstruction.

**Signal processing**

The application of MWIR/LWIR combs to molecular spectroscopy or ranging strongly relies on novel signal processing routines tailored to non-stationary sources. In some cases these are inspired by electronic intelligence (ELINT) algorithms, including Doppler or CW radar, or pulse compression. Such algorithms are essential for the prolonged integration (averaging) of spectroscopic signals. Still, many of them are in infancy or show weak compatibility with noise-rich signals obtained with chip-based comb sources.

Not only does time-domain signal processing need new ideas or adaptations of state-of-the art algorithms known in the signal processing community, the analysis of frequency-domain data can also be problematic in molecular DCS [119]. Due to excessive phase noise relating to the finite (or technically broadened) optical linewidth, lines of the multiheterodyne spectrum do not look discrete. They instead take Lorentzian-like shapes, often with multiple peaks [79,80]. How to unambiguously extract the modal phases and amplitudes from a smeared-out or multi-peaked (modulated) spectrum is still a challenging question, particularly for novices. Rapid peak interpolation, coherent demodulation, or lock-in-based [120] approaches can be used to solve some of these problems. Nevertheless, there is hope for artificial-intelligence-assisted algorithms that provide less computationally-demanding solutions for multi-frequency extraction of the phase and amplitude.

While some of these problems can be mitigated using computational phase correction techniques [79–82], the resulting corrected spectra are still subject to the same problems of sample response extraction as in uncorrected data.

Other areas of signal processing that will greatly improve the applicability of MWIR/LWIR devices are algorithms for single-comb spectroscopy. This is because high-rate or even real-time DSP processing of data requires significant computational resources. The first technique is sub-nominal Fourier transform spectroscopy [67], which enables MHz-resolution measurements using compact (mm-sized) interferometers. By digitally extracting the offset frequency from the measured interferogram combined with the knowledge of $f_{rep}$, it can enhance the measurement resolution by more than two orders of magnitude. Elegant techniques to accurately estimate the frequency offset and provide artifact-free adaptive resampling of acquired data will be needed in the future. Similar problems concern the Vernier spectroscopy technique [84], which utilizes a tunable optical cavity to simultaneously enhance the analyte interaction path and filter one comb line at a time. Due to dynamic optical feedback from the cavity, the line spacing in the measured spectrum is not uniform despite the equidistance between lines in the optical domain. Consequently, fast and efficient algorithms may be needed to equalize the line spacing (resample the data) and align multiple measurements to facilitate averaging in the frequency domain.

### *Concluding Remarks*

The unique features of chip-based MWIR/LWIR OFC sources strongly motivate further development and improvement of their performance. Of course, all problems faced by this technology cannot be tackled at once. Priority should be given to optimization of the SWAP aspects, followed by the emission profile, waveguiding, bandwidth, and reproducibility. There are still many unexplored phenomena, such as quantum correlations [121] between the modal intensities and phase originating from four-wave mixing responsible for frequency pulling in FM combs. It may be expected that such phenomena can be universal, irrespective of the active region design. Future efforts should also focus on making MWIR/LWIR comb spectroscopy easier and more practical to implement. Currently, the number of DCS practitioners in this spectral region is relatively small compared to the fiber and solid state laser communities. Finally, in the context of a recent work by Roy et al., MWIR/LWIR combs may need to be redefined or reclassified in the language of the liquid comb theory [122]. Liquid combs, in contrast to regular combs, are noisy but preserve intermodal coherence. States resembling this mode of operation are routinely observed in ICL or diode devices, suggesting that the next revolution of OFC sources may be around the corner.

**Back matter**

Funding
Content in the funding section will be generated entirely from details submitted to Prism.
Acknowledgments

Ł. A. Sterczewski acknowledges funding from the European Union (ERC Starting Grant, TeraERC, 101117433). Views and opinions expressed are however, those of the authors only and do not necessarily reflect those of the European Union or the European Research Council Executive Agency. Neither the European Union nor the granting authority can be held responsible for them. The author would like to thank Dr. Jerry Meyer, and Dr. Igor Vurgaftman at NRL for helpful suggestions.

Disclosures

The author does not declare a conflict of interest.

5. Mid- and Long-Wave Infrared Photonic Crystal Surface Emitting Lasers (PCSELs)

WEIDONG ZHOU[*], SEUNGHYUN LEE, AND MINGSEN PAN

**Department of Electrical Engineering, Photonics Center, University of Texas at Arlington, Arlington, Texas, TX 76019, USA**
*wzhou@uta.edu*

*Overview*

The photonic crystal surface-emitting laser (PCSEL) is a new generation of semiconductor light source technologies, featuring high power density from a broad emission area. In recent years, research and investigations of PCSELs have gained popularity due to their great potential for high power applications. Compared to other semiconductor laser structures, such as the edge-emitting lasers (EELs) and vertical cavity surface-emitting lasers (VCSELs), PCSELs utilize more flexible 2-dimensional (2D) in-plane optical feedback to implement single mode lasing over mm- or even over cm-surfaces with nearly diffraction-limited output beams. Here we briefly review the current status of PCSEL development in semiconductor light sources. We discuss the challenges and opportunities of implementing mid- and long-wave infrared (MWIR and LWIR) PCSELs, followed by a future outlook.

### 3. Current Status

In recent decades, semiconductor light sources have become ubiquitous in many fields, due to their advantages of size, power, speed, and efficiency. Driven by a market totaling billions of dollars, commercial models of semiconductor lasers are widely used in applications of data centers, light detection and ranging (LiDAR) [1], optical sensing [2], communications [3, 4], material processing [5], etc.

**Cavity design of semiconductor lasers.** The common EEL and VCSEL semiconductor laser structures are shown in **Fig. 1**(a) and (b), respectively. EELs emit light from a side of the chip that is parallel to the optical feedback in the waveguide structure. Thus the beam emitted from the relatively small aperture is asymmetrical and diverging. The output power of EELs can be scaled from mW to kW by increasing the bar length from tens of microns to a few centimeters. VCSELs have an optical cavity perpendicular to the chip surface, and optical confinement is realized by a pair of distributed Bragg reflector (DBR) mirrors that sandwich the active region, which is typically created by epitaxial growth. VCSELs can produce a symmetric, circular beam, although the divergence angle is again quite large if the cavity diameter is small. A common challenge for both EELs and VCSELs is to maintain single-mode operation at high power levels. As the size increases to support power scaling, the cavity becomes multimode, which limits the laser beam quality and brightness.

The PCSEL is another type of semiconductor laser configuration (**Fig. 1**(c)) that was first proposed and demonstrated 25 years ago [6-8]. PCSELs utilize a 2D photonic crystal (PC) layer's in-plane band-edge modes that evanescently couple with the active region [8, 9]. The 2D optical feedback forms a resonance cavity in which the lasing thresholds of the band-edge modes can be designed to have a single dominant mode. Thus, the PC cavity design can implement single-mode lasing and power scaling in a broader area. Due to the mode dispersion engineering capabilities in the 2D PC lattice, higher order modes can be suppressed to support only fundamental laser operation. A surface-emitting beam then forms from the evanescent waves of the lasing mode. Since the mode operates at the band edge (k = 0) and is leaky (above the light line), as shown in **Fig. 2**, the output beam has extremely low divergence in the surface-emitting direction. Thanks to the structural advantages of a PC cavity, the PCSEL's output beam can be orders of higher brighter even without an external collimation lens. The implementation of surface emission by the in-plane optical feedback in a PCSEL also eliminates the need for thick DBR mirrors. This is especially advantageous at longer wavelength where the required DBR mirror thickness can become excessive.

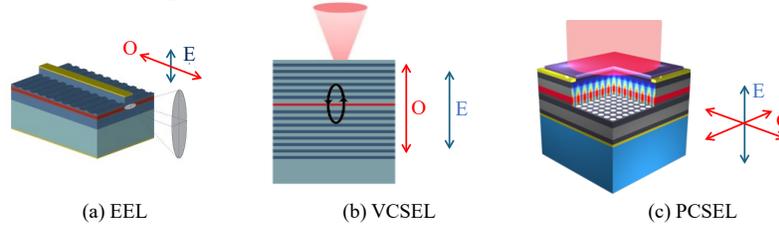

(a) EEL          (b) VCSEL          (c) PCSEL

**Fig. 1.** Three semiconductor laser architectures: (a) EEL: In-plane optical Fabry-Perot (FP) cavity for in-plane (edge) mission; (b) VCSEL: Vertical optical FP cavity for surface-emission; (c) PCSEL: In-plane photonic crystal lasing cavity with a leaking mode for surface-emission. Red arrows: optical feedback directions. Blue arrows: electrical injection directions.

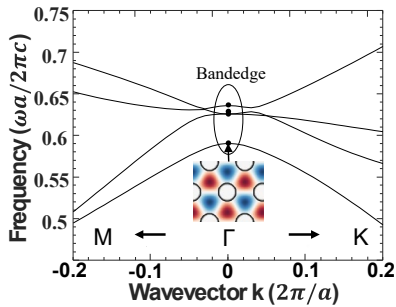

**Fig. 2.** Photonic band structure of a PC near the $\Gamma$ point. A triangular lattice with circular air holes is illustrated. Standing-wave patterns form in a PCSEL cavity due to the in-plane feedback, and emission can be formed from the surface-normal direction.

**Performance achievement of diode PCSELs.** Based on the epitaxial growth of PIN heterostructures, PCSELs have been demonstrated in different wavelengths in the infrared (IR) region, as shown in **Fig. 3**. By engineering the optical feedback in a PCSEL cavity, high power high brightness lasers have been demonstrated especially by S. Noda and collaborators.[10] Single-mode operation in large sized PCSELs can be achieved by introducing double-lattice PC designs [11]. Output powers of 50 W and brightnesses of 1 GWcm$^{-2}$sr$^{-1}$ have been achieved by GaAs heterostructures emitting at 940 nm [12]. In recent years, Watt-class GaAs-based PCSELs have also been demonstrated in the 1040 nm range [13]. For longer optical communication wavelengths of 1300 nm and 1550 nm, PCSELs have been realized with InP-based heterostructures with active InGaAsP [14, 15] and AlInGaAs [16] multiple quantum wells (MQWs). SUNY has reported 2 μm PCSELs based on a GaSb-based type-I diode structures that operate in continuous wave (CW) mode at room temperature [17]. Although the initial devices have low output power on the order of 10 mW, it should be possible to substantially enhance the output powers and brightnesses of GaSb-based PCSELs emitting from 2 to 3.5 μm [18].

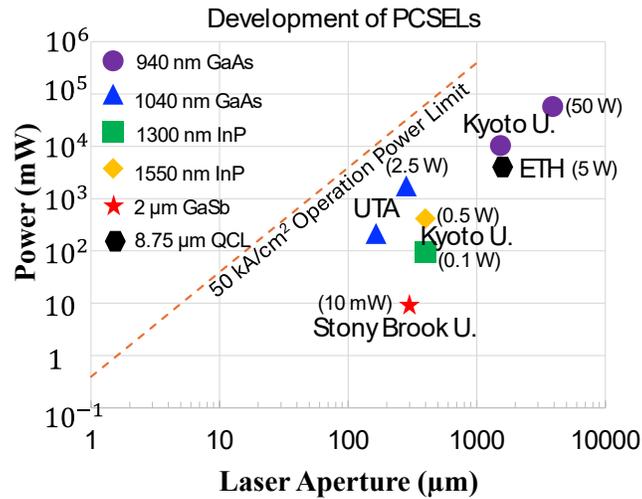

**Fig. 3.** Maximum output powers of PCSELs operating at different wavelengths. The records are indicated in parentheses. Refs for diode lasers: 940 nm GaAs (circles)[12, 19]; 1040 nm GaAs (triangles) [13, 20]; 1300 nm InP (square) [14]; 1550 nm InP (diamond) [15]; 2 μm GaSb (star) [17]; 8.75 μm pulsed quantum cascade laser (QCL) PCSEL (hexagon) [21]

**Broader aspects of PCSEL technologies.** PCSELs have been investigated and addressed in broader aspects. In the field of free space optical communications, the PCSEL's narrow beam divergence can transmit the laser beam over a longer distance. Moreover, PCSEL modes have larger optical confinement factors in the active region compared to that for VCSELs. Thus, PCSELs of the same size can achieve higher speed than VCSELs. Hamamatsu reported a commercial PCSEL with bandwidth of 2.32 GHz and power of 25 mW. [22] Several-GHz modulation bandwidth has been achieved with Watt-class output power. [23] PCSELs based integrated on heterogeneous platforms have also been demonstrated, such as Si/III-V PCSELs [24], and 2D materials [25, 26].

PCSELs with narrow spectral linewidth and high-power operation have been developed. We have reported spectrum linewidths of about 330 kHz for GaAs-based PCSEL devices.[27] Another report demonstrated an intrinsic linewidth of 70 kHz.[28] However, the laser output power efficiency decreases when the cavity Q is too high, and it remains challenging to combine higher power with narrower linewidth. The solution requires larger cavities where the Q is high and the emission area is large, which simultaneously supports high power and narrow linewidth. For 1mm emission area, Ref. [29] demonstrated PCSELs with 1kHz-class linewidth and 5-W output power.

Semiconductor laser arrays are important in the applications of power scaling [30], beam combining [31], and modulation [32]. However, it is challenging to achieve coherent laser arrays with the EEL or VCSEL architecture due to the weak in-plane evanescent coupling. The lateral coupling can be controlled by the lateral leakage of the in-plane feedback of PC cavities in the PCSELs [33]. By placing individual PCSEL cavities within lateral evanescent coupling range, Ref. [20] investigated the feasibility of passively coupled PCSEL array device on a monolithic mesa structure. The combined beam from a PCSEL array exhibits similar coherence and linewidth as the single PCSELs.

**MWIR and LWIR PCSELs.** Gas and vapor molecules have signature absorption features in the mid-infrared spectrum. Optical sensing of greenhouse/toxic gases can be traced down to the sensitivity of parts per billion for both environmental and industrial monitoring applications. [34] In these wavelength regimes, conventional PIN diode lasers have strong non-radiative processes, specifically high Auger recombination and free carrier absorption (FCA).[35-37] Thus, most recent laser development in the MWIR and LWIR spectral ranges has been driven by cascade-type lasers, i.e., QCLs that employ intersubband radiative transitions and interband cascade lasers (ICLs) [38, 39]. QCLs can be engineered to exclude the use of p-type materials and thus avoid intervalence band absorption.[40] ICLs with type-II designs improve the energy conversion efficiency by introducing cascading and reducing the Auger non-radiative recombination rate [37, 41]. Ref. [42] demonstrated MWIR PCSELs based on type-II "W" active regions emitting at wavelength of 3.7 μm with optical pumping.

**QCL-based PCSEL (PC-QCLs).** InP-based QCLs have emerged as the dominant laser platform for the MWIR and LWIR ranges from approximately 4 μm to 12 μm. Due to the intersubband selection rules governed by Fermi's Golden Rule, the radiative transition in QCLs requires the electric field of the optical mode to be polarized perpendicular to the QW plane (i.e., TM-polarized). Theoretical analyses and numerical simulations have shown the feasibilities of TM modes in a photonic crystal cavity for surface emitting lasers [43-46]. Ref. [47] demonstrated QCL-based PCSELs at 4.3 μm. Beam divergence less than 1 degree and 50 mW pulsed output power were obtained at 77 K. Ref. [48] achieved 8.5 μm PCSELs with pulsed output power 176 mW at RT. With 1.5 mm device size, 5 W 8.75 μm PC-QCLs were demonstrated at pulsed operation with 10 degree divergence [21]. These results suggest that aligning photonic designs of cavities with the laser emission structures offer a compelling route to realize high-power high brightness MWIR and LWIR lasers.

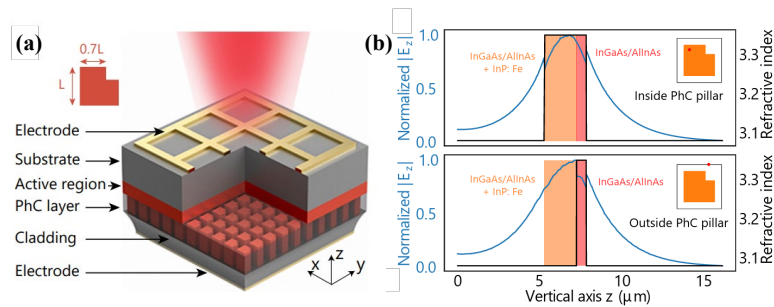

**Fig. 4**. QCL-based PCSEL design: (a) Schematic of PCSEL structure with QCL active region. (b) Mode intensity distribution in layers. Inside PC pillar (top) and outside PC pillar (bottom). Shaded are 2 μm thickness of active regions, with 1.4 μm etched PC patterns and filled with semi-insulating InP. Adapted with permission from [21] © Optica Publishing Group.

**Fig. 4** shows the design of the 8.75 μm PCSEL in Ref. [21] where the recorded power of 5W was characterized at pulsed operation. This advance brings QCL PCSELs operating at RT to the Watt-class.

*Challenges and opportunities*

## 5. Regrowth interface

Similar to the common design of PCSELs, the structure of the QCL PCSEL consists of layers for metal contact, cladding, active region, and PC layers, as shown in **Fig. 4**(a). The patterned PC layer in a PCSEL device functions as an optical cavity that is close to the gain region for strong optical overlap. In the design and fabrication of a PCSEL device, in order to embed the photonic crystal layer the epitaxial growth is generally separated into two steps. For diode lasers, as for a VCSEL the first growth on the n-type substrate is usually the n-cladding layer, after which the MQW/PC layers are grown on top. Then the PC cavity is patterned on the PC layer using reactive ion etching (RIE). Then the second growth (or regrowth) of the p-type cladding and contact layers are carried out. QCL PCSELs also employ regrowth, although all the materials are usually n-type.

Regrowth interface engineering is important, since the regrowth profiles determine the refractive index contrast that forms the PC cavity. Low index contrast can induce issues of weak lateral optical feedback and thus higher light leakage from the side, and the mode operates within a larger offset from the k=0 point in the momentum space. This results in larger angular divergence of the output beam.

The shape of the air hole voids formed by the regrowth affects the final device performances. The common challenge is that if the air hole becomes too small, or is even completely refilled by the regrowth, the result is reduced gain threshold contrast between the different band edge modes. Not only is the hole shape opened differently with tool-based recipes, but the etching also plays an important

role. The etching depth has been reported to impact the mode coupling between the PC and the MQWs. [49]

Such material refilling during regrowth has more impact on larger air holes. Sublattice PC designs with smaller holes overcame this challenge when researchers first demonstrated double-lattice PCSELs [19] and triple-lattice PCSELs [16] operating at both 940 nm and 1550 nm. The air void shapes can be more stably controlled with regrowth on smaller air holes. For example, record performance was demonstrated with active regions grown on PC layers with double lattice designs. [15] Airhole-retaining-regrowth has been studied for GaAs- [50], InP- [51], and GaSb- [52] based materials for PCSELs below 2.1 μm. Nonetheless, for longer wavelengths the air holes scale to larger sizes, and it has remained challenging to regrow air voids that implement a large index contrast. Further investigations into the PC index contrast for the MWIR and LWIR PCSELs are desired to improve the mode quality and laser performance.

## 6.  Electrical injection

An efficient electrical injection design controls the performance optimization, and is important if the PCSEL is to implement the functionalities of direct modulation, integration, and compact design. In Ref. [53], we reported gain-guided PCSEL mode control and optimization by applying different sizes of the p electrode to control the carrier injection region. Optimal single mode operation was obtained by maximizing the gain overlap for the fundamental mode.

One challenge is to maintain lateral uniformity of the injected carrier density when scaling to large devices [54]. For example, the size of the ring contact opening on the substrate backside is similar to that of the device, so current spreading to the cavity center becomes weaker if the device is much larger than the substrate thickness.  For device diameters from 0.5 mm to several mm, the current spreading can be enhanced by finger contacts on the substrate side. For cm-size devices, more complex designs for current spreading in the substrate must be considered.

Another challenge is the carrier injection efficiency. Similar to oxide aperture confinement in VCSELs, unique PCSEL designs can be implemented for high-speed operation.

## 7.  Optical confinement

Optical confinement in the active region determines the coupling between the optical mode and the gain region. The optical confinement factor is governed by the overlap of the mode intensity with the active QWs. In PCSELs, the photons emitted from the active layers will be coupled into the band edge modes of the PC, where the mode with the lowest lasing threshold will start the lasing action.

In diode lasers, PCSEL modes generally have higher active region confinement factors of about 0.05 to 0.07 than in the corresponding VCSELs. As is also the case for ICL and QCL PCSELs, the vertical positioning of the gain QWs relative to the edge mode intensities in the patterned PC structure must be designed carefully. The PCSEL edge mode profile usually has a single vertical peak, as shown in **Fig. 4**(b), which decays exponentially away from the PC/active region. This sets a trade-off of the number of the cascade stages vs. laser efficiency and performance, because active QWs that are positioned too far away from the mode peak will provide a relatively smaller contribution to the laser gain.

### *Future developments to address challenges*

Record performances of 5 W output power have been demonstrated with QCL-based PCSELs at 8.75 μm. [21] However, those demonstrations are still underperforming when compared to the diode PCSELs at 1 μm, in terms of beam divergence, RT operation, CW operation, etc. Future developments are expected to solve the current challenges in MWIR and LWIR PCSELs.

### 1. Optical cavity design

Optical cavity design is critical for overall laser performances. Cavity modes become less different in lasing threshold when the cavity is as large as a few hundred periods of PC. In the shorter wavelength PCSEL structures, different cavity designs such as the double-lattice, and triple lattice PC are considered to address such scaling challenges. [54] However, it is not clear how the performance degrades in long wavelength PC cavities by the complete material refilling into the PC. Future development could focus on how to implement high index contrast PC cavities.

### 2. Optical confinement

It is also feasible to design higher confinement of the cavity modes with the ICL or QCL structures. There are opportunities to utilize the staged active regions, as demonstrated in recent years the multi-junction VCSELs [55]. In Ref. [56], PCSELs with two cascaded junctions are demonstrated to have higher efficiency with the suppression of fundamental waveguide modes. The design has a cavity mode that has amplitude peaks at each junction region. Although the research has not considered odd mode and even mode competition in the first-order waveguide modes, the results of power and efficiency are impressive. Similar vertical waveguide mode design of PCSELs in ICL and QCL can be carried out to optimize the lasing efficiency and output power.

### 3. Electrical injection design

Uniform carrier injection is desired for efficiently couple optical gain to the PCSEL cavity modes. Watt-class semiconductor surface-emitting lasers usually have mm-scale emission area with current light emitting structures. With increasing the aspect ratio (device size to chip thickness), the current spreading to the contact center decreases. Electrical contact mesh or fingers on the emission window can efficiently help resolve this challenge with a few percent of output power blocked from the beam.

Development of novel materials with conductive transparent properties in the MWIR and LWIR wavelength can further simplify the electrical design for PCSELs.

With a thin layer of conductive transparent materials functioning as current injection and anti-reflective coating (ARC), the optical output power can be improved. Future investigations into different aspects of electrical and optical designs and optimizations are expected.

## 4.   Lasing efficiency

The wall-plug efficiency (WPE) of demonstrated PCSELs is around 20% for GaAs- and InP-based diode structures [57]. However, thus far the WPEs for PCSELs operating in the MWIR and LWIR regions remain muchlower, typically around 1%. Multiple approaches can be used to tackle the low efficiency, including optimization of the electrical injection design, optical confinement design, and optical mode properties. Current ridge waveguide QCL structures exhibit WPEs as high as20-30%, and continued efforts are being made to further improve their performance [58, 59].

PC-QCL design can adopt similar strategies to optimize the electrical injection design, such as enhancing the voltage efficiency. In addition, PCSELs uniquely benefit from tailored optical designs. The optical confinement of the selected band edge modes within the active layers can be maximized by carefully engineering the PC lattice and adjusting the layer thicknesses. Further consideration of eliminating the mode competition in a PCSEL cavity helps to improve the laser efficiency. The gain threshold difference between modes from different bands can be increased by the spatial patterns in the PC design, as the filed distribution of these modes usually occupies different locations within the unit cell. Similarly, higher-order modes from the same band can be suppressed by the PC cavity designs.

Below 2 μm, it has been reported that triangular, double-lattice, and triple-lattice air hole designs achieve high-order mode suppression [16, 19, 60]. We recently found that the Q-factor contrast ratio between the fundamental and first higher-order mode can be designed to be independent of cavity size by utilizing the bound states in the continuum (BIC) properties. The experimental investigations into this exotic behavior are currently underway.

Ultimately, single-mode operation in QCL-based PCSEL structure can be accomplished by these advanced photonic designs. These measures can significantly improve the lasing efficiency while achieving single mode lasing from the broad area devices.

To overcome the weakness in index contrast of current PCSELs, Ref. [21] adopted strong coupling between PC and active layers by partial etching of active region. Etching on the active layers may cause issues of reduced material gain, loss of uniformity, and non-radiative recombination, which is not desirable for high-power lasers. Further improvement of the lasing efficiency is highly possible by overcoming the issues of integrating PC cavities with stronger lateral feedback, and by improving the index contrast and uniformity of the air holes.

### *Concluding Remarks*

Remarkable development and contribution have been made by PCSEL designs to the field of semiconductor light sources. Due to its extraordinary ability to modulate the mode field of cavity resonance, PCSEL designs are pushing the laser technology towards smaller size and higher brightness.

Three major challenges and opportunities are discussed, including regrowth profile impact, electrical injection design and optical confinement optimization. Future developments are expected to further improve performances of PCSELs in MWIR and LWIR. Addressing the regrowth challenges for a stronger refractive index contrast in the PC layer can potentially improve mode properties and beam properties.  Laser efficiency can be further improved by the optimal alignment between optical modes and gain layers. PC cavity design helps scale to larger emitters with single-mode operation capabilities.

In conclusion, the lasing efficiency for MWIR and LWIR PCSELs is currently lower than those in the NIR. The lasing efficiency can be further improved by the optimal alignment of active layers with the mode field distribution. Multi-junction PCSEL designs are potential candidates for optimal utilization of the thick active regions consisting of the cascading stages. On the other hand, improvement of the semiconductor design to reduce the intrinsic optical loss of the materials can enhance the slope efficiency for high power applications.

PCSELs have advantages due to more flexible photonic modulation to control and couple emitted photons from semiconductor structures. Therefore, realizing high-power, high-brightness lasers in CW operation in MWIR and LWIR wavelengths using PCSEL designs are very promising.

6. Single-Mode Ring Cascade Lasers

## ROLF SZEDLAK,[1,*] BENEDIKT SCHWARZ,[1] AND GOTTFRIED STRASSER[1]

**[1]Institute of Solid State Electronics, TU Wien, Gußhausstraße 25, 1040 Vienna, Austria**

*\*rolf.szedlak@tuwien.ac.at*

*Introduction*

Mid-infrared light sources are essential for a wide range of applications, including gas sensing, spectroscopy, environmental monitoring, and medical diagnostics, due to the presence of strong molecular absorption features in this spectral region. Quantum cascade lasers (QCLs) [1] and interband cascade lasers (ICLs) [2] have emerged as the leading electrically pumped semiconductor lasers for mid-infrared emission, owing to their wavelength flexibility, compactness, and continuous-wave operation capabilities. These device classes have enabled the development of compact and powerful mid-infrared systems, making them indispensable in both scientific and industrial contexts. Among them, Fabry-Pérot (FP) distributed feedback (DFB) lasers [3] represent the current standard for electrically pumped single-mode mid-infrared sources. Their popularity stems from several key advantages: straightforward fabrication using well-established ridge waveguide geometries, high optical confinement, robust performance, and the potential for high output power.

However, a fundamental limitation of these devices lies in their reliance on cleaved facets for optical feedback and light emission. The facets are typically formed at the final stage of fabrication, meaning that the device's performance is critically dependent on the cleave quality and location. Consequently, full optical characterization is only possible after complete fabrication, precluding on-chip testing during earlier process steps. This limitation not only hampers yield, but also leads to inefficient utilization of fabrication resources. Moreover, facet-emitting QCLs and ICLs inherently produce highly divergent and asymmetric beam profiles. This is due to the small and often rectangular emission aperture defined by the waveguide cross-section. The resulting beam shapes complicate optical alignment and beam shaping, and can hinder integration into sensing or imaging systems that benefit from more symmetric and collimated output profiles.

To address these challenges, ring QCLs [4] and, more recently, ring ICLs [5] have emerged as promising alternatives. These devices are based on ring-shaped waveguides, which eliminate cleaved facets entirely. Optical feedback and vertical emission is achieved using second-order DFB gratings embedded in the top metal or semiconductor cladding layers. The second-order DFB grating refracts light out of the waveguide according to the Bragg condition, enabling vertical emission. This emission occurs both upwards and downwards, meaning that light from a ring QCL or ring ICL is emitted in two opposite directions perpendicular to the chip surface, as indicated in Fig. 1(a). This geometry offers several compelling advantages. First, the absence of facets decouples the optical performance from the cleaving process, allowing for wafer-scale optical testing and improved fabrication yield. Second, vertical emission through a grating increases the emitting area and provides a more symmetric, collimated beam with reduced divergence, which is crucial for efficient free-space coupling and simplified optical integration.

Compared to other single-mode mid-infrared laser architectures, ring QCLs and ICLs offer a unique combination of integration potential and beam quality. Vertical-cavity surface-emitting lasers (VCSELs), while offering excellent beam circularity and on-chip testing capabilities in the near-infrared, are incompatible with QCL technology due to the intersubband selection rule that prohibits vertical emission [6]. Photonic crystal QCLs (PhC-QCLs) [7] can also provide high-power surface emission and high beam quality, but they rely on complex two-dimensional periodic patterning and deep etching, which impose strict fabrication requirements and potential yield limitations. In contrast, ring QCLs and ICLs use simpler grating designs and standard planar processing techniques.

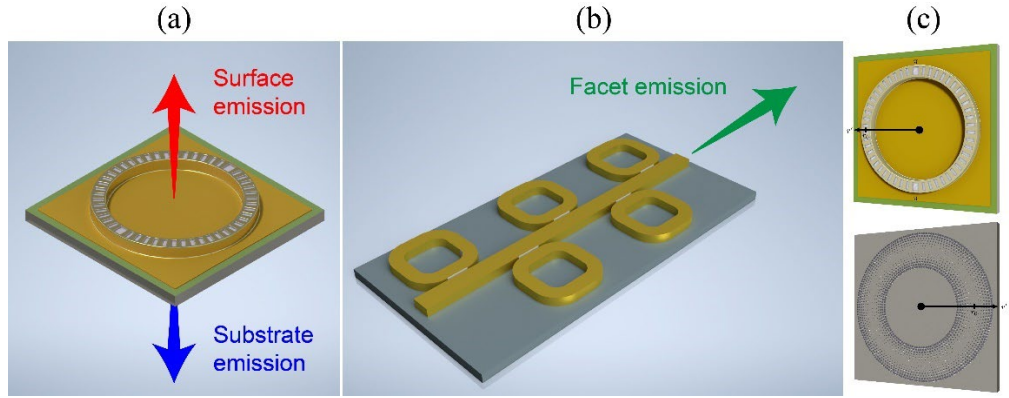

Figure 5: (a) Sketch of surface and substrate emission from a ring QCL with a π-phase shift in the DFB grating. (b) Sketch of five ring lasers evanescently coupled to a collector waveguide. The latter combines the emission from all the rings and emits the light via its facet. (c) Sketch of a π-shift ring QCL (top) with a collimating metalens fabricated directly into the other side of the substrate.

External-cavity QCLs (EC-QCLs) [8] offer a high degree of spectral tunability and can deliver narrow linewidths over broad wavelength ranges, but their external-cavity design makes them inherently bulky and less suited for on-chip integration. Random lasers [9] have recently attracted attention for their unconventional feedback mechanism, which relies on multiple scattering within a disordered gain medium. While they typically operate as broadband sources, it has been shown that introducing a reconfigurable optical field, capable of modifying the permittivity of the active region, can lead to tunable single-mode operation. However, implementing such a reconfigurable field remains technically challenging, especially in compact and integrated configurations.

In this context, ring QCLs and ICLs stand out by combining the fabrication simplicity and integration potential of planar devices with the beam quality and testing flexibility needed for scalable mid-infrared sensing applications. In addition, the ring geometry allows coupling to adjacent waveguides, opening new avenues for power extraction, spectral engineering, and photonic integration beyond what is feasible with traditional facet- or surface-emitting designs. Fig. 1(b) illustrates a configuration in which five rings evanescently couple to a central collector waveguide. Unlike conventional ring QCLs that rely on distributed feedback gratings for single-mode operation, this approach achieves spectral selectivity through alternative feedback mechanisms, such as curvature variations and controlled coupling between the rings and bus waveguide. This design exemplifies how ring geometries can be adapted for novel functionality, while remaining compatible with DFB concepts in combinations that further enhance the mode control and spectral purity.

This roadmap article explores the current status, challenges, and future directions of single-mode ring QCLs and ICLs, with a particular focus on their potential for compact, high-performance mid-infrared light sources in sensing and integrated photonics.

**Current status**

A key advantage of ring QCLs and ICLs is their compatibility with on-chip optical integration. Optical components such as lenses and polarizers can be directly fabricated on the backside of the semiconductor chip using standard planar techniques. Although similar elements have also been realized for facet-emitting devices, their fabrication often involves significantly more complex processes, such as ion beam milling [10], which is not suited for large-scale production, or manual lens alignment [11], that results in less efficient and less versatile components. In contrast, ring QCLs have enabled the integration of compact and lithographically defined on-chip polarizers [12] and collimating metalenses [13] composed of subwavelength holes, offering a simpler and more scalable fabrication procedure. A collimating metalens design is sketched in Fig. 1(c). The metalens on the back side of the chip is directly fabricated into the substrate and aligned with the ring QCL on the front side. These metalenses have also been used to azimuthally twist the wavefront, enabling the emission of beams that carry orbital angular momentum (OAM) [14]. Additionally, grating-based phase engineering, such as introducing a $\pi$-phase shift in the distributed feedback grating, enables transformation of the emission profile from a pattern with a central minimum to one with a central intensity maximum [12]. These on-chip beam modifications have proven to be a powerful tool for tailoring the emission characteristics of ring QCLs, which is essential for sensing applications that demand highly collimated and versatile beams.

Ring QCLs and ring ICLs have already demonstrated impressive performance across the mid-infrared and terahertz spectral regions, with lasing observed from 3.7 μm to 94 μm [5, 15]. Notably, the highest continuous-wave (CW) power recorded to date is 0.5 W from an overgrown mid-infrared ring QCL operating at 4.6 μm [16]. A SEM image of a device that has been cleaved through the center of the ring to show the circular ridge waveguide is depicted in Fig. 2(a). An alternative approach based on a top grating in combination with electroplating ensured a collimated and concentric beam profile as illustrated in Fig. 2(b) [17].

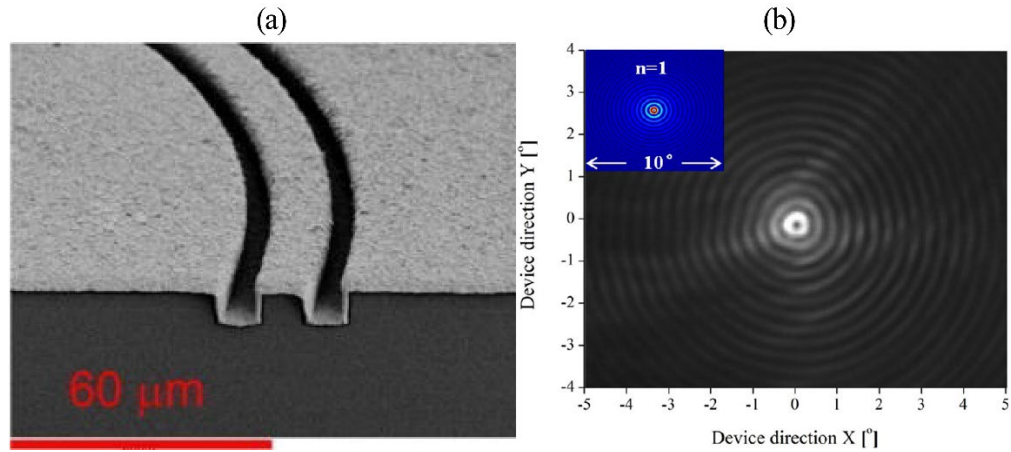

Figure 6: (a) SEM image of the cleaved cross section of an overgrown CW ring QCL [16]. (b) Farfield pattern of a CW ring QCL with a top grating operating at the fundamental mode. The inset shows the corresponding farfield simulation [17].

In general, it is possible to design the DFB grating of a ring QCL such that the losses introduced by the grating are lower than the mirror losses in a facet-emitting ridge QCL. When ring QCLs and facet-emitting ridge lasers were fabricated from the same heterostructure material, the ring QCL exhibited a threshold current density of 1.8 kA/cm² compared to 2.75 kA/cm² for the ridge laser [18]. It is noteworthy, however, that reduced grating losses correspond to less outcoupling of the light and therefore lower optical output power. This reflects an inherent trade-off in ring QCLs between low threshold current density when the out-coupling is weak and higher output power when it is strong. On the other hand, the DFB grating of a ring laser offers greater flexibility than a facet-emitting design, in that it can be tailored to suit the specific application needs. High-performance CW ring QCLs have achieved threshold current densities as low as 1.06 kA/cm², with wall-plug efficiency of 6.25% and maximum output power up to 0.5 W [16]. For comparison, CW single-mode facet-emitting DFB ridge QCLs have demonstrated higher threshold current densities around 1.92 kA/cm², combined with higher wall-plug efficiencies up to 16.6% and output powers reaching 5.1 W [19]. Facet-emitting multimode CW ridge QCLs without DFB gratings have achieved even higher wall-plug efficiencies of 22% with threshold current densities of 1.22 kA/cm² and output powers of 5.6 W [20]. The threshold currents of ICLs are generally lower than in QCLs. Room-temperature CW ring ICLs have demonstrated a threshold current density of around 0.6 kA/cm² and a wall-plug efficiency of approximately 0.5% [21]. In contrast, facet-emitting single-mode ICLs have achieved lower threshold current densities of about 0.33 kA/cm² [22]. Other single-mode designs demonstrated significantly higher wall-plug efficiencies of up to 9% [23], with even higher values up to 18% recorded in multi-mode devices [24].

A current research focus is the miniaturization of ring QCLs and ICLs, motivated by the need for lower power consumption and higher chip-level integration. Smaller devices also allow higher yield per wafer, reducing cost per laser and supporting large-scale deployment. For ring ICLs, radial sizes down to 80 μm have shown no degradation in performance, with typical power dissipation at threshold and maximum output power of 100 mW and 170 mW, respectively. The lowest recorded power dissipation at threshold for these rings was found to be 68 mW for a CW ring with radius 80 μm and threshold current of only 15 mA at room-temperature. An array of these devices is depicted in Fig. 3(a) [25].

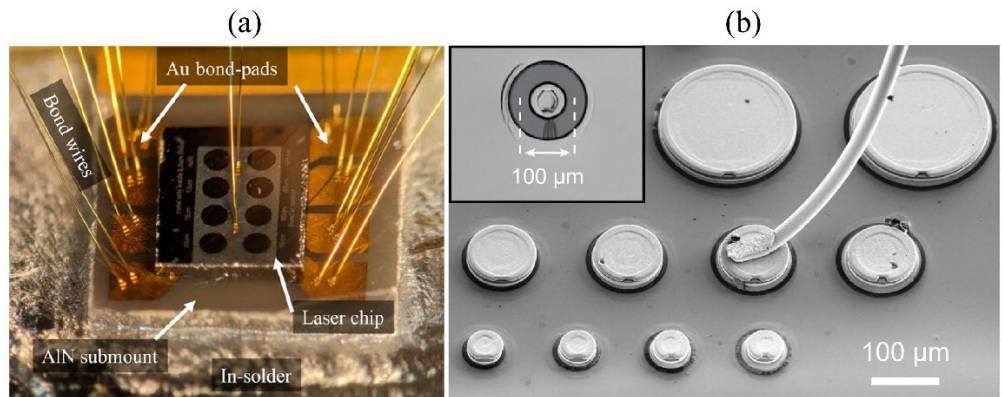

Figure 7: (a) Array of flip-chip bonded low-dissipation ring ICLs onto an AlN submount [25]. (b) Low-dissipation ring QCLs with different sizes featuring arcuate gratings (inset) [26].

For microring QCLs with arcuate grating sections, as illustrated in Fig. 3(b), a linear relationship between ring radius and threshold current density was observed, though rings consistently outperformed linear cavities in terms of threshold current density. The curved gratings of arcuate grating sections are deliberately kept relatively short—typically covering only 5–7 grating periods—and occupy just a small segment of the total ring circumference rather than encircling it completely, as is common in traditional ring QCLs. The lowest power dissipation was measured for a 50 μm radius ring with 267 mW at threshold and 374 mW at maximum output power. Lasing was still achievable for rings as small as 25 μm [26].

The inherently symmetric and collimated emission of ring QCLs and ICLs, which is superior to that of facet emitters, makes them excellent candidates for mid-infrared sensing. Their 2D integrability also enables multi-wavelength arrays, which have been realized in the form of 16 ring QCLs covering a gapless tuning range of 180 cm$^{-1}$ [27]. These arrays offer a more compact and robust alternative to external cavity QCLs, eliminating the need for moving parts. Moreover, ring QCLs have demonstrated superior spectral tunability compared to their linear counterparts [28]. These devices have been successfully implemented in various sensor platforms. For example, ring QCLs have been combined with substrate-integrated hollow waveguides (iHWGs) [29], significantly reducing the required gas volume and overall sensor size. Ring ICLs have enabled $CO_2$ isotope analysis with record-low power dissipation among surface-emitting mid-infrared ring lasers, marking a crucial step toward handheld sensing devices [25]. Additionally, heterodyne phase-sensitive dispersion spectroscopy (HPSDS) has been demonstrated using ring QCLs, showing their compatibility with advanced spectroscopic techniques [30]. An important development has been the integration of laser and detector on a single chip using bi-functional quantum cascade heterostructures [31], leading to monolithic gas sensing platforms that reduce alignment complexity, device footprint, and manufacturing cost. The ability to interchange the roles of the laser and detector within the same device further enhances its versatility, enabling flexible system configurations tailored to specific sensing needs [32].

### Challenges and opportunities

For many applications in sensing and spectroscopy, CW operation at room temperature is a critical requirement. The primary obstacle to achieving this goal is efficient heat dissipation. High-power CW ring QCLs operating at room temperature have been demonstrated using an overgrowth approach that integrates a buried second-order DFB grating directly above the active region. This architecture enabled high optical output power of 0.5 W at 4.6 μm. However, the resulting beam exhibited broad divergence, attributed to overcoupling of the grating and subsequent lasing in higher-order modes [16]. A promising strategy to address this issue has been the introduction of a surface grating placed at a greater distance from the active region, thereby reducing the grating coupling strength. To maintain thermal conductivity, the surface grating was electroplated. This design yielded a first-order beam profile with narrow divergence, although at the cost of a reduced output power of 0.2 W. The performance parameters also degraded, including threshold current density, slope efficiency, and wall-plug efficiency [17]. A key challenge remains: achieving high-power emission with a collimated first-order beam while preserving overall device efficiency.

In parallel, devices with low power dissipation are essential for handheld sensing platforms that depend on minimal power consumption. Achieving such low power usage is closely linked to minimizing the physical size of the laser, as smaller structures generally require less current and generate less heat. In this regard, ring ICLs with standard second-order DFB gratings have shown robust performance down to a radius of 80 μm, without degradation in output characteristics. Lasing has also been demonstrated in rings as small as 50 μm, although further size reduction is currently limited by higher losses. The study showed that these increased losses cannot be fully explained by enhanced absorption in the SiN isolation layer due to stronger mode bending and increased modal overlap in smaller rings. Therefore, these losses are presumed to originate not only from absorption in the dielectric isolation layers but also from interface roughness and residuals from fabrication processes on the sidewalls [25]. The challenge is to identify fabrication and design strategies that mitigate such losses, enabling even smaller rings with lower power dissipation.

For microring QCLs incorporating arcuate grating sections, a key limitation lies in the beam divergence. The combination of a short arcuate grating and a small emission aperture leads to a highly divergent output beam [26]. This is particularly problematic in applications where beam quality is essential. Increasing the aperture and the grating length could address this issue, but must be achieved without compromising the low power dissipation already demonstrated. Moreover, further size reduction and improved heat dissipation of these ring lasers remain desirable yet challenging goals.

A broader challenge facing both ring QCLs and ring ICLs, as well as mid-infrared sources in general, is the issue of beam alignment in multi-wavelength laser arrays. Each laser emits a different wavelength along a distinct optical axis, causing spatial separation of the output beams. For applications like multi-wavelength spectroscopy, this misalignment requires additional beam-combining optics to spatially overlap the emission from different lasers [33]. These external components add complexity, increase the system size, and reduce robustness. As a result, this requirement partially offsets the key advantages of 2D-integrated ring laser arrays, namely their compactness and mechanical stability, compared to EC-QCLs, which also rely on bulk optical components for wavelength tuning.

Despite these challenges, ring QCLs and ring ICLs continue to offer substantial opportunities in sensing and spectroscopy. Their compact, surface-emitting geometry offers clear advantages in scenarios requiring integration, beam shaping, and low power consumption. Beyond existing implementations, further opportunities arise from the unique beam engineering capabilities of ring lasers. One such example is the application of a ring QCL in combination with a metalens generating an OAM beam for detection of chiral molecules, leveraging the beam-shaping capabilities unique to ring geometries.

### *Future developments to address challenges*

*To overcome the current limitations in ring QCLs and ring ICLs, several targeted strategies can be pursued. These developments aim to improve thermal management, reduce losses, optimize beam quality, and enable compact multi-wavelength sources. In the following, we outline key directions for future research and development.*

*One challenge in ring QCLs lies in achieving a collimated, first-order beam while avoiding degradation in performance metrics such as efficiency and output power. This degradation has so far been a consequence of heat dissipation constraints and grating overcoupling. To address this, the overgrowth approach can be further optimized by increasing the separation between the active region and the DFB grating, thereby reducing the grating strength and minimizing higher-order mode operation. Alternatively, surface grating designs remain promising, especially when combined with electroplated metals to maintain efficient thermal conduction. These surface gratings could be tailored to provide the required feedback while ensuring minimal optical loss and sufficient thermal pathways. The continued refinement of grating placement and coupling strength could reconcile the trade-off between beam quality and device performance.*

*The challenge for low-dissipation ring ICLs is to reduce the optical and electrical losses that limit the miniaturization and efficiency, especially for portable sensing platforms. One promising path forward is optimization of the dry-etching process during fabrication. A refined etching recipe that ensures smooth, clean, and residue-free outer ring sidewalls could significantly reduce optical scattering and absorption. Additionally, eliminating the outer sidewall isolation and the associated electrical contact could help reduce absorption losses in the dielectric materials. However, this modification must be carefully evaluated, as it may negatively affect thermal management. A detailed trade-off analysis between improved optical performance and reduced heat dissipation is necessary.*

*The major limitation in arcuate-grating-based microring QCLs is the poor beam quality due to small apertures and short grating lengths. To overcome this, future device designs should incorporate wider apertures and extended arcuate grating sections. This will likely require new lateral contact schemes, enabling current injection without obstructing the emission aperture. Additionally, modifying the grating geometry itself could increase optical output power and improve beam shape. Enhanced heat dissipation can be achieved through flip-chip bonding onto AlN submounts and electroplating of the device sidewalls. Beyond improving beam quality, these developments would enable new architectures such as 2D arrays of arcuate ring QCLs, facilitating compact multi-wavelength platforms. Extending the concept of microring resonators with arcuate gratings to ICLs could establish a new class of ultra-*

*efficient, low-dissipation mid-infrared sources with power consumption as low as a few tens of milliwatts suitable for handheld devices and mobile sensing platforms.*

*A key challenge for multi-wavelength spectroscopy is the lack of alignment among the optical axes of individual lasers in an array, which necessitates bulky beam-combining optics and undermines system compactness. Future designs could resolve this by exploring novel device concepts that ensure co-axial beam emission. One promising idea is the implementation of concentric ring laser arrays, resembling a Matryoshka configuration as presented in Fig. 4(a). These rings share the same optical axis but naturally differ in beam divergence due to their different sizes. Another approach is to integrate a collector waveguide that couples to multiple ring lasers as sketched in Fig. 1(b) and outlined in Fig. 4(b) [34]. This waveguide unifies the output into a single beam with shared divergence and alignment. Although such beams may be broader due to waveguide geometry and facet emission, the collector waveguide can also serve as an integrated amplifier, boosting the combined output power and enabling efficient coupling into downstream optics.*

(a)                                                    (b)

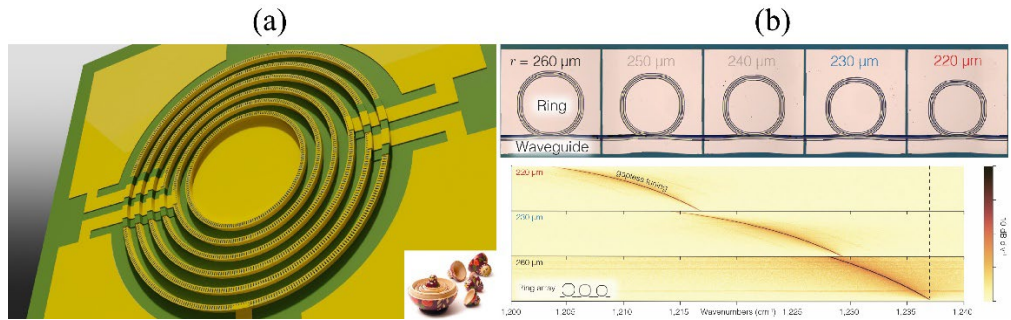

Figure 8: (a) Sketch of a concentric ring QCL array with 6 ring lasers sharing one common optical axis. (b) Microscope image (top) and spectral tuning behavior (bottom) of an array of single-mode ring lasers coupled to a collector waveguide. Light from all the rings is emitted via the facet of the collector waveguide [34].

*The realization of ring lasers with a common optical axis also opens the door to mid-infrared phased arrays, which represent a transformative opportunity for beam steering and multiplexed sensing. Optical phased arrays are already well-established in the near-infrared for applications such as Light Detection and Ranging (LIDAR) and free-space optical communication. In the mid-infrared, they could unlock new possibilities in chemical imaging, remote sensing, and environmental monitoring. While phased arrays have typically relied on linear cavity lasers [35], demonstrations of coherent coupling between ring QCLs suggest that a 2D phased array architecture based on ring geometries is feasible [36]. Such an array would combine the inherent advantages of ring lasers—compact footprint, on-chip beam shaping, and robust integration—with the directional control offered by phased array principles. This concept holds great potential to surpass current QCL-based phased arrays in performance, scalability, and functionality.*

### Concluding Remarks

Ring quantum and interband cascade lasers represent promising platforms for compact, monolithic mid-infrared light sources. Their inherent ability to produce surface-emitting, single-mode beams from small, planar devices makes them highly attractive for integrated sensing and spectroscopy. While significant progress has been made in demonstrating high-power operation, low-power dissipation, and advanced emission concepts such as orbital angular momentum, several key challenges remain. These include optimizing the heat management, minimizing optical losses at small scales, improving beam quality, and enabling scalable and efficient multi-wavelength systems emitting along a common axis.

Looking ahead, innovations in device architecture, fabrication processes, and system-level integration are expected to further enhance the performance and functionality of ring QCLs and

ICLs. Novel grating designs, improved thermal handling, and array-based concepts such as concentric lasers or phased arrays could transform these sources into versatile components for next-generation infrared applications. Continued research in these directions will lead to full exploitation of the advantages of the ring geometry, and push the limits of miniaturized mid-infrared photonics.

# Nicolas Schäfer[1], Johannes Koeth[1], and Robert Weih[1,*]

**[1] Nanoplus Advanced Photonics Gerbrunn GmbH, Oberer Kirschberg 4, 97218 Gerbrunn, Germany**
*\*robert.weih@nanoplus.com*

### Overview

Mid-wave and long-wave infrared (MWIR/LWIR) light-emitting diodes are emerging as compact, low-power, and spectrally tailorable alternatives to conventional thermal emitters, which suffer from high power consumption, poor directionality, and limited modulation speeds. Advances in heterostructure engineering—most notably interband cascade LEDs (ICLEDs)—have enabled significant gains in internal quantum efficiency, optical extraction, and spectral control. Strategies such as cascade stage optimization, substrate engineering, and surface structuring have improved room-temperature output power into the multi-milliwatt range and extended emission coverage from 1.8 to over 10 µm. Resonant cavity designs and multi-wavelength architectures now allow both narrowband precision and broadband coverage for applications in gas sensing, spectroscopy, and scene projection. Despite challenges such as light-extraction bottlenecks, Auger recombination, and reduced wave-function overlap in LWIR devices, ongoing innovations in material growth, photonic structuring, and device integration promise further efficiency gains and expanded application reach in infrared photonics.

### Current status

Thermal emitters have been the default non-coherent light sources of choice within the MWIR and LWIR, commonly used in FTIR spectroscopy. However, despite their broad spectral coverage and low cost, thermal emitters suffer from several serious limitations. These include high power consumption, low efficiency, poor directionality, slow modulation capability, limited flexibility for spectral shaping, high operating temperatures, and an intrinsic coupling between center wavelength and total output power.

Hence, it is desirable to overcome these limitations and gain more flexibility for other applications including gas sensing or infrared scene projection. LEDs offer high modulation bandwidth, room temperature operation, small footprint, low power consumption, and the capability of spectral engineering to suit the applications requirements.

In contrast to their counterparts in the visible portion of the spectrum, MWIR and LWIR LEDs have not yet gathered the same attention despite their great potential.

The first MWIR LEDs, based on bulk InAs, were demonstrated in 1966 [1,2], achieving room temperature emission at 3.7µm. Later, in 1993, bulk InAsSb LEDs were introduced, emitting around 4.2µm – within the prominent $CO_2$ absorption band.

In the following years, heterostructure-based approaches such as single quantum wells [4], multiple quantum wells [5,6] and superlattice structures [7,9,10,11] led to ongoing improvements in device performance. In particular, strained type-II quantum wells proved effective in suppressing Auger recombination. Additionally, the implementation of carrier blocking barriers made of AlSb [12], AlAsSb [13] or InAlAs [14] led to a multifold increase in quantum efficiency. Liquid-Phase-Epitaxy-fabricated LEDs based on InAs/InAsSbP emitted more than 3mW of output power in pulsed mode at 300K at a wavelength of 3.3 µm [15]. Nevertheless, the maximum continuous wave (cw) output power was still limited to roughly 150 µW.

**Performance Optimization**

To improve performance characteristics of MWIR and LWIR LEDs such as total output power and wall plug efficiency, two fundamental parameters must be optimized: internal quantum efficiency and optical extraction efficiency.

The first parameter, internal quantum efficiency, is primarily addressed through the layer design of the epitaxial stack. As early as 1996, the interband cascade process was employed to generate MWIR electroluminescence (EL) [16], leading to ICLEDs in the 5–8 μm spectrum region in early 1997 [17]. ICLEDs employ active stage designs similar to those used in interband cascade lasers, benefiting from the same carrier recycling and tailored band structure concepts. The cascade process within ICLEDs was verified [18] by showing the proportional relation between the number of active stages (each incorporating at least one InAs/GaInSb/InAs QW structure) and the optical power. Utilizing the cascading scheme by connecting several InAs/GaSb SLs in series Koerperick et al. realized a 520 x 520 μm mesa that emitted a room temperature optical output power of more than 800 μW at a wavelength around 4 μm [19]. As a result of further improvement of the active stage design and material quality, Abell et al. achieved room temperature cw output power as high as 1.6 mW from a 15 stage ICLED in 2014 [20], showing the potential of the interband cascade concept.

Optical extraction is the second key parameter that needs to be optimized to realize high-efficiency LEDs. In substrate-emitting ICLEDs, a major challenge for photon extraction is absorption in the highly-Te-doped GaSb substrate. Free carrier absorption (FCA) by electrons is a dominant sub-bandgap absorption mechanism that significantly reduces the number of photons reaching the substrate-to-air interface.

In the IR regime, several variations of the chip design have been shown to improve the proportion of light outcoupled from the LEDs. In the case of substrate emitting devices, Das et al. attributed a 6-fold increase of output power to substrate thinning and another 50% to texturing of the substrate [21]. However, the pronounced loss due to below-bandgap absorption within GaSb remains even in a thinned substrate. Simply omitting the intentional doping is not a viable option, as GaSb remains intrinsically p-type due to gallium vacancies and antisites. This results in a residual carrier concentration between $1\times10^{16}$ cm$^{-3}$ and $1\times10^{17}$ cm$^{-3}$, depending on the growth technique and conditions.

One approach to mitigating substrate-related absorption is to grow the epitaxial layers on a mismatched GaAs substrate, as demonstrated by Provence et al. in 2015 [22]. In their study, cascaded InAs/GaSb LEDs were grown on both GaSb and GaAs substrates for comparison. The GaAs-based structure required a 3000 nm GaSb buffer layer to accommodate the lattice mismatch. At 77 K and 50% duty cycle, the peak radiance was measured at 0.69 W/cm²·sr for the GaSb-based device and 1.06 W/cm²·sr for the GaAs-based counterpart. The improved performance on GaAs was attributed to the higher transparency and better thermal conductivity of the GaAs substrate. However, high transparency can also be achieved with GaSb substrates. This typically requires the suppression of free-carrier absorption, primarily by reducing native acceptor concentrations. It can be accomplished through lithium diffusion or tellurium compensation doping, the latter first demonstrated by Milvidskaya et al. [23]. While tellurium doping effectively mitigates substrate absorption, it renders the substrate semi-insulating, making it necessary to grow a highly doped GaSb contact layer to maintain reliable ohmic contact formation. Substrate-emitting ICLEDs grown on doping-compensated substrates achieved a 70% higher output power compared to those on Te-doped substrates (n > 5 × $10^{17}$ cm$^{-3}$) when thinned to 150 µm. A 500-nm-thick Te-doped interstitial layer was implemented for the substrate-side contact formation [24].

Epitaxial growth on silicon is another approach to reduce undesired substrate absorption. Besides inherent CMOS compatibility, it also allows for selective growth on pre-processed wafers, supporting integration with downstream silicon-based photonic circuits and system-on-chip (SoC) manufacturing. In 2021, Canedy et al. demonstrated ICLEDs grown on 4-degree-offcut silicon [25]. Despite a 12% lattice mismatch, the devices showed good performance, benefiting from improved heat dissipation in an epi-up configuration. A GaSb buffer layer was essential to accommodate the 12% lattice mismatch between silicon and the GaSb-based ICLED structure, enabling high-quality growth; however, despite reducing dislocation density, the buffer could not fully eliminate defects, leading to slightly rougher surfaces, higher leakage currents, and moderate efficiency losses compared to devices grown on native GaSb.

Three years later, Ince et al demonstrated ICLEDs grown on GaSb-on-Si wafers with improved threading dislocation densities (~$10^7$ cm⁻²), achieving in some cases device performances comparable to those for growth on native GaSb [26]. Compared to earlier work, they optimized the AlSb nucleation process, identified strain effects during high-temperature ICLED growth as the cause of dislocation multiplication, and proposed defect filter layers for further improvement.

In LEDs, the photons emitted during spontaneous recombination are nearly isotropic. This results from the high crystal symmetry of the semiconductor and the distribution of electron and hole states, which average out the directional dependencies. Consequently, light is emitted uniformly in all directions.

One significant bottleneck in light extraction for LEDs is the trapping of MWIR photons by total internal reflection (TIR), which limits the external quantum efficiency (EQE). The high refractive index of GaSb (n $\approx$ 3.7) substantially contrasts that of air, enhancing the likelihood of TIR. When dealing with a smooth substrate surface, surface effects such as TIR, must be considered. For the GaSb-Air interface, the critical angle $\theta_c$ for TIR calculated using geometric optics is 16°. This means that only light within a very narrow cone of angles can escape the semiconductor. As a result, 91% of the emitted light is trapped and reflected back into the material, severely hindering light extraction. There are two conceptually different approaches to enhancing the light extraction efficiency. One focuses on modifying the conditions for TIR to alter the escape cone. This can be achieved through various techniques, including antireflection coatings, lens-shaped materials, surface roughening and plasmonic metastructures. Another approach involves narrowing the internal emission profile and effectively steering the photons trajectories to increase the number of photons within the escape cone. Strategies for this approach include incorporating a structured or unstructured backside mirror, placing the active region in a cavity with mode enhancement, sidewall reflectors, stage positioning in the vicinity of a backside mirror, and achieving extremely high internal quantum efficiency (IQE) to facilitate photon reabsorption and re-emission. A concept called stage positioning was first introduced by Kim et Al [27]. An ICLED's 22 active stages were divided into four groups that were strategically placed at antinodes of the optical field. This arrangement ensured that the emitted light interferes constructively when reflected at near-normal incidence from the metal contact of the epitaxial-side-down mounted device, leading to a wall plug efficiency of 0.4% at low powers and cw room temperature optical output power of 2.9 mW.

Combining the concept of stage positioning and epitaxial growth on doping-compensated substrates enabled record performance of ICLEDs in the 3.4 μm emission wavelength regime. Figure 1 presents the temperature-dependent L-I-V characteristics of an ICLED with square-shaped mesa structure (d = 640 μm). These devices were fabricated on semi-transparent GaSb substrates and thinned to a final thickness of 150 μm. Individual emitters were diced and flip-chip mounted onto AlN heat spreaders using AuSn solder and then affixed to a single-sided aluminum printed circuit board (PCB). The PCB was subsequently mounted on a thermoelectrically-cooled measurement station for characterization. A room temperature cw output power of more than 10 mW could be reached while at low driving current, and the wall plug efficiency was improved to ≈1.8%.

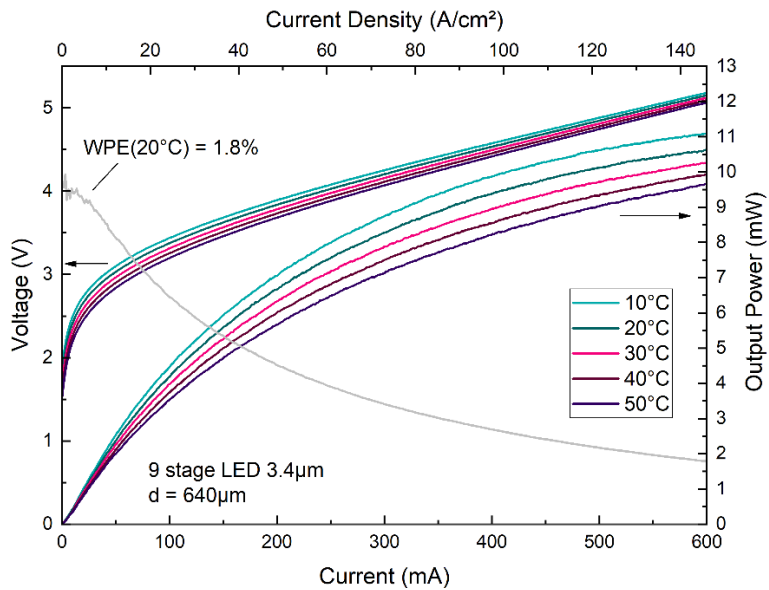

Figure 9: cw L-I-V characteristics for a flip-chip mounted ICLED with 9 active stages and 640x640 μm² square shaped active area

Figure 2 shows that the entire spectral region between 1.8 and 6.5 μm can be covered utilizing either conventional InGaAsSb/AlGaAsSb QWs or W-QWs in a cascaded configuration. These technologies currently represent state-of-the-art performance in terms of optical output power in the MWIR spectral region.

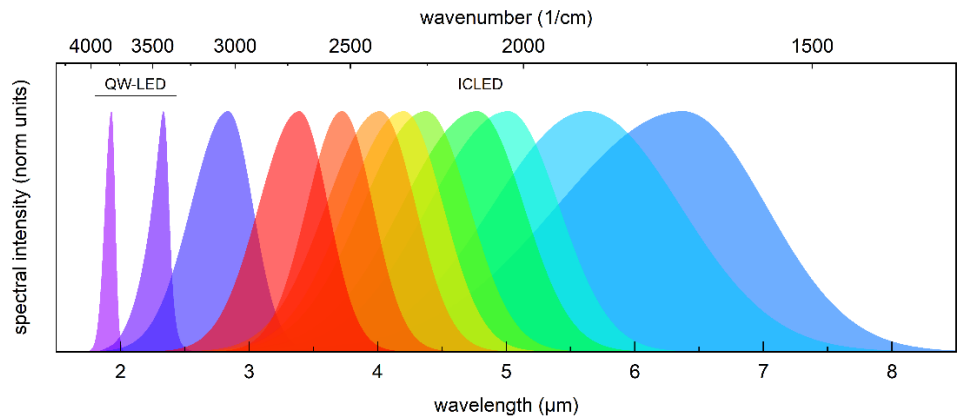

Figure 10: Spectral coverage of LEDs emitting in the MWIR range. GaSb LEDs utilizing InGaAsSb/AlGaAsSb quantum wells are suitable for wavelengths between 1.8–2.6 μm. For wavelengths beyond 2.6 μm, LEDs based on interband cascade structures represent the highest-performance GaSb-based technology, emitting at room temperature up to 6.5 μm.

**Spectral Shaping:**

**RCLEDs**

One key challenge for applications requiring high optical intensity or spectral precision is the inherently low spectral power density of ICLEDs, especially when compared to lasers. Although stage positioning allows partial tuning of the spectral linewidth, it offers only limited control over the optical mode. Greater mode control can be achieved by leveraging microcavity effects, such as integrating the active region within a Fabry–Perot cavity. In 2020, Al-Saymari et al. reported an RCLED emitting at 4.5μm. The structure design used a 1-λ-thick microcavity formed from high-index-contrast AlAsSb/GaSb distributed Bragg reflector (DBR) mirrors [28]. With AlInAs/InAsSb active MQWs between a 5-pair top DBR and 13.5-pair bottom DBR, the RCLED exhibited 85× higher peak intensity, 13× higher integrated output power and 16× narrower full-width at half maximum spectral linewidth (FWHM=70nm) compared to a reference LED without the DBRs. In the same year, Diaz-Thomas et al. demonstrated an IC-RCLED with resonance peak emission at 3.3 μm [29]. The 7-stage active region was positioned in a 3-λ-thick optical cavity sandwiched between a 14-pair GaSb/AlAsSb bottom DBR and a 2-pair ZnS/Ge dielectric top DBR. They also implemented a technique that is common in AlGaAs/GaAs VCSELs: A 20-nm-thick $Al_{0.98}Ga_{0.02}As$ layer was deposited near the interface to enable electrical confinement through oxidation, to direct current laterally for improved device efficiency. The oxidation also reduces surface recombination at the mesa edges, which is typically high in GaSb-based structures. Both approaches rely on a DBR back mirror to achieve the high reflectivity (>95%) required for efficient emission. However, the relatively small refractive index contrast between GaSb (n ≈ 3.9) and AlAsSb (n ≈ 3.2) at 3 μm necessitates a relatively large number of DBR pairs, making the epitaxial growth thick and time-consuming. Moreover, the added electrical resistance from the mirror can contribute to self-heating, which in turn may enhance non-radiative recombination. Another limitation of typical device architectures is the use of annular anode contacts for current injection, which tends to introduce radial inhomogeneity in the current distribution. Although the high lateral conductivity of the quantum wells in ICLEDs promotes uniform current flow at later stages, non-uniformities may still arise in the initial stages of injection.

In 2024, Schäfer et al. [30] introduced a novel design approach for Resonant Cavity ICLEDs that utilizes only a single DBR. In this configuration, which has its origins in VCSEL development and was later successfully adapted to RCLEDs [33], the anode DBR is replaced by a metal mirror to realize a substrate-emitting device. This design offers several advantages, including a simplified layer structure and reduced epitaxial growth time. Uniform electrical pumping is achieved via the metal mirror, which simultaneously serves as the p-side contact. The structure also provides high reflectivity (R > 0.94) across the entire spectral range of interest, and features low voltage drop due to the incorporation of an intracavity contact layer, which eliminates the need to electrically pump the cathode-side DBR. An 89% reduction in the emission spectrum width, as indicated by the FWHM of 70 nm, was achieved. The single-DBR configuration enables tuning of the cavity length by applying a refractive index-matched material on top of the epitaxial structure after growth, which effectively shifts the cavity's spectral response as illustrated in Figure 3. This approach was used to fabricate emitters with two cavities of different lengths, producing two distinct and independently controllable spectral emission bands. The dual-wavelength functionality allows one emission band to act as a reference channel, making these LEDs highly suitable for cost-effective gas-sensing applications.

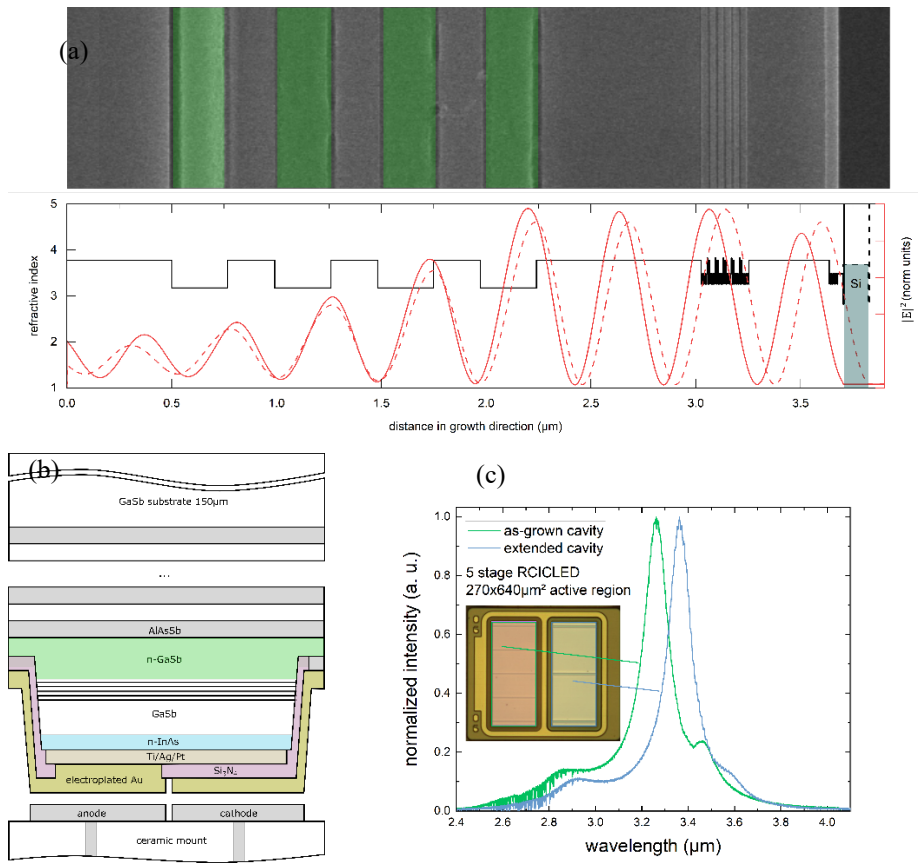

Figure 11: Epitaxial layer structure and simulated optical field intensity of a Resonant Cavity ICLED with 4 DBR pairs (a) and schematic after LED processing (b). Dual-wavelength emission after cavity length tuning with refractive-index-matched material on top of the epitaxial structure after growth (c).

**Broadband ICLEDs**

In certain applications, it can be advantageous to narrow the spectral emission to achieve high intensity at specific target wavelength. However, other use cases benefit from covering a broad spectral range. This can be realized by combining active quantum well structures with different emission wavelengths within the same epitaxial stack. Figure 4 shows an example of such a multilayer structure with three distinct emission peaks. By placing these active regions at the antinodes of the optical field and carefully selecting the number of cascade stages, a broad emission spectrum with FWHM of 2650 nm was achieved.

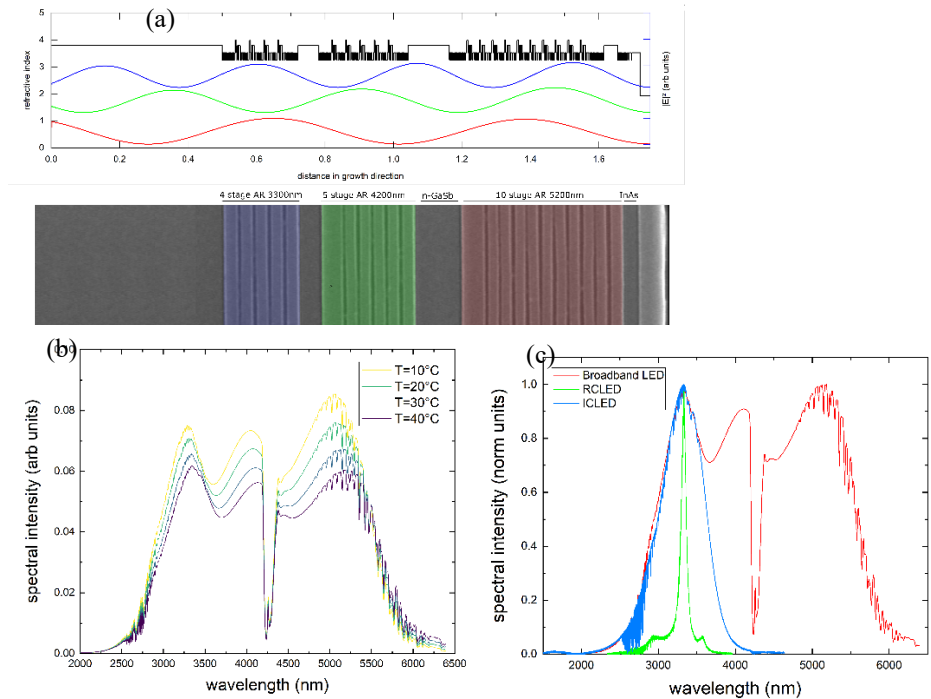

Figure 12: (a) Simulation and SEM cross section image of a quantum well structure designed for broadband emission. (b) Several active regions, each engineered to emit light at a different wavelength, resulting in three emission peaks. (c) Spectral comparison of a reference ICLED (blue), an RCLED (green) and a broadband ICLED (red).

**LWIR LEDs**

In 1997, Zhang et al. reported EL in the 10–15 µm range from 10-stage ICLEDs and the measured optical output power reached 50 nW without accounting for divergence loss [31], which is much higher than the maximum power reported for intersubband QC LEDs in the 8-13 µm wavelength region [32]. In 2008, Das et al. reported EL in the 6–12 µm range from antimony-based type II interband cascade structures comprising 30 stages [34]. The emitted optical power of 100-µm-diameter top-emitting devices varied from a few µW at room temperature to as much as 22 µW at 77 K. This is higher than the results published in 1994 for QC LEDs with emission wavelengths near 5 µm [35]. In 2010, the group extended their concept to develop a two-color infrared LED that emitted in both the 3–4 µm and 5–10 µm spectral regions [36]. The device was grown on an n-type GaSb substrate and incorporated 30 cascaded stages for LWIR emission and 15 cascaded stages for MWIR emission. In 2011, Koerperick et al [11] reported cascaded InAs/GaSb superlattice LEDs for LWIR applications, achieving a peak emission wavelength of 8.6 µm and output power up to 600 µW at 77 K. The emission wavelength was primarily tuned by adjusting the thickness of the individual InAs and GaSb layers within the superlattice. This tunability enabled the integration of multiple superlattices to design dual-color LEDs [10] or even broadband LEDs [37] by combining emissions from different wavelength ranges.

Recent efforts have focused on extending ICLED performance into the LWIR range. Figure 5 shows an ICLED with a peak emission at 7.8 µm and output power of 1.8 mW at 77 K from a 640 × 640 µm² substrate-emitting mesa that was flip-chip mounted on a ceramic carrier. The device's active region is composed of a single stack of 18 cascaded stages that is positioned at an antinode of the optical field intensity within the optical cavity to maximize light extraction. This cascaded design enables efficient carrier recycling and multiple photon emissions per injected carrier, making ICLEDs fundamentally more efficient than superlattice-based LEDs in the LWIR range. Surface roughening was also applied to enhance optical outcoupling. For room temperature operation, we observed a redshift in the emission spectrum toward 10 µm with an output power of 370 µW. Furthermore, pixel-scale devices with dimensions of 18 × 18 µm² yielded output powers of 8 µW at 77 K, supporting their suitability for high-resolution focal plane array integration.

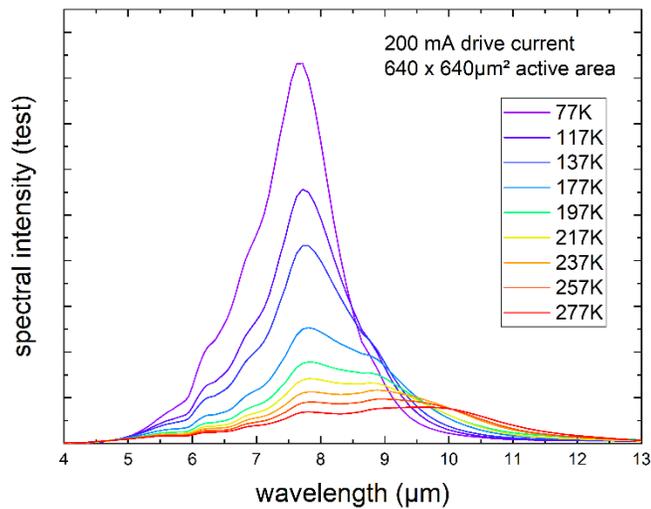

Figure 13: Emission spectra of an LWIR ICLED at different temperatures at a driving current of 200mA.

### *Challenges and opportunities*

One of the central challenges in the development of LWIR ICLEDs lies in the poor spatial overlap between the electron and hole wave functions within the quantum wells. This limited overlap reduces radiative recombination efficiency, making interband transitions less effective at longer wavelengths. As the target emission wavelength increases, the transition energies decrease, requiring precise engineering of quantum well and barrier widths to define states with the desired energy separation. However, at lower transition energies, the electron and hole states tend to become more weakly confined and more spatially separated, especially in type-II band alignments. This leads to more delocalized wave functions and reduced spatial overlap, ultimately limiting the radiative efficiency. Long-wavelength photons also suffer from increased reabsorption within the device layers, requiring careful substrate and waveguide design to minimize losses.

Another significant challenge in ICLED development still lies in overcoming light extraction bottlenecks. While early efforts involving surface texturing and photonic structures have shown promise in enhancing light outcoupling, overall light emission remains suboptimal. Furthermore, aspects such as emission directivity and beam shaping have yet to be adequately addressed. These optical inefficiencies present a critical obstacle for practical applications. To mitigate these limitations, integration of optical elements such as microlens arrays or other micro-optical structures offers a promising route.

For broadband ICLEDs, engineering multiple cascade stages with staggered transition energies enables spectrally wide emission. However, this approach introduces difficulties such as carrier leakage between non-resonant states and imbalance in carrier distribution across stages. Additionally, maintaining high radiative efficiency over a broad spectrum is challenging due to the trade-off between energy selectivity and wave function overlap.

### *Future developments to address challenges*

To address limitations from the inherently weak spatial overlap between electron and hole wave functions in type-II quantum wells in LWIR, alternative architectures such as cascaded superlattice LEDs—particularly those based on InAs/GaAsSb or InAs/GaSb—are being explored, as they allow better control over confinement and wave function overlap.

Auger recombination is a key non-radiative loss mechanism in ICLEDs and becomes increasingly significant at high carrier densities due to its superlinear dependence on carrier concentration. One strategy to mitigate this effect is to distribute the injected carriers across multiple quantum wells within each cascade stage. For instance, employing a triple W-QW design reduces the carrier density per well, thereby suppressing Auger processes while maintaining overall radiative output.

Surface-engineering strategies are relevant for the integration of MWIR LEDs into miniaturized optoelectronic platforms, where directional emission, beam shaping, and efficient outcoupling are essential for system-level performance. As such, continued development of structured surface designs represents a critical pathway toward practical and scalable light sources. Approaches such as microlens arrays, photonic crystals, sub-wavelength gratings, and metasurfaces can be engineered to break the symmetry of light propagation and redirect photons into the escape cone, thus mitigating TIR. These techniques not only improve light outcoupling efficiency but also allow tailoring of the far-field emission pattern. In 2021, Montealegre et al. [38] demonstrated a significant enhancement in radiance—from 7.6 W/cm²·sr to 13.4 W/cm²·sr by thinning and roughening the GaSb substrate of 18×17 μm² ICLEDs, measured at 80 K. This underscores the potential of relatively simple surface treatments in improving optical performance. Analogous strategies are well-established in visible-light LEDs, where patterned substrates (e.g., textured sapphire with backside reflectors) are commonly employed to redirect internally trapped light. Adapting similar design concepts to the MWIR and LWIR spectral ranges could likewise yield substantial improvements.

Furthermore, hybrid approaches combining photonic crystals with surface plasmon effects have shown promising results in enhancing efficiency in near-infrared (NIR) LEDs [39] and are now being investigated for longer wavelengths. More recently, metasurfaces—ultrathin nanostructured layers capable of controlling phase, polarization, and wavefront—have emerged as powerful tools not only for increasing extraction efficiency but also for enabling beam shaping without sacrificing the active area. Periodic or disordered structures can also induce diffuse scattering, helping to homogenize the emission profile.

## *Concluding Remarks*

Continued advances in optical cavity design, multi-wavelength active region engineering, and device-level optimization are driving ICLEDs toward mainstream adoption in portable sensing, environmental monitoring, and medical diagnostics. The synergy of materials innovation, nanofabrication, and device modeling promises higher efficiencies, broader spectral coverage, and new applications in the MWIR photonics landscape. Meeting these challenges will unlock the full potential of ICLEDs as versatile, efficient, and compact MWIR sources for the next generation of photonic technologies.

### Disclosures

The authors declare no conflicts of interest.

# 8. Infrared focal plane arrays

## ANTONI ROGALSKI

*Institute of Applied Physics, Military University of Technology, 2 Kaliskiego Str., 01-934
Warsaw, Poland; e-mail: antoni.rogalski@wat.edu.pl*

### Overview

The paper presents the trends in the development of infrared detector arrays in the last decade. In the beginning, the important mechanisms of detector operation are briefly discussed. More attention is paid to the advantages and disadvantages of materials used in active areas of infrared detectors; especially HgCdTe alloys, type-II superlattices, quantum wells, and lead salts. The second part of the article deals with the performance of hybrid infrared detector arrays with particular emphasis on their development (pixel scaling) and the associated challenges regarding hybridization techniques, system optics, and signal readout circuitry.

### Introduction

Focal plane array (FPA) technology has revolutionized many types of imaging, from $\gamma$ radiation to terahertz and even radio waves. While in the early 1980s, the technology of infrared (IR) detector arrays with dimensions of 64×64 pixels was developed; currently, arrays containing about $10^8$ pixels are already being produced. For example, the Euclid Near Infrared Spectrometer and Photometer (NISP) array, consisting of 16 H2RG devices (each with a 2k×2k 18-μm pixel size) and operating in space (with 67 million pixels) with a long-wave sensitivity limit of 2.3 μm [1] – was launched on July 1, 2023. The largest array currently under development, consisting of a 10-μm pitch 300-megapixel HgCdTe FPA, is shown in the lower left corner of Fig. 1. This array is the heart of the multiband visible and near-infrared (NIR) camera incorporated into the Roman's Wide Field Instrument and will be launched into space in the mid-2020s [2].

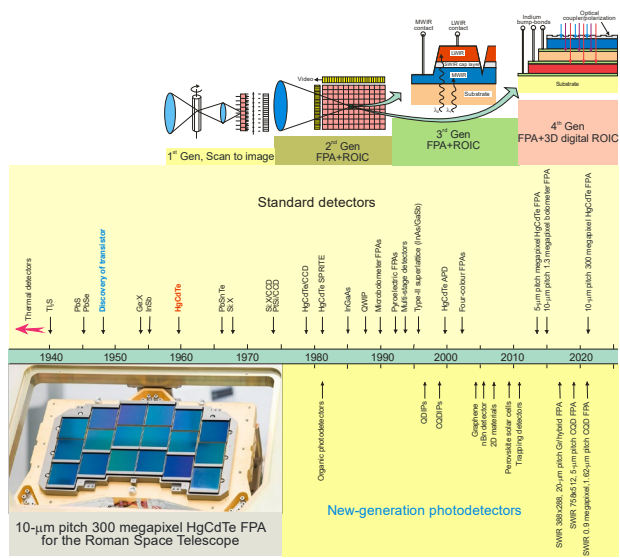

Fig. 1. Development of IR detectors and history of imaging systems: first generation (scanning systems), second generation (staring systems), third generation (with very high pixel counts), and fourth generation (with very high pixel counts, multicolor detection, 3D ROIC, and other on-chip features). The approximate dates of the appearance of various IR detector materials are indicated at the bottom of the figure, covering both the standard detectors that dominate the commercial market and those incorporated into new detector generations over the past thirty years. The lower left corner shows the largest IR detector array consisting of a 10-μm pitch 300-megapixel HgCdTe FPA.

IR arrays are nominally at the same rate of development as dynamic random-access memories (DRAMs), but are lagging 10 years behind the latter. The size of FPAs will continue to grow but at a slower rate. However, the market trends that previously forced the development of larger arrays are no longer as strong now that the megapixel sensor barrier has been crossed.

## Material systems used in infrared detector technology

Most IR detectors can be classified into two broad categories: thermal and photon detectors [3-7]. Among thermal detectors, monolithic microbolometer arrays are the most popular. Today, microbolometer arrays are produced in larger volumes than all other IR array technologies together. More information on their performance can be found in Ref. 8.

The most important criteria for selecting a detector are its spectral range, sensitivity, and operating temperature (Fig. 2). New emergency low-dimensional solid (LDS) photodetectors have been omitted from our considerations. These are discussed in a newly published monograph [9].

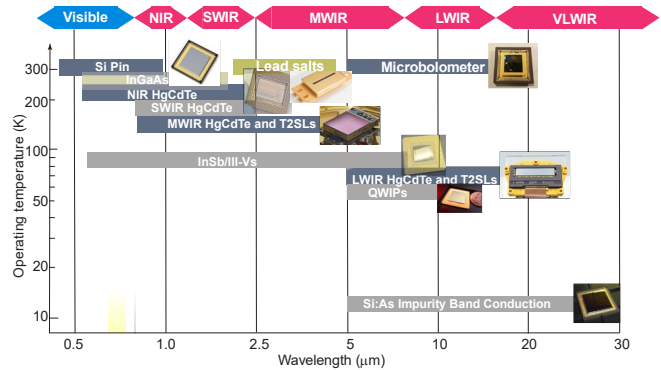

Fig. 2. The most important semiconductor materials used to design commercial infrared detectors.

Piotrowski and Rogalski showed that the IR photodetectors performance can be estimated by the formula $D^* = k(\lambda/hc)(\alpha/G_{th})^{1/2}$ [10], where: $\lambda$ – wavelength, $h$ – Planck's constant, $c$ – light speed, $\alpha$ – absorption coefficient, $G_{th}$ (in cm$^{-3}$s$^{-1}$) – thermal generation rate in the active detector's region, and $k$ – coefficient dependent on radiation coupling to detector including e.g. antireflection coating, microcavities or plasmonic structures. A transformation of this equation shows that the detectivity depends on the square root of the absorption coefficient and the carrier lifetime, $D^* \propto \sqrt{\alpha\tau}$.

Among narrow-gap semiconductors, the ternary HgCdTe alloys and lead salts have the highest absorption coefficient (above 10$^3$ cm$^{-1}$ at the absorption threshold) [6]. Similar values $\alpha$ are characterized by InAs/GaSb T2SLs. The InAs/InAsSb superlattices absorption coefficient is approximately half as large as that for InAs/GaSb T2SLs [11]. In quantum well-infrared photodetectors (QWIPs), the $\alpha$ is considerably modified by low-dimensional phenomena, and its typical value in the LWIR region is between 600 and 800 cm$^{-1}$.

Generally, the influence of ionic bonds in II-VI materials is stronger than in III-V ones (with strong covalent bonds), making the electron wave functions around the lattice atoms more confined. II-VI compounds are more resistant to the formation of lattice defects caused by deviations from crystal perfection. Consequently, SRH lifetimes in the low-doped HgCdTe material (about 10$^{13}$ cm$^{-3}$) are more than three orders of magnitude higher than those observed for III-V alloys with similar band gaps. Ref. 12 showed that the upper values of SRH lifetimes in HgCdTe are 10 ms for MWIR detectors and 0.5 ms for LWIR detectors. Interband Auger recombination mechanisms in HgCdTe affect long carrier lifetimes, enabling high operating temperature (HOT) photodetectors [12]. The SRH mechanism determines carrier lifetimes in InAs/GaSb T2SLs and is below 100 ns [13]. The absence of gallium in the MWIR InAs/InAsSb superlattices affects much longer lifetimes, up to 10 μs.

In LWIR QWIP devices, LO phonons play a decisive role in the intersubband (IS) relaxation process, and for this reason, carrier relaxation times are typically around 10 ps. Consequently, QWIPs are not competitive with HgCdTe photodiodes, especially in the operating range above 70 K. Potential advantages of QWIP over HgCdTe arrays include: the use of standard fabrication MBE growth technologies, high efficiency, and thus lower cost, higher thermal stability and external radiation hardness. The drawback of QWIP detectors is relatively low quantum efficiency, typically below 10%. For these reasons, the scientific community's acceptance of QWIP technology has been slow and difficult to implement. Therefore, funding for QWIP technologies has now been significantly reduced [14].

The interband quantum cascade detectors (IB QCDs) based on T2SL materials have complicated detector architecture. Their design is particularly pronounced for HOT operations with performance comparable with HgCdTe [15]. However, their challenging technology (with multiple interfaces and stressed thin films) and high cost have resulted in further funding being abandoned.

Based on the fundamental photoelectric properties of semiconductors, we will estimate $\sqrt{\alpha\tau}$ values for materials operated in three IR spectral ranges: NIR ($\lambda_c$ = 1.6 µm), MWIR ($\lambda_c$ = 5 µm), and LWIR ($\lambda_c$ = 10 µm) – see Table 1. The values of $\sqrt{\alpha\tau}$ for the NIR material systems (InGaAs, HgCdTe) are similar. However, for the longer wavelength regions (both MW and LW), the HgCdTe figure of merit, $\sqrt{\alpha\tau}$, is better than for the other material systems. Furthermore, the $\sqrt{\alpha\tau}$ metric indicates that standard materials (InGaAs, HgCdTe, and T2SL) appear to be superior to emerging new-generation photodetectors, such as transition metal dichalcogenides (TMDs), perovskites and other LDS photodetectors [16].

**Table 1. Estimated $\sqrt{\alpha\tau}$ Figure of Merit for Infrared Material Systems at Room Temperature and Wavelengths $\lambda$ = 1.6, 5, and 10 µm (after Ref. 16)**

| $\lambda$ [µm] | Material system | Material parameters | | | $\sqrt{\alpha\tau}$ [(s/cm)$^{1/2}$] |
|---|---|---|---|---|---|
| | | Doping [cm$^{-3}$] | $\alpha$ [cm$^{-1}$] | $\tau$ | |
| 1.6 | InGaAs | $5\times10^{14}$ | $1\times10^{4}$ | 200 ms | 1.4 |
| | HgCdTe | $5\times10^{13}$ | $1\times10^{4}$ | 1 ms | 3.2 |
| 5.0 | InAs$_{0.91}$Sb$_{0.09}$ | $1\times10^{15}$ | $3.2\times10^{3}$ | 10 ms | 0.18 |
| | HgCdTe | $5\times10^{13}$ | $3.2\times10^{3}$ | 1 ms | 1.8 |
| | InAs/GaSb SLs | $5\times10^{14}$ | $2.4\times10^{3}$ | 20 ns | $6.9\times10^{-3}$ |
| | InAs/InAsSb SLs | $5\times10^{14}$ | $1.2\times10^{3}$ | 2 µs | $4.9\times10^{-2}$ |
| | PbSe | $1\times10^{17}$ | $3\times10^{3}$ | 10 ns | $2.0\times10^{-2}$ |
| 10.0 | HgCdTe | $5\times10^{13}$ | $2.2\times10^{3}$ | 0.1 ms | 0.47 |
| | InAs/GaSb SLs | $5\times10^{14}$ | $1.6\times10^{3}$ | 10 ns | $4.0\times10^{-3}$ |
| | InAs/InAsSb SLs | $5\times10^{14}$ | $8.0\times10^{2}$ | 0.2 µs | $1.3\times10^{-2}$ |
| | GaAs/AlGaAs QWIP | $5\times10^{17}$ | $7.0\times10^{2}$ | 10 ps | $9.0\times10^{-5}$ |

## Infrared focal plane arrays

The typical design of an IR camera is similar to a digital video camera. As shown in Fig. 3, the detector of an infrared camera is a focal plane array (FPA) consisting of μm-sized pixels made of different materials depending on the spectral range of the imaging. The system optics (lenses and filters) are matched to the required spectral range and type of application.

Figure 3(b) shows the radiant power collection by optics of the imaging system. Here $A_s$ and $A_d$ are, respectively, the surfaces of the object and the detector, $r$ is the distance of the object to the lens (system optics), $A_{ap} = \pi D^2/4$ and $D$ are the surface and the diameter of the lens (aperture, entrance-pupil). The detector is placed in the system's focal plane in the distance $\approx f$ to the entrance pupil. The optic system is characterized by the so-called F-number, i.e., $f/\# = f/D$.

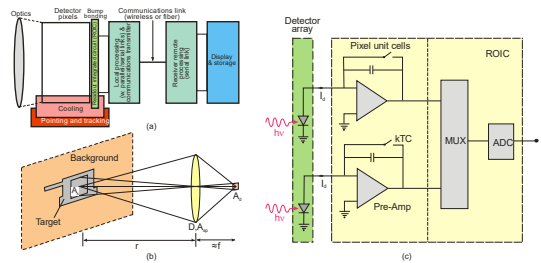

Fig. 3. Imaging system: (a) configuration of important sub-systems, (b) radiant power collection by an optical system, and (c) readout integrated circuit (ROIC) for two pixels of an FPA.

So-called "staring arrays" scanned electronically by circuits integrated into the arrays are now commonly used. The architecture of the readout circuits is diverse and is described in detail in Ref. 3–7,17,18, among others. The readout integrated circuits (ROIC) types include pixel deselecting functions, anti-blooming at each pixel, sub-frame imaging, output preamplifiers, and many others.

Different types of FPA architectures are used in thermal imaging systems [19–22]. They are generally classified into monolithic and hybrid. However, this distinction is not generally decisive. This is because of the decisive issues related to evaluating the benefits of a specific architecture and its manufacturability. Each target application imposes specific technical requirements, implementation costs, and implementation schedules.

Although efforts have been made over the past 40 years to develop monolithic structures using various IR photodetector materials, only a few have reached the practical application level. These include PbS and PbSe, and in the last decade, also colloidal quantum dot (CQD) arrays. However, the large temperature coefficient of expansion of lead salts mismatched with Si has an influence on the fabrication of only small arrays. Currently, the production of low-cost CQD FPAs is not yet widely used. Other IR material systems [InGaAs, InSb, HgCdTe, T2SLs (InAs/GaSb and InAs/InAsSb), GaAs/AlGaAs QWIPs, and extrinsic silicon] are fabricated in hybrid configurations.

Hybrid technology allows both the detector material and the readout (multiplexer) chip to be optimized independently. These arrays enable a large fill factor (FF) close to 100% and simultaneously allow an increase in the signal processing area on the multiplexer chip. Most hybrid arrays use photodiodes with their typically low power dissipation, inherently high impedance, negligible $1/f$ noise, and easy multiplexing via ROICs. In addition, the photodiodes can be polarised in the reverse-biased direction to increase impedance, allowing a better electrical match to the compact, low-noise silicon readout preamplifier circuits. It is also worth noting that the response of photodiodes remains linear for much higher excitation levels than photoconductors, mainly due to higher doping levels in the active region.

The development of hybrid technology began in the late 1970s, and it took another decade to reach mass production. Fully two-dimensional (2D) IR detector arrays were commercialized in the early 1990s. The commercialization of multispectral arrays has also been successfully initiated, the most mature of which is represented by megapixel two-color arrays operating in MW/LW spectral rangers.

Various hybridization techniques are currently in use [23–26]. The most common technique is flip-chip bonding (see Fig. 4). In this technique, indium bumps are formed on both the detector array and the ROIC chip. The array and ROIC are aligned, and then a force is applied to cold weld the indium bump. In the second approach, indium bumps are formed only on the ROIC; the detector array is aligned and close to the ROIC, the temperature is raised to melt the indium, and contact is established by reflow.

Large FPAs with fine-pitched bumps require considerable force to ensure a reliable connection. Its use can damage detector layers, cause misalignment associated with the bonding process, and cause "hybrid slippage". The fusion bonding (or direct bonding) process requires less than 10 lbf (compared to > 1,000 lbf for an indium-based process). It is particularly optimal in fabricating very large arrays up to 10k×12k and thin detector active layers (< 10 μm). The disadvantage of this method is its extreme sensitivity to surface flatness, surface roughness, and the presence of messy particles.

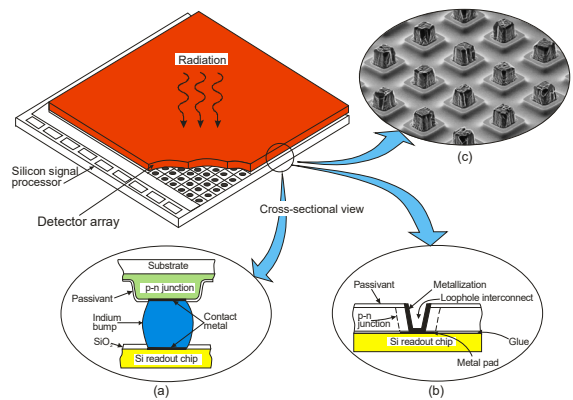

Fig. 4. IR FPA interconnect techniques between a detector array and silicon multiplexer: (a) indium bump technique, (b) loophole technique, and (c) SEM photo shows mesa photodiode hybrid array with indium bumps. The smallest pixel pitch of hybrid arrays is 5 μm.

Hybrid arrays are also fabricated using loophole connections – see Fig. 4(b). In this case, the metallic contact of the photodiode is connected downstream to the silicon readout circuit through a hole of a few microns in diameter [27,28]. The detector and multiplexer chips are glued/bonded together before the detector is fabricated.

Typically, hybrid arrays are backside illuminated (photons fall on a transparent substrate). To increase bonding, epoxy resin is injected into the space between the readout processor and the detector array. The thickness of opaque substrates is abraded to less than 10 μm to achieve higher quantum efficiency and reduce crosstalk. Increasingly, however, the substrates are being completely removed.

The size of traditional hybrid systems is typically limited to 1-megapixel arrays. The hybridization process limits the pixel size to 5 μm because the solder bumps require sufficient volume for reliable bonding, which in turn is limited by the aspect ratio and pixel spacing.

## Performance of infrared detector arrays

Considerable efforts are being made to reduce the imaging system's size, weight, and power (SWaP) consumption — the ultimate goal is to reduce system costs. The SWaP criterion influences the shrinking of pixels in the arrays. Pixel reduction is also indicated to improve resolution and IR system reliability.

For IR thermal imaging systems, the basic parameters of their quality are thermal sensitivity and spatial resolution. In order, they are determined by the noise equivalent difference temperature ($NEDT$) and the modulation transfer function ($MTF$). Thermal sensitivity defines the minimum temperature difference that can be observed above the array noise level. The $MTF$, refers to the spatial resolution and answers the question – how small can an object be imaged by the system?.

The $MTF$ is influenced by several factors, the most important of which are determined by the optics, detector, and display, and can be cascaded by simply multiplying the $MTF$ components — $MTF = MTF_{optics} \times MTF_{detector} \times MTF_{display}$ [29]. In practice, the resulting $MTF$ of the whole detection system is mainly determined by the optics and detector, and therefore $MTF_{od} = MTF_{optics} \times MTF_{detector}$. Other blurs (aberrations, crosstalk, diffusion, etc.) are considered less insignificant.

Figure 5 summarizes the different $MTF$ behavior of the system [30], where the primary parameters are the detector size, $d$, optics F-number ($f/\#$), and wavelength $\lambda$. The detector-limited region occurs when $F\lambda/d \leq 0.41$, and the optics-limited occurs when $F\lambda/d \geq 2$ [see Fig. 5(a)]. When $F\lambda/d = 0.41$, the Airy disk equals the detector size. A transition in the region $0.41 \leq F\lambda/d \leq 2.0$ is large and indicates a change from detector-limited to optics-limited performance. When $F\lambda/d = 1.0$, the spot size equals 2.44 pixel sizes (Airy disk determined by diffraction). In the past, most systems were designed to avoid aberration at the pixel — the resulting blur was less than 2.5 pixels ($\sim F\lambda/d < 1.0$).

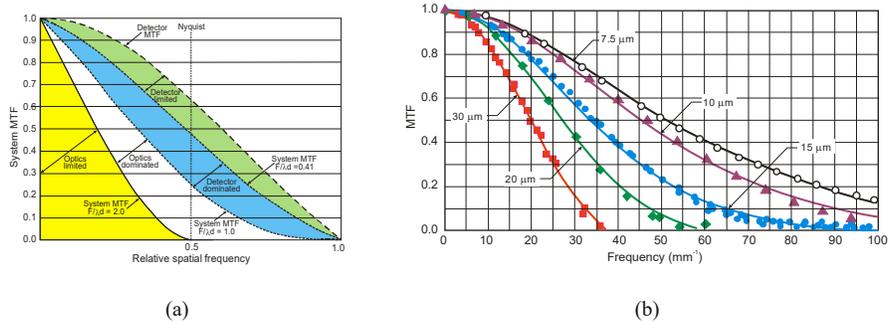

(a)                                    (b)

Fig. 5. Modulation transfer function: (a) system *MTF* curves illustrating the different regions with the design space for various *Fλ/d* conditions. Spatial frequencies are normalized to the detector cutoff (after Ref. 30); (b) *MTF* curves of n-on-p HgCdTe structures from 30 μm to 7.5 μm pixel pitch (after Ref. 31). Both theoretical simulations (solid lines) and experimental data (different points) are shown.

The development of large-format hybrid IR arrays with small pixel pitch (less than 10 μm) requires overcoming many difficulties caused by electrical and optical crosstalks. Berthoz *et al.* [31] conducted an experimental and theoretical comparative study of the effects of different pixel designs (planar, loop, mesa, and depleted p-i-n HgCdTe photodiodes) on detection range and image quality of IR imaging system. As an example, Fig. 5(b) shows the improvement of the *MTF* function for the n-on-p HgCdTe photodiode by reducing the pixel size from 30 μm to 7.5 μm (keeping other parameters constant). As can be seen, good agreement was obtained between experimental measurements and simulations.

A very good detector choice for large arrays containing small pixels is the P-i-N photodiode. Due to the very low doped region, full depletion of the i-region can be achieved at low reverse voltage. The electric field in the depleted structure eliminates electrical crosstalk. The lowest possible level of doping of the photodiode active i-region enables improved performance.

It is predicted that in the near future it will be possible to achieve room-temperature operation of IR arrays with high pixel density (small pixels) fully compatible with the background and diffraction-limited performance due to system optics [12,32]. As previously noted, the advent of small pixels also results in better spatial and temperature resolution of imaging systems.

Most current IR imaging systems have the classic view shown in Fig. 6, where the pixel size ranges from 5 to 50 μm. Long-range identification systems use high *f/#* optics (for a given aperture) to reduce the angular view of the detector. On the other hand, wide field of view (WFOV) systems typically have low *f/#* and short focal lengths. However, it turns out that long-range identification systems do not have to be limited to high *f/#* systems, as very small pixels allow high performance with a smaller housing size [30,33–36].

The fundamental limit to pixel size is diffraction. However, it turns out that some degree of oversampling beyond the diffraction limit is desirable (this is possible due to nanofabrication performance). Reducing the focal plane in proportion to the detector size does not change the detector's field of view, so in the optics-limited region, smaller detectors do not affect the system's spatial resolution.

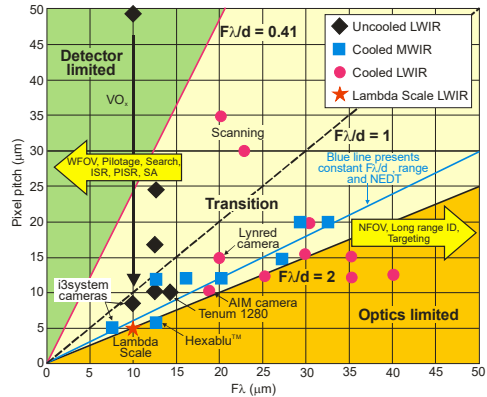

Fig. 6. $F\lambda$-$d$ space for IR system design together. Straight lines represent constant $NEDT$. There are an infinite number of combinations that provide the same range. The experimental data points are adapted after DRS Technologies [35] and other manufacturers (as marked).

The fundamental limit to pixel size is diffraction. However, it turns out that some degree of oversampling beyond the diffraction limit is desirable (this is possible due to nanofabrication performance). Reducing the focal plane in proportion to the detector size does not change the detector's field of view, so in the optics-limited region, smaller detectors do not affect the system's spatial resolution.

Considerations carried out in Ref. 33 indicate that for MWIR imaging systems with $f/1$ optics, the smallest size of the detector should be 2 μm, while in the LWIR systems – 5 μm. Taking the $f/1.2$ systems, these limits should be 3 μm and 6 μm for the MWIR and LWIR ranges, respectively.

The lines in Fig. 6 represent a constant $F\lambda/d$ and, at the same time, denote the constant detection range and $NEDT$ of the thermal system (both detection range and $NEDT$ is proportional to $F\lambda/d$ [12]) . For a given aperture $D$ and wavelength $\lambda$, the detection range is determined by the condition of optimal resolution $F\lambda/d = 2$ and the minimum value of $NEDT$ for a given $\tau_{int}$. Fig. 6 also includes experimental data points for various classes of thermal imaging systems developed by DRS Technologies [35] and other manufacturers operating in the MWIR and LWIR bands. The first uncooled imaging sensors produced in the early 1990s featured large pixel sizes of about 50 μm and fast optics to achieve useful system sensitivity. LWIR imaging systems typically come close to meeting the condition $F\lambda/d = 2$. This tendency has recently been observed also for MWIR systems. As pixel dimensions decrease over time, the thermal imaging systems gradually evolve from the "limited detector" regime to the "limited optics" regime.

Currently, the infrared market is dominated by megapixel FPA arrays with an area in the order of 1 cm². The hybridization process limits the pixel size to 5 μm because the solder bumps require sufficient volume for reliable bonding, which in turn is limited by the aspect ratio and pixel spacing. The oversampling of the diffraction spot may provide some additional resolution for smaller pixels, but this effect saturates quickly as the pixel size decreases. However, with the current state of technology, it appears that the cost of arrays (which includes the array, semiconductor wafer, cryogenic cooler, and other subcomponents) with a pixel size of 8 μm is approximately 9% and 19% lower compared to arrays with pixels size of 5 μm and 15 μm, respectively [37]. Simply put, system-level costs reach a point of diminishing returns and become increasingly expensive when the smallest pixel sensors are used.

As pixel size is reduced, the requirements for "bump-bonding," ROIC, signal-integrating capacitor, and high signal-to-noise ratio become difficult to reconcile. Electronic signal reading systems have a great influence on IR FPA performance. Then their efficiency is often limited by readout circuits (ROIC storage capacity). In this case $NEDT = \left( \tau C \eta \sqrt{N_w} \right)^{-1}$ [38], where $C$ is the scene contrast, $\eta$ is the quantum efficiency, and $N_w$ is the number of photogenerated carriers integrated for one integration time, $\tau_{int}$. The number of integrated photogenerated carriers can be described by $N_w = \eta A_d \tau_{int} \Phi_B$, where $A_d$ is the detector area and $\Phi_B$ is the background flux. So *NEDT* is inversely proportional to the square root of the integrated charge – that is, the larger the charge is collected in the reading capacitor during signal integration, $\tau_{int}$, the greater the sensitivity of the array is obtained. The size of the pixel has a large impact on the well capacity. In large arrays with small pixels, the photon flux collected by the pixels is reduced while the integration time is increased.

Table 2 presents the typical performance of megapixel hybrid IR FPAs offered by major manufacturers. Despite the numerous advantages of III-V compounds [mainly type-II superlattices (T2SLs) and barrier detectors] over HgCdTe (such as lower tunneling and surface leakage currents, as well as less efficient Auger recombination), the expectation of better detector performance was not realized due to the low SRH lifetimes [13]. In general, III-V semiconductors are more chemically stable than their II-VI counterparts. For this reason, III-V-based FPAs stand out for their greater functionality, spatial homogeneity of composition, temporal stability, scalability, manufacturability, and affordability — so-called "ability" advantages.

**Table 2. Typical Performance of Megapixel Hybrid IR FPAs Offered by Major Manufacturers**

| Spectral range (μm) | Detector material | Size/ Architecture | Pixel size (μm) | Oper. temp. (K) | $D^*(\lambda_p)$ (cmHz$^{1/2}$/W)/ *NETD* (mK) |
|---|---|---|---|---|---|
| 0.7–1.7 | InGaAs | 640×512 | 12.5×12.5 | 300 | $2.9 \times 10^{13}$ |
|  | InGaAs | 1280×1024 | 12.5×12.5 | 300 | $2.9 \times 10^{13}$ |
| 3-5 | InSb | 1280×1024 | 15×15 | 80 | —/20 |
| | InAsSb XBn | 1280×1024 | 15×15 | 150 | —/30 |
| | HgCdTe | 1280×1024 | 15×15 | 110 | —/20 |
| | T2SL | 1280×1024 | 15×15 | 130 | —/25 |
| 8-10 | HgCdTe | 640×512 | 15×15 | 80 | —/25 |
| | InAs/GaSb T2SL | 640×512 | 15×15 | 80 | —/30 |

To achieve high-sensitivity LWIR FPAs (sey < 30 mK), a large amount of integrated charge is required to be accommodated in a very small unit. The state-of-the-art storage density of conventional (based on analog electronics) ROIC design rules is about $2.5 \times 10^4$ e⁻/μm² [34]. For a 5-μm planar unit cell, the charge capacity in standard ROIC technology is less than 1 million electrons, whereas 8 to 12 million electrons are required for good sensitivity.

The main goal of IR FPA development is to achieve BLIP operating conditions at room temperature. This goal also necessitates the development of suitable ROIC circuits. It follows that the charge collected from the full field of view $2\pi$ integrated during the frame is effectively stored in the detector node, even with reduced pixel spacing. For this reason, a much larger well capacity per μm² is required (above $1 \times 10^7$ e⁻/μm² for LWIR), compared to the values available in current ROIC designs [39].

The limitation of charge accumulation in small pixels can be addressed by using MEMS capacitors adapted for 3D ROIC design in a so-called digital readout integrated circuit (DROIC). Currently, typically three CMOS layers are stacked and vertically bonded, which increases the capabilities for charge collection and processing within a single pixel [40].

The use of DROIC overcomes the limitations of conventional FPAs by digitizing the signal in pixels. Digitization of the readout enables increased dynamic range, faster low-noise digital readouts, and on-chip processing. In a typical analog readout, the depth of the well allows for a charge collection of about 10 million electrons, while digital readout circuits provide more than a billion electrons. In some cases, this is even over 1 trillion electrons. This means that by using a digital circuit, we can integrate the signal over a longer period of time.

A novel 3D integration technology implemented as an alternative to bump bonding is reviewed in a number of articles; see, for example, Ref. 41–44. One example of this solution is Cu micro bumps technology for interconnection of a 5-μm pitch in megapixel arrays. 3D ROICs-stacks with analog and digital Si circuit layers connected via metal-filled silicon vias (TSVs) are increasingly being used. New techniques provide low-temperature bonding and alignment accuracy in the 1–2 μm range.

### Concluding remarks

In the last decade, we have witnessed rapid progress in the development of the IR FPA. Among the spectacular successes, we can include several megapixel sensors with pixel dimensions of less than 6 μm [45]. Other achievements should also be mentioned. One of these is the development of high-quality p-i-n HgCdTe photodiode technology, which has influenced the formulation of "Law 19" in 2019 [32]. The formulation of this law represents a significant advance towards fulfilling Kinch's [34] prediction (SWaP criterion) that significant reductions in IR system production costs can be achieved under high operating temperature (HOT) conditions with pixel arrays that are fully background-compatible and diffraction-limited performance due to the system optics.

The current preference in imaging systems is to oversample beyond the diffraction limit, using sub-wavelength scale optics. It appears that pixel sizes can be reduced to around 5 μm for cooled and uncooled LWIR applications with $f/1$ optics. However, with the current state of technology, the choice of 8-μm pixel spacing in IR imaging cameras is the most cost-optimal.

Small-pitch IR FPAs require more advanced readout circuits with large ROIC well capacitance per unit area. This challenge is addressed by fabricating MEMS capacitors tailored for 3D ROIC design; in the so-called digital readout integrated circuit (DROIC). Meeting the SWaP criterion also comes down to faster system optics than $f/1$, improved hybridization technologies, or broader implementation of monolithic technologies. In the latter case, impressive progress has been made using III-V antimonide-based detectors [45].

For over 60 years, HgCdTe has successfully addressed serious challenges from various detector families. Recently, a new generation of LDS-based photodetectors (mainly 2D materials) has emerged, with performance approaching that of HgCdTe photodiodes. Thus, HgCdTe now has more competitors than ever before [16]. Thus, the question: *will the discovery of other materials affect the status of HgCdTe?* is very timely. Partially, the answer to this question is given in Ref. 9.

# 9. HgCdTe Detectors

ADAM PIOTROWSKI[*], JAN SOBIESKI, PAWEŁ LESZCZ AND JÓZEF PIOTROWSKI

*Vigo Photonics, Ożarów Mazowiecki, PL*

\* apiotrow@vigophotonics.com

**Overview**

The mid infrared (MWIR) spectral region, defined here as the range between 3 and 14 μm, holds immense significance across diverse fields. Initially, its importance stemmed from its crucial role in fundamental scientific measurements. Applications such as Fourier Transform Infrared (FTIR) spectroscopy, precise temperature sensing, and other scientific measurements, alongside defense, security, and industrial process control, capitalize on the unique molecular "fingerprint" vibrational signatures present in this range, enabling highly specific material identification and analysis. This capability drives the demand for high-performance MWIR detection technologies. The increasing availability of fast semiconductor lasers, such as interband cascade lasers (ICLs) [1] and quantum cascade lasers (QCLs) [2], perfectly matched to uncooled HgCdTe (MCT) detectors, is creating new opportunities for next-generation chemical sensors, including cavity-enhanced spectroscopy, open-path sensors, and cavity ring-down spectroscopy. A key direction in this field is the development of affordable, small-sized laser-based systems with broad wavelength tunability. This trend further emphasizes the need for faster detectors.

Mercury cadmium telluride (HgCdTe) stands as the premier material system for infrared detection [3]. HgCdTe is a semiconductor alloy formed from mercury telluride (HgTe) and cadmium telluride (CdTe). A key attribute of HgCdTe is its tunable bandgap, which can be precisely engineered by adjusting the mercury-to-cadmium ratio. This tunability allows for the fabrication of detectors optimized for specific infrared wavelengths with high quantum efficiency [4]. These properties have established HgCdTe as crucial for applications requiring high sensitivity and spectral selectivity in the MWIR.

The development of HgCdTe detectors spans several decades, marked by continuous advancements in material growth and device fabrication techniques. From early bulk growth methods to sophisticated epitaxial techniques, the pursuit of improved material quality, device performance, and reduced cost has been a driving force. The monolithic integration of multiple functionalities on a single semiconductor chip, including enhanced absorption, suppression of thermal noise generation, elimination of parasitic impedances, and integration of optical concentrators, is a key strategy for achieving the best possible signal-to-noise ratio in these devices [5-7].

This paper aims to provide an overview of the status of HgCdTe infrared detector technology, with a particular focus on emerging trends in material growth. We will explore recent advancements in low-temperature molecular beam epitaxy (MBE) with post-growth annealing for effective doping activation, metal-organic chemical vapor deposition (MOCVD) on alternative substrates (such as GaAs) using lattice-matched buffer layers, and interdiffused multilayer processing. These techniques address critical challenges in HgCdTe material growth, such as achieving high material quality, precise composition control, and controlled doping.

## Current status

A particular advantage of MCT is its minimal change in lattice constant with varying cadmium content [4]. This property allows for the growth of complex heterostructures that are necessary for various types of photodetectors operating at various temperatures and spectral ranges. The material is characterized by a high ratio of the absorption coefficient to the thermal generation rate ($\alpha/G$), which determines the ultimate detectivity of the detectors [8]. The easy tunability of the bandgap also enables the fabrication of multispectral detectors.

The fabrication of HgCdTe detectors was initially based on the growth of bulk crystals from melts [4]. The present devices rely on epitaxial growth of HgCdTe from the liquid (LPE) and gas (MOCVD, MBE) phases. Epitaxial techniques allow for low temperature growth of high-quality HgCdTe heterostructures on CdTe and alternative low-cost GaAs, Si and other substrates. [9-11]

LPE enables the fabrication of high-quality crystalline layers [4] but does not allow for sharp compositional and doping interfaces in the HgCdTe heterostructures. Another drawback is the limited uniformity of the resulting layers.

MOCVD is an efficient technique for large-scale growth of epitaxial heterostructures on CdTe and GaAs substrates with up to 4-inch sizes [11]. The growth occurs at ~360°C. The structures can be doped with well-behaved donor (I) and acceptor (As) dopants. The residual doping is mid-$10^{14}$ cm$^{-3}$ range, and dislocation densities are $>10^6$/cm$^2$. The heterostructures are characterized by relatively soft grading of the interfaces (~0.2 to 0.6 μm).

The most common growth method for HgCdTe is MBE, which provides a perfect crystalline structure with exceptionally low residual doping levels (low $10^{13}$ cm$^{-3}$ range) [12], sharp interfaces in heterostructures and a low dislocation density. However, since MBE growth occurs at a low temperature (~180°C), the activation of arsenic acceptor dopants remains low. Post-growth activation annealing at ~350°C and prolonged ~180°C vacancy annihilation annealing is required.

Photoconductors (PC) are the simplest and most reliable HgCdTe devices [3]. They consist of a narrow gap semiconductor layer deposited on CdTe (or CdTe-buffered GaAs substrate), surface passivation layer and two electric contacts (Fig. 1). Long wavelength infrared (LWIR) photoconductors with cutoff wavelength up to 25 μm and operating at 77K achieve near background radiation detectivity limit, D*(300 K, 180°) [13]. They are commonly used for broadband IR spectroscopy [12]. They are typically housed in long-lasting vacuum dewars. Photoconductors have relatively slow responses that are determined by the recombination time. An exception is uncooled LWIR devices with sub-nanosecond response time [5], although unfortunately the sensitivity of those devices is very poor.

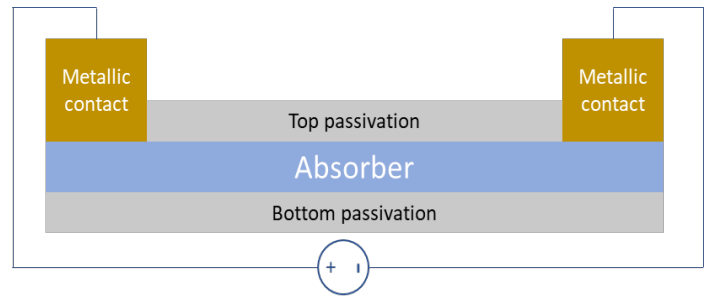

Figure 1. Schematic design of a heterostructure photoconductor.

Most high-end applications require photovoltaic (PV) detectors. These detectors offer high response speed and high detectivity, with the option of zero bias flicker-free operation.

Typically, the devices are various modifications of N⁺iP⁺ heterostructures [15] with lightly doped (i) narrow gap absorbers and two N⁺ and P⁺ heavily doped wide bandgap layers (Fig 2).

- The main part of the device is an absorber of IR radiation. Its bandgap determines the cutoff wavelength of the device.

- The N⁺ layer is the electron contact and hole barrier. It usually serves as an optical filter that determines the cut-on wavelength

- The P⁺ layer is the hole contact and electron barrier.

Proper design of the layers and their interfaces reduces the dark current generation volume to the neutral absorber and prevents the generation of Shockley-Read-Hall (SRH), tunnel and surface leakage currents. Cryogenically cooled HgCdTe photovoltaic devices achieve extreme detectivities, much better than those of any competing material. Low-noise avalanche photodiodes and single-photon detectors can be realized with some modification of the devices.

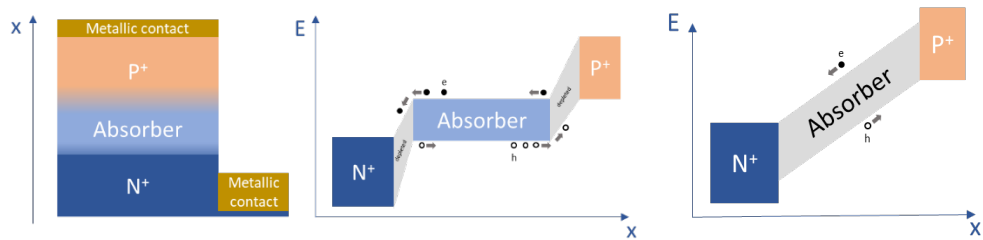

Figure 2. Schematic design of a mesa structure PV detector (left); band diagram of an N+iP+ heterostructure under bias free operation (center); and a reverse-biased non-equilibrium device operated with fully depleted absorber (right).

Unfortunately, the thermal generation rate of charge carriers in HgCdTe, similarly to other semiconductors, increases exponentially with the product of the operating temperature and the cut-off wavelength. This results in a rapid deterioration of the detectors' sensitivity with increasing temperature or wavelength. Auger thermal generation (impact ionization) is the dominant mechanism of thermal generation at elevated temperatures [16]. Additional problems of MWIR/LWIR detectors operating at elevated temperatures are the short diffusion length of charge carriers in the absorbers and low parallel resistance of the devices. This results in low quantum efficiency and high noise of preamplifiers coupled with low-resistance detectors. However, there are alternative ways to increase sensitivity other than cooling. The most important of them are:

- Optimized design of the device and use materials with the best figures of merit. Lightly doped MCT with a low concentration of SRH recombination centers is a material of choice due to its low SRH generation rate.

- Decreasing the absorber volume without limiting the device field of view and quantum efficiency (QE). This can be done by reducing the surface area with optical concentrators and decreasing absorber thickness while enhancing the absorption using optical resonant cavities [17] or plasmonic antennas [18].

- Use of cascade photovoltaic detectors with absorber thickness less than the diffusion length, to ensure perfect collection of the photogenerated charge carriers and increase the resistance of the devices [19].

- Suppressing the Auger generation by the extraction of charge carriers from the lightly (low 10¹³ cm⁻³) absorber. This can be realized by the non-equilibrium mode of operation of reverse-biased heterostructural photodiodes [7]. HgCdTe uniquely enables the efficient extraction of carriers (right schematic in Fig. 2) to a very low level that allows radiative-limited operation [20].

"Rule 07" is a heuristic measure of IR detector quality that has been used for over two decades [21]. It was recently refined as "Rule 22" [22], which defines the dark current as a function of cutoff wavelength and temperature (Fig. 3). This rule serves as a benchmark for evaluating the quality of HgCdTe photodiodes and provides a state-of-the-art reference point for comparing the performance of detectors made from emerging materials, such as bulk and superlattice AIII- BV compounds. The ultimate performance of a photodiode is determined by incoming signal or background photon noise. "Law 19" considers only current generation by emission from the neutral regions of epilayer [20]. It represents the limit that can be achieved when the carriers are extracted with a bias to suppress Auger recombination as illustrated on the right side of Fig. 2. This enables D* twice higher than BLIP(300K, π) with one- stage thermoelectric cooling (Fig. 3).

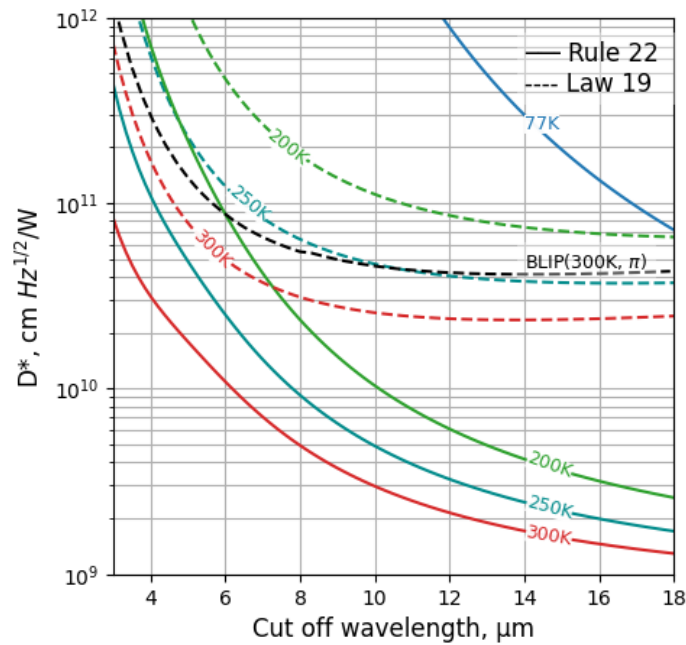

Figure 3. Normalized detectivity for a detector in which the only noise is the dark current as defined by Rule 22 or Law 19 for various operating temperatures.

According to "Rule 22", BLIP(300 K, 180°) performance can be achieved for cutoff below 6 µm at 200 K (achievable with TEC cooling). For longer wavelengths, reaching BLIP performance without Auger suppression requires much lower operating temperatures—for instance, a 20 µm cut-off would necessitate liquid nitrogen cooling.

## Challenges and opportunities

The major disadvantage of MCT is its weak Hg-Te bonds, which make the material vulnerable to environmental exposures during the fabrication, storage and exploitation of the devices. A related limitation is the need to use low-temperature manufacturing processes due to the weak mercury bonds. Due to the material's fragility, the standard industrial semiconductor production methods must be modified for use in fabricating MCT devices. This hinders the scaling of the technology for mass production, as it relies on semi-handmade processes. Furthermore, Restriction of Hazardous Substances (RoHS) regulations are pushing detector manufacturers to replace cadmium and mercury-based materials with more environmentally friendly alternatives.

## Future developments to address challenges

To fully unlock the potential of MCT material, several critical challenges must be addressed.

- The high cost of MCT detectors imposes a major limitation on their application. It results from the use of expensive CdTe substrates and low fabrication yield. This can be overcome by optimizing the epitaxial processes and using widely- available, inexpensive substrates (e.g., GaAs or Si) buffered with thin CdTe layers.

- MCT's fragility presents significant difficulties during production, storage, and operation. The deposition of CdTe layers provides a degree of surface protection, but further refinement is necessary to extend the device lifespan. This can be achieved with the improved deposition of hard dielectric coatings.

- The need to cool infrared detectors to low temperatures is a significant obstacle to a wide range of applications. This applies to cooling to liquid-nitrogen temperature, but also thermoelectric cooling. Multi-stage thermoelectric coolers consume substantial electric powers and need large heat sinks that considerably increase the size of the packaged detector.

- A unique opportunity is to construct devices that perform close to the fundamental limit using only a one-stage thermoelectric cooler [20]. The advantages are low cost, low power consumption, miniature size, and low heat generation. A particular advantage is stabilization of the detector parameters as the ambient temperature changes.

## Concluding Remarks
HgCdTe remains the most important variable bandgap semiconductor for MWIR photodetectors. The most important reasons for this are:

- The extreme flexibility of HgCdTe, which is the only detector material that can cover the whole IR spectral range. Having an early composition-independent lattice parameter, it can be tailored for optimized detection at any region of IR spectrum. Thus, heterostructural photodetectors can be constructed easily.
- None of the new competing materials can offer fundamental advantages like HgCdTe's high $\alpha/G$ figure of merit.
- The very low concentration of SRH centers and uncontrolled impurities in HgCdTe enables the effective suppression of both SRH and Auger generation which leads to the highest detectivities for low and elevated operating temperatures.

- The detector surfaces can be effectively passivated by applying wide-gap HgCdTe layers [4, 23].

However, there have been numerous attempts to replace HgCdTe with alternative materials:

- This is motivated by limited material stability and requirements of the RoHS directive aimed at eliminating the use of highly toxic heavy metals including Hg and Cd.

- The current development of detectors based on the RoHS-compliant A3B5 semiconductors, particularly with InAs/GaSb and InAs/InAsSb superlattice absorbers, has led to the improvement of their properties [24] and the increase of their market share.

- However, it should be assumed that HgCdTe will be used for many years to come, especially in applications requiring the highest sensitivity at low and elevated operating temperatures, including avalanche, single photon and hyperspectral devices.

## 8. Back matter

### Disclosures

The authors declare no conflicts of interest.

# 10. Single Element Lead-Salt Detectors

## MILAD RASTKAR MIRZAEI,[1,2] RICHARD KIM,[1,*] AND JEUNG HUN PARK[3]

*1Laser Components Detector Group Inc., Chandler, AZ, 85225, USA*
*2ECE Department, University of Oklahoma, Norman, OK 73019, USA*
*3University of Southern California, Marin☐ Del Rey, CA, 90292, USA*
*\*rkim@laser-components.com*

**Introduction**

The mid-wavelength infrared (MWIR), spanning 3-5 µm, and the long-wavelength infrared (LWIR), covering 8-14 µm, are critical spectral regions for a multitude of applications such as gas sensing and analysis, thermal imaging, surveillance, spectroscopy, environmental monitoring, industrial process control, and medical diagnostics[1-4]. The enduring importance of these spectral windows is fundamentally linked to specific atmospheric transmission windows and the characteristic vibrational and rotational absorption/emission spectra of many molecules, as well as thermal signatures of objects at or near terrestrial temperatures that fall within these ranges. These needs continually drive the demand for detectors operating in these bands[5-7].

The lead chalcogenides, specifically lead sulfide (PbS), lead selenide (PbSe), lead telluride (PbTe), and the ternary alloy lead-tin telluride (PbSnTe), represent a historically significant and still highly relevant class of semiconductor materials for detection in the MWIR and LWIR regions. The prominence of PbSe and PbS detectors largely stems from their excellent alignment with SWaP+C (size, weight, power, and cost) considerations. They exhibit high sensitivity and can operate effectively at or near room temperature, particularly within the MWIR spectral range, thus substantially reducing system complexity and operational power demands[8, 9]. Moreover, the inherent cost-effectiveness of PbSe and PbS detectors, especially for single-element configurations, distinguishes them from more complex cooled detectors such as mercury cadmium telluride (MCT) [10, 11] (covered in a separate sub-topic article of this Roadmap).

The performance and applicability of lead salt detectors are intrinsically linked to their fundamental material properties. These IV-VI narrow-gap semiconductors share a common rocksalt crystal structure, and a direct bandgap located at the L-point of the Brillouin zone[12-14]. A distinctive feature is their anomalous positive temperature coefficient of the bandgap ($dEg/dT > 0$). The shift to increasing bandgap, and thus decreasing cutoff wavelength, with increasing temperature has significant implications for the temperature stability of uncooled or thermo-electrically stabilized detectors[15, 16]. Another common characteristic is their very high static dielectric constant. This contributes to high carrier mobilities but also induces high device capacitance that limits high-frequency operation[17-19]. Table 1 provides a comparative summary of these properties.

Table 1. Fundamental Properties of Lead Salt Detector Materials (All have the cubic rocksalt crystal structure)

| Property | PbS | PbSe | PbTe | $Pb_{1-x}Sn_xTe$ |
|---|---|---|---|---|
| Typical Spectral Range (μm) | 1-3.3 | 1-5.2 | 1.5-5.5 | 8-14 |
| Bandgap at 300K (eV) | 0.41 | 0.278 | 0.31 | Tunable |
| Electron Mobility at 300K, bulk ($cm^2V^{-1}s^{-1}$) | ~600-800 | ~1000-1020 | ~6000 | Varies with $x$ |
| Hole Mobility at 300K, bulk ($cm^2V^{-1}s^{-1}$) | ~700-900 | ~900-1000 | ~4000 | Varies with $x$ |
| Static Dielectric Constant ($\epsilon_s$) at 300K | ~170-175 | ~200-280 | ~400 | - |
| High-Freq. Dielectric Constant ($\epsilon_\infty$) | ~17-18 | ~23-25 | ~33-38 | - |

The journey of lead salt detectors began in the early 20th-century. Significant acceleration leading up to, during, and after World War II included Kutzscher's 1933 discovery of PbS photoconductivity [20] and Cashman's subsequent work on PbS, PbSe, and PbTe[21]. Early applications were predominantly military, such as missile seekers[22]. As the technology matured, a gradual shift towards commercial and industrial applications was driven by improving cost-performance ratios[5]. This transition underscored the increasing need for cost reduction, enhanced long-term stability, improved reproducibility, and more rigorously defined performance metrics, which remain key drivers for contemporary research.

This article focuses specifically on the present status and future prospects of single-element lead salt photodetectors, namely PbS and PbSe for MWIR applications, and PbTe and its alloy PbSnTe for LWIR applications. The discussion covers their fundamental material properties, operating principles including sensitization, current fabrication techniques, performance benchmarks, key challenges, and promising research directions with associated milestones. The discussion will not focus on lead-free alternatives or detector arrays.

## Current Status

Single-element lead salt detector operation is based predominantly on the photoconductive effect, where the electrical conductivity of a polycrystalline semiconductor film changes upon illumination. A critical aspect is the sensitization process, since as-grown films exhibit very low photoresponse. Activation of the infrared sensitivity requires thermal treatment in a specific atmosphere, usually involving oxygen and/or iodine[23]. This complex process induces chemical and structural modifications that lead to the formation of various lead oxides, oxysulfides, or oxyselenides such as $PbSeO_3$ at grain surfaces and boundaries[24, 25]. For PbSe detectors, iodine exposure is often critical. It potentially acts as an n-type dopant, forms compounds like $PbI_2$, or serves as an oxidation enhancer that facilitates recrystallization[26-29]. To date, there is ongoing debate about the dominant operation mechanism. Proposed models include barrier mechanisms, where illumination changes the carrier mobility by modulating the potential barriers at grain boundaries[30]; charge-separation mechanisms, particularly for PbSe, where sensitization may form p-n junctions at crystallite surfaces to create n-type shells around p-type cores to spatially separate the photogenerated carriers, reduce recombination, and enhance lifetime and responsivity, with oxygen passivating the defects or facilitating separation[11, 31-33]; and trap-assisted conduction mechanisms, where some researchers suggest oxygen-related traps are central, with iodine acting as an oxidation enhancer[27, 34, 35].

Chemical Bath Deposition (CBD) has been a widely adopted as a relatively low-cost, solution-based method for depositing lead salt thin films, particularly for photoconductive (PC) devices, which can be adapted for large-area deposition[4]. However, CBD faces challenges in reproducibility and achieving highly uniform films (Figure 1a-c) since its properties are sensitive to substrate conditions[36, 37], although it maintains appeal due to its simplicity and high performance of the resulting detectors. Alternatively, Vapor Phase Deposition (VPD) techniques, including thermal evaporation and sputtering, generally provide better control over film uniformity and reproducibility (Figure 1d-f), and are more readily compatible with standard semiconductor manufacturing processes[38]. While potentially more expensive for small-scale single-element production, VPD can be more cost-effective in mass production.

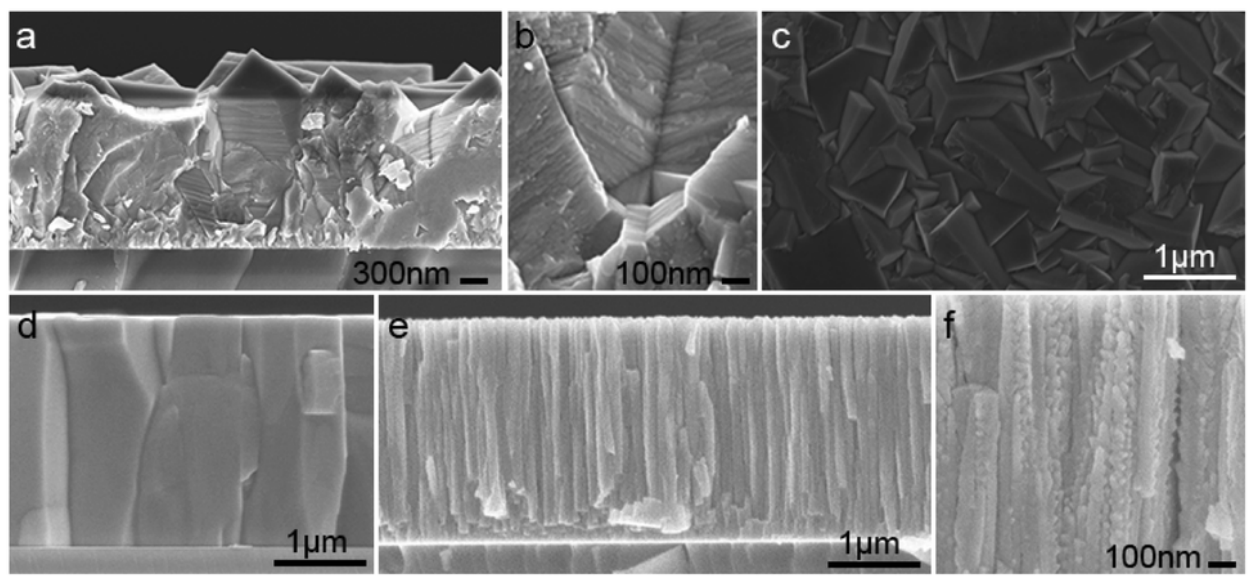

Figure 1 Typical microstructural features of as-grown PbSe films. (a) Low-magnification; (b) high-magnification side-view; and (c) top-view FESEM images of CBD-PbSe films grown 80 °C for 3.0 h. (d) Low-magnification side-view FESEM image of a traditional VPD-PbSe film fabricated at 300 °C in $2 \times 10^{-4}$ Pa.; (e) Low- and (f) high-resolution side-view FESEM images of modified VPD-PbSe film fabricated at 25 °C substrate temperature in $2 \times 10^{-2}$ Pa. Reprinted  from Ref.[39].

The key performance metrics of single-element lead salt detectors include detectivity (D*), responsivity (R), noise equivalent power (NEP), and response time (τ). For PbS at room temperature (295K), D* values typically range from $10^9$ to $10^{11}$ Jones. Uncooled detectors can achieve D* around $10^{10}$ Jones, and cooling improves it beyond $10^{11}$ Jones. For example, Laser Components' uncooled PB25 series offers typical D* values of $3.5\times10^{10}$ Jones at 90 Hz, increasing to $1.1\times10^{11}$ Jones at 650 Hz with responsivity of $8\times10^5$ [40]; time constants are typically in the hundreds of microseconds to milliseconds range[41]. PbSe detectors are typically faster than PbS, with time constants less than 10 μs[42]. Laser Components lists a typical D* of $1.8\times10^{10}$ Jones at 1 KHz for uncooled CBD-grown PbSe, which cooling to -50 °C improves to $3.6\times10^{10}$ Jones[42]. The highest reported detectivity for a CBD-grown PbSe detector, attained by incorporating an antireflection layer, is $4.2 \times 10^{10}$ Jones[43]. Although CBD-grown PbSe detectors are low-cost and high-performance, challenges with uniformity and reproducibility have led researchers to explore alternatives like VPD, which offers uniform films and the capability for monolithic integration with silicon readout circuits[39, 44]. New Infrared Technologies (NIT) LEPTON series (VPD PbSe) quotes typical uncooled Peak D* of $2\times10^9$ Jones, and $1\times10^{10}$ Jones for cooled versions[45]. Compared to the more popular PbSe and PbS materials, fewer papers have been published on PbTe photodetectors. Resonant cavity enhanced (RCE) PbTe detectors have shown D* = $0.72\times10^{10}$ Jones at a resonance wavelength of 3.5 μm[46]. For PbSnTe operating in the LWIR, D* values around $10^{10}$ Jones are typically achieved at 77K[22, 47]. At 85K and 10.2 μm cutoff wavelength, D* = $5.8\times10^{10}$ Jones has been reported for MBE-grown PbSnSe on CaF2-BaF2 on a silicon substrate. While PbS and PbSe perform well near room temperature in the MWIR, PbTe and PbSnTe require cryogenic cooling to attain competitive D* in the LWIR, which reduces their SWaP+C advantages over alternatives like microbolometers. The best reported D* for room-temperature lead-salt photodetectors as well as competing material systems are depicted in Figure 2.

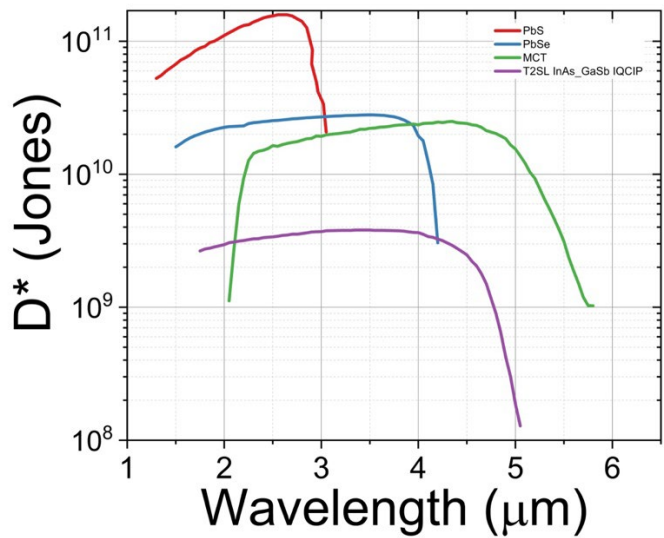

Figure 2. D* Comparison between best reported room-temperature MWIR photodetectors. Data gathered from Refs [48-51]

Single-element lead salt detectors, particularly PbS and PbSe, are widely used in non-military applications due to their low cost and reliable MWIR performance. Major applications include non-dispersive infrared (NDIR) gas sensing for gases such as $CO_2$, CO, and $CH_4$[52-54]. Their high-frequency response also makes them ideal for flame and spark detection in industrial safety systems[55, 56]. In pyrometry, they enable non-contact temperature measurements[57]. Additional uses span industrial process monitoring, medical diagnostics, and environmental monitoring[58-61], which all benefit from MWIR spectral sensitivity for selective measurements without the complexity or cost of cryogenics.[39]

## Key Challenges & Bottlenecks

A central challenge arises from the polycrystalline nature of films typically fabricated via CBD or VPD. While grain boundaries are crucial for sensitization, they also introduce carrier scattering and trapping that degrade the mobility and increase the noise, with variations compromising performance and reproducibility[62, 63]. Structural defects, including point defects and dislocations, serve as recombination centers. Achieving uniformity in film thickness, grain structure, doping, and sensitization remains particularly challenging, especially for CBD-grown films[39]. The sensitization step itself is a significant reproducibility challenge, due to its high sensitivity to process parameters.

Long-term stability and reliability are critical. Lead salt films, especially if not adequately passivated or if the sensitization is incomplete, can be susceptible to environmental factors like oxygen and moisture that lead to performance drift and eventual failure[64-66]. Nanostructures like PbSe CQDs are notably prone to oxidation. The inherent exposure of the active region during sensitization creates a fundamental tension between device activation and long-term environmental stability.

Noise fundamentally limits the detector sensitivity. Dominant types in lead salt detectors include 1/f noise (flicker noise), which is prominent at low operating frequencies and is generally attributed to fluctuations in carrier number or mobility due to trapping/detrapping processes at grain boundaries, surface states, or defects[30, 67-70]. Generation-recombination (G-R) noise, which arises from statistical fluctuations in the carrier generation and recombination rates, is particularly significant in photoconductors and often limits high-performance devices [73,74]. Johnson-Nyquist noise (thermal noise) is fundamental to any resistive element at a finite temperature[26]. Lastly, shot noise arises from the discrete nature of charge carriers and is associated with photocurrent fluctuations[71]. The interplay of these noise sources determines the overall noise floor of the detector. For instance, CBD-fabricated PbSe detectors have been noted to exhibit large 1/f noise. In PC devices, 1/f noise and shot noise (due to the random arrival of carriers at the contacts) are often dominant under bias, while in photovoltaic (PV) devices operating at zero bias, 1/f and Johnson noise are typically the main interference signals.

The D* of a lead-salt photoconductor is strongly influenced by both the applied bias voltage and the chopping frequency, due to the shifting dominance of underlying noise mechanisms. At low bias voltages, noise is primarily limited by dark current shot noise or Johnson noise, with the former typically dominating when the dark current is non-negligible. As the bias increases to a moderate or high level, G-R noise becomes the primary limiting factor, driven by increased carrier generation and gain-dependent fluctuations. Additionally, 1/f noise—associated with traps, grain boundaries, and contact imperfections—dominates at low chopping frequencies (typically below 1 kHz), particularly in polycrystalline films. However, its influence diminishes significantly at higher chopping frequencies, where G-R and shot noise become the prevailing noise sources. Therefore, for optimal D* performance, PbSe detectors should be operated at sufficiently high frequency to suppress 1/f noise, and the bias conditions must be carefully balanced to manage G-R noise without inducing excessive dark current or secondary excess noise phenomena. Identifying the dominant noise regime under specific operating conditions is crucial for tailoring the material quality, device structure, and operating parameters to maximize performance. The performance of lead-salt photoconductors is fundamentally constrained by the trade-off between gain and bandwidth, which is governed by the gain-bandwidth product (GBP). While carrier generation occurs almost instantaneously (~10 fs for 1 μm absorption depth)[8], the practical response time is limited by the carrier recombination lifetime ($\tau_{lifetime}$), transit time ($\tau_{transit}$), and RC time constant. Typical room-temperature $\tau_{lifetime}$ values for sensitized PbSe films range from 1 to 100 μs, while $\tau_{transit}$ spans 5–100 μs for interelectrode spacings of 100–1000 μm and carrier mobilities of 30–100 cm²/V·s[4]. This yields photoconductive gains of 0.3–100, but limits the bandwidth from a few kHz to tens of kHz, especially in high-resistance configurations. Since the gain scales as G = $\tau_{lifetime}$ / $\tau_{transit}$, longer lifetimes improve the responsivity but reduce speed. Conversely, short lifetimes favor high-speed operation at the expense of gain and sensitivity. The low mobility in polycrystalline films exacerbates this trade-off, and RC time constants can also limit the bandwidth of high-resistance devices. Therefore, the optimization of lead-salt photoconductors requires a balancing of material quality, geometry, and biasing conditions to meet the application-specific needs—whether to prioritize high responsivity (e.g., NDIR gas sensing) or high speed.

Another challenge faced by lead-salt materials is environmental regulation. Lead salt photodetectors are increasingly scrutinized by global environmental regulations, particularly the EU RoHS and REACH directives, which restrict the use of hazardous substances in electronic devices[72, 73]. Laser Components Detector Group is working with relevant industrial associates to obtain a further extension of the exemption for medical technology and industrial safety technology beyond the year 2024. The renewal of such exemptions is in process[74, 75] and hinges on the availability of viable lead-free alternatives. As non-toxic materials like InAs/GaSb T2SLs and heavy-metal-free CQDs advance, the case for continuing lead use becomes harder to justify. This creates regulatory and market instability for lead-based technologies. Long-term viability will depend not only on technical innovation but also on the development of robust encapsulation strategies, improved end-of-life recycling, and maintaining a clear exemption pathway through compelling performance and safety justifications.

## Future Research - Materials Science & Fabrication Technologies for Single-Element Lead Salt Detectors

The advancement of single-element lead salt photodetectors is closely tied to progress in materials science, with future efforts focusing on nanostructured materials such as colloidal quantum dots (CQDs), nanowires (NWs), and two-dimensional (2D) materials. These forms offer pathways to the exploitation of quantum confinement, increased sensitivity via high surface-to-volume ratios, and novel device architectures. PbS, PbSe, and PbTe CQDs have gained attention for their size-tunable bandgaps and solution-processability, with PbSe CQDs demonstrating high responsivities in the near- and short-wave infrared[76]; however, challenges remain in synthesizing large, high-quality MWIR CQDs, particularly concerning poor air stability and surface passivation issues for larger PbSe QDs[77-79], as well as difficulties in the cation exchange processes for shell formation[80]. Lead salt NWs offer advantages like efficient charge transport and high photoconductive gain[81], with synthesis methods including template-assisted electrodeposition and colloidal techniques[82-84]; yet, NWs are highly sensitive to surface oxidation and require robust passivation, good doping control, and high-quality contacts[81, 83, 84]. Research into 2D lead salts holds promise for quantum well effects and layer-dependent tunability, but the synthesis of high-quality, stable 2D lead salt sheets remains underdeveloped[85]; current efforts often focus on hybrid heterostructures that combine lead salt CQDs with established 2D materials for transport, underscoring the importance of interface engineering. For instance, PbSe CQDs on $MoS_2$ have demonstrated room-temperature responsivity of 137.6 A/W and detectivity of $7.7 \times 10^{10}$ Jones at 2.55 μm, enabled by effective charge transfer at the interface[86]. Alternative synthesis methods for nanostructured PbSe films, such as CBD with antioxidants[87], spray pyrolysis[88], photoelectrodeposition [89], molecular ink deposition (achieving D* up to $10^9$ Jones)[9], and VPD of PbSe/CdSe nanostructures (achieving D* of $2.17 \times 10^{10}$ Jones and EQE >100%)[11], also show promise for improved performance and functionality.

The long-term stability of lead salt photodetectors, particularly nanostructures, hinges on effective surface engineering, primarily through passivation and encapsulation. Passivation strategies aim to minimize surface traps and environmental degradation, employing techniques like ligand exchange with various organic or inorganic species[90]. Atomic Layer Deposition (ALD) offers conformal, pinhole-free coatings such as $Al_2O_3$ or PbS that are suitable for complex nanostructures[91]. Emerging 2D materials and traditional coatings like chlorides or $As_2S_3$ also offer promising routes[92, 93]. Future advances will likely focus on multifunctional, stable layers.

Resonant Cavity Enhanced Detectors (discussed in a separate sub-topic article of this Roadmap) boost the quantum efficiency and enable narrowband detection by embedding a thin lead salt absorber within an optical resonator[94, 95]. This configuration increases the optical path length and absorption in an ultrathin layer, allowing reduced dark current, faster response, and enhanced D* with high spectral selectivity; PbEuSe RCEDs have achieved <1% spectral linewidth and QE up to 50%. Tunability via MEMS-based movable mirrors is also an area of interest. However, RCEDs demand precise control of the layer thickness and refractive index, typically requiring techniques like MBE.

Integrating plasmonic nanostructures or metasurfaces (discussed in a separate sub-topic article of this Roadmap) with lead salt photodetectors enhances the light absorption, spectral selectivity, and polarization control, which are crucial for boosting the performance of ultrathin active layers[96]. Localized surface plasmon resonances and metasurface-engineered resonators concentrate the electromagnetic field near the absorber, thereby increasing absorption. For instance, PbSe Mie metasurfaces have achieved over 99% absorption[97], and PbSe/PbS CQDs combined with metallic metasurfaces showed substantial responsivity enhancement that reached 375 A/W D* ≈ $10^8$ Jones[37], while other integrations have yielded D* up to $1.22 \times 10^{11}$ Jones[96]. These enhancements permit thinner devices to realize lower noise and faster response, although precise fabrication is required because the performance is highly sensitive to nanostructure geometry and material properties.

Achieving high-performance infrared detection at or near room temperature is a central goal, particularly for LWIR detectors where thermally-generated noise, primarily due to Auger and SRH recombination, is significant. The near symmetry of the conduction and valence bands at the L points of the lead salt band structure offers inherently lower Auger rates than in III-V and II-VI materials[98]. Device-level strategies, including unipolar barrier architectures (e.g., nBn, XBp) that block majority carriers and suppress G-R dark currents[99, 100], are now being explored for epitaxial PbSe[101]. Reducing the detector volume and ensuring high material quality are also critical. Implementing these solutions in polycrystalline or CQD lead salts is challenging.

Mitigation of 1/f noise in lead salt photodetectors can be approached through material, device, and system-level strategies. Surface passivation is critical for suppressing surface-state fluctuations, while contact engineering ensures low-resistance and stable interfaces[102]. Improving the material quality via defect reduction and grain control also reduces carrier fluctuation noise. In terms of device architecture, photovoltaic operation at zero bias can the avoid large dark currents associated with 1/f noise in photoconductors[103]. System-level techniques such as optical chopping and lock-in amplification shift the signal above the 1/f noise knee, although they add complexity and cost[11]. The primary contributors to 1/f noise often lie at surfaces, grain boundaries, and interfaces.

### Roadmap Milestones & Vision for Single-Element Lead Salt Photodetectors:

The sustained advancement and relevance of single-element lead salt photodetectors will rely on a strategic combination of material innovation, device engineering, scalable manufacturing, and responsiveness to environmental regulations. This will position them favorably for applications requiring uncooled or TE-cooled MWIR sensitivity, low cost, and mature processing.

To remain competitive, a phased roadmap is essential. Short-term efforts (1–3 years) should prioritize improving the material stability and device reproducibility, with key goals including enhanced sensitization uniformity in polycrystalline films, development of robust ligand exchange protocols for CQDs to achieve significant EQE with minimal degradation, and demonstration of nanowire-based detectors with $D^* > 5 \times 10^{10}$ Jones and rapid response. Reduction of the 1/f noise is also critical. Medium-term (3–5 years) objectives involve implementing advanced architectures and achieving performance breakthroughs. Priorities include fabricating large-area, flexible CQD arrays with high pixel yields and $D^* > 1 \times 10^{11}$ Jones, developing heterostructure detectors with responsivity >100 A/W, and achieving tunable bandgaps in lead salt alloys. Device-level targets include uncooled LWIR detectors with $D^* > 5 \times 10^{9}$ Jones using Auger suppression and RCEDs with >70% QE. Long-term (5–10+ years) goals should center on disruptive innovation and broader commercialization, including the development of highly stable lead salt nanostructures with near-theoretical QE for LWIR detection near room temperature, as well as the scalable, low-cost manufacturing of flexible, transparent detectors. Continued progress will require breakthroughs in surface passivation, Auger suppression, and defect management. To address lead toxicicty, efforts should aim to reduce the lead content, implement effective encapsulation, and support recycling. Lead salt detectors should focus on niches where their unique advantages are unmatched, such as industrial sensing, environmental monitoring, medical diagnostics, and emerging consumer applications. Integration with readout electronics and on-chip AI will enhance the system-level functionality.

## Concluding Remarks

Single-element lead salt photodetectors remain highly relevant in the infrared sensing landscape, particularly for MWIR and LWIR applications where uncooled or thermoelectrically cooled operation, compact design, and low cost are prioritized. Despite advanced lead-free technologies, lead salt detectors offer a compelling value proposition due to mature processing and favorable SWaP+C characteristics. Ensuring their long-term viability will require a concerted, phased strategy grounded in materials innovation, device architecture advancement, and manufacturing scalability. Short-term objectives include improving the reproducibility, enhancing stability, and reducing 1/f noise. Medium-term goals emphasize high-performance device structures such as flexible CQD arrays and resonant cavity-enhanced devices, supported by scalable fabrication. Long-term success will hinge on breakthroughs in HOT device architectures, Auger suppression, and nanoscale material stability. Future detectors should target specific application niches where the unique strengths of lead salts can be leveraged, while addressing environmental concerns through lead reduction, encapsulation, and recycling to ensure sustainable deployment. This roadmap envisions the evolution of single-element lead salt photodetectors as key enablers of next-generation infrared technologies, which retain a vital role by aligning research with emerging application demands.

# 11. Type-II superlattice mid- and long-wavelength infrared detectors

DAVID Z. TING*

*NASA Jet Propulsion Laboratory, California Institute of Technology, Pasadena, CA 91109, USA*
*David.Z.Ting@jpl.nasa.gov

## Overview

MCT is the most successful infrared photodetector material to date, benefitting from remarkable properties such as the very long Shockley-Read-Hall (SRH) lifetimes, and from over 80 years of development (reviewed in a separate sub-topic article of this Roadmap). MCT grown on nearly lattice-matched CdZnTe (CZT) substrates offers continuous cutoff wavelength ($\lambda_{cutoff}$) coverage from the short-wave infrared (SWIR) to the very long wavelength infrared (VLWIR), while providing high quantum efficiency (QE) and low dark current for high-performance applications. On the other hand, MCT is a II-VI semiconductor with relatively weak ionic Hg-Te bonds and high Hg vapor pressure; it is soft and brittle and requires extreme care in growth, fabrication, and storage, particularly for long wavelength infrared (LWIR) and VLWIR alloys that contain high Hg fraction. In general, III-V semiconductors are more robust than their II-VI counterparts due to stronger, less ionic chemical bonding. III-V semiconductor-based infrared focal plane arrays (FPAs) excel in operability, spatial uniformity, temporal stability, scalability, producibility, and affordability (the so-called 'ility' advantages). InGaAs FPAs with $\lambda_{cutoff}$ ~1.7 μm dominate the SWIR market because they perform nearly at the theoretical limit. And despite a significantly lower operating temperature than MCT of comparable cutoff wavelength, InSb FPAs ($\lambda_{cutoff}$ ~5.3 μm) dominate the mid-wavelength infrared (MWIR) market in volume due to superior manufacturability, yield, and affordability.

As illustrated in Figure 1, traditional bulk III-V semiconductor detectors grown on (nearly) lattice-matched substrates are limited by the lack of wide cutoff wavelength adjustability found in MCT. Other than InGaAs and InSb, which have achieved wide success respectively in the SWIR and MWIR, InAs ($\lambda_{cutoff}$ ~ 3.4 μm) on InAs substrate and $InAs_{0.91}Sb_{0.09}$ ($\lambda_{cutoff}$ ~ 4 μm) lattice-matched to GaSb substrate are also possible. However, they had not been practical due to surface leakage problems (until introduction of the unipolar barrier detector architecture). The past two decades have seen accelerated progress in III-V semiconductor infrared photodetector technology. Advent of the unipolar barrier device architecture such as the nBn [1], the XBn [2], the complementary barrier infrared detector (CBIRD) [3,4], the double heterostructure (DH) [5], and the pMp [6] has in many instances greatly alleviated the generation-recombination (G-R) and surface-leakage dark current issues that had been problematic for many III-V photodiodes. Meanwhile advances in a variety type-II superlattices (T2SLs) such as InGaAs/GaAsSb, InAs/Ga(In)Sb, InAs/InAsSb, and InGaAs/InAsSb, as well as in bulk III-V material such as InGaAsSb and metamorphic InAsSb, have provided continuously adjustable cutoff wavelength coverage from the SWIR to the VLWIR. Together these developments have led to a new generation of versatile, cost-effective, high-performance infrared detectors and FPAs based on robust III-V semiconductors, providing a viable alternative to MCT. As illustrated in Figure 1, alloy and T2SL detector structures that can be grown lattice-matched to, or pseudomorphically on, GaSb substrates now provide continuous and wide-ranging cutoff wavelength coverage. This section briefly discusses mid- and long-wavelength III-V T2SL infrared detectors; more detailed information can be found in review articles in the literature [7-13].

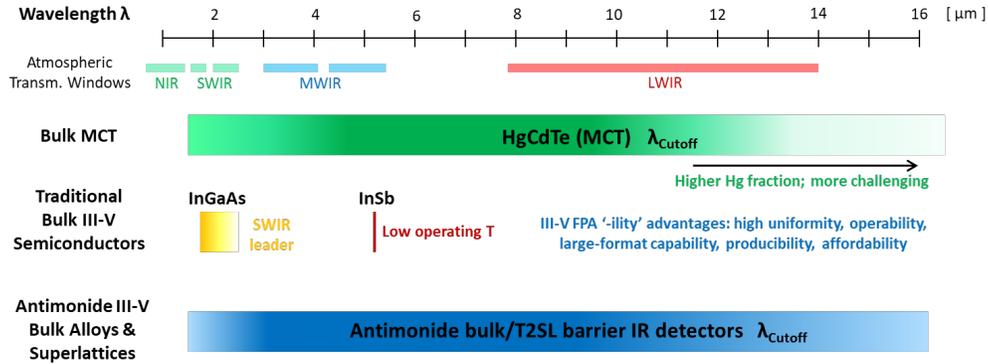

*Figure 14. . Illustrations of cutoff wavelength coverage of HgCdTe, bulk InGaAs and InSb, and antimonide alloys and superlattices.*

## Current status

A key contributing factor to the current success of T2SL infrared detectors has been the development of the unipolar barrier device architecture. Although III-V type-II superlattices can  be less susceptible to band-to-band tunneling than bulk semiconductors [14] and can achieve reduced Auger recombination in properly designed structures [15], they have also had to contend with a higher SRH generation rate (and hence larger depletion dark current) than in MCT. However, the introduction of nBn [1] and XBn [2] structures showed that unipolar barriers can substantially suppress depletion dark currents in III-V infrared detectors. A unipolar barrier is a special heterostructure construct that blocks one carrier type (electron or hole), but allows unimpeded flow of the other. The unipolar barrier concept was originally introduced in the context of double- heterostructure (DH) lasers [16], although they were not referred to as such at the time. Unipolar barriers can be used in many ways to improve MCT and III-V infrared detector performance. A unipolar barrier inserted into a photoconductor can block the flow of majority carrier dark current without impeding minority carrier transport [17]. Similarly, a unipolar-barrier detector design can reduce the diffusion dark current in a photodiode [18]. The non-equilibrium high operating temperature (HOT) detector [19] can use the DH architecture with unipolar electron and hole barriers for carrier exclusion and extraction, as can the fully-depleted HOT detector [20]. The successful demonstration of nBn/pBn detectors [1,2] with MWIR InAsSb absorbers led to wide adoption of the unipolar barrier device architecture for T2SL based infrared detectors, initially for InAs/GaSb T2SL absorbers [3,21], and later with InAs/InAsSb T2SL and InGaAs/InAsSb T2SL absorbers.

InSb has been one of the most widely used infrared photodetector material. It is a robust III-V semiconductor with favorable FPA manufacturability, and its band gap is ideally suited for covering the 3–5 μm MWIR atmospheric transmission window. The InSb FPA offers the III-V 'ility' advantages and has dominated the MWIR FPA market in volume. Despite its success, however, InSb has notable drawbacks compared to MCT, the major competing MWIR FPA technology. For example, the SRH lifetime of ~400 ns for InSb is short compared to that of MWIR MCT [22]. Unlike MCT, which can be passivated effectively using the high-bandgap CdTe, InSb lacks heterostructure capability, and achieving good surface passivation is more challenging. As a result, InSb FPAs operate at much lower temperatures (~80 K for ion-implant planar InSb; 95–100 K for MBE grown epi-InSb) than MWIR MCT FPAs [23].

In 2012, unipolar barrier detectors with InAs/InAsSb T2SL absorbers were reported in the MWIR [24] and the LWIR [25]. This turned out to be a breakthrough for MWIR photodetectors. The InAs/InAsSb T2SL has some notable contrasting advantages over InSb. For example, MWIR InAs/InAsSb T2SL samples have been reported to have SRH lifetimes of ~10 µs [26]; which is much longer than those for InSb. A variety of heterostructures, including electron and hole unipolar barriers, can be formed for InAs/InAsSb T2SL absorbers for device performance enhancement. Furthermore, surface passivation is a less demanding issue for InAs/InAsSb T2SL unipolar barrier detectors, since the high band gap barrier material can preclude exposure of the narrow gap absorber at the detector mesa top surface. The cutoff wavelength of the InAs/InAsSb T2SL absorber can be adjusted from ~4 µm to the VLWIR. MWIR FPAs made from unipolar barrier detectors with InAs/InAsSb T2SL absorbers also operate at 40 to 50 K higher temperatures than InSb FPAs [27], while retaining the same III-V semiconductor manufacturability and affordability benefits. The higher operation temperature enables the use of smaller cryocoolers for compact sensor engines. As a result, InAs/InAsSb T2SLs have replaced InSb MWIR FPAs in many applications.

The InAs/GaSb T2SL infrared detector technology is well-established, with FPAs and sensor engines routinely manufactured by, for example, Fraunhofer IAF/AIM [28], SCD [29], IRnova [30], and i3systems [31]. The emergence of InAs/InAsSb T2SL infrared detectors has added to the capability and versatility of T2SL infrared detector technology. Research investigations of InAs/InAsSb T2SLs are now ubiquitous, with results reported from many groups worldwide. SCD from Isreal reported modeling results of InAs/InAsSb superlattice detectors [32]. Collaborating groups from China reported results on MWIR detector grown on GaSb [33], the Military University of Technology and Vigo of Poland and National Taiwan University demonstrated microlens enhanced LWIR detectors grown on GaAs [34], Lancaster and Warwick in the UK demonstrated MWIR detectors grown on Si substrates [35], as have Groups from Montpellier and Grenoble in France [36]. Researchers from i3system and collaborators in Korea reported high operating temperature MWIR FPA results [37]. Fraunhofer IAF, Germany, reported MWIR InAs/InAsSb T2SL nBn detectors with low diffusion-limited dark current and good QE [38]. ASELSAN, Turkey, demonstrated MWIR FPA and integrated detector dewar cooler assembly (IDDCA) with >99.5% operability at 110 K [39].

While most of the more recent InAs/InAsSb T2SLs have been grown on (100) GaSb substrates by MBE, growths by a variety of other modes have also been demonstrated. MWIR and LWIR detectors have been grown by MOCVD [40]. Growths on GaAs [34], and Si [35,36] have been reported. Growth of MWIR and LWIR detectors on (211)A and B, and (311)A and B GaSb substrates have also been reported [41]. The influence of proton radiation on photoluminescence, minority carrier lifetime, and device characteristics in InAs/InAsSb superlattice structures has been investigated [42]. Deep levels in InAs/InAsSb superlattice were revealed by analysis of forward-bias tunneling current [43]. Different optical field enhancement techniques have been applied to InAs/InAsSb T2SL detectors, including 1D resonant cavity [44], Fabry-Pérot cavity with metallic nano-antennas [45], and plasmonic detector architecture [46].

## Challenges and opportunities

In a T2SL, the wavefunctions of the electron and hole band-edge states are largely located in separate layers. As a result, the band gap of a T2SL can be smaller than those of its constituent semiconductors. While this enables T2SLs to achieve the small band gaps needed for thermal infrared applications, it comes with inherent disadvantages. Having the band-edge electron and hole wavefunctions located separately reduces the oscillator strength and hence the absorption coefficient. Figure 2(a) shows the cutoff wavelengths (derived from band gaps calculated using the effective bond orbital model [47]) for sets InAs/GaSb T2SL (solid blue) and InAs/InAsSb (dashed red) T2SLs. In general, larger superlattice periods are needed to achieve the longer cutoff wavelengths. Also, because InAs/GaSb has a stronger type-II band offset than InAs/InAsSb, the InAs/GaSb T2SL can achieve the same LWIR cutoff wavelength with a shorter period than an InAs/InAsSb T2SL. Figure 2(b) shows selected bulk and superlattice absorption coefficients calculated using the NRL MULTIBANDS® modeling software [48]. In MWIR, the calculated absorption coefficients of InAs/GaSb and InAs/InAsSb T2SLs are somewhat smaller but comparable to that of InSb. In the LWIR, the calculated absorption coefficients of T2SLs are significantly smaller than that for the bulk $InAs_{0.4}Sb_{0.6}$, although we should keep in mind that while the superlattices can be grown pseudomorphically on GaSb substrates, there is no lattice-matched substrate for bulk $InAs_{0.4}Sb_{0.6}$. The relatively weak absorption is a challenge for (V)LWIR T2SL.

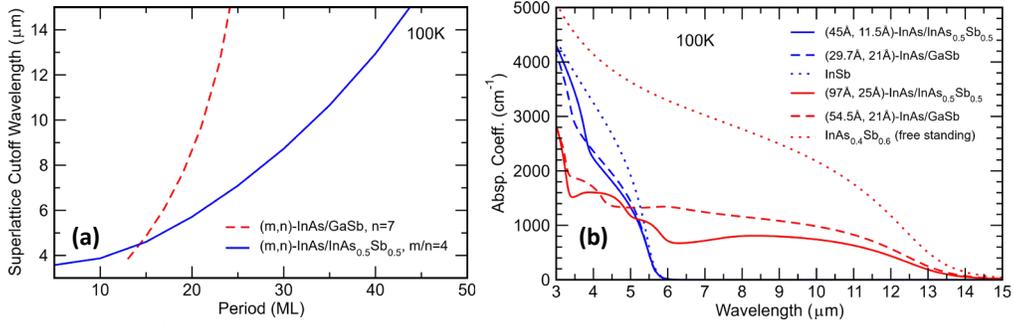

Figure 15. **Error! Reference source not found.** (b) Calculated absorption coefficients for MWIR and LWIR bulk and superlattice materials.

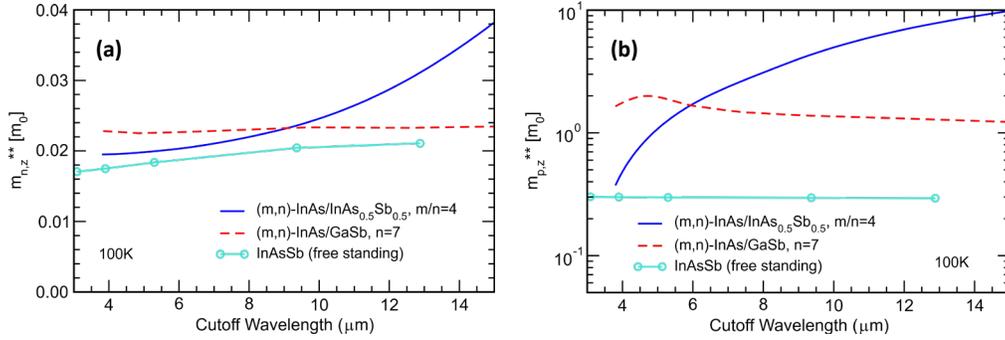

Figure 16. (a) Calculated growth-direction electron conductivity effective mass $m_{n,z}$** as a function of cutoff wavelength for selected bulk and superlattice materials, and (b) the corresponding growth-direction hole conductivity effective mass $m_{p,z}$**.

Because of the smaller absorption coefficient, a thicker LWIR T2SL absorber is needed to achieve the same absorption QE as a bulk material. In addition, to achieve good external QE, the diffusion length needs to be sufficiently long to accommodate the needed absorber thickness. The diffusion length depends on the conductivity effective mass of the minority carrier. Figure 3 shows calculated electron and hole conductivity effective masses [49] along the growth (transport) direction. While the T2SL and bulk electron conductivity effective masses are generally comparable, the T2SL hole mass can be significantly larger than for a bulk semiconductor material, especially for a InAs/InAsSb T2SL in the (V)LWIR. This strongly favors electron transport (in a *p*-type absorber) over hole transport (in an *n*-type absorber) in these T2SLs.

Although *p*-type T2SL absorbers have more favorable diffusion lengths and QEs, surface dark current considerations complicates the picture. For example, in InAs/InAsSb T2SLs the exposed surface is always degenerate *n*-type, regardless of whether the InAs/InAsSb T2SL itself is doped *n*-type or *p*-type. Near the *n*-type absorber surface, the accumulated surface potential repels the minority carriers (holes) from the etched absorber sidewall surface, which consequently is relatively benign. However, for a *p*-type absorber the surface band bending is problematic. The degenerate *n*-type surface of the *p*-type absorber attracts minority carriers (electrons); the inverted surface also creates a surface *p-n* junction with a sub-surface depletion region, and is subject to various surface dark current mechanisms [50]. Therefore, good passivation is crucial for devices with exposed *p*-type absorber surfaces.

Having sufficient QE is especially important in minimizing spectral crosstalk in dual-band detectors [51]. In a dual-band detector containing a larger band gap ($E_{g,1}$) material Absorber 1 and a smaller band gap ($E_{g,2}$) material Absorber 2, Absorber 1 serves to detect photons with $h\nu > E_{g,1}$ (Band 1), while Absorber 2 is designated to detect photons with $E_{g,2} < h\nu < E_{g,1}$ (Band 2). Ideally Absorber 1 should absorb nearly all the incident Band 1 photons in one pass, leaving Absorber 2 to subsequently absorb Band 2 photons. If there is insufficient absorption QE and Absorber 1 cannot absorb all the photons in Band 1 in a single pass, the surviving Band-1 photons can still be absorbed by Absorber 2, and creating a false signal in Absorber 2. This spectral crosstalk (i.e., Absorber 2 detecting photons in Band 1) needs to be minimized in order that we can clearly distinguish Band 1 and Band 2 photons.

We have briefly discussed the primary challenges for T2SL infrared detectors in this subsection; some additional details have been reported in the literature [49,50,52].

### Future directions and outlook

We conclude this subsection by identifying a few topics for future development in T2SL infrared detectors, and point out some promising approaches that merit further investigation. Improving QE and reducing dark current are always of interest for infrared photodetectors, as they lead to increased signal to noise ratio and/or detector operating temperature. As discussed earlier, compared to bulk material, LWIR T2SL have noticeably weaker absorption, which poses a challenge for achieving high QE. Improved understanding of carrier transport [11,53,54] may lead to T2SL absorber with longer minority carrier diffusion length, allowing the use of thicker absorbers for higher QE. Alternatively, instead of increasing absorber thickness, optical field concentration into a smaller detector volume using immersion lenses [55], metalenses [56], plasmonics, metastructures and resonant cavities [44-46,57] can enhance QE; concentrating light into a smaller volume also reduces dark current generation. One of the problems in raising photodiode operating temperature is the reduction of minority carrier diffusion length at higher temperatures due to increased phonon scattering. Multi-junction cascade detectors (discussed extensively elsewhere in this paper) make use of a series of shorter absorber segments connected by tunnel junctions, and offer an effective approach for circumventing this problem. T2SL cascade detector can be and has been used in combination with immersion lens for increased operating temperature [58]. T2SL surface passivation has been an active research area [59]; continued development in the effective and consistent surface passivation is critical for LWIR T2SL detector dark current suppression. Surface dark current can also depend strongly on the device structure design and the accompanying processing scheme, as demonstrated with the inverted device structure [60] and the reticulated shallow etch mesa isolation (RSEMI) two-step etch scheme [61,62]. Despite the relatively long SRH lifetimes reported for MWIR InAs/InAsSb T2SL, in general T2SL minority carrier lifetimes are still significantly shorter than those of high-quality low-background HgCdTe [20,63]; the fundamental reasons for this are not well understood. It may be worthwhile to explore whether T2SL SRH lifetimes could be increased significantly. If so, T2SL would be able to take advantage of the fully depleted device architecture [20] for very low dark current operation.

Other topics of interest include very large format FPAs, advanced low-power consumption high-dynamic range readout integrated circuits (ROICs), and improved radiation tolerance for space applications.

The outlook for T2SL infrared detectors is quite positive. There is already a good understanding of the basic properties of infrared T2SLs, and the effective use of heterostructures for improving detector performance. Benefitting from the horizontal integration model established under the Vital Infrared Sensor Technology Acceleration (VISTA) U.S. government program [64], the infrastructure for manufacturing antimonide-based T2SL infrared detectors is well established. GaSb substrates in 2", 3", 4", and 5" diameter formats are available commercially, while 6" is available for sampling. Material growth foundries have established multi-wafer growth capabilities, demonstrating high-uniformity growths of antimonide detectors on 6" GaSb substrates [65], as well as on 8" GaAs, Ge, and Si substrates via metamorphic buffers [66]. The availability of large-diameter substrates is beneficial for the economy of scale, and for very large format FPA development. The growth foundries have also established material characterization, and device fabrication and characterization capabilities to facilitate rapid feedback on material quality. This infrastructure has enabled the rapid commercialization of T2SL infrared detector technology.

### Acknowledgements

The author thanks A. M. Fisher, B. J. Pepper, C. J. Hill, S. A. Keo, S. B. Rafol, A. Khoshakhlagh, A. Soibel, S. D. Gunapala, M. Bouchet, J. N. Schulman, X. Cartoixà, E. S. Daniel, S. V. Bandara, and L. Höglund. A part of the research described in this publication was carried out at the Jet Propulsion Laboratory, California Institute of Technology, under a contract with the National Aeronautics and Space Administration (80NM0018D0004).

# 12. Quantum-engineered interband cascade infrared photodetectors

**RUI Q. YANG,**[1,*] **MICHAEL B SANTOS,**[2]

[1]*School of Electrical and Computer Engineering, University of Oklahoma, Norman, OK 73019, USA*
[2]*Department of Physics and Astronomy, University of Oklahoma, Norman, OK 73019, USA*
*\*Rui.q.Yang@ou.edu*

## Overview

Conventional photodetectors are based on electronic transitions across the bandgap of a single-stage continuous semiconductor layer by absorbing incident photons and collecting the photogenerated carriers in an external circuit. The thermal noise (Johnson noise) limited detectivity $D^*$ of a photodetector, which essentially characterizes the signal-to-noise ratio (S/N), is determined by the quantum efficiency (QE) $\eta$, representing the signal, and the thermal noise current. The QE increases with the absorber thickness ($d$) and absorption coefficient ($\alpha$), while the thermal noise current also increases with the absorber thickness. Hence, the benefit to $D^*$ decreases with further increase of the absorber thickness after a certain value of $d$. In the case of an infinite diffusion length ($L$), the thermal-noise-limited detectivity $D^*$ reaches an ultimate peak value $D_p^*$ at an optimal absorber thickness $d_o = 1.26/\alpha$, and is given by [1,2]

$$D_p^* = (0.319)\frac{\lambda}{hc}\sqrt{\frac{\alpha}{g_{th}}}, \qquad (1)$$

where $h$ is the Planck constant, $c$ is the speed of light, $\lambda$ is the wavelength of the incident photons, and $g_{th}$ is the thermal generation rate. At this optimal absorber thickness $d_o$ for the peak detectivity $D_p^*$, the QE is constrained to 71%.

In real devices, the diffusion length $L$ is finite and decreases with operating temperature T when T is relatively high. Particularly in narrow bandgap materials, the absorption coefficient $\alpha$ is small and $L$ is typically shorter than $1/\alpha$ at high temperatures so that the $\alpha L$ product is less than one. In this case, many of the photogenerated carriers will recombine before they can be collected in the external circuit. Consequently, to reduce the noise and achieve a maximum detectivity $D_M^*$, the optimal absorber thickness needs to be reduced [3]. As shown in Fig. 1, in a single-stage detector this value of $D_M^*$ with a finite $L$ is smaller than $D_p^*$ and depends on $\alpha L$ [3-4], especially when $\alpha L<1$. The diffusion length limitation can substantially affect the performance of a conventional photodetector at elevated temperatures.

However, this diffusion length limitation can be circumvented in the quantum-engineered interband cascade infrared photodetector (ICIP) architecture [5], as schematically shown in Fig. 2, which originated from interband cascade lasers (ICLs) [6-7]. The first experimental demonstration was reported based on ICL structures in 2005 [8], and with InAs/GaSb type-II superlattice (SL) absorbers in 2010 [9].

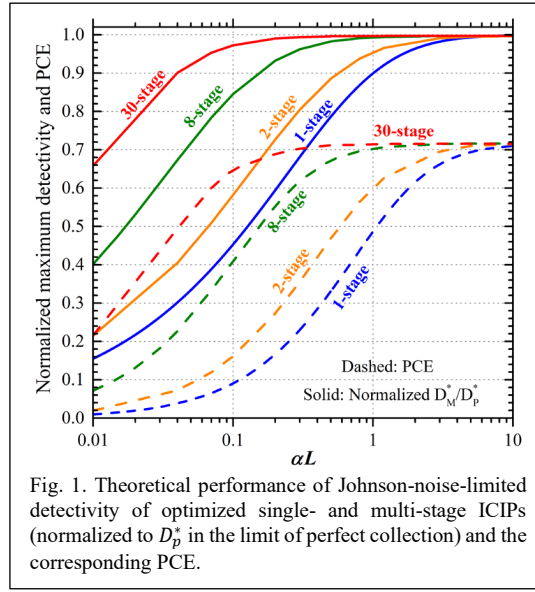

Fig. 1. Theoretical performance of Johnson-noise-limited detectivity of optimized single- and multi-stage ICIPs (normalized to $D_p^*$ in the limit of perfect collection) and the corresponding PCE.

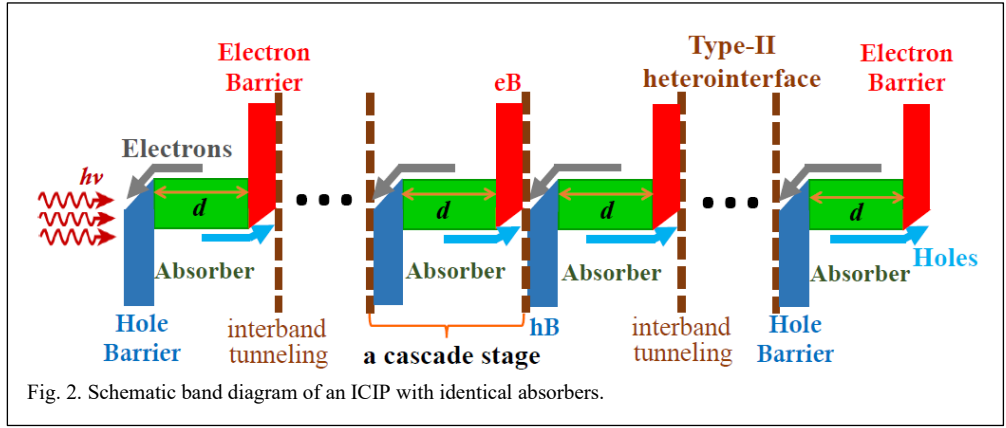

Fig. 2. Schematic band diagram of an ICIP with identical absorbers.

An ICIP structure comprises thin discrete absorbers that are separated by unipolar barriers – an electron barrier (eB) and a hole barrier (hB), to form multiple cascade stages. Adjacent stages are connected in series through smooth interband tunneling facilitated by type-II broken-gap heterointerfaces in the InAs/GaSb/AlSb material system [10-11]. The unipolar barriers have inclined band-edge profiles (Fig. 2) to induce fast extraction of the photogenerated carriers. When electron-hole pairs are created by photoexcitation in one stage, the electrons and holes diffuse toward opposite edges of the stage and then recombine with carriers from adjacent absorbers at the type-II heterointerface between the eB and hB as shown in Fig. 2. Hence these interfaces, together with contact layers at the two ends of the ICIP structure, act as equivalent collection points for photo-generated carriers. The particle conversion efficiency (PCE) [3], which is defined as the sum of the ratio of collected carriers in every stage to the incident photons, is a more appropriate figure of merit for ICIPs instead of the conventional external QE. These distinct features in ICIP structures enable the carriers to move quickly toward a collection point that is at most only a single cascade stage distance away, a distance designed to be shorter than the diffusion length. This design allows the total absorber thickness, the sum of all the individual absorber thicknesses, to be significantly thicker than the diffusion length, thereby circumventing the diffusion length limitation. The thin discrete absorbers and series connections make the device resistance high, so that noise is suppressed in an ICIP structure, and the photocurrent flows more smoothly to an external circuit. As such, the PCE and the attainable maximum detectivity are universally higher in ICIPs than in the conventional single-stage detectors when the diffusion length is finite as shown in Fig. 1. For example, the ultimate value of $D_M^*$ for the ICIP with 30 identical discrete absorbers is nearly a constant from the highest number until $\alpha L$=0.1, with the PCE maintaining approximately the optimal value of 71%.

These theoretical results and features indicate that the ICIP architecture should provide feasible advantages for high-temperature and high-speed operation with optimal device performance. Each cascade stage in an ICIP structure is like a double heterostructure unit with complementary unipolar barriers [12] and retains all the advantages of barrier structures. Hence, the discrete absorber architecture of ICIPs provides a great deal of flexibility for quantum engineering of cascade stages and manipulating carrier transport to achieve high-temperature and high-speed operation without compromising the absorption QE. The ICIP architecture would be especially beneficial to long wave (LW) and very LW infrared photodetectors based on absorber materials such as Ga-free InAs/InAsSb SLs, where a small absorption coefficient and short diffusion length (particularly for holes) are typical concerns.

## Current status

The experimental development of ICIPs based on InAs/GaSb SL absorbers was initiated at the University of Oklahoma (OU) in 2008. The peak responsivities of initial 7-stage ICIPs with cutoff wavelengths in the mid-wave (MW) region (4-7 μm) at 300 K exceeded 200 mA/W, and were relatively insensitive to the operating temperature [9,13]. This verified the advantage of efficiently collecting photogenerated carriers in ICIPs with relatively thin absorbers (0.15-0.25 μm). Afterward, a group at the University of New Mexico (UNM) reported the operation of a 7-stage ICIP based on the same InAs/GaSb SL absorber design as in [9] at temperatures up to 420 K with a cutoff wavelength 7 μm [14], the highest operating temperature ever reported for a MW infrared (IR) photodetector. This validated the advantages of ICIPs for high temperature operation. Subsequently, Pusz et al. [15] used a device made from the same ICIP structure to achieve a Johnson-noise-limited $D^*$ of $3 \times 10^9$ Jones at room temperature (RT). Their reported response time of this device at RT was 5 ns at zero bias and decreased to ~1 ns with increasing reverse bias voltage, which verified the short transit time of carriers in an ICIP with thin absorbers. Also, the resistance-area products ($R_0A$) of all early ICIPs were substantially higher than those of state-of-the-art HgCdTe (MCT) photodiodes when compared at RT [15]. The higher values of $R_0A$ in ICIPs provided solid evidence of reduced noise compared to conventional detector structures with a single absorber.

Later, additional groups began to research and develop ICIPs based on the InAs/GaSb/AlSb material system, including teams in China, Poland, Germany, Austria, France, and JPL/NRL [16-22]. ICIPs were made with encouraging results for a wide range of cutoff wavelengths from the short-wave (SW) to the very LW IR [23-24], although they were not optimized for device performance. Significant advances were achieved in understanding and demonstrating the advantages of ICIPs based on InAs/GaSb type-II SLs and ICL structures [17-21, 25-32]. These include, for example, the operation of LW ICIPs above RT with a detectivity exceeding $10^8$ Jones at 300 K [25], demonstration and interpretation of multiple negative differential conductance features and their unusual temperature dependence [26], discovery and explanation of electrical gain [27-29], extraction of thermal generation rate and carrier lifetime [30], focal plane arrays (FPAs) at high operating temperatures (up to 180 K) [31] and with noise equivalent temperature difference (NETD) of 27 mK at 125 K [32], and RT ICIPs on a GaAs substrate displaying high detectivity [17]. The high frequency operation of ICIPs has also been demonstrated, with a 3-dB bandwidth of 1.3 GHz at zero bias [33] and 7.04 GHz at a reverse bias [19-20]. When measured with a microwave probe, a 50 μm×400 μm ICL-based ICIP device exhibited a flat microwave response region over 10 GHz, demonstrating the viability of high bandwidths following optimization of the packaging and impedance matching [21]. Furthermore, a monolithically integrated mid-IR ICL and ICIP exhibited a record-high detectivity (~2×$10^{10}$ Jones) for RT operation [34], paving the way for important applications such as on-chip miniaturized sensors, spectrometers, optical communication and optical processing.

Recently a 3-stage ICIP with a cutoff wavelength near 5 μm at RT was used for high-speed free-space optical (FSO) communication applications, and demonstrated a data rate of 12 Gbit/s with an on–off keying scheme and 14 Gbit/s with a 4-level pulse amplitude modulation scheme [22]. The system with the ICIP had a much lower noise floor (<-45 dB) than the system (>-30 dB) with a quantum cascade detector (QCD) [discussed in a separate sub-topic article of this Roadmap]. The wide spectral response, broad bandwidth, and increased sensitivity of ICIPs make them highly suitability for FSO communication. It is expected that RT ICIPs with a cutoff wavelength in the LW atmospheric transmission window (8-12 μm) should also have similar advantages in terms of low noise, high speed and enhanced sensitivity.

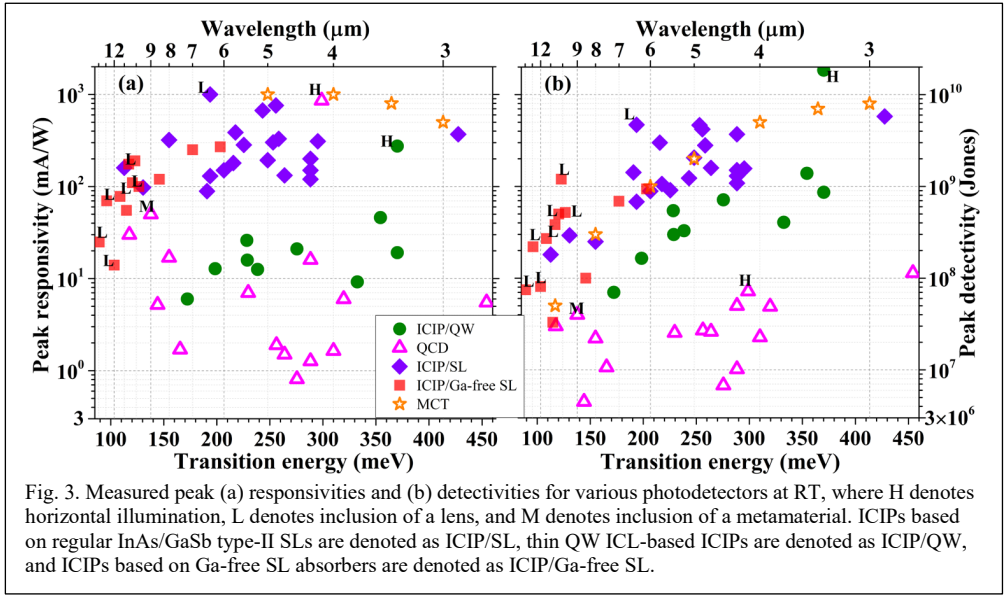

Fig. 3. Measured peak (a) responsivities and (b) detectivities for various photodetectors at RT, where H denotes horizontal illumination, L denotes inclusion of a lens, and M denotes inclusion of a metamaterial. ICIPs based on regular InAs/GaSb type-II SLs are denoted as ICIP/SL, thin QW ICL-based ICIPs are denoted as ICIP/QW, and ICIPs based on Ga-free SL absorbers are denoted as ICIP/Ga-free SL.

To compare with MCT photodetectors and multi-stage QCDs based on intersubband transitions within the conduction band, Fig. 3 presents the measured responsivity and detectivity values reported in the literature [5,35-42] and discussed above. The metrics are plotted as a function of transition energy $\Delta E$, which is the bandgap for ICIPs and MCT or the energy separation of the two relevant conduction subbands for QCDs. Compared to RT MCT photodetectors that are commercially available from Vigo [36], Fig. 3 shows that ICIPs have already demonstrated comparable responsivities and higher detectivities in a relatively long wavelength ($\geq 5$ μm) region, even though the reported ICIPs were not optimized. This suggests that ICIPs have a high potential to overtake MCT photodetectors across a wide wavelength spectrum at RT and above.

Owing to a substantially longer carrier lifetime $\tau$ in Ga-free InAs/InAsSb SLs compared to InAs/GaSb SLs, it is expected that ICIPs with Ga-free SL absorbers will have improved device performance. Equation (1) predicts higher detectivity, since $g_{th}$ is inversely proportional to $\tau$. Preliminary ICIPs with Ga-free InAs/InAsSb SL absorbers were reported with RT cutoff wavelengths of ~8.5 μm [37], as well as 10.8 μm, 12 μm and 7 μm [38-41]. Very recently, RT operation of Ga-free SL-based ICIPs was extended to longer wavelengths (e.g. a 50% cutoff wavelength of 13.9 μm) [42]. Some devices were fabricated with an optical immersion lens on the GaAs substrate [39-42]. However, we see from Fig. 3 that without an optical lens, the reported Ga-free SL ICIPs have not yet achieved better performance than regular InAs/GaSb type-II SL ICIPs. This was partially attributed to the relatively low absorption coefficient $\alpha$ in Ga-free SLs, which reduces responsivity and partially cancels the benefit of a longer lifetime for enhanced $D^*$ implied in Eq. (1).

Compared to QCDs, which cannot be illuminated vertically, the responsivity of an ICIP is generally higher, especially when a SL absorber is used to enhance the absorption. The QCD's lower responsivity is partially attributed to the low escape probability that is proportional to the carrier lifetime [43]. This value is close to 100% for ICIPs, due to a much longer lifetime (~ns or more) [9] compared to a short intraband scattering time (~ps). Another factor is related to the polarization selection rule for intersubband transitions within a QW's conduction band, which prohibits the absorption of vertically-incident light in a QCD. To accommodate the incident light, the QCD facets are typically made by polishing at an angle of 45° to the growth direction. Alternatively, the light illumination can be along a horizontal direction to achieve a high responsivity as shown in Fig. 3(a) with a value of 860 mA/W from a single-stage QCD near 4.15 μm [44]. As indicated in [4], the maximum QE in a QCD or ICIP is inversely proportional to the number of cascade stages ($N_s$). Hence, in Fig. 3(a) with horizontal illumination the responsivity of the one-stage QCD is higher than for the 6-stage ICIP/QW. The responsivity can also be improved by incorporating a metamaterial that enhances the light-matter interaction. This was demonstrated in a QCD with peak photo-response near 9 μm at RT [45], in which the responsivity of ~50 mA/W was comparable to those in ICIP/SLs as shown in Fig. 3(a). However, due to significant noise with a high dark current density, the detectivity $D^*$ (e.g. ~4×10$^7$ Jones with a metamaterial) of QCDs is several times to an order of magnitude lower than that in ICIPs with a similar ΔE as shown in Fig. 3(b).

## Challenges, opportunities and future developments

ICIPs based on type-II SL absorbers share similar advantages and disadvantages with single-stage type-II SL photodetectors, which are reviewed in a separate sub-topic article of this Roadmap. They can cover a wide spectral range from SWIR to VLWIR by adjusting the layer thicknesses of the SL without changing the constituent materials. However, because the electron and hole wavefunctions are located primarily in different layers of the SL, the absorption coefficient arising from interband transitions across the bandgap is small compared to that for bulk material or a type-I SL. This results in a reduced ultimate peak detectivity, as indicated by Eq. (1). Nevertheless, ICIPs are not limited to type-II SL absorbers, since bulk material absorbers can be used as demonstrated initially with a quaternary GaInAsSb alloy [46]. Other challenges with type-II SLs include a relatively short carrier lifetime compared to MCT materials and surface passivation in the device fabrication, which result in a high thermal generation rate and high dark current density. For example, a p-type type-II SL absorber is preferred in terms of longer minority carrier (electron) diffusion length. However, surface band bending in a p-type absorber causes accumulation of electrons on the surface and formation of a surface *pn* junction, resulting in a substantial surface dark current density [47-48]. Besides the requirement of good surface passivation, increasing the electron barrier thickness may help to reduce the dark current density [13, 25-26], which has not been systematically explored. Another issue related to operation of ICIPs at high temperatures is the lack of systematic and reliable material parameters such as absorption coefficient and diffusion length for type-II SLs at high temperatures, as many past studies focused on the operation of type-II SL detectors at relatively low temperatures. This likely affects the optimization of device design, but should be resolved with more efforts in the future.

With the development and availability of efficient semiconductor mid-IR light sources such as ICLEDs, ICLs [49-50] and quantum cascade lasers (QCLs) [51], the demand for ICIPs is expected to increase for applications such as FSO communications, heterodyne detection, and photonic integrated circuits (PICs). Because ICIPs can simultaneously have a high sensitivity and operate at high speed, and are compatible with efficient ICLEDs and ICLs, they will be more desirable in applications that require low power consumption.

Although ICIPs have been demonstrated with advantages and flexibility to circumvent the diffusion length limitation, they are still in a relatively early phase and the ultimate performance projected by theory has not been achieved yet. The research and development of ICIPs that incorporate optical structures such as metamaterials or a resonant cavity (reviewed in separate sub-topic articles of this Roadmap) will enhance the responsivity and reduce the noise, resulting in further improvement of the device performance. By manipulating the individual absorber thickness and number of cascade stages in a resonant cavity [52], optimal device performance can be achieved for meeting the specific requirements of a given application, which should be especially beneficial at high temperatures where the suppression of noise is important.

## Concluding Remarks

In summary, ICIPs represent a new type of IR photodetector that can suppress noise and circumvent the diffusion length limitation of conventional single-stage III-V devices, which makes them capable of operating at higher temperatures with higher speed. Experimental investigations have demonstrated their advantages and flexibility, indicating their promising future. However, continued and expanded efforts are required to realize their full potential for integration into practical systems with expanded applications.

### *Acknowledgments*

The authors are grateful for support received in the past from NSF and AFOSR and to their collaborators and former students R. T. Hinkey, J. F. Klem, M. Johnson, H. Ye, H. Lotfi, L. Lei, W. Huang, L. Li, S. M. S. Rassel, Y. Jiang, Z. Tian, J. A. Massengale, J. C. Keay, T. D. Mishima, P. R. Larson, J. A. Gupta, H. C. Liu, D. Lubyshev, Y. Qiu, J. M. Fastenau, and W. K. Liu for their contributions.

# 13. Background, development and future for Quantum Cascade Detectors

**VIRGINIE TRINITÉ,[1] SALVATORE PES,[1] AND JEAN-LUC REVERCHON[1]**

[1]*III-V Lab, 1 avenue Augustin Fresnel, 91767 Palaiseau Cedex, France*
*virginie.trinite@3-5lab.fr*

## Overview

Quantum cascade detectors (QCDs) are photodetectors in which the electronic band structure is engineered at the quantum level to enable the detection of photons with energies below the semiconductor bandgap, as illustrated in Fig. 1. The combination of long-wavelength detection capability and high material quality makes QCDs a compelling technology for mid-infrared (mid-IR) photon detection. QCDs emerged in the early 2000s [1,2] inspired by two key mid-IR photonic technologies: the quantum well infrared photodetector (QWIP) [3,4] and the quantum cascade laser (QCL, which is discussed in a separate sub-topic article of this Roadmap) [5]. Together, QCDs, QWIPs, and QCLs form a family of intersubband (ISB) optoelectronic devices, which operate based on intra-band transitions between quantum-confined states within the conduction band of semiconductor hetero-structures. These ISB components are unipolar, majority-carrier devices. Because the active optical transitions occur far from the semiconductor band edge, typical bandgap-related defects such as mid-gap Shockley-Read-Hall (SRH) traps have activation energies much larger than the photon energies being detected. As a result, the dark current is primarily governed by the small effective bandgap within the conduction band, while other material defects remain electrically inactive. This leads to a natural form of "self-passivation" of ISB devices, allowing structures such as mesas or small resonators to be defined through etching without introducing parasitic currents.

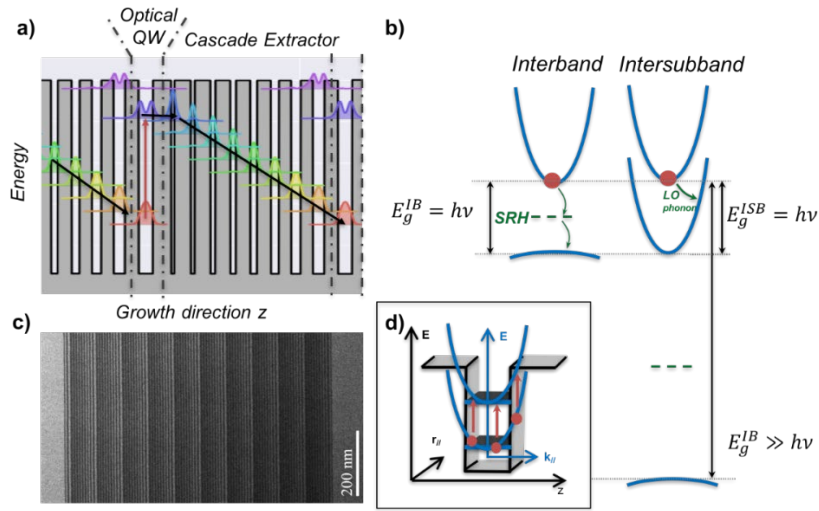

Fig. 17: (a) Band structure of an InGaAs/AlInAs QCD on InP, designed to detect around a 5 μm wavelength (here just one period of the complete structure is shown for simplicity). The optical quantum well (QW) is Si-doped with a typical sheet carrier density of a few $10^{11}$ cm⁻², ensuring that the lower energy state (marked in red) is populated with electrons. The red arrow indicates photon absorption. This excites a photoelectron to the upper level of the cascade extractor, where it is scattered to the next period via efficient longitudinal optical (LO) phonon emission (shown with black arrows). (b) Comparison of the mechanisms limiting photoelectron lifetime in interband versus ISB detectors. For interband detectors, mid-gap Shockley–Read–Hall (SRH) defects act as extrinsic, technology-limiting centers for electron–hole recombination. In contrast, ISB detectors are primarily limited by intrinsic LO phonon emission as the dominant

recombination process. (c) High-resolution transmission electron microscopy (TEM) image of a 10-period QCD structure corresponding to the QCD design shown in (a). In the image, InGaAs layers appear bright, while AlInAs layers appear dark. (d) Hybrid ($E, k_\parallel, z$) representation of photon absorption within the optical QW of a QCD. (Adapted from [6])

**Current status**

## 10. Quantum Cascade Detectors origin and imaging applications

QCDs were originally developed for infrared imaging and are directly derived from QWIPs. The format of detectors used today in thermal imagers depends heavily on the sophistication of their respective fabrication technologies. Indeed, the first thermal imagers were based on a single detector that was relatively easy to manufacture. However, this configuration required a double scanning system to reconstruct the observed scene. Progress in material science and fabrication processes has since enabled the development of linear detector arrays, simplifying the imager architecture while increasing its sensitivity. The natural trend was ultimately toward so-called staring arrays, which eliminate the need of any opto-mechanical scanning mechanism. The use of multi-detector arrays increases the integration time per pixel, which consequently enhances the signal-to-noise ratio.

As previously mentioned, QWIPs operate based on intraband optical transitions between quantum subbands in quantum heterostructures. This approach benefits from the powerful tool called band-gap engineering. QWIPs based on AlGaAs/GaAs heterostructures were among the first technologies to produce large 2D arrays in the early 2000s. Only a few suppliers have actively developed QWIP-based infrared imagers. Notable contributors include III-V Lab and Sofradir for Thales in France [7,8], and IRNova in Sweden [9]. III-V Lab developed QWIP arrays and produced more than 1000 from 2006 to 2012. The detector format included full TV resolution (640x512) with a 20 µm pixel pitch, as well as 1280x1024 SXGA resolution using micro-scan (µScan) technology. III-V Lab also developed polarimetric capabilities (see Fig. 2a). Additionally, more complex architectures were developed, including dual-band configurations as presented in Fig. 2b. In 2013, III-V Lab's QWIP imaging technology was transferred to Sofradir (now Lynred), though this transition led to limited follow-up development. Currently, IRNova remains the only company still offering QWIP-based solutions [10,11].

QWIPs are photoconductive detectors. A DC bias must be applied to the structure to produce a photocurrent, which is then superimposed on a relatively large dark current. A typical ratio of 10 between photocurrent and dark current is tolerated due to the QWIP's high dark current & photo-response uniformity. In contrast, interband detectors such as InSb or HgCdTe require much higher ratios (typically 100 or more) to achieve comparable performance. However, a side effect of accepting a significant dark current is rapid saturation of the readout circuit's integration capacity. Different solutions were proposed to mitigate this issue. One notable concept is the "skimmed QWIP", in which the dark current is effectively suppressed [7]. Historically, the high dark current level of QWIPs motivated the development of QCDs as an alternative for imaging applications. Indeed, unlike QWIPs, QCDs operate in the photovoltaic regime, which prevents the readout capacitor from being filled by the dark current during integration. The QCD also has a reduced dark noise, which could eventually increase the operating temperature.

Although QCDs can achieve better detectivity than QWIPs, the benefits are not large enough to justify the overall costs and disruption associated with transitioning the mid infrared imagers fabrication. For these reasons, and despite the original motivation behind developing the technology, QCDs are not expected to enter the mid-infrared imaging market in the near future.

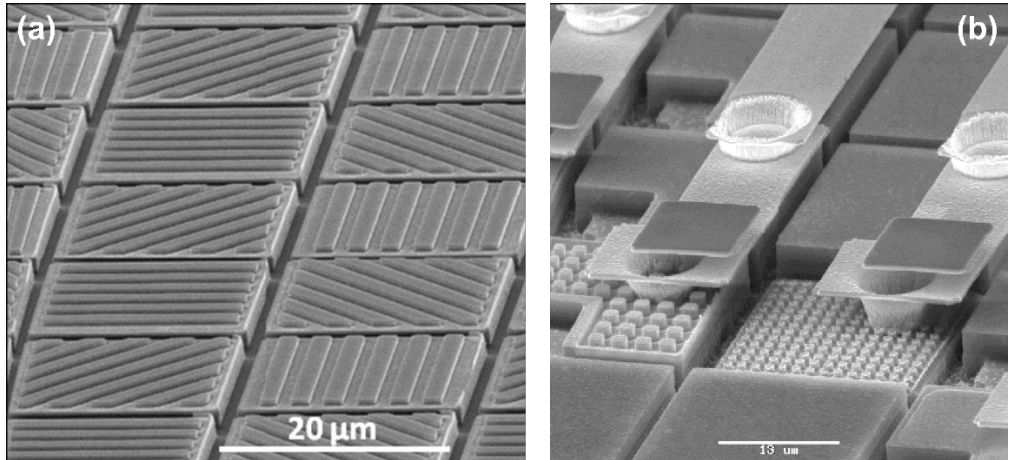

Fig. 18: SEM pictures of (a) a QWIP polarimetric array (2×2 pixels) (reproduced from [7]) and (b) a QWIP bi-spectral configuration (reproduced from [10]).

Since 2000 [12–14], the demonstration of QWIPs coupled to patch antenna resonators (PARs) has drastically redefined the perspective of the ISB pathway. Thanks to the antenna effect, this geometry provides a photonic collection area much larger than the detector's electrical surface, i.e., the one generating dark current. In addition, it fulfills the ISB polarization rules to allow efficient absorption under normal incidence. This leads to a significant increase of signal-to-noise ratio [13], making ISB detectors and PARs relevant altogether. Room temperature operation of both QWIPs and QCDs detectors embedded in metal-metal PARs have been reported in heterodyne detection schemes. It is the object of the next chapter.

QWIPs still remain the ISB preferred solution, since they offer proven stability and reliability, and are already fully integrated into established industrial production lines. That said, it is worth noting that QCDs in pixel-array configurations have been proposed and demonstrated, albeit in a limited number [15]. The integration of QCDs in patch configurations for imaging have been also considered at the patent level [16]. In the short term, the development of QCD technology appears to be more focused on single-element devices rather than imaging architectures. In this context, Hamamatsu began commercializing a high-speed 4.65μm QCD detector with 20 GHz bandwidth in 2021 [17]. The ability of QCDs to operate at room temperature, combined with their wide bandwidth, makes them key enabling components for high-speed applications such as free-space optical communications, dual-comb spectroscopy, LiDAR, and heterodyne interferometry. This point will be discussed further below.

## 11. Evolution of the QCD performance

QCDs have been demonstrated over a broad spectrum of operating wavelengths (from 1 to 20 μm, and even in the THz range [23,24]), operating temperatures (from 10 to 300 K), and illuminating conditions. We describe here how the QCD performance has evolved by updating the figures of the work done in [6]. The typical metrics used to compare the performances are also briefly reviewed.

The most common metric for comparing detectors is the responsivity $\mathcal{R}$ (expressed in A/W), which is defined as the value of the photocurrent density $J_{opt}$ divided by the optical flux $\phi$, allowing samples of different sizes to be readily compared. Although intuitive and easy to use, this metric has two inherent biases. Firstly, devices operating at longer wavelengths are favored due to a greater number of photons for the same incoming power. Secondly, since in first approximations the responsivity is proportional to the number of flowing electrons it increases the responsivity with higher doping level, at the cost of increased dark current and consequently higher dark noise.

$$\mathcal{R} = \frac{J_{opt}}{\phi} = \frac{J_{illuminating} - J_{dark}}{\phi} \tag{1}$$

We can correct the first drawback by substituting the external quantum efficiency (EQE) $\eta$, defined as the number of photoelectrons flowing in the external circuit per incident photon. The EQE should not be confused with the internal quantum efficiency (IQE), which is defined as the number of photoelectrons flowing in the external circuit per absorbed photon (ignoring the loss of impinging photons that are not absorbed in the optical quantum wells). The EQE is related to the responsivity $\mathcal{R}$ by:

$$\eta = \mathcal{R} \frac{h\nu}{e} \tag{2}$$

where $h\nu$ is the photon energy and $e$ the elementary charge. The second drawback can be corrected by using the specific detectivity $D^*$. $D^*$ is a normalized signal over noise ratio, and is defined as the inverse of the noise equivalent power (NEP, the minimal optical input power that generates a photocurrent equal to the noise current $i_n$), normalized by the detector area $A$ and the measurement bandwidth $\Delta f$.

$$D^* = \frac{\sqrt{A\Delta f}}{NEP} = \frac{\mathcal{R}\sqrt{A\Delta f}}{i_n} \qquad \text{in } Jones = cm.W^{-1}Hz^{-1/2} \tag{3}$$

As discussed in detail in [6], the main shortcoming of this metric is that it strongly depends on the illuminating conditions. So here for simplicity we limit the discussion to dark detectivity, when $i_n$ is measured in dark conditions. This metric depends merely on the detector's intrinsic performances and gives the upper limit to its sensitivity.

Another drawback of using $D^*$ is that in most cases, people do not measure the device's noise but instead infer it from simpler current measurements. Even more concerning, the literature contains misleading formulae that treat shot noise and Johnson noise as independent and additive contributions, an assumption that is not physically accurate (see for instance [6,25]). We note here that some approximate expressions for the noise based only on I(V) measurements can be used for QCDs, because their noise is generally frequency-independent over the measurement bandwidth. The correct expressions for $D^*$ are:

$$D^* = \mathcal{R} \Big/ \sqrt{2e\frac{1}{N}|J| + \frac{4k_BT}{dV/dJ|_{V=V_0}}} \text{ for } V > V_0 \tag{4a}$$

$$D^* = \mathcal{R} \Big/ \sqrt{2e\frac{1}{N}|IJ| + \frac{4k_BT}{dV/dJ}} \text{ for } V < V_0 \tag{4b}$$

where $J$ is the total current density flowing through the device (no distinction between optical and dark contributions), and $N$ is the number of QCD periods. $V_0$, the bias for which the total current is zero, corresponds to a direct bias under illumination, and $V_0 = 0$V in dark conditions. $dV/dJ|_{V=V_0}$ is the static resistance times the area of the detector, and $dV/dJ$ is the differential resistance times the area. These formulae are in good agreement with experimentally measured dark $D^*$. In particular, reduction of noise at low reverse bias, which is of particular interest experimentally, is well described (see Fig. 4a).

The figure of merit (FOM) used in Fig 3 is defined as:

$$FOM = \frac{D^*}{\lambda} \sqrt{T} \exp\left(-\frac{hc}{2\lambda k_B T}\right) \qquad (5)$$

Here the division by $\lambda$ normalizes the responsivity to an EQE, for a fair comparison with respect to the wavelength. The $\sqrt{T}$ and the exponential terms normalize the detectivity by the thermal activation of the dark noise, to account for the different temperatures used to report the detectivity. Note that this term supposes that the activation energy of the dark current is equal to the detected photon energy. It is rarely the case in practice, however, since the activation energy is lowered by the optical well Fermi energy as well as by potential diagonal transitions from the optical well to the extractor. That means that the FOM can slightly vary with temperature for a given device. However, the FOM will be even more affected if the activation energy changes notably with temperature.

Figure 3a shows the impact of the underlying system material on the FOM. InP indicates the InGaAs/AlInAs material system lattice-matched on InP, which is most often used for QCDs operating in the 4.5 μm to 10 μm wavelength range. For the longer wavelength and THz windows, the GaAs/AlGaAs system (here briefly indicated as GaAs) is preferred, although other systems with greater conduction band offset (CBO) could in principle be used.

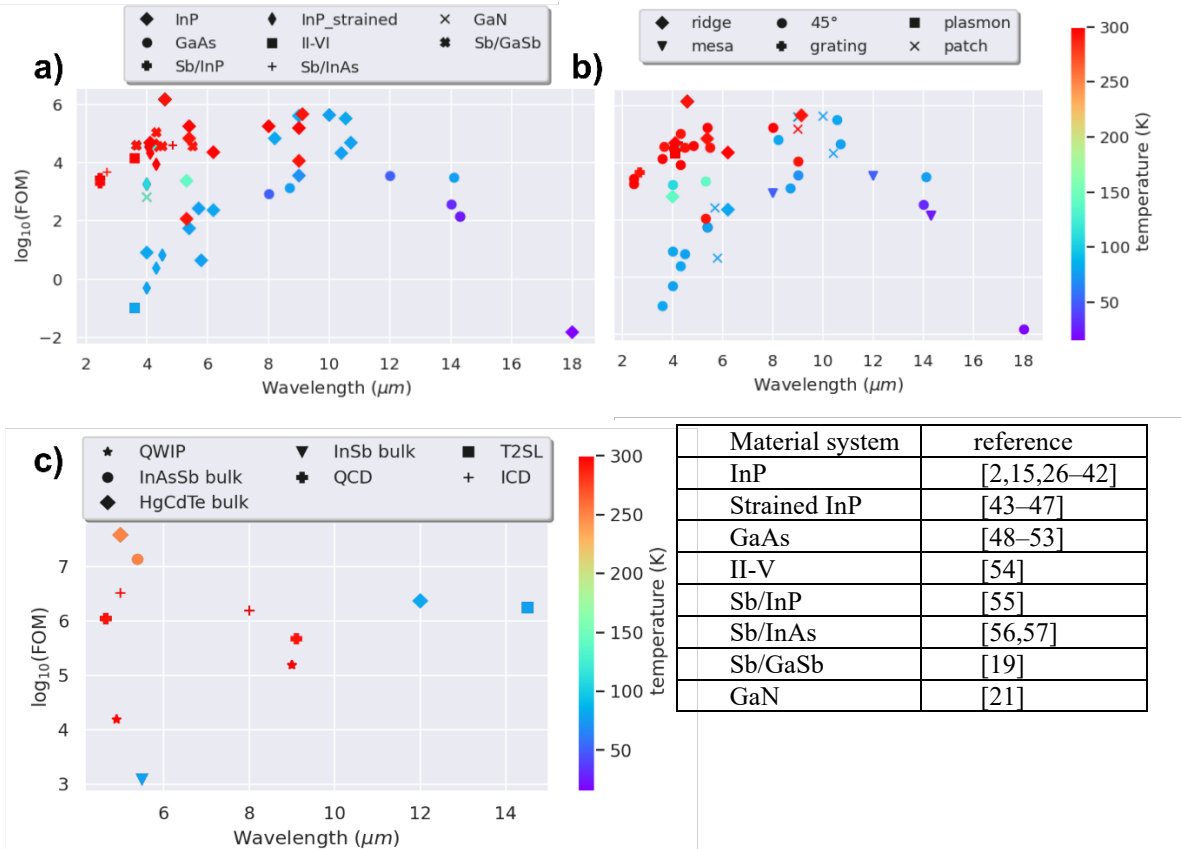

Fig 19:Comparison of the QCD performance using the figure of merit defined in Eq. (5): a) Impact of different material systems on the performance (see the text for definitions of the materials); b) Impact of the optical coupling technology used; c) Comparison with other detection technologies.

For shorter wavelengths, the need for a higher CBO definitely drives the choice of material system. Strained InGaAs/AlInAs on InP (indicated as InP strained on the figure) can cover wavelengths down to 4 μm, even if the performance is not as good as the lattice matched counterpart. InAs/AlAsSb on InAs (Sb/InAs) [23,24] can reach even shorter wavelengths down to 2.7 μm, but this material system suffers from lower crystal quality. QCD devices on GaSb substrates currently give better results for Sb-based designs, since the GaSb substrate has a significantly higher level of maturity [18,19]. InGaAs/AlAsSb on InP (Sb/InP) was also tried and gives similar results to the QCDs on GaSb. For these wavelengths, the available CBO for devices grown lattice-matched on InP ranges from 1.48 eV with $In_{0.53}Ga_{0.47}As/AlAs_{0.56}Sb_{.44}$ to 520 meV with $In_{0.53}Ga_{0.47}As/Al_{0.52}Ga_{0.48}As$. More exotic materials were also used, but with very few demonstrations: Group III-Nitrides also offer a large range of available CBOs, for example, 1.9 eV for AlN–GaN. ISB transitions were reported at 4.2 μm in the GaN/AlGaN material system grown on sapphire [20,21], and at 3.25 μm with ZnO/MgZnO on ZnO [22].

Figure 3b shows the impact of light coupling scheme on the FOM. QCDs in a ridge configuration, where the light is injected on the ridge facet, show the best results. This coupling is in fact very efficient, because it naturally copes with the selection rule of ISB transitions, and it is also beneficial for the long cavity to efficiently absorb the incoming optical signal. The patch geometry also shows very interesting results, due to the efficient coupling offered by the patch resonator and the antenna effect. For this coupling scheme we found only one reported result for the detectivity of a QCD operating at 9 μm [26], although another detectivity value was reported for a dual-color QCD [27,28]. In that case, simultaneous optimization of the performance in both bands is not trivial, which can explain a degraded FOM for the latter device. Other coupling schemes show similar FOMs. 45° coupling is the most commonly used (given the simplified fabrication process), and provides fairly favorable results. However, we emphasize that the 45° mesa configuration requires challenging optical alignment within the optical systems. The same is true for the ridge configuration, although ridge QCDs are in principle meant for integration into photonic integrated circuits (the subject of a separate sub-topic article in this Roadmap), so alignment is typically defined during the lithography fabrication steps.

Table 1 and Fig. 3 c) compare the performances of QCDs to that typical of other commercial devices. If devices based on bulk materials still present the best performances, the advantage of QCDs lies in the lack of saturation and the very high frequency and high-speed carrier dynamics. These characteristics make them particularly suitable for heterodyne detection in the mid-infrared region, as elaborated further in Section 5.

**Table 1. Performance comparison between QCD and other technologies.**

| | MWIR | | | | | | LWIR | | | | |
|---|---|---|---|---|---|---|---|---|---|---|---|
| | InAsSb[a] | HgCdTe[b] | InSb[a] | QCD [29][a] | QWIP [58] | ICIP [59] | T2SL[b] | HgCdTe[b] | QCD [40] | QWIP [60] | ICIP [59] |
| λc (μm) | 5.4 | 4-5 | 5.5 | 4.6 | 4.9 | 5 | 14.5 | 12 | 9.1 | 9 | 8 |
| Cooling | 2 stages | 2-4 stages | $LN_2$ | RT | RT | RT | $LN_2$ | $LN_2$ | RT | RT | RT |
| Detectivity (Jones) | $1,5.10^{10}$ | $6.10^{10}$-$2.10^{11}$ | $1.6.10^{11}$ | $10^9$ | $10^7$ | $2.10^9$ | $1,6.10^{10}$ | $5.10^{10}$ | $6,5.10^7$ | $2.10^7$ | $2.5.10^8$ |
| $Log_{10}$ (FOM) | 7.1 | 7.6 | 3.1 | 6.0 | 4.2 | 6.5 | 6.2 | 6.4 | 5.7 | 5.2 | 6.2 |
| Typical rise time/bandwidth | 15ns | 80ns | 30ns | 20GHz | | | 150ns | 300ns | ~1->25GHz | ~1->100 GHz | |

InAsSb, HgCdTe, and InSb are bulk interband technologies. T2SL (type 2 superlattices) and ICIP (interband cascade infrared photodetectors) are interband devices with engineered quantum level as the QCDs. (a https://www.hamamatsu.com/) and (b https://vigophotonics.com) indicate data in direct detection operation from the Hamamatsu and Vigo catalogues, respectively. Detectors based on HgCdTe, T2SLs, and ICIPs are discussed in separate sub-articles of this Roadmap.

## 12. Modeling state of the art

On a fundamental level, a QCD is a quantum cascade laser (QCL) operated near thermal equilibrium. QCD theory benefits from all the work on electronic transport in QCLs. Several levels of modeling are now available for description of the electronic transport. An empirical and simple model treats a single period of the QCD as a Schottky junction between a metal (the ground state of the optical well) and a semiconductor (the adjacent thermalized cascade, up to the upper level of the optical well). Indeed, the states of the cascade are well connected to avoid any detrimental series resistance, and a single Fermi-level can be applied to the whole cascade (see Fig. 4a) and b)). Voltage drops occur between the cascades. This so-called thermalized cascade picture of transport [61] is very useful for quick modeling, as in [59].

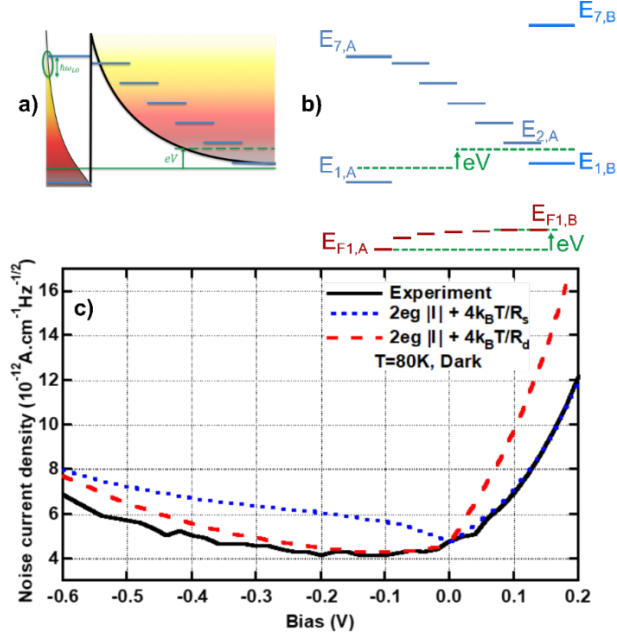

Fig. 20: a) Illustration of the analogy between a QCD cascade and a Schottky diode. The autumn colormap shows the electronic density distribution. b) Schematic representation of the thermalized cascade model that allows the Schottky analogy: Under a bias per period V, each quantum state in blue (here 7 states) of the A and B period has its own Fermi level in red. A good approximation is that the states of the cascade have a common Fermi level $E_{F1,B} \simeq E_{F2,A} \simeq E_{F3,A} \ldots \simeq E_{F7,A}$. c) Comparison between experimental dark current noise for a QCD @ 8µm and shot & Johnson noise contribution. The latest depends on static or differential resistance as explained in [6,25].

A well-known model is the rate equation model in the Wannier-Stark hopping, which has successfully computed the performances of mid-IR QCDs [62,63]. It was also applied on a much larger scale to QCLs [64]. It defines the mental framework in which QCD and QCL designers conceive the device's active region. Its main shortcomings are well known. First, the framework supposes that the structure is periodic: the transport model applies for infinite structures, which is typically a good approximation for devices that contain more than 10 periods. Secondly, the assumption of thermalized subbands loses track of the distribution function in the in-plane momentum ($k_{//}$). Monte-Carlo methods are implemented when the in-plane distribution is needed. However, most importantly, it fails to deal with quantum states that are resonant, that is, when the detuning $\Delta$ between two levels is smaller than the ISB linewidth $\Gamma$. In this case, the coupling of states in resonance will create an artifact short circuit in the device. This makes it problematic to model the transition from the upper state of the QCD's optical well to the first state of the extractor.

A better alternative is the mixed density matrix model, which treats non-resonant parts of the structure in Wannier-Stark hopping while resonant parts are treated using coherent second order tunneling processes. This approach has been developed successfully for QCLs [65–67] and applied to QCDs [68]. The level of approximation represents a good compromise between quantitative agreement with experiment and cumbersome calculations. It allows our team to easily design new QCDs, because the fast simulations can rapidly try several configurations. Figure 5 shows an example of the level of agreement we obtain with this model. The agreement is excellent for both dark current and spectral responsivity. Absolute comparison for the responsivity is not shown because of uncertainties concerning the optical part of the devices. At negative bias, the current is dominated by transitions between the fundamental level and first excited level in the quantum well (E1,A and E7,A in the illustration of Fig. 4b). We also observe peaks in the current that are the signature of alignment of the fundamental level with the other levels in the cascade that open a parasitic resonant channel. In this case, the QCD can no longer be modeled as a Schottky junction, and all the equations above eventually fail.

Our team has also used this framework to develop a complete theory for the noise, which allows the detector's noise to be calculated without any additional assumptions [62,69]. To our knowledge, QCDs are the only IR detectors whose signal-to-noise ratio can be calculated completely without any external fitting parameters.

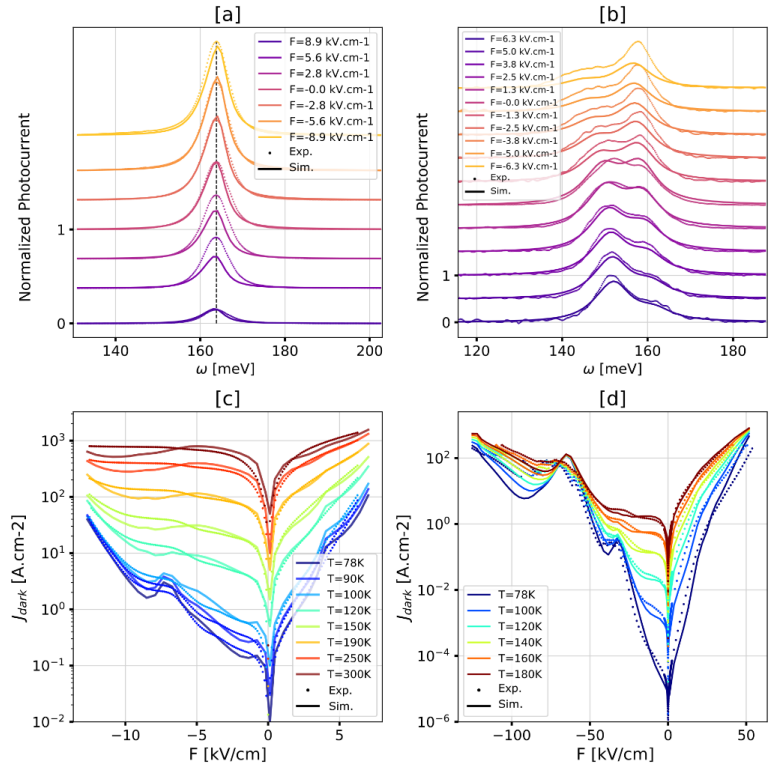

Fig 21 : Comparison of simulations (solid lines) versus experiment (dotted lines): [a][b] Photocurrent is measured at 78K at several biases and normalized by the maximum value at maximum negative bias. [c][d] Evolution of dark current density vs. bias at a series of temperatures. [a][c] is a QCD@8μm and [b][d] a QCD@9μm. Data are from [70], where details of the QCD structure can be found

More advanced simulation of the transport has more recently been performed using non equilibrium-Green's function (NEGF) models [71], which retain an energy-resolved description of the transport at the cost of additional numerical burden. This formalism is now also available in the commercial software NextNano.NEGF [72,73].

Recent modeling developments have accounted more realistically for graded interfaces. This combines a smooth potential to describe the interfaces of finite width with a more realistic description of interface roughness scattering in the transport [74,75]. Another interesting development has gone beyond mixed density matrix models by using a real density matrix approach without an "arbitrary" separation between resonant and non-resonant coupling of the states. Density matrix models, not designed for QCDs/QCLs, are quite common but are configured to describe discrete quantum levels, and do not account for the in-plane $k_{//}$ momentum. The work done in [76] recast a generalized, $k_{//}$ defined, density matrix into one with effective discrete levels. It paves the way to a numerically-efficient tool that precisely describes the electronic transport coupled with the optical field though Maxwell-Bloch-type equations. However, one limitation of this model is that it cannot quantitatively account for correlations between more than two states.

## 13. QCDs for free space optical communication and heterodyne detection

For free space optical (FSO) links subject to variable or degraded weather conditions, signal transmission in the MWIR or LWIR atmospheric window offers significant advantages compared to the Short Wavelength Infrared (SWIR) domain, by providing more reliable availability of the communication channel during unfavorable atmospheric conditions such rain or fog [77–79]. Indeed, longer wavelengths are less prone to the effects of aerosols and air turbulences to provide more stable signal transmission. This major advantage is combined with an unrivaled stealth potential, thanks to low Mie scattering and the presence of incoherent infrared photons. Furthermore it allows the optical signal to be hidden in the surrounding thermal background in contrast to SWIR, where off-axis beam detection is possible [80]. The MWIR optical carrier frequencies are also compatible with the high-speed data transfer required by telecom systems. Thanks to the peculiar advantages against typical impairments affecting classical optical channels, fast MWIR photonic devices are approaching new fields of application that are now dominated by telecom-graded components, such as FSO communications networks as well as light detection and ranging (LiDAR) optical systems.

Although pioneering works on $CO_2$-laser-based communications can be traced back to the late 1970s [81], interest in FSO in the mid-infrared region regained momentum with the development of QCL sources. QCLs have by now reached a very high level of technological maturity and have demonstrated great performance across the entire MWIR spectral range. Widely tunable, narrow-linewidth QCL sources operating at room temperature are now commercially available, producing continuous wave output power ranging from some hundreds of milliwatts in the distributed-feedback (DFB) configuration, up to several watts when Fabry-Perot (FP) cavities are employed. The high degree of freedom offered by QCLs makes them ideal for integration into high-speed optical systems operating in the mid-infrared. In particular, devices based on the InGaAs/AlInAs material system lattice-matched on InP substrates can directly benefit (with some adjustments) from concepts and fabrication processes developed for the 1.55 µm telecom semiconductor industry.

Among all detectors suitable for addressing the MWIR and LWIR optical windows, QCDs have an advantage (shared with QWIPs) that makes them particularly attractive in photonic applications such as FSO communications, dual-comb spectroscopy, and LiDAR systems, namely their inherently high speed. This is due to the ultrafast ISB carrier dynamics that leads to intrinsic cut-off frequencies expected to exceed 100 GHz [82,83]. In contrast to imaging applications (where maximizing the carrier lifetime helps to increase the operating temperature), the fast carrier dynamics of ISB detectors, driven by the efficient LO phonon scattering, becomes a significant asset when high-speed performance is essential.

Beyond the high frequency response, the ultrafast carrier dynamics also confers another major advantage to ISB detectors: their response is virtually insaturable, providing saturation levels reaching several MW/cm² [84]. The combination of high speed and high linearity is particularly advantageous for exploitation in heterodyne detection schemes, where an external optical signal is coherently mixed with a high-purity local oscillator acting as a signal amplifier. This process generates an AC electrical signal with intermediate frequency (IF) typically in a range spanning several tens of GHz.

Additionally, the unipolar nature of carrier transport in ISB devices ensures self-passivation, a significant advantage that offers fabrication flexibility that is rarely achieved with conventional interband detectors. Notably, self-passivated ISB devices can be structured into sub-wavelength optical cavities to enhance light-matter interactions and boost the detector's performance (as discussed below). Furthermore, the Lorentzian (or quasi-Lorentzian) spectral response of ISB devices is narrow enough to avoid integrating the photonic noise over the entire LWIR band, but wide enough to guarantee a flat response over the entire IF band of interest (the full width at half maximum of the spectral response is typically of the order of ~10 meV, i.e. ~2.5 THz).

To appreciate the gain in sensitivity that a coherent detection scheme offers in contrast to direct detection, one can look at the signal-to-noise ratio (SNR) of both cases. In a direct detection scheme (cf. Fig. 6a), the SNR can be expressed by considering the average current $I_{signal}$ generated by an optical signal with power $P_{signal}$ impinging on the detector, along with the various noise contributions (photocurrent $i_{signal}$ and dark noise $i_{dark}$), as:

$$SNR_{direct} = \frac{I_{signal}}{i_{signal} + i_{dark}} = \frac{\mathcal{R}P_{signal}}{\sqrt{2e\mathcal{R}P_{signal} + \frac{2e}{N}|I_{dark}| + \frac{4k_BT}{R}}} \sqrt{\frac{1}{\Delta f}} \qquad (6)$$

where $\Delta f$ is the effective measurement bandwidth. In the formula above, $R$ should be taken as the static resistance defined as $R_s = dV/dI|_{V=V_0}$ for V>V0 (see eq.4a), or the dynamic resistance $R_d = dV/dI$ for V<V0 (see eq.4b). For the $\frac{2e}{N}|I_{dark}|$ term, here we used the noise gain factor $1/N$ that is typical of QCDs, but the expression can easily be adapted to the case of QWIPs by considering the photoconductive gain g.

Optimal performance requires that $I_{signal} \gg I_{dark}$, a condition that typically requires cooling the ISB detector to reduce its dark current that is exponentially activated with temperature. This thermodynamic limit is clearly visible in the expression for dark current, which depends on the ratio between the activation energy $E_a$ and the thermal energy $k_BT$ of the system, according to:

$$I_{dark} \propto I_0(V,T) \, e^{-\frac{E_a}{k_BT}}$$

$I_0$ is a pre-factor that depends on the applied voltage and temperature.

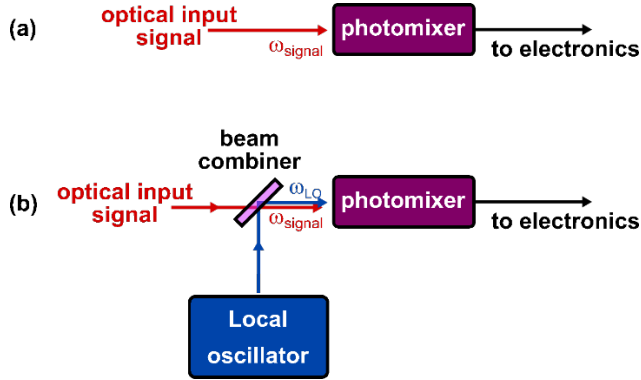

Fig. 22. Schematic illustration of a) a direct detection and b) a heterodyne detection scheme.

Although the different mid-infrared detection technologies are in competition to reduce the pre-factor $I_0$, it is clear that the direct detection scheme itself has a fundamental limit, which often makes ambient temperature operation elusive, in particular for devices operating in the LWIR. However, the approach based on heterodyne detection is radically different [4]. In a coherent detection scheme (cf. Fig. 6b), the signal-to-noise ratio can be written as:

$$SNR_{heterodyne} = \frac{I_{AC}}{i_{photon} + i_{dark}} = \frac{2\mathcal{R}\sqrt{P_{signal}P_{LO}}}{\sqrt{2e\mathcal{R}P_{LO} + \frac{2e}{N}I_{dark} + \frac{4k_BT}{RS}}}\sqrt{\frac{1}{\Delta f}} \tag{7}$$

where $I_{AC}$ is the average current resulting from the beating of the optical signal and the local oscillator (LO) at $\omega_{IF} = \omega_{photon} - \omega_{LO}$. In practical cases, $P_{LO} \gg P_{signal}$, and the signal contribution in the photonic shot-noise term can be neglected. Thus, if a sufficiently powerful local oscillator is employed at the receiver side, the receiver noise is dominated by the LO shot-noise, and even the dark noise contributions can be neglected. Therefore, the main advantage of the heterodyne scheme is the possibility of achieving photon-shot-noise limited detection, for which the signal-to-noise ratio becomes basically independent of temperature. The SNR and noise equivalent power (NEP) then assume a simplified form:

$$SNR_{heterodyne} = \frac{\eta P_{signal}}{h\nu \Delta f} \tag{8}$$

$$NEP_{heterodyne} = \frac{h\nu}{\eta}\Delta f \tag{9}$$

Early published work on heterodyne mid-infrared detectors began with the development of $CO_2$ lasers as local oscillators, and employed MCT detectors [85]. For a long time these systems were confined to niche applications such as astronomy [86], and suffered from the low saturation levels of the MCT technology. Mid-infrared QCLs have proven their relevance as high quality LO sources for heterodyne receivers, thanks to their narrow emission linewidth (<1 MHz), continuous wave operation, high-frequency stability and emitted output power [87]. On the detector side, the parameters needing optimization for heterodyne photomixing are:

i. Sensitivity, which determines the required LO power and therefore, ultimately, the operating temperature.
ii. Photocarrier extraction/recombination time, which determines the receiver's ultimate bandwidth.
iii. Device impedance, which should be matched with the external circuit/load. It also limits the power and cut-off frequency of the output signal via the RC product.
iv. Saturation level and thermal dissipation, which limit the LO power injected.

MCT components such as avalanche photodiodes combine record sensitivities with proven bandwidths of a few GHz [88–90]. However, in addition to the limited maximum bandwidth and linearity, their distribution to civilian markets faces significant difficulties due to high cost and use restrictions (for example the European REACH regulation [91]). This does not apply to the GaAs and InP material systems from which ISB detectors are manufactured.

There have already been notable demonstrations of heterodyne detection based on ISB detectors. The pioneering work of H.C. Liu et al. demonstrated heterodyne mixing of several tens of GHz at 77 K on QWIPs by heterodyne beating of $CO_2$ lasers [92]. This very promising proof of concept was nevertheless limited by a high RC time constant due to a large mesa dimension (75 μm-diameter), limited extraction of the RF signal due to the impedance mismatch, and inefficient optical coupling. Some of these problems were subsequently overcome by implementing mesas with smaller surface areas [93], but the unresolved issue of low ISB detector response has always limited their use in telecom applications. However, the integration of QWIP and QCD structures inside different types of plasmonic resonators and photonic crystals has shown promising potential to improve the performance [94–97]. A plasmonic antenna, in the form of a double metal patch resonator is illustrated in Fig. 7c. In this case, a QCD active zone is inserted between an upper contact and a ground metallic plane. The patch antenna resonator (PAR) is in practice a $\lambda/2$ cavity in the direction parallel to the surface, while in the perpendicular direction the confinement is typically sub-wavelength. The radiation pattern of such a resonator corresponds to a quadrupole antenna ($\lambda/2n_{eff}$), where $n_{eff}$ is the effective index of the optical mode. Its optical response is controlled (among other things) by the geometrical parameters $s$ (lateral size) and $p$ (distance between neighboring patches) of the antenna array (Figs.5b).

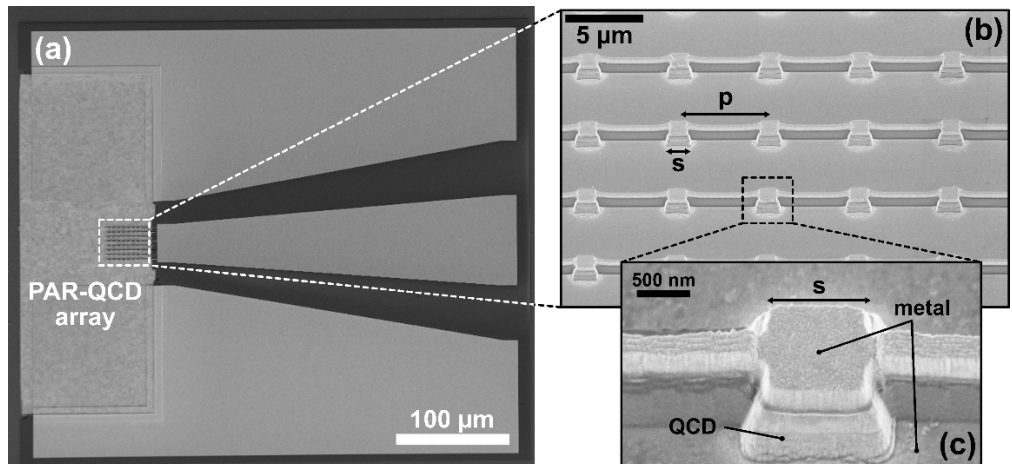

Fig. 23 – (a) Scanning electron microscopy picture of a high-speed heterodyne detector that integrates an 8x8 PAR-QCD array (~25x25 μm² area) and a 50 Ω-matched coplanar waveguide; (b) zoom on the PAR-QCD array connected with metallic nanowires; and (c) detail of a single patch antenna resonator embedding the QCD detector region. Adapted from Quinchard et al. [98].

When coupled to ISB detectors, this type of resonator offers several advantages for high-speed operation or heterodyne detection. The use of a patch antenna resonator greatly simplifies the geometry (Fig.5a) and allows the incident light, polarized parallel to the surface at normal incidence, to be coupled to the patch antenna mode, which is TM polarized. The rotation of the polarization by 90° satisfies the ISB selection rule and thus solves the problem of limited of QCDs (or QWIPs) response by improving the absorption under normal incidence. Another advantage relates to the sub-wavelength nature of the resonator, which facilitates single-mode operation and therefore assures strong spatial overlap between the signal and the LO modes for high mixing efficiency. Furthermore, the angular acceptance of the resonator allows the use of high-numerical-aperture lenses to strongly focus the LO and signal beams. More importantly, the effective optical collection area at resonance is much larger than the geometric (and therefore electrical) surface of the detector. This ensures good light collection despite the small device area, and thus minimizes the device's capacitance without limiting its response or bandwidth. Recent applications of the patch antenna resonator approach to both QWIP [60,99,100] and QCD detectors [26,98] permitted impressive demonstrations of room temperature heterodyne detection.

### Future developments to address challenges

We note that although QWIPs and QCDs can achieve comparable performance when fully optimized, approaching the ultimate heterodyne noise-equivalent power (NEP), QCDs have few notable advantages compared to QWIPs. First, the photovoltaic operation of QCDs guarantees better thermal behavior compared to a polarized structure such as the QWIP. In addition, the QCD's lower dark noise suggests an easier transition to the LO-limited regime at high temperatures. Secondly, the design of the active zone is more versatile and more robust in QCDs on InP than in QWIPs on GaAs. This is particularly the case for very thin structures with a small number of periods, which are naturally suited for integration into sub-wavelength antennae resonators. Furthermore, the response of QCDs decreases only modestly with temperature (unlike QWIPs). Finally, the entirely 2D nature of electronic transport in a QCD allows much more accurate quantitative modeling of the component characteristics [70,101,102], allowing shorter development cycles (which can be advantageous for future industrialization).

Finally, unlike QWIPs that usually employ the GaAs/AlGaAs material system, QCDs are primarily developed in InGaAs/AlInAs on InP, which is also the material system of choice for mid-infrared QCLs. Similarly, fast transistors on InP for telecom are much more efficient than those on GaAs. This opens the way to developing high-speed coherent mid-infrared transceivers with trans-impedance amplifier stages, where a powerful QCL acting as the active source and/or local oscillator, passive routing waveguides, and a QCD heterodyne photomixer are monolithically integrated onto the same photonic platform [103–107]. Nonetheless, although this perspective is very appealing, a trade-off of the laser and detector performance must be anticipated, as in general the two operations are symmetrically different. A different approach that avoids compromising the performance of both building blocks may be heterogeneous integration, for instance employing flip-chip [108], wafer bonding [109] or μ-transfer printing technology [110]. Optical waveguides for low losses and high-quality factor configuration are not clearly identified.

## 14. Conclusion

Since the late 2010s, the integration of QWIPs and QCDs with patch antenna resonators has significantly reshaped the development trajectory of ISB detectors [12,14]. By leveraging the antenna effect, this architecture enables a photonic collection area that far exceeds the detector's electrical footprint. As a result, it supports high-speed operation without compromising sensitivity, and it remains compatible with room-temperature performance. This configuration inherently satisfies the polarization selection rules of ISB transitions, enabling efficient light absorption even under normal incidence. This leads to a substantial enhancement of the signal-to-noise ratio, making ISB detectors and patch antenna resonators (PARs) highly complementary and particularly well suited to implementation in heterodyne detection schemes [13]. Additionally, the approach exhibits strong tolerance for variations in the angle of incidence, further underscoring its robustness and versatility for optical system integration. The future of QCDs relies on heterodyne detection and benefits from the lack of saturation. Commercial QCDs have been available since 2021 [17]. The room temperature operation and bandwidth larger than 20GHz make them a key component for free space optical communication, high speed MWIR spectroscopy, and optical heterodyne detection.

# 14. Resonant cavity infrared detectors

## J. R. Meyer* and I. Vurgaftman

*Naval Research Laboratory, Washington, DC 20375, USA*
*jerry.r.meyer.civ@us.navy.mil*

## Overview

Generally, the active absorber region of a conventional broadband midwave or longwave infrared (MWIR or LWIR) detector must be several microns thick if it is to attain high external quantum efficiency (EQE). However, for some applications it may be preferable to position a very thin absorber between two highly-reflecting mirrors that provide multiple passes, in order to produce high EQE within a narrow spectral bandwidth [1]. This resonant cavity infrared detector (RCID) can be advantageous if the light to be detected is from a laser rather than a broadband thermal source, or when limited bandwidth is otherwise desired as in spectroscopic chemical sensing [2,3] or hyperspectral imaging [4]. To minimize the noise currents associated with thermal background radiation, the RCID should also simultaneously minimize the parasitic detection of wavelengths outside the narrow spectral band of interest. Furthermore, the resonant cavity architecture can provide enhanced frequency response, since photocarriers generated within the thin absorber traverse a far shorter distance before being collected [1,5]. Enhanced frequency bandwidth combined with high sensitivity is critical for applications such as free space optical communication [6] and heterodyne detection [7]. An RCID may also be beneficial when the minority-carrier diffusion length is much shorter than the absorption depth, due to low vertical mobility or short photocarrier lifetime. This is again because the distance over which the photogenerated carriers must be collected is much shorter in a broadband detector.

The concept has been developed extensively at shorter wavelengths [1,8], where resonant cavity photodiodes that provide enhanced frequency response are relatively mature. MWIR and LWIR RCIDs with lead salt and III-V absorbers were reported over 20 years ago [9-12], with resonance wavelengths ($\lambda_{res}$) ranging from 3 to 8 μm and linewidths ($\delta\lambda$) as narrow as ≈ 40 nm. EQEs up to 90% were reported for the lead-salt RCIDs [13], although accompanied by relatively low resistance-area products ($R_0A$). Consequently, the specific detectivities ($D^*$) were at least 2 orders of magnitude lower than the *Rule '07* trend [14] for state-of-the-art (as of 2007) broadband HgCdTe photodiodes with cut-off equal to the resonance wavelength ($\lambda_{res}$).

Another way to induce resonant response is to etch a high-contrast grating into an epilayer above the absorber, which couples IR input into in-plane waveguide modes [15-17]. Light is then confined to the waveguide by a heavily-doped cladding layer grown below the absorber. For example, U. of Texas demonstrated devices with $\lambda_{res}$ between 4 and 5 μm that yielded up to 60% EQE with 50-60 nm linewidth [16]. Specific detectivities have reached 65% of *Rule '07* at room temperature, although riding on a large non-resonant background and for only one polarization of the incoming light relative to the grating orientation.

## Current status

Only recently has the performance of resonant cavity detectors operating at $\lambda_{res} \geqslant 3$ μm become favorable enough to generate practical interest [18-21]. For example, U. Lancaster reported EQE up to 84% at 500 mV bias for $\lambda_{res} \approx 4.4$ μm and $\delta\lambda = 46$ nm [19], and EQE up to 69% at 1.3 V bias for $\lambda_{res} \approx 3.7$ μm and $\delta\lambda = 42$ nm [**Error! Bookmark not defined.**]. However, this was accompanied by typical EQEs of $\geqslant 5\%$ at non-resonance wavelengths.

All of the MWIR RCIDs discussed above included bottom distributed Bragg reflectors (DBRs) and absorbers that were grown or deposited as a single structure. For the higher-performance RCIDs grown by III-V molecular beam epitaxy (MBE) on a GaSb [**Error! Bookmark not defined.-Error! Bookmark not defined.**] or InAs [**Error! Bookmark not defined.**] substrate, this required the challenging growth of a thick Ga(As)Sb/AlAsSb DBR while maintaining precise control over each layer thickness. For example, due to roughly linear scaling with wavelength, the device reported in Ref. [22] with $\lambda_{res} \approx 7.7$ µm required that the bottom mirror be grown to a thickness of 14 µm.

However, Praevium, NRL, and Intraband simplified the GaSb-based growth considerably, by heterogeneously bonding a Ga-free *nBn* detector with 100-nm-thick absorber to a commercially-purchased GaAs/Al$_{0.92}$Ga$_{0.08}$As DBR with reflectivity > 99% [23]. This approach, which is illustrated schematically in Fig. 1, provides a larger quality factor (*Q*) for the cavity, along with the potential for higher growth and processing yields. Figure 2 illustrates simulated profiles of the refractive index and optical mode intensity within the cavity. Note that the active absorber is centered on an anti-node of the optical field.

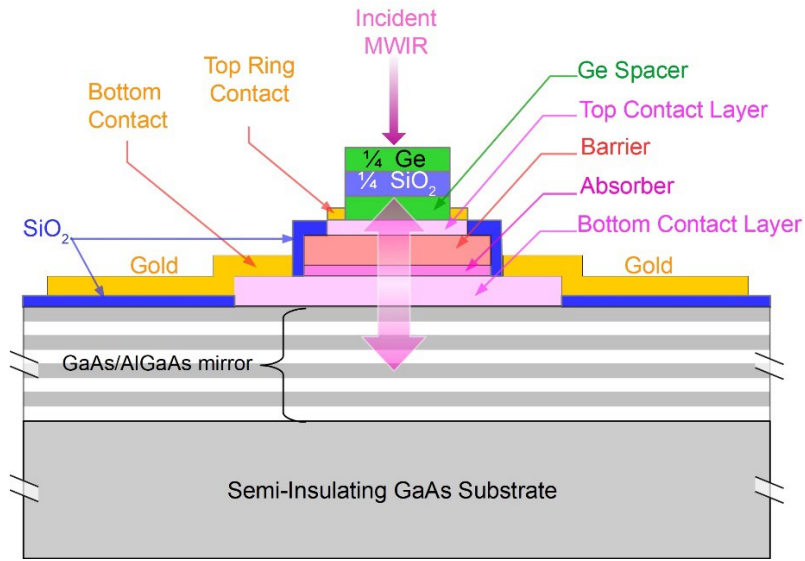

Fig. 1. Schematic of a typical RCID structure that comprises: (1) a GaAs/AlGaAs bottom DBR grown on a GaAs substrate; (2) an *nBn* detector; (3) a Ge spacer layer; and (4) a 1½-period deposited Ge/SiO$_2$ top DBR.

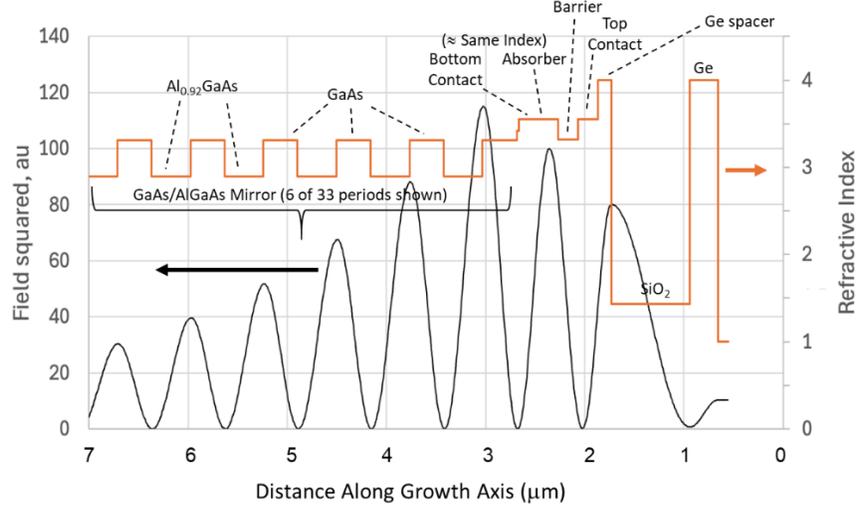

Fig. 2. Simulated profiles of the refractive index (orange) and optical mode intensity (black) along the vertical axis of the resonant cavity illustrated in Fig. 1. Reproduced with permission from Ref. [24].

A recent RCID fabricated in this manner with 100-nm-thick absorber, $\lambda_{res} \approx 4.6$ μm, and only 150 mV operating bias ($V_b$) combined EQE up to 58%, resonance linewidth as narrow as 19 nm, and EQE < 1% at all non-resonance wavelengths between 4 and 5 μm. A subsequent RCID formed with the same GaAs/AlGaAs bottom DBR and somewhat thicker absorber (300 nm) produced EQE = 61-74% with spectral bandwidth 20-40 nm and responsivity > 2.2 A/W at $V_b$ = 150 mV for all temperatures between 125 K and 300 K [24]. Figure 3 illustrates the EQE spectra near resonance for one of these devices. The resonance wavelength shifts due to gradual variation of the refractive index with temperature. Like all previously-reported RCIDs, the dark current for this device was somewhat above *Rule '07*, and scaling of the dark current with absorber thickness was not observed. Therefore, at higher temperatures $D^*$ fell short of that attainable by a broadband HgCdTe device. However, the RCID is especially advantageous at lower temperatures where thermal background photocurrent rather than dark current dominates the noise. When the thermal background for a realistic system scenario with f/4 optic that views a 300 K scene was derived from the RCID's experimental EQE spectrum at a given operating temperature, the resulting $D^*$ of $7.5 \times 10^{12}$ cmHz$^{1/2}$/W at 125 K was 4.5× higher than for a state-of-the-art HgCdTe device with 4.6 μm cut-off wavelength [24]. We emphasize that this enhancement relies on the RCID's nearly total suppression of thermal background noise at wavelengths outside the resonance band of interest.

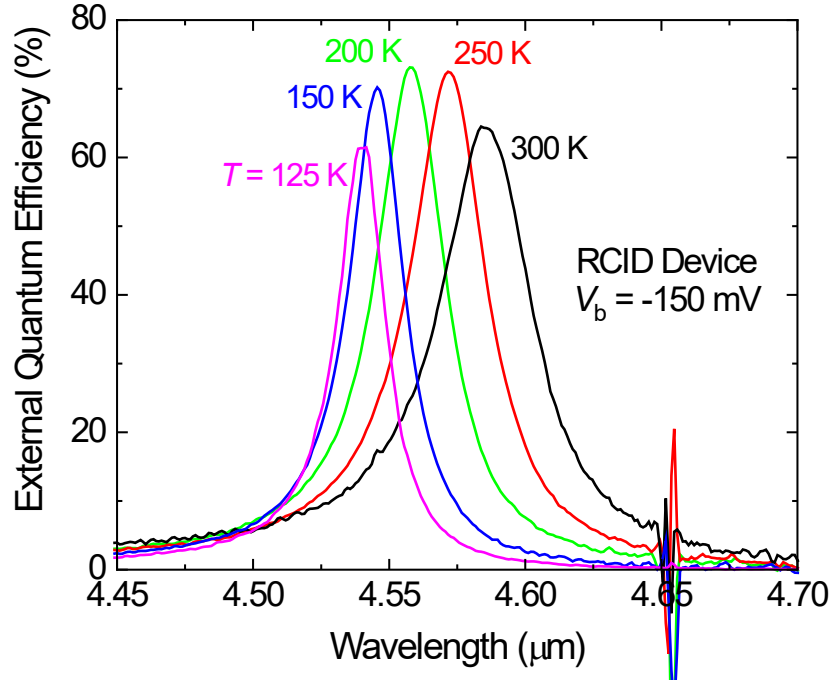

Fig. 3. External quantum efficiency near the resonance peak of an RCID device at a series of temperatures. The FWHM linewidth gradually increases from 22 nm at 125 K to 42 nm at 40 K. Reproduced with permission from Ref. [24].

It was mentioned above that the RCID architecture can also provide enhanced frequency response due to its rapid collection of photocarriers from the very thin absorber. However, this also requires minimization of the device capacitance, for example, by limiting the diameter ($D$) of the detector's mesa. Intraband, Praevium, and NRL recently reported an RCID with $D = 30$ μm and $\lambda_{res} = 4.6$ μm that provided 20-dB larger link budget than any previous MWIR result for ~ 5 Gb/s operation at room temperature [5]. The device combined high responsivity (1.97 A/W) with a low noise equivalent power (NEP) of 0.7 pW/$\sqrt{Hz}$ and high frequency bandwidth of 6.7 GHz at 3-dB. Table II of Grillot *et al.* [6] indicates that this responsivity is already > 3× higher than that of any other MWIR detector technology with frequency response exceeding 1 GHz.

## Challenges and opportunities

The primary characteristics and figures of merit that will determine the value of RCIDs in future MWIR and LWIR sensing systems include resonance linewidth, dark current, peak EQE, parasitic EQE at non-resonance wavelengths, tunability of the resonance wavelength, operating bias, frequency response, performance of arrays *vs.* individual devices, manufacturability, and cost, all as functions of the operating temperature and peak resonance wavelength. Other relevant properties that may be derived in part from these characteristics are $D^*$, responsivity, and noise-equivalent power. Many of the likely applications will require operation at room temperature or with thermoelectric cooling.

The best recent MWIR RCIDs have displayed resonance linewidth down to ≈ 20 nm [23], peak EQE up to 80% [19], non-resonant EQE < 1-2% over broad spectral ranges on both sides of the resonance peak [23,24], operating bias 150-500 mV [19,24], and frequency response up to 6.7 GHz [5]. These characteristics are already attractive for some real-world systems (even though no single device has displayed all of the listed properties), and they can potentially be improved further. For example, EQE > 90% appears quite feasible [24], and the resonance bandwidth can be reduced further by increasing the mirror reflectivities for higher cavity Q. The optimal $\delta\lambda$ may not be ultra-narrow, however, depending on the application. Regardless of the reflectivities and absorbances in the active layers, the cavity should be designed for "critical coupling" to maximize the EQE [25].

Of the remaining characteristics that require further improvement, perhaps the most challenging will be to minimize the dark current for ambient or thermoelectrically-cooled operation. Although dark current is less relevant at lower temperatures where thermal photocurrents dominate the noise, many important applications will require compact and inexpensive systems that do not require cooling. Furthermore, the low dark currents of state-of-the-art broadband HgCdTe devices set a high bar for RCIDs to surpass [3,4]. Type-II III-V MWIR detectors have rarely matched the *Rule '07* standard [26,27].

The dark current of a diffusion-limited photovoltaic detector [28,29] scales linearly with the thickness of the absorber. Thus, the RCID's very thin absorber will be highly advantageous when diffusion processes dominate. However, the dark current densities reported in Ref. [24] actually increased slightly when the absorber thickness was decreased from 300 nm to 100 nm. Furthermore, the dark currents for RCID devices with thin absorbers were not significantly lower than those of the best conventional broadband *nBn* detectors with thick InAsSb-InAs superlattice absorbers and similar cut-off wavelengths [27,30]. The dominant mechanism producing these dark currents has not yet been identified, although impact ionization is expected to play a role at higher temperatures [23], while tunneling and *g-r* processes are likely to dominate at low temperatures [29].

The dark current in a HgCdTe detector decreases even further, to a dependence now known as *Law 19* [31], when Auger recombination is suppressed by a strong bias that fully depletes both the majority and minority carrier population in the absorber. A telling signature in HgCdTe photodiodes is a dramatic negative differential resistance (NDR) that occurs at a bias marking the onset of full depletion [32]. In principle, carrier depletion should similarly reduce the dark currents in III-V type-II detectors [2]. However, to date at most a very slight NDR is observable at higher temperatures in the *I-V* characteristics of NRL's *nBn* detector structures. Until the mechanism(s) that induce the excessive dark currents have been identified and mitigated, it may be advantageous to employ bulk HgCdTe, rather than a III-V superlattice, as the thin absorber region [33] of an RCID.

In 2007-2009, ETH Zurich demonstrated dynamic electrical tuning of the detector resonance wavelengths, via piezoelectric- [12] or MEMs (micro-electron-mechanical system)-induced [34] variation of the cavity length. Figure 4 schematically illustrates a typical cavity for a MEMS-tunable RCID. More recently, U. Lancaster statically varied the resonance wavelengths of different devices on the same chip with InGaAsSb absorbers ($\lambda_{res}$ = 2.0-2.1 µm) by spatially chirping selected epitaxial thicknesses during the MBE growth [35]. We note that one can alternatively vary the cavity lengths of different RCIDs on a chip by depositing chirped spacer layer thicknesses on individual devices, or by depositing a spacer with fixed thickness and then etching to different depths. Although EQEs for the dynamically-tuned RCIDs from Refs. [12,32] were well below those reported more recently, the protocols for fabricating such devices with voltage-controlled height of a reflecting surface [36] are sufficiently mature that much better performance may be expected once the cavity designs are fully optimized. Similarly, means for statically varying the $\lambda_{res}$ for different devices on a chip during growth or subsequent processing are well established and suitable for future optimization. The maximum tuning range for both dynamically- and statically-tunable RCIDs will be limited by the range over which the cavity length can be varied in practice, as well as the spectral bandwidths of the top and bottom mirrors.

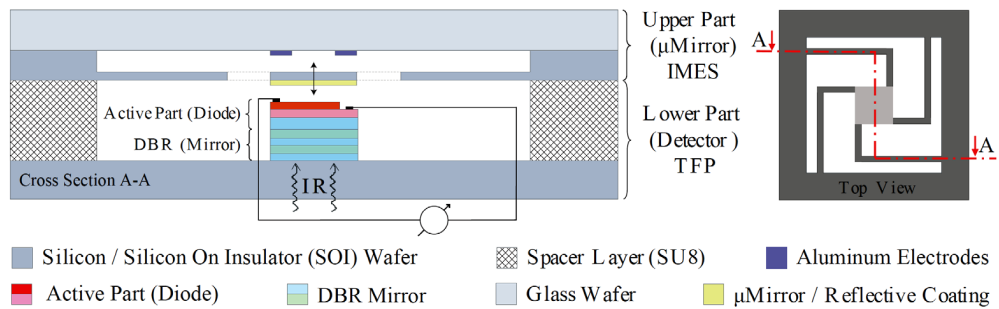

Fig. 4. Cross sectional schematic of a resonant cavity detector with cavity length tunable by MEMS. The side view is shown at left, and the top view at right. Reproduced with permission from the authors of Ref. [34].

An RCID's frequency response is limited primarily by: (1) the time required for collecting photogenerated minority carriers and (2) the RC time constant of the detector circuit. Due to its very thin absorber, it is unlikely that photocarrier collection limited the 7 GHz 3-dB bandwidth that was recently reported in Ref. [5]. The capacitance of that device was minimized by reducing the mesa diameter to 30 μm, and by employing a semi-insulating GaAs substrate for the GaAs/Al$_{0.92}$Ga$_{0.08}$As bottom mirror that was bonded to the detector. To further increase the speed, future devices may employ even smaller mesa diameters. The net signal-to-noise ratio can then be enhanced by separately monitoring multiple small devices, and then combining or averaging their responses.

Imaging by a focal plane array (FPA) [37] of RCIDs will be valuable in a number of applications such as remote chemical sensing and hyperspectral imaging. Conventional fabrication protocols that indium bump bond the array to a read-out integrated circuit (ROIC) should be compatible, following removal of the substrate for illumination from the bottom. Dynamically-tunable two-dimensional (2D) arrays should also be feasible, although it may be challenging to maintain the same variation of cavity length for all array elements. A simpler first step may be to process a series of linear one-dimensional (1D) arrays on the chip, with chirping of the cavity lengths for the different columns introduced during the growth or subsequent processing. Imaging at a given $\lambda_{res}$ could then be realized by scanning a selected linear array.

The preceding comments concerning the status of RCID technology apply primarily to MWIR devices, whereas LWIR RCIDs are currently far less mature [11,22]. The cavity fabrication becomes somewhat more challenging in the LWIR due to scaling of the cavity and mirror layer thicknesses with wavelength, and it may become more challenging to minimize dark currents relative to state-of-the-art broadband detectors. Parasitic absorption within the cavity may also become more significant. Nonetheless, thermal background radiation is more pervasive at longer wavelengths, so there is even stronger motivation to minimize the non-resonant photocurrents if the signal of interest falls within a restricted bandwidth.

Both the yield and cost of manufacturing RCIDs relative to conventional broadband IR detectors will be affected somewhat by the additional requirements for growing, bonding, or depositing top and bottom mirrors to form the cavity. However, this will be offset somewhat by less demanding growth of the much thinner absorber. Naturally the introduction of means for tuning the resonance wavelength will have further impact.

**Future developments to address challenges**

If the challenges discussed above can be overcome successfully at both ambient and cryogenic temperatures, we may expect MWIR and LWIR RCIDs to become the sensors of choice for a number of applications. For example, RCIDs that combine high EQE with low dark current and a strong suppression of thermal background photocurrents will be advantageous in most any scenario involving the detection of an MWIR or LWIR laser signal. At high temperatures, $D^*$ will then exceed that of state-of-the-art broadband detectors if lower dark current related to the thin absorber and carrier depletion can be realized. At lower temperatures, $D^*$ will be enhanced by the RCID's inherently strong suppression of thermal background photocurrents at wavelengths outside the narrow spectral band of interest, which has already been demonstrated experimentally [24]. Applications include free space optical communication [5,6,38], spectroscopy-based chemical sensing [3,39-41], on-chip spectroscopy [42,43], and hyperspectral imaging [4], along with the generic detection of other weak, narrow-band signals in the field or in the laboratory.

Whereas conventional laser spectroscopy combines a narrow-band laser source with a broadband detector, optimized RCIDs may reverse that by combining an inexpensive broadband IR source, such as an interband cascade LED [44,45] or even a thermal emitter, with a narrow-bandwidth detector whose resonance wavelength targets a specific fingerprint spectral feature. Moreover, tunability of the resonance over a sufficient spectral range will enable flexible multi-species sensing that can span the absorption features of a broad variety of chemicals.

Besides minimizing the bit error rates in high-speed data links [5], the RCID's combination of high frequency response with high responsivity will also be advantageous in applications such as real-time kinetic observations of complex chemical reactions [41] and optical heterodyne detection [46]. Heterodyne spectroscopy with dual frequency combs is realized by beating the spectral responses of two combs in a chemical fingerprint region [47]. This down-converts the IR spectroscopic information to the radio-frequency (RF) domain, where it can be readily detected as photocurrent using a fast detector such as an RCID.

Dynamic piezoelectric [12] or MEMS [34,36] tuning of the cavity length, or static chirped spatial tuning of the resonance wavelength [35] will further expand the RCID functionality, *e.g.*, in multi-species chemical sensing or in military applications such as hyperspectral imaging. Flexible adjustment of the spectral peak will nearly always be advantageous, if only to assure optimal matching of the laser wavelength with a narrow RCID resonance.

Some applications will require linear or 2D RCID arrays that provide FPA imaging. These include remote [48] or microscopic [49] chemical sensing and hyperspectral imaging, as well as many other civil and military scenarios. Most of these will further benefit from dynamic or chirped tuning of the array's resonance wavelength.

## Concluding Remarks

Each of the applications discussed above is driven by its own specific set of performance metrics. Nonetheless, if future RCIDs can succeed in combining high detectivity with narrow spectral bandwidth, tunable resonance wavelength, and high speed, they will be advantageous over broadband detectors in almost every system that requires spectral selectivity rather than accumulated integration over a broad spectral bandwidth.

### *Acknowledgments*

We thank Drs. Kevin Leonard and Richard Espinola of ONR for supporting this research, Vijay Jayaraman and Robert Marsland for valuable discussions, and the scientists at Praevium Research, Intraband LLC, and NRL for essential contributions to the RCID development.

### *Disclosures*

The authors declare no conflicts of interest.

# 15. Mid-Wave and Long-Wave Infrared Avalanche Photodiodes

NATHAN GAJOWSKI,[1] SANJAY KRISHNA,[1,*]

[1]*Department of Electrical and Computer Engineering, The Ohio State University, Columbus, Ohio, 43210, USA*
**krishna.53@osu.edu*

## Current status

Avalanche Photodiodes (APDs) are solid-state detectors that absorb photons and subsequently impact ionize the photogenerated carriers to increase the signal. This process allows a small number of incident photons (low optical power) to produce many electrons transiting the device (large current). The resulting electronic gain makes APDs useful for detecting signals of extremely low optical intensity. In the mid- and long- wave infrared (MWIR and LWIR), relevant applications include passive and active imaging, sensing, free space communication, and other medical and military applications. Detectors operating in these wavelengths are of particular interest due to atmospheric transmission windows resulting in long range under adverse weather conditions.

APDs offer performance improvement over unity gain detectors only when the circuit noise limits the overall system signal-to-noise ratio (SNR). This can be seen in Eq. (1) as the average gain, $M$, multiplies both the photo and dark currents of the detector, but not the circuit noise contribution [1].

$$SNR = \frac{I_{ph}M}{\sqrt{2q(I_{ph} + I_{dark})BF(M)M^2 + \sigma_{circuit}^2}} \qquad (1)$$

Here $q$ is the fundamental electron charge, $B$ is the bandwidth in Hz, $I_{ph}$ is the primary photocurrent, $I_{dark}$ is the primary dark current including that from background radiation, and $\sigma_{circuit}$ is the input-referred circuit noise, in Amps, from the transimpedance amplifier (TIA) or readout integrated circuit (ROIC). The excess noise factor [2], $F(M)$, which is a function of gain, is also introduced, and often imposes an upper limit on the effectiveness of an APD's gain in improving the system SNR. F(M) is the excess noise that results from the stochastic nature of the impact ionization, with $F$=1 corresponding to noiseless gain [3]. For this reason, for most materials there is a single gain point, or operating voltage, that maximizes the SNR of an APD detection system.

The excess noise of an APD can often be analytically characterized by McIntyre's local field model of multiplication noise [4].

$$F(M) = kM + (1-k)\left(2 - \frac{1}{M}\right) \qquad (2)$$

This model makes several simplifying assumptions, including the shape of the probability distribution function. However, the closed form expression is often useful for understanding the physics of the impact ionization process and characterizing $F(M)$ using the ratio of hole to electron impact ionization coefficient, $k=\beta/\alpha$. Since $\beta$ and $\alpha$ are field dependent material parameters, $k$ is often quoted for specific materials and alloy compositions. To achieve low noise multiplication, one carrier type should dominate the impact ionization, which corresponds to $k \sim 0$ for electron- and $k \sim \infty$ for hole-dominated gain. It is also critical to inject the dominate impact ionizing carrier from the photogeneration region for low noise gain. In the case of this single carrier ionization limit, $F(M)$ saturates at a value of 2. However, realistic factors such as dead space, the probability distribution of impact ionization events, and multiplier thickness result in a device specific effective $k$ value. Additionally, some materials such as HgCdTe and InAs have band structures conducive to single carrier impact ionization and demonstrate excess noise factors below 2 even at high gains, displaying sub-local noise behavior. These materials are sometimes referred to as eAPDs as they exclusively multiply electrons.

The field of MWIR and LWIR APDs is quite young, with a limited number of experimentally demonstrated devices. The main material systems used are compound II-VI and III-V semiconductor alloys. The dominant technology for MWIR and LWIR APDs is currently the II-VI HgCdTe. Due to its direct band gap, bulk HgCdTe can achieve high Quantum Efficiency (QE) at wavelengths from 1 to 30 μm by varying its composition. The band structure of this alloy is also ideal for achieving high gains ($\sim 10^2$-$10^4$) with low excess noise ($>1.5$), because the impact ionization is almost exclusively by electrons. The combination of these properties results in high performance detectors utilizing simple structures like p-n and p-i-n. HgCdTe APDs have been demonstrated in both the MWIR and LWIR [5-7]. The main drawbacks of the HgCdTe material system are the low manufacturing yields, lack of applications outside of IR leading to increased cost and decreased robustness when compared to III-V materials that can be grown in commercial foundries [8].

In comparison, III-V materials have stronger chemical bonding, making them more robust, and have broad applications resulting in a mature and maintainable manufacturing base. However, no single III-V alloy has both the extreme band gap tunability and favorable impact ionization properties of HgCdTe. Since different materials are optimal for the absorber and multiplier sections, the result is called a separate absorber and multiplier (SAM) APD. MW/LWIR III-V absorbers include the bulk materials InAs ($\lambda_c \sim 3.2\ \mu m$) [9], InAsSb ($\lambda_c \sim 3.2 - 12.5\ \mu m$) [10], and InSb ($\lambda_c \sim 5.4\ \mu m$), as well as superlattices such as InAs/GaSb [11-13] and InAs/InAsSb that can span the MW/LWIR. Some of which have been incorporated into SAM APD heterostructures. While these are the most reported III-V absorbers, other superlattices [14] and quantum dot absorbers [15] have also been employed in APDs. The reported devices often utilize wider band gap AlAsSb multipliers, which are sometimes alloyed with In or Ga to achieve high gains and low excess noise factors without tunneling [16, 17]. Additionally, InAs and InSb have themselves displayed favorable electron-initiated impact ionization properties, with peak gains of $\sim$330 [18] and $\sim$10 [19] as well as low excess noise factors of $\sim$1.6 [20] and $>2$ (predicted) [21], respectively.

The noise equivalent power (NEP) of a receiver system offers a fair comparison of APDs for a given application since it incorporates TIA or ROIC noise, which can be the limiting noise source of a system. NEP also includes the dark current, QE, gain, and excess noise of the device. NEP is often discussed in the units of Watts, which is more relevant for conventional thermal detectors. For quantum detectors, it is more insightful to present the incident power in units in photons per second and use noise equivalent photons (NEPh) as the relevant figure of merit. This removes the wavelength dependance and is convenient when discussing the extremely small optical signals that APDs are often called on to detect. Figure 1 shows a comparison of NEPh for reported MW and LWIR APDs operating at their highest gain. It should be noted that the highest gain does not necessarily result in the lowest NEPh or highest SNR. For devices with $k>0$, where the excess noise increases with gain, there is an operating point that optimizes these figures of merit for a system. Alternatively, devices with $k=0$ or sub-local noise will provide the greatest enhancement of these figures of merit at their maximum gain. However, it may not be desirable to operate at the highest gain due to bandwidth degradation.

To compile this comparison, numerous assumptions and extrapolations were necessary. Table 1 shows a summary of the reported MW/LWIR APDs and the calculated and assumed values that were employed. NEP was calculated using eq. 2.223 from [1] and then converted to NEPh assuming a TIA noise of 10 nA and bandwidth of 200 MHz.

**Table 1.** Comparison of reported MWIR and LWIR APDs. Bolded values were calculated, extracted from figures, based on standard values in literature, or a combination thereof. Due to limited data, values may also have been taken from other operating temperatures, biases, or device sizes in the same report. Excess noise was calculated with the local model [4] when the excess noise was not given at the required gain. The gain values were not optimized for minimizing noise for these calculations, rather the maximum reported gain was used.

| Author | Absorber | Multiplier | Temp. (K) | $\lambda$ ($\mu$m) | Max gain, $M$ | $F(M)$ | QE | $\langle I_{dp} \rangle$ (A) | Responsivity (A/W) | NEPh (Photons s$^{-1}$) |
|---|---|---|---|---|---|---|---|---|---|---|
| Dadey [16] | Al$_{0.05}$InAsSb | Al$_{0.7}$InAsSb | 100 | 3 | 850 | **35.91** | 0.58 | **3.54·10$^{-3}$** | **1192.89** | **3070** |
| Ramirez [15] | QW/InAs QDs | GaAs | 77 | 5 | 14 | **4.95** | 0.0084 | **4.5·10$^{-12}$** | **0.47** | **531** |
| Huang [9] | InAs | AlAs$_{0.13}$Sb$_{0.87}$ | 295 | 3.27 | 13.1 | **1.98** | **0.23** | 1·10$^{-3}$ | 8.09 | **9494** |
| Dehzengi [14] | 11/2 ML InAs/InSb | | 150 | 3.75 | 6 | **2.96** | 0.82 | 5·10$^{-5}$ | **14.94** | **738** |
| | | | 77 | 3.75 | 7.4 | **3.36** | 0.77 | 5·10$^{-5}$ | **17.17** | **844** |
| Li [10] | InAs$_{0.9}$Sb$_{0.1}$ | InAs$_{0.9}$Sb$_{0.1}$ / (5/2 ML AlAsSb/GaSb) | 200 | 3.9 | 29 | **4.59** | 0.86 | 2.76·10$^{-5}$ | **78.88** | **650** |
| | | | 150 | 3.9 | 121 | **13.54** | 0.54 | 2.76·10$^{-5}$ | **206.91** | **1776** |
| Beck [5] | HgCdTe ($\lambda_c$=4.34) | | 77 | 4.34 | 1270 | 1.35 | ~0.9 | 1.4·10$^{-13}$ | **4001.01** | **0.06** |
| | HgCdTe ($\lambda_c$=9.7) | | 77 | 9.7 | 114 | **1.35** | ~0.9 | 2.38·10$^{-8}$ | **802.70** | **10** |
| Yan [11] | 14/12 Å InAs/GaSb | AlAs$_{0.06}$Sb$_{0.94}$ | 77 | 3 | 32.1 | **2.12** | 0.067 | 2.02·10$^{-9}$ | **5.15** | **59** |
| | | | 200 | 3.25 | 6.1 | **1.86** | 0.053 | 5.95·10$^{-5}$ | **0.85** | **9886** |
| Banerjee [12] | 35/3/20/13/1 Å InAs/GaAs/Ga$_{0.6}$In$_{0.4}$Sb/AlSb/GaAs | | 77 | 6.3 | 105 | ~4 | - | 9.52·10$^{-4}$ | - | - |
| Abautret [21] | InSb | | 77 | 5.45 | 3 | - | - | 3.46·10$^{-6}$ | - | - |
| White [18] | InAs | | 200 | ~3.25 | 330 | ~1.6 | ~0.5 | 1.11·10$^{-6}$ | **432.51** | **42** |
| Ker [22] | InAs | | 77 | ~3 | 26.5 | ~1.6 | ~0.5 | 1.02·10$^{-8}$ | **32.06** | **5** |

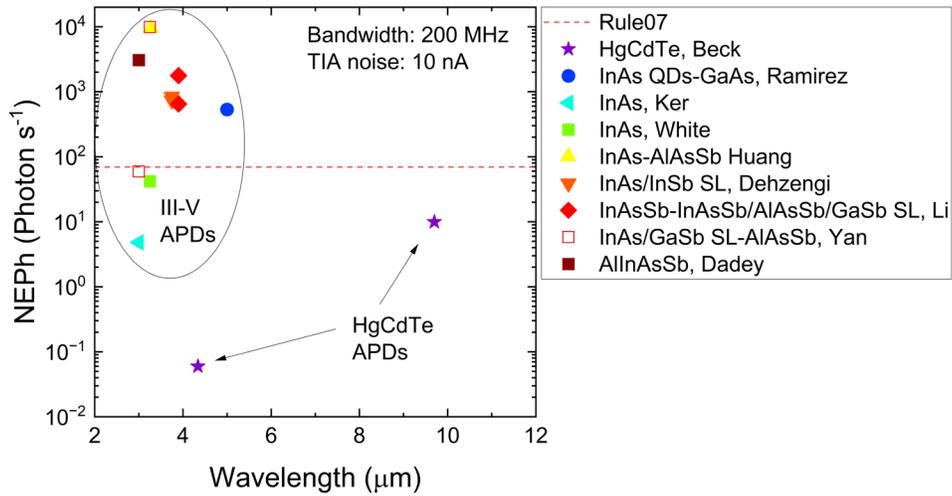

**Figure 1.** Calculated NEPh for an APD receiver system comparing reported MW/LWIR APDs and Rule '07 [23] of a unity gain device for reference. For SAM devices, the absorber and multiplier materials are separated with a hyphen (Absorber-Multiplier). Reports include Beck [5], Ramirez [15], Ker [22], White [18], Huang [9], Dehzangi [14], Li [10], Yan [11], and Dadey [16].

For this noise level and bandwidth, both MWIR and LWIR HgCdTe APDs display lower NEPh than a unity gain Rule '07 detector (dashed line) by roughly 3 and 1 orders of magnitude, respectively. This can be attributed to their high unity gain QE in conjunction with extremely low noise even at high gains. Some III-V material systems also display lower NEPh than Rule '07, but cannot fully cover the MW and LWIR spectral regions. From this analysis it appears that InAs is the most promising III-V multiplier due to its eAPD properties.

## Challenges and opportunities

The central challenge facing MWIR and LWIR APDs is the competing electric field requirements for useful gain vs. low dark current. Useful gain requires fields in the tens or hundreds of kV/cm, while tunneling can occur in the required narrow band gap materials at just a few kV/cm or less (depending on the cutoff wavelength), resulting in severe dark current penalties.

The SAM APD is often implemented effectively in III-V short-wave infrared (SWIR) devices. SAM APDs can implement separate materials for each layer, with different band gaps to meet the needs of the layer. A charge layer is often added to the device to form a SACM, which allows for a greater field offset between the absorber and multiplier for reduced dark current from the absorber at equivalent gains. Figures 2a and 2b show the structure of an SACM device and the corresponding ideal field profile. A band gap grading layer is often required and implemented in tandem with the charge layer to facilitate transport of the photogenerated carriers from the absorber into the multiplier as shown in Figure 2d. In reality, the fields in both the absorber and multiplier have non-zero slopes, and the slope directions and magnitudes are determined by the doping polarity and doping concentration of the layers. Additionally, the field levels can spike near doping level and material interfaces, as shown in Figure 2c. A high performance SACM device must be designed such that the absorber field is below the absorber tunneling threshold while the multiplier field is above the impact ionization threshold but below the Zener tunneling threshold. The optimal gain level for linear-mode APDs is dependent on the material, device structure, and system noise.

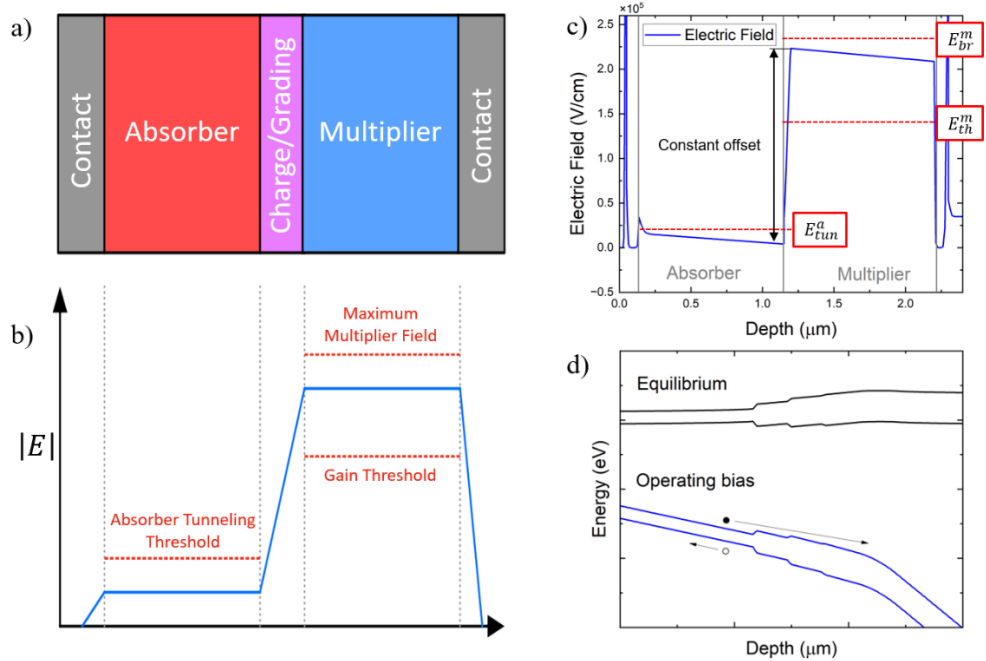

**Figure 2.** (a) SACM device structure; (b) electric field profile; (c) simulated MWIR SACM field profile at the operating bias, which illustrates field spiking; and (d) simulation of the band gap step grading under the equilibrium condition at the operating bias.

While the SACM approach works well for SWIR APDs, both the absorber band gap and tunneling threshold decrease as the absorption wavelength is extended to the MWIR and LWIR. This inherently limits the device operating range as the charge layer sets a static field offset between the absorber and multiplier. However, since an APD has a single gain point where the device maximizes the SNR, there is the opportunity to improve the SNR through co-design by the system and APD device designers.

Another challenge faced by MW/LWIR APDs is the current lack of custom ROICs designed for these devices. To achieve peak system SNR, APDs often operate at biases above the capabilities of commercially available ROICs. Therefore, we need to either develop ROICs capable of handling high voltage traces or decrease the operating bias of the APDs to increase the SNR of the overall imaging or sensing system.

Additionally, MW/LWIR APDs will face the same challenges as other detectors operating at these wavelengths. These include high dark current that results not only from the narrow band gap, but also from the high surface component that often plagues narrow band gap materials (especially those containing InAs).

Compared to HgCdTe, III-V materials face challenges of their own. III-V multipliers generally have higher unintentional background doping than that of highly optimized HgCdTe (in the range of $10^{13}$ cm$^3$), limiting the gain in multiplication regions due to the peak field. III-V materials generally have more degrees of freedom, including the availability of ternaries, quaternaries, and quintenary alloys, and quantum engineered structures like superlattices and metamorphic structures. This leads to a wide number of parameters to optimize the material quality. Moreover, these superlattices and other low dimensionality absorbers have lower quantum efficiencies than bulk materials due to their lower density of states.

**Future developments to address challenges**

To become technologically relevant, MWIR and LWIR APDs must offer superior performance to the state-of-the-art detectors operating at unity gain. To achieve high performance, a MW/LWIR APD must exhibit low dark current and noise at the operating bias. Key advancements to achieving an APD with SNR enhancement in this wavelength range include:

1. Reduction of dark current: One way to reduce the bulk dark current of an MW/LWIR device is to minimize depletion in the narrow band gap absorber material. This is leveraged by both barrier [24] and junction detectors that operate at extremely low biases. Additionally, in some barrier detector designs depletion is intentionally confined to the wider bandgap unipolar barrier to reduce the dark current resulting from depletion in the narrow band gap absorber. However, to function as an APD the device must operate with an intentionally high field in the multiplication region of the device.

2. Optimization of carrier transport: Carrier transport from the absorber to the multiplier is critical. Careful device design should enable the demonstration of an APD with no depletion in the MW/LWIR absorber, and a high enough field to achieve gain in the wider band gap multiplier, while enabling carrier transport between the two layers.

3. Heterostructure Engineering: The excess noise in the multiplier must be low to attain a noise advantage over unity gain devices. The excess noise in the multiplier depends mainly on the material used, and the relation between the electron and hole impact ionization coefficients. Both device band diagram engineering [3, 25] and superlattice band structure engineering [26] have been proposed and demonstrated to improve multiplier excess noise properties.

4. Suppression of surface leakage current: A key dark current reduction tool is the implementation of a barrier to reduce the major surface current component at these wavelengths. The addition of a wider band gap unipolar barrier may be utilized to increase the surface shunt resistance and therefore reduce the dark current and noise of an APD at operating bias [27]. Robust passivation techniques that can suppress the surface leakage current at the operating bias also need to be developed.

One compelling solution may be an SAM device in which the primary photogenerated carrier *sees a zero or negative band offset when transiting from the absorber to the multiplier*. In this device, no grading layer is needed as the carriers can travel into the multiplier unimpeded. One possible material combination that satisfies these conditions is a T2SL InAs/GaSb absorber and bulk InAs as the multiplier grown on an InAs substrate, as shown in Fig. 3. *In MWIR or LWIR InAs/GaSb T2SLs, the quantum confined conduction miniband will always be above the bulk band edge of InAs, which allows the easy transit of photogenerated electrons from the T2SL into the InAs multiplier.* A charge layer can also be added to the device to provide fields high enough to induce impact ionization in the InAs. This device structure would leverage an undepleted T2SL MW or LWIR absorber design, low noise gain made possible by the favorable band structure of InAs and implement an AlAsSb barrier to reduce the surface current. Due to the relatively simple design and binary materials, this device stack should be relatively easy to grow. Strain-balanced InAs/GaInSb or InGaAs/Ga(In)Sb could also be used for T2SL absorbers on InAs. The barrier could consist of AlAsSb or a wider band gap T2SL.

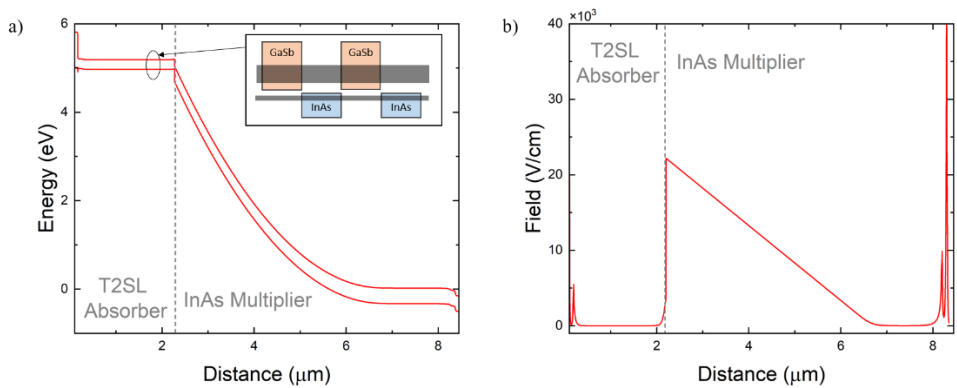

**Figure 3.** Simulated (a) band diagram and (b) field profile of an MWIR SACM using InAs/GaSb T2SL absorber and InAs multiplier at operating bias. As shown in the insert, the quantum confined T2SL miniband is above the band edge of bulk InAs, resulting in the easy transport of photogenerated electrons from the absorber to the multiplier with no grading required.

Similarly, a T2SL absorber grown on GaSb can be paired with an AlGaSb multiplier, which is reported to exhibit impact ionization coefficients that are favorable for hole-initiated impact ionization. This device would exhibit an n-i-n-i-p design that would easily allow photogenerated holes to transit into the AlGaSb multiplier due to the negative band offset.

In other material systems, a path to creating high gain is to design a conventional linear-mode SACM and operate it at only one gain point. The design of this device would be critical. The operating gain should be selected theoretically using knowledge of the selected material's excess noise. The charge layer should then be designed such that the multiplier reaches the field required to achieve the selected operating gain at the same bias the grading is lowered, and carriers from the absorber can transit unimpeded into the multiplier. In this way, the field will not penetrate the absorber, limiting the tunneling current. However, increasing the bias beyond this operating voltage will likely result in increased dark currents from tunneling. This is a reasonable design choice, as the peak SNR occurs at one gain for a system with a set amplifier/ROIC noise floor. A similar device could be designed to operate in Geiger mode by setting the charge in the charge layer such that the field in the multiplier slightly exceeds the avalanche breakdown voltage at punch through. Geiger mode APDs cannot detect a signal amplitude, only the signal timing. This makes these devices useful for ranging applications.

In HgCdTe, a similar device that separates the absorption and multiplication regions could be realized using compositional grading. This is especially important in LWIR devices as the presence of tunneling at the fields required for impact ionization likely increases the noise, nullifying any enhancement of the SNR. Implementing an SAM design could likely suppress the tunneling current in the narrow band gap absorber and enable APDs with SNR enhancement at LWIR wavelengths. However, in this material system the compositional grading required to increase the band gap will also change the lattice constant of the material, which will likely degrade material quality. When designing these devices, care must be taken to minimize the field throughout the narrow band gap absorber, including field spikes near interfaces. This has been achieved at doping interfaces by grading the doping.

Additionally, thinning the absorber may provide an appealing way to enable detectors operating at elevated temperatures with comparable dark current levels. Resonant enhancement could be implemented to reduce the dark current by reducing the absorber thickness while maintaining high QE using plasmonic [28] or distributed Bragg reflectors [29]. Bismuth containing materials are also under investigation, as they could offer increased absorption over Ga-containing or Ga-free superlattices in these wavelength rages to provide similar QE with a thinner absorber [30].

Regardless of design, disruptive III-V devices will display a receiver NEPh at least an order of magnitude below a unity gain Rule '07 device for a given system noise and bandwidth. Ideally, these devices will operate at biases attainable by current commercial ROIC technology, namely below 6V. Additionally, reducing the receiver NEPh below 1 will enable single photon detection and accurate photon counting. Realizing these devices within the III-V material system will reduce the cost of these detectors by leveraging the mature manufacturing base due to the versatility of these materials.

## Concluding Remarks

Research on MWIR and LWIR APDs is still in its early stages, with a limited number of demonstrated devices. Further development of the device design and processing is needed to realize a performance advantage over unity gain detectors throughout this wavelength range. When realized effectively, MW/LWIR APDs have the possibility to enhance the detection of extremely low light signals, opening the door to new capabilities for imaging, sensing, and communication systems.

## Acknowledgement

The authors thank Dr. Pradip Mitra for helpful discussion and resources on HgCdTe and Dr. Seunghyun Lee for discussion and simulation data on SACM APDs.

Research was sponsored by the United States Air Force Research Laboratory and the United States AFRL Regional Hub and was accomplished under Cooperative Agreement No. FA8750-22-2- 0501. The views and conclusions contained in this document are those of the authors and should not be interpreted as representing the official policies, either expressed or implied, of the United States Air Force or the U.S. Government. The U.S. Government is authorized to reproduce and distribute reprints for Government purposes notwithstanding any copyright notation herein.

Professor Sanjay Krishna acknowledges support from the George R. Smith Chair of Engineering at The Ohio State University. Nathan Gajowski acknowledges support from the Science, Mathematics, and Research for Transformation (SMART) Scholarship-for-Service Program within the Department of Defense (DoD).

# 16. SiGeSn Materials for Midwave Infrared Optoelectronics

**SHUI-QING YU,[1]\* WEI DU,[1] AND RICHARD SOREF[2]**

[1]*Department of Electrical Engineering and Computer Science, University of Arkansas, Fayetteville, AR 72701, USA*
[2]*Department of Engineering, University of Massachusetts at Boston, Boston, MA 02125, USA*
*\*syu@uark.edu*

## 1. Introduction of the SiGeSn material system

Si, Ge and their alloys have been the miracle materials driving the digital revolution for the electronics industry.[1-4] The rapid "Moore's law" miniaturization of device sizes has yielded an ever-increasing density of fast components integrated on Si, which has pushed feature sizes down close to their ultimate physical limits. At the same time, there has been a parallel effort to broaden the reach of the materials by expanding their functionalities well beyond electronics, which is evident in the development of Group-IV photonics. Simply by simply introducing another group-IV element, Sn, a new material platform SiGeSn has been created with tremendous new electrical, optical, and mechanical properties that may dramatically change the landscape of future microelectronics/photonics.[5,6] Figure 1 summarizes the increased publication numbers on this material system versus time for the past 35 years, representing dramatic increases in community interest and research progress. The inset pie chart shows that material research has accounted for close to 65% of the total publications, indicating that SiGeSn is still in the material development rather than device development stage. Readers are directed to a few extended reviews for more details concerning this material.[7]

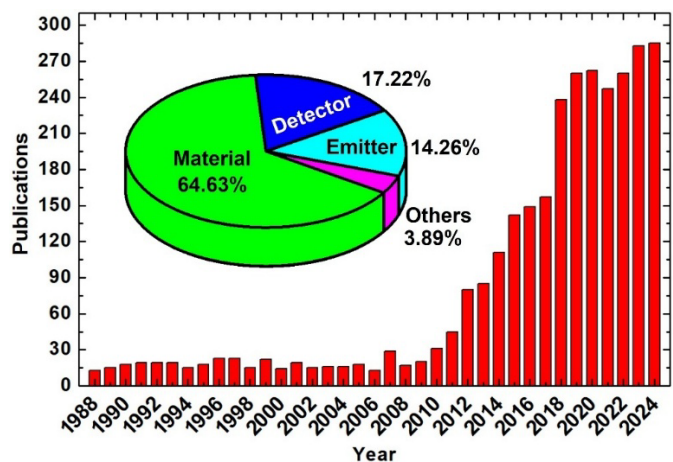

Fig. 1 SiGeSn-related yearly publications according to data collected from Web of Science. (Inset: The distribution of research subjects for this material).

**Some of the strategic attributes offered by SiGeSn are**: i) Ability to independently tune the lattice constant and bandgap by simultaneously varying the compositions of Si, Ge, and Sn (Fig. 2); ii) Potential to achieve a true direct bandgap by incorporating only a few percent of Sn; iii) Possibility of forming a desirable type-I band alignment to provide favorable quantum confinement for optoelectronics designs; vi) Potential to cover mid-IR wavelengths up to 12 μm through band-to-band transitions, and to cover all wavelengths beyond 12 μm through intersubband transitions; v) Compatibility with CMOS processing at a low growth temperature (below 400ºC); vi) Feasibility of Selected Area Growth (SAG), which is highly desirable for optoelectronic integration; vii) Potential for compliant virtual substrates consisting of fully-relaxed GeSn layers on Si for material integration; viii) Potential to obtain high thermoelectric figure of merit by adding Sn into a SiGe superlattice to reduce the thermal conductivity and increase the electrical conductivity; ix) Potential for high electron and hole mobilities through strain engineering for future CMOS devices; x) Potential for high- performance intersubband devices due to the long phonon scattering lifetimes in this non-polar material; xi) Potential to enhance the device performance due to a low Auger recombination coefficient.

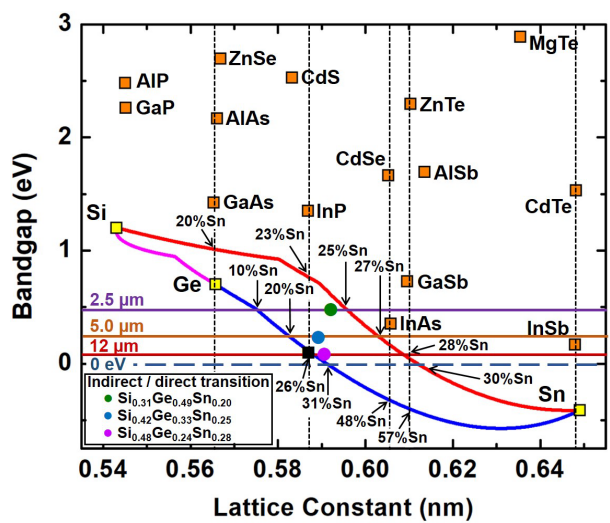

Fig. 2 Bandgap energy versus lattice constant for different compounds including SiGeSn

## 2. SiGeSn Material Growth

From the material growth perspective, it is known historically that the major technical challenges of incorporating Sn into the Ge lattice are the large (17%) lattice mismatch between the elements and the instability of α-Sn's diamond-cubic structure above 13ºC.[5] As a result, the thermodynamic solubility of Sn in Ge is less than 0.5%, while that of Ge in Sn is zero.[8] GeSn alloys are thus unstable, and their synthesis requires conditions that are far from equilibrium. The synthesis of SiSn and SiGeSn alloys suffers a very similar thermodynamic solubility issue, even though the SiGe alloy can achieve any arbitrary composition. SiGeSn alloys (GeSn, SiSn, and SiGeSn) have been grown successfully by both physical vapor deposition (PVD) and chemical vapor deposition (CVD) techniques. Due to the intrinsic growth challenges, the composition for successful Sn incorporation has been used as an important figure of merit in the early stages of the material development. Along with the evolution of growth techniques, recent research has gradually shifted towards focusing on substantial improvement of the material quality, as this is essential to the realization of desirable optical and electronic material properties. Figure 3 shows a composition contour plot for materials with certain compositions that were grown by different techniques, and for which a large range of ternary alloy compositions has not yet been explored. With α-Sn demonstrating topological quantum properties, there is now increased interest in studying random or digital semi-metal SiGeSn alloys (compositions indicated as the shaded region) for Group-IV topological quantum alloys. This represents a paradigm shift from mainstream topical quantum material studies, which have searched for new exotic materials that possess significant new topological quantum properties. Group-IV topological quantum alloys will demonstrate different quantum properties by using a single material platform that tunes the Si, Ge, and Sn compositions for the integration of topological quantum devices on a Si-CMOS platform, by analogy to the early vision of integrating SiGeSn optoelectronics on the Si-photonics platform.

## 2.1 MBE (PVD) grown SiGeSn

Many PVD techniques, including molecular-beam epitaxy (MBE), sputter deposition, solid phase recrystallization, and solid phase epitaxy, have been used to synthesize both epitaxial and polycrystalline GeSn thin films. MBE has a long history of use for SiGeSn material growth and has demonstrated a large range of Sn compositions in GeSn, SiSn, and SiGeSn (Fig. 3). However, a major problem encountered with MBE is the propensity of Sn to segregate toward the film surface.[8] To counteract this effect, MBE growth of GeSn must be conducted at very low temperatures (~100-200ºC), which places a severe limitations on the film critical thickness and material quality. To overcome this effect, the (Si)GeSn must be grown at low temperatures and be assisted by *in-situ* characterization,[9] along with the growth of superlattice structures and surfactants to help stabilize the materials.[10,11] While MBE-grown GeSn with low Sn composition has achieved good quality, as evidenced by device demonstrations, only recently has photoluminescence (PL) characterization confirmed the MBE material composition and quality to have reached the critical milestone of true direct bandgap,[12] which is 11 years behind the same milestone for CVD-grown GeSn. However, the PL intensity is much lower than for a CVD benchmark sample with similar composition and structure, indicating the need for substantial further improvement of the material quality.

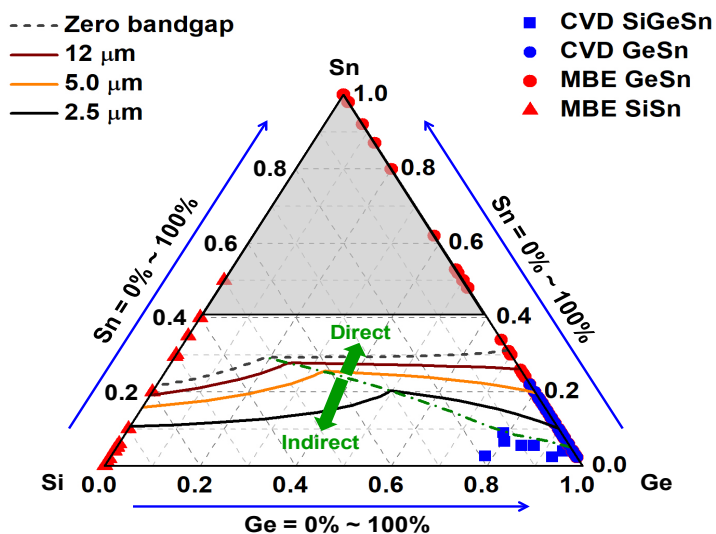

Fig. 3 Composition diagram showing constant bandgap contours, indirect-direct bandgap partition, existing compositions, and shaded area (Semi-metal).

## 2.2 CVD-grown SiGeSn using the SnD₄ precursor

A significant material breakthrough was made by the Kouvetakis' group at Arizona State University using UHV-CVD.[13-15] This success originated from the discovery of a viable Sn precursor (SnD₄) that is remarkably stable at 22ºC for extended time, thereby enabling the synthesis of SiGeSn alloys using the CVD technique with Si, Ge, and Sn-based hydrides.[14] For a decade in early 2000s, this growth approach had dominated a significantly advance of material development by growing a variety of materials such as GeSn, SiSn, SiGeSn, and their heterostructures such as quantum wells (QWs). This led to the demonstration of doping and consequently PN junctions and protype LEDs and photodiodes, and contributed to the early understanding of the material band structure. The solid foundation laid by these developments led to the blooming of SiGeSn research.[16] Nonetheless, the approach has major limitations due to precursor instabilities and availability, making it extremely difficult to scale-up for follow-up investigations by the broad research community. GeSn lasers, another important material development milestone, have not been demonstrated by this growth approach, while many teams using $SnCl_4$ as precursor for CVD growth have broadly demonstrated both optically-pumped and electrically-injected GeSn lasers.

## 2.3 CVD-grown SiGeSn using the $SnCl_4$ precursor in industry-CVD reactors

A new era of GeSn growth began in 2012, when several groups almost simultaneously committed to using industrial grade Group-IV CVD reactors made by major tool vendors for the material development. It was also spurred by the discovery that the new $SnCl_4$ precursor can work with different Si ($SiH_4$, $Si_2H_6$, or $Si_3H_8$) and Ge hydrides ($GeH_4$ and $Ge_2H_6$) to produce device quality GeSn and SiGeSn,[17-22] which was originally motivated by the material's application to transistors. This wave of strategic investment has been responsible for the demonstration of material critical properties; for example, the true direct bandgap signature and eventual attainment of the key GeSn laser milestone has enabled the transition of SiGeSn research from pure material studies to device development. It also serves as a testing vehicle to provide feedback for material growth with different compositions, to obtain sophisticated structures, to improve the material quality, and to eventually demonstrate desirable device performance. So far, almost all of the best GeSn device results have been obtained using materials grown by this approach. Therefore, the discussions in the following sections will assume $SnCl_4$-based CVD growth, unless a different growth approach is specified.

## 2.4 Materials demonstrating critical properties: A pathway from composition, to quality, and then intrinsic properties

While exploring the growth of different compositions of SiGeSn material remains the focus for material development, it is equally important that the material quality be sufficient for obtaining desirable critical properties. From the device application perspective, this ensures maximum design freedom in the material "toolbox", which can supply the needed building blocks for deploying bandgap engineering of different device structures.

For bulk GeSn, one of the most important properties is to have a true direct bandgap with efficient light emission. This is often obtained by growing relaxed GeSn bulk material with Sn composition beyond 8%. Since the first demonstration of true direct bandgap GeSn, bulk GeSn can now be grown routinely with a clear direct-bandgap PL signature and dramatically-enhanced intensity at cryogenic temperatures. This starkly contrasts the intensity decrease with decreasing temperature that occurs when the bandgap is indirect and the PL is photon-assisted. Figure 4 shows a summary of composition- dependent absorption and PL spectra for a series of compositions. As common in other semiconductor materials, the achievable bandgap is limited by the material composition, and by the strain condition that is further constrained by the layered material compositions.

At present, the primary driver of SiGeSn development is its use as the barrier material in GeSn QW lasers. Indeed, SiGeSn offers unique material properties that are yet to be explored. Examples are the feasibility of tuning the composition to achieve direct and indirect bandgap transitions that significantly enhance the carrier lifetime (See section 4.3.4), or to attain short-range ordering for bandgap engineering (See section 4.2.4).

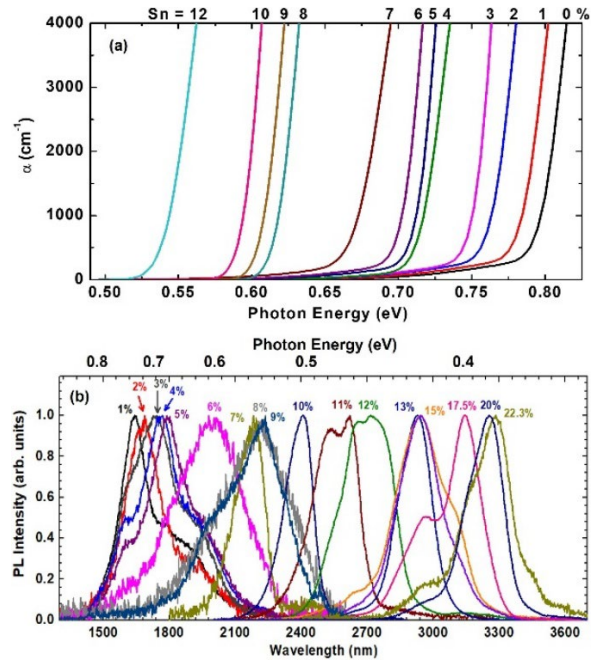

Fig. 4 Sn compositional dependence of (a) absorption coefficient α, and (b) PL spectra

As an example of state-of-art work to develop GeSn with different compositions for building structures with desirable properties, in the following we highlight the work to obtain: i) direct bandgap type-I alignment with high injection efficiency and ii) high mobility in the quantum transport regime.

As the optical properties of bulk GeSn have been studied extensively, increasing attention has turned to quantum well (QW) structures, driven by the well-established understanding that quantum confinement can significantly enhance device performance. To support the design and optimization of GeSn-based devices, QW studies aim to identify and engineer the following key properties: 1) Bandgap directness in the well region; 2) Type I or Type II band alignment; and 3) Barrier heights for the Γ and L valleys in the conduction band (CB), as well as for the heavy-hole (HH) and light-hole (LH) states in the valence band (VB).

In early studies, relaxed Ge was commonly used as the barrier layer.[23] However, due to significant compressive strain experienced by the GeSn well layer (when pseudomorphically grown on Ge), the resulting band structure typically remains indirect, and the L-valley in the CB does not exhibit a type-I band alignment. Moreover, insufficient carrier confinement in the Ge/GeSn QWs limits their effectiveness in light-emitting applications. To mitigate compressive strain in the QW, researchers later introduced a relaxed GeSn buffer layer prior to the QW growth. This buffer, with a lower Sn composition, serves as a barrier layer relative to the GeSn QW with higher Sn content.[24,25] However, the Sn composition difference between the barrier and well must be carefully controlled, and is typically limited to around 6%. A larger mismatch would introduce excessive compressive strain in the well that not only fails to promote further bandgap directness, but may also degrade carrier confinement. This constraint becomes particularly problematic in long-wavelength applications, where the Sn content in the well is relatively high (e.g., 15%). In such cases, the barrier layer, which also contains a significant amount of Sn, may itself become a direct bandgap material, enabling radiative recombination in the barrier that competes with the QW emission. To overcome these limitations, recent studies have explored the use of ternary SiGeSn alloys as barrier materials.[26-28] By carefully tuning the Si and Sn compositions, the lattice constant and band structure can be engineered independently. This approach provides sufficient conduction and valence band offsets for effective carrier confinement, while also reducing the strain imposed on the GeSn well layer. As a result, SiGeSn emerges as a promising barrier material for high-performance GeSn QW lasers. Another critical figure of merit for device performance is the carrier collection efficiency. To enhance this property, a dual-barrier QW structure consisting of a GeSn well sandwiched between a GeSn inner barrier and SiGeSn outer barrier was recently demonstrated.[29] This configuration showed clear enhancement of the QW emission, indicating improved carrier injection and collection dynamics.

However, it is important to note that material growth remains a major challenge, particularly in achieving precise and independent control over the Si and Sn compositions in the SiGeSn alloys while maintaining high crystalline quality. These limitations currently constrain the available design space for optimized QW structures that use SiGeSn barriers. Therefore, the future development of GeSn-based QW devices will depend heavily on continued advances in epitaxial growth techniques. As more high-quality SiGeSn alloy compositions become accessible, a wider range of band structure and strain engineering options will become viable, enabling more efficient and versatile QW device designs.

GeSn heterostructures show significant promise, not only for optoelectronic applications but also for emerging spintronic and quantum electronic technologies.[30,31] Their high carrier mobility and strong spin–orbit coupling (SOC) make them particularly attractive for spintronic devices. Among the two main types of SOC, the Rashba effect offers a key advantage over the Dresselhaus effect due to its tunability via external gating—a critical feature for device design. Unlike group III–V materials, where Rashba and Dresselhaus SOC effects coexist, due to the inversion symmetry of group IV materials, GeSn alloys exhibit only the Rashba effect. This makes high-quality GeSn particularly well-suited for detailed studies of carrier transport, especially for holes. In 2021, a Hall mobility of 20,000 cm²/V·s was demonstrated for a two-dimensional hole gas (2DHG) in $Ge_{0.94}Sn_{0.06}$/Ge QWs.[30] A subsequent study reported a record-high hole mobility of 80,000 cm²/V·s in shallow and undoped GeSn/Ge QWs. Given the advantages of ternary SiGeSn alloys for carrier confinement in heterostructures, the electronic transport properties of 2D holes in GeSn/SiGeSn QWs have also been investigated, with peak mobilities of 9,000 and 19,000 cm²/V·s reported. Notably, at low carrier densities, devices incorporating SiGeSn barriers exhibited mobilities several times higher than those in comparable GeSn QWs with Ge barriers.[32,33]

Furthermore, applying tensile strain to n-type GeSn epitaxial films has been shown to enhance band-to-band tunneling, enabling high tunneling rates in GeSn tunnel field-effect transistors (TFETs).[34] In such devices, electron Hall mobilities of up to 5,000 cm²/V·s were measured at lower doping concentrations. Additionally, when the strain was varied from 1% compressive to 0.5% tensile, the electron mobility increased by a factor of seven. These transport studies underscore the considerable potential of GeSn for future quantum devices, particularly those that integrate quantum optical and quantum electronic functionalities on a monolithic platform.

## 3. Development of GeSn Based Emitter and Detectors

This section summarizes the status of GeSn based device development, with highlight on some key milestones and important device specifications. We will focus on: i) GeSn lasers for future optoelectronic integration, ii) GeSn infrared detectors for imaging applications, and iii) GeSn avalanche photodiodes (APDs) for LiDAR and sensing applications. The selection of those devices is due to their potential for significant impact on dominant research efforts in the GeSn research community. Other devices such as LEDs, waveguide detectors, and high-speed detectors will be discussed only briefly because their key application scenarios are less clear.

### 3.1 Current status of GeSn laser development

Silicon-based lasers have long been sought due to their potential for monolithically integrating photonic components with silicon microelectronics.[35] However, silicon's indirect bandgap fundamentally limits its suitability as a laser material. As a result, III-V materials have dominated the semiconductor laser market and remain the subject of extensive research and development.[36] While Ge-based lasers have been demonstrated, the inducement of a direct bandgap typically requires tensile strain or heavy n-type doping—approaches that significantly reduce the laser efficiency.[37] More recent studies have shown that GeSn alloys, when alloyed with sufficient Sn content, undergo a transition to direct bandgap that is a key requirement for efficient light emission. Importantly, GeSn can be grown directly on silicon substrates, offering a promising pathway for seamless integration with existing complementary metal-oxide semiconductor (CMOS) compatible platforms. Following theoretical predictions that a relaxed GeSn alloy with approximately 6% Sn should exhibit a direct bandgap, the first experimental confirmation of such a material was reported in 2014.[38] This marked a major milestone in the development of group-IV light sources.

The first optically-pumped Fabry-Perot (F-P) lasers were demonstrated in 2015,[18] employing a direct-bandgap GeSn active region with approximately 13% Sn. These devices exhibited a maximum lasing temperature of 90 K, and threshold pump intensity of 325 kW/cm$^2$ at 20 K. Since then, optically-pumped GeSn lasers have been studied extensively as an intermediate step toward electrically-injected lasers, with the goal of understanding lasing behavior and the underlying optical gain mechanisms. Significant efforts have been directed toward performance enhancement, including: 1) Material development, based on strategies such as increasing the Sn content and applying tensile strain to improve the direct-gap radiative recombination efficiency while maintaining high material quality. 2) Layer structure engineering: The use of double heterostructures (DHS) and multiple quantum wells (MQWs) to enhance carrier confinement and facilitate population inversion. Concurrently, SiGeSn ternary alloy growth has been pursued to provide more favorable band offsets for carrier confinement. 3) Advanced optical cavity designs: Beyond traditional F-P cavities, microdisk and photonic crystal (PhC) cavities have been employed to increase optical confinement and reduce the lasing thresholds.

As a result of these advancements, F-P lasers using 20% Sn have operated to 270 K, with lasing at 3.5 μm and a reduced threshold of 47 kW/cm$^2$ at 77 K;[39] With a SiNx stressor and microdisk cavity, continuous-wave (CW) lasing was achieved at 70 K, with a threshold of 1.1 kW/cm$^2$ at 25 K;[40] A pedestal-supported microdisk structure was developed to enhance thermal dissipation, leading to room temperature lasing from a 17% Sn device, with a higher threshold of 3.27 MW/cm$^2$ at 305 K.[41] These progressive improvements highlight the immense potential of GeSn as a group-IV laser gain material, particularly for future photonic platforms that are monolithically integrated on silicon.

A deep understanding of the intrinsic material properties, gained through optical pumping studies, has enabled a major breakthrough: demonstration of the first electrically-injected GeSn F–P laser in 2020.[42] This device achieved a maximum lasing temperature of 100 K, a peak output power of 2.7 mW per facet at 10 K, and a threshold current density of 1.4 kA/cm$^2$ at 77 K. Subsequent work, which focused on identifying the loss mechanisms and optimizing the device structure,[43] led to improved performance with a maximum lasing temperature of 140 K and a reduced threshold of 0.76 kA/cm$^2$ at 77 K.[44] In parallel, an alternative electrically-injected microring cavity was explored to promote strain relaxation. This structure demonstrated lasing to 90 K, with a threshold of 25 kA/cm$^2$ at 5 K.[45] In 2024, a significant further advance was achieved by combining a SiGeSn/GeSn MQW structure with a microdisk cavity, which resulted in CW lasing to 35 K.[46] This device emitted at 2.32 μm, with a threshold current density of 6.2 kA/cm$^2$ at 5 K. Figure 5 summarizes key milestones in the historical progression of GeSn laser development.

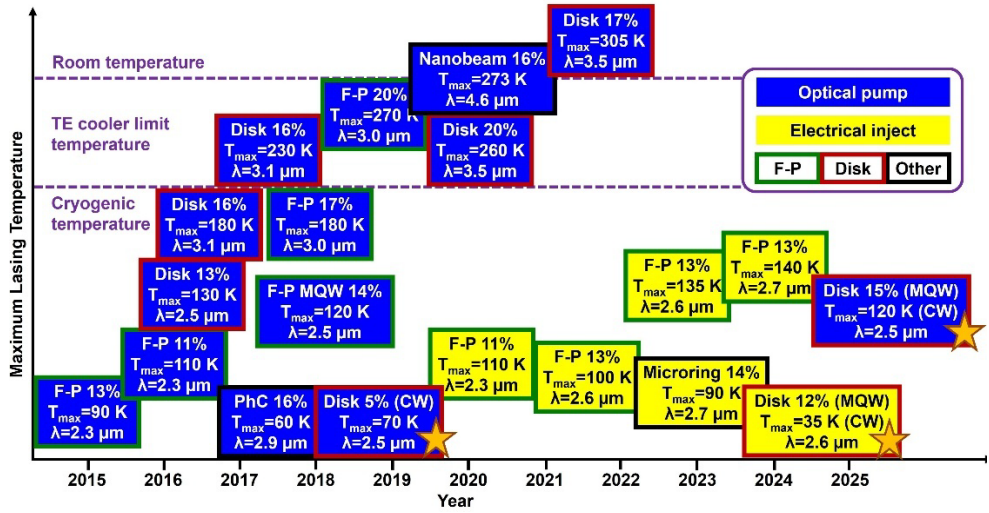

Fig. 5 Historical progression of key milestones in GeSn laser development.

## 3.2 Development of GeSn Based Infrared Detectors for IR imaging applications

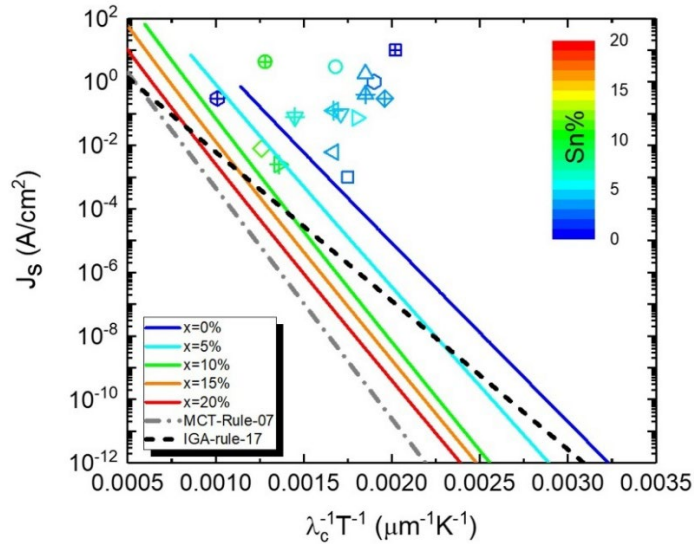

Fig. 6 Calculated GeSn-rule-23 performance by means of dark current density vs. reciprocal of the product cutoff wavelength and temperature (solid lines), compared with reported experimental data (scatters) from the literature. MCT rule-07 (dashed-dotted line) and IGA-rule-17 (dashed line) are also depicted for comparison.

Low-cost, high-performance short-wave and mid-wave infrared (SWIR and MWIR) photodetectors hold significant promise for both defense and civilian applications, particularly in night-vision technologies.[47] Targeted markets include advanced video surveillance systems, automotive night-vision cameras, and integrated infrared imaging in smart mobile devices such as smartphones, tablets, and other portable or wearable electronics. Due to their compatibility with standard Si CMOS processes and the availability of high-quality materials, (Si)GeSn alloys represent a potentially disruptive technology for next-generation SWIR and MWIR sensing and imaging. These alloys offer a pathway to reduced size, weight, and power (SWaP), while enabling monolithic integration with electronics on the same silicon substrate.[4]

Over the past two decades, GeSn-based photodetectors have been extensively developed worldwide. It is well acknowledged that key IR detector figures of merit, such as dark current, cutoff wavelength, responsivity, and specific detectivity (D*), are strongly influenced by the operating temperature, active layer thickness, and Sn composition. To reasonably assess the device performance, the "GeSn Rule-23" guideline was introduced in 2023.[48] This rule provides a framework for estimating the performance potential of GeSn photodiodes (PDs), specifically normal-incidence p-i-n homojunction structures grown on GeSn virtual substrates. According to "GeSn Rule-23", the calculated dark current density at 300 K under a -1 V bias ranges from $10^{-6}$ to 10 A/cm$^2$ as the Sn composition increases from 0% to 20%. Current state-of-the-art GeSn PDs demonstrate dark current densities of » 0.02 A/ cm$^2$, 0.03 A/cm$^2$, and 0.3 A/cm$^2$ for devices with 5%, 10%, and 15% Sn, respectively.[49-51] It is important to note that because the active layer thicknesses vary among these devices, their dark current densities, when normalized to a 3000-nm-thick GeSn layer, are approximately 0.04 A/cm$^2$, 0.09 A/cm$^2$, and 1.88 A/cm$^2$, respectively, for a more consistent comparison with the "GeSn Rule-23" benchmark. In fact, in practical device design the active layer should be sufficiently thick to maximize light absorption and to mitigate tunneling-related dark current, both of which are critical for achieving high detection performance.

For practical photodetection applications, the overall device performance must be considered rather than just the dark current density. A longer-wavelength spectral cutoff requires a higher Sn composition, which in turn reduces the bandgap energy and increases the intrinsic carrier concentration. This results in a lower resistance–area product ($R_0A$) and higher dark current density. Additionally, at elevated operating temperatures thermal excitation further increases the intrinsic carrier concentration, compounding the rise in dark current. Therefore, plotting dark current density as a function of the reciprocal of the cutoff wavelength-temperature product provides a more comprehensive and illustrative metric for evaluating the device performance. Cutoff wavelengths of 2.7 μm and 3.3 μm have been demonstrated with devices containing 11% and 15% Sn, respectively.[52,53] Moreover, reported responsivity values range from 0.1-0.65 A/W at 1.55 μm to 0.35 A/W at 2.0 μm.[54]

It is worth noting that in the well-established "MCT Rule-07" for HgCdTe and "IGA Rule-17" for extended InGaAs infrared PDs,[55,56] the measured device data align closely with the corresponding empirical models, reflecting the maturity and optimization of these material systems. In contrast, many reported performance metrics for GeSn PDs deviate significantly, often by orders of magnitude, from "GeSn Rule-23" (Fig. 6). This disparity highlights the considerable room for improvement in GeSn device performance. Advancements such as optimized doping profiles, reduced background doping concentrations, and the implementation of effective anti-reflection coatings and surface passivation strategies are expected to significantly accelerate the development and competitiveness of GeSn photodetectors.

Furthermore, a prototype 320 × 256 GeSn focal plane array (FPA) has been demonstrated. Each pixel consists of a p-i-n Ge/GeSn/Ge heterostructure with Sn compositions of 2.5% and 9.5%,[57,58] respectively. The pixel pitch is 30 μm, and each pixel has an area of 27×27 μm$^2$. The I–V characteristics exhibit typical diode behavior, and the measured dark current density is comparable to that of standalone photodetectors. Notably, instead of using conventional thin-film epitaxy to grow the diode layer structures, a state-of-the-art lithographically-defined aspect ratio trapping (ART) growth technique was employed.[59] This approach significantly improves the crystalline quality of the GeSn by effectively filtering threading dislocations (TDs). Moreover, this growth is inherently compatible with the pixelated architecture of area image sensors, making it particularly promising for future FPA development.

### 3.3 GeSn APD

Infrared (IR) avalanche photodiodes (APDs), which offer internal gain through avalanche multiplication, are widely used in applications requiring high sensitivity. These include light detection and ranging (LiDAR), fiber-optic telecommunications, eye-safe imaging, and medical sensing.[60,61] Traditional IR APD technologies based on III-V compound semiconductors or HgCdTe can achieve excellent performance, but often come with high material costs.[62] Moreover, most III-V-based APDs suffer from relatively high excess noise due to their large impact ionization coefficient ratio. A major breakthrough in APD design was development of the separate absorption, charge, and multiplication (SACM) structure. In this architecture, the light absorption and carrier multiplication regions are spatially separated, and a charge control layer is introduced between them to shape the electric field profile across the device.[63] Leveraging this design, GeSn-on-Si APDs have been realized by using GeSn as the IR absorber and silicon as the multiplication layer, which offers a low impact ionization coefficient ratio ($k$) and thus significantly-reduced excess noise. This makes GeSn-on-Si SACM APDs a promising and cost-effective solution for high-performance IR detection.

Most earlier demonstrations of GeSn APDs employed MBE growth, incorporating a relatively thick Ge buffer layer between the Si charge layer and the GeSn absorption layer.[64-66] While the thick Ge buffer improves the crystalline quality of the GeSn layer, it significantly compromises the photocarrier collection efficiency. Specifically, carriers generated in the front-end GeSn absorber must travel across the thick Ge layer before reaching the Si multiplication region, resulting in carrier recombination losses and thus limited avalanche gain. To address this issue, GeSn-on-Si APDs without a Ge buffer have also been reported.[67] However, the large lattice mismatch between Si and GeSn introduces a high density of defects at the Si/GeSn interface, which can act as trap states that capture photo-generated carriers and degrade the APD performance. Devices operating around 1550 nm have shown maximum responsivity values ranging from 0.3 to 4 A/W, and avalanche gain values around 10 to 20. However, many of these reports lack clear evidence of punch-through behavior or saturation in responsivity prior to avalanche breakdown, which raises uncertainty about whether the reported gains fully reflect the true avalanche performance.

A systematic study of GeSn APDs grown by chemical vapor deposition (CVD) was reported in 2025.[68,69] A relatively thin Ge buffer layer (~100 nm) was employed to facilitate efficient carrier transport while still maintaining high material quality, enabled by a uniquely-developed growth technique. The device exhibited a well-defined punch-through voltage at −15 V, and photoresponse saturation prior to avalanche breakdown at −17 V. At 1550 nm and 77 K, the device achieved an avalanche gain of 4.5 and a peak responsivity of 1.12 A/W, measured with respect to the saturated primary responsivity of the same device. This provided a reliable and accurate gain assessment. This work represents a comprehensive investigation, integrating both material and device characterizations, and clearly establishes the correlation between material quality and APD performance. Although a device with 2% Sn was demonstrated in this work, the reported APD design is fully scalable to higher Sn compositions (e.g., 10%) to enable full coverage of the SWIR spectrum.

When operated in Geiger mode, the GeSn APD becomes a single-photon avalanche diode (GeSn SPAD) detector. Simulations of a waveguided GeSn/Si SPAD integrated on a silicon photonics platform predicted high-performance SP detection at room temperature in the 1550 to 2000 nm wavelength range.[70] In essence, the GeSn/Si SPAD will extend the performance of the experimentally-proven Ge/Si SPAD.[71] It is envisioned that the GeSn SPAD will facilitate room-temperature photonic-quantum-computing,[72] and will enable several practical sensing and communications applications.

### 3.4 Other GeSn devices

### 3.4.1 GeSn light-emitting diodes (LEDs)

Like III-V infrared light-emitting diodes (LEDs, which are discussed in Section 7 of this Roadmap), GeSn LEDs are being developed for chemical and biomedical sensing applications, as many gas molecules including water vapor, nitrogen-containing species, and hydrocarbons exhibit characteristic absorption fingerprints in the mid-IR spectral range. Both double heterostructure and MQW configurations have been employed to fabricate GeSn LEDs.[23,73-75] Room-temperature electroluminescence (EL) has been demonstrated, with emission at wavelengths extending up to 2.3 μm under pulsed current injection of current densities ranging from a few hundred to several thousand A/cm$^{-2}$. More recently, semiconductor grafting technology has realized a GaAs/GeSn/Ge n–i–p heterojunction structure.[76] This approach provided a high-quality heterointerface comparable to that of lattice-matched epitaxy, while also enabling favorable band alignment for enhanced carrier confinement. Notably, room-temperature EL under direct current (DC) injection was achieved with an emission wavelength centered at 1.9 μm.

### 3.4.2 GeSn waveguide photodetectors

Planar waveguide GeSn photodetectors (WGPDs) featuring both laterally- and vertically-stacked p–i–n junctions have been demonstrated.[77-80] Light is coupled into the waveguide through either end-fire or evanescent-wave coupling, and subsequently absorbed within the photodiode to generate photocurrent. The waveguide geometry provides an extended optical path length, which enables efficient photon absorption and thereby enhances the responsivity. In addition, WGPDs offer more convenient integration and communication with other photonic components, making them highly suitable for integration in electronic–photonic integrated circuits (EPICs). Although the dark current in WGPDs tends to be on the higher end compared to normal-incidence GeSn photodetectors with similar Sn composition, impressive performance has been reported. A detection range extending up to 1950 nm and responsivity of 0.292 A/W at 1800 nm were achieved. By utilizing a thick graded GeSn absorption layer with Sn compositions of 4.85% and 10.41%, a cutoff wavelength of 2.61 μm and specific detectivity (D*) of $3.05 \times 10^6$ cmHz$^{1/2}$W$^{-1}$ at 2 μm were demonstrated.

### 3.4.3 GeSn high-speed photodetectors

Frequency bandwidth is another critical figure of merit for some applications. A notable example is integrated microwave photonics (IMWP), which seeks to incorporate microwave photonic functionalities within photonic integrated circuits. Early demonstrations of GeSn PDs reported bandwidths of a few gigahertz at a wavelength of 2.0 μm.[81,82] Subsequent advances achieved bandwidths exceeding 10 GHz using Ge/GeSn MQW structures.[83] More recently, high-speed GeSn PDs operating near 2.6 μm demonstrated a 3 dB bandwidth of 7.5 GHz,[84] while devices with cutoff wavelength extended to 2.8 μm reached bandwidths of up to 12 GHz.[85] Furthermore, by employing a GeSn buffer layer with a graded Sn composition, a GeSn PD achieved a bandwidth exceeding 50 GHz at 1630 nm.[86] It is worth noting that despite the relaxed GeSn active region in these high-speed devices, their responsivity remains relatively high. It typically ranges from 0.2 to 0.4 A/W, and does not degrade significantly at wavelengths beyond 2.0 μm.

### 3.4.4 GeSn photovoltaic photodetectors

Utilizing large-area "low speed" GeSn photodetectors operated in photovoltaic (PV) mode, it is envisioned that practical "laser energy harvesting" can be realized in the future. The energy harvesting system, known as wireless power transfer, begins with the propagation of a collimated beam from a powerful CW laser through the earth's atmosphere (or through space) to a remotely located receiver comprised of a GeSn PV "panel" (an interconnected array of large PV diodes). Laser radiation incident on this panel then generates considerable DC electric power because the infrared light is converted efficiently into voltage and current. The laser wavelength for this app may be within the 2 to 11 μm range, for example, 4 μm for a quantum cascade laser and 10.6 μm for a $CO_2$ laser.

### 3.4.5 Longer-wave applications of GeSn light emitters and detectors

While Sections 3.2, 3.3 and 3.4 above discussed significant applications of SWIR and MWIR GeSn photodetectors and LEDs, longer-wave applications also hold promise. Because high-Sn-content GeSn (roughly 20% Sn) extends the GeSn device operation to the LWIR (defined here as 8-14 μm ), two examples of enabled applications are: (1) LWIR FPAs that can in principle compete successfully with MCT and QWIP cameras, and (2) high performance LWIR chemical-biological-medical on-chip sensors that utilize co-integrated GeSn PDs and LEDs.

## 4. Challenges and opportunities

### 4. 1 Limited material growth capabilities

The success of SiGeSn material development relies on access to optimal-quality material growth. While it is not a scientific barrier, SiGeSn research currently faces the challenge of limited access to the most advanced growth techniques and equipment. The initial SiGeSn research followed a route similar to that of III-V material development, with slow progress that explored different competing growth techniques on research MBE and MOCVD systems. This led to better understanding of the material properties, with the goal of transferring successful techniques to manufacturing. However, a dramatic change occurred due to the great success of group-IV device demonstrations achieved on industry-standard manufacturing systems, while meanwhile the existing tools in research institutions were left far behind. While the continuing success of SiGeSn device development relies on the exploitation of new approaches implemented on the most advanced industrial systems, it is ironic that most US research institutions have given up investment in group-IV reactors to avoid direct competition with industry. Although many opportunities for developing SiGeSn are emerging, it is awkward that the research community is inhibited from pursuing them due to a lack of in-house infrastructure. Moreover, even though many existing group-IV CVD reactors (some of them obsolete) are available in CMOS foundries, concerns about potential contamination by introducing Sn precursors into the reactors, and consequently the entire process line, prohibits the commitment of those resources to SiGeSn research.

Fortunately, this situation is likely to change in the next few years thanks to the global interests of microelectronics and associated huge infrastructure investments. One near-term solution that could dramatically increase the industry CVD growth capabilities is to re-purpose the existing reactors in foundry with minor modifications to grow SiGeSn. One clear advantage of this approach will be a smooth future transition when the material is sufficiently mature for adoption by the foundry, as it is directly developed in the foundry using CMOS-compatible procedures. In order to seize this opportunity, material researchers should work closely with the foundries to develop new operational models that overcome the challenges, for example, how to coordinate effective material characterization that provides feedback for the new growth yet avoids long idling times for the foundry facilities. Although this would require careful design of the experiments and detailed planning, it will significantly reduce the "cost-to-own" of one-time capital investment and the cost of maintenance that are already covered by the foundry. In the last decade, researchers in U.S. have used this operational mode for material development to obtain excellent results such as high-quality materials, lasers, and detectors.[22,27,39,42-44,68,69,87-90] Recently, there are reports of new commercial foundry capabilities that demonstrate extraordinary results shortly after the tool commission.[87,91]

Compared with the significant progress made by CVD, the advantages of MBE growth for SiGeSn work lie in i) its capability to explore compositions far beyond what can be obtained by CVD, which is limited by the thermal chemistry to decompose the precursors and ii) its capability to control sharp interfaces in superlattice and QW structures, for example, GeSn/Ge, SiGe/GeSn, SiGeSn/Ge… However, most MBE growth of SiGeSn has used old SiGe MBE machines that added Sn effusion cells. Those machines, which are designed for SiGe growth at high temperatures, are not well suited for growing GeSn and SiGeSn alloys at much lower temperatures. This is partly because high-temperature thermal radiation from the cell causes the substrate temperature to overshoot, leading to instability and drift from the optimized growth conditions. It is worth pointing out that combining the controlled growth of (Si)GeSn alloys at a low temperature with the high-temperature cells required for Ge and Si poses a major challenge not only for SiGeSn growth, but also to the whole MBE community. It will require completely new engineering designs, such as more aggressive active substrate temperature control at the low temperature range for both heating and cooling that must be compatible with the UHV environment, special cell design and machine configuration (for example, using a "dummy cell"), and incorporating a special thermal camera that can accurately monitor the substrate temperature in the low temperature regime. Of course, these advances will depend on the high capital investment associated with long development time for both tools and materials.

## 4.2 Future Material Development Challenges and Opportunities from Scientific Perspective

We describe the future challenges and opportunities of SiGeSn material optimization by following four aspects of how a material is generally developed and then subsequently matured: i) Obtaining new SiGeSn compositions (more compositions will provide more building blocks for future devices); ii) Improving the material quality (the foundation for future device applications); iii) Measuring the intrinsic material properties (critical for future design and performance optimization, but only possible when high quality material is available); and iv) Multi-disciplinary approaches that simultaneously address the challenges of compositions, quality, and properties.

### 4.2.1 Growth to extend the achievable compositions

**Using of new precursors**: It has been discovered that the CVD growth conditions, such as substrate temperature, chamber pressure, gas flow rate, and local film strain, can all significantly affect control of the Si, Ge, and Sn compositions. This applies particularly to the Sn incorporation, a major figure of merit for success of growth. While the growth conditions can be fine-tuned to push the boundary of accessible compositions, a more fundamental challenge that limits the maximum composition range attainable with CVD is the thermal chemistry characteristics of the precursors. It has been shown that a lower growth temperature combined with higher growth rate tends to incorporate more Sn. However, effective CVD requires an elevated substrate temperature to decompose the corresponding Si, Ge, and Sn precursors that facilitate the growth. Therefore, the growth process stops when the substrate temperature drops to a range where the precursors will not decompose. This challenge can potentially be mitigated by using high-order hydrides such as $Si_2H_6$, $Si_3H_8$, $Ge_2H_6$, $Ge_3H_8$, $SnD_4$, and even $SnH_4$ that could decompose at much lower temperature. The initial adoption of this approach has led to successful SiGeSn material development. Industry reactors tasked with delivering high-quality materials for device applications have shifted to the use of standard precursors that are relatively cheap and much easier to handle (more stable for shipping and storage). However, the alternative precursors could be revisited since most have already been used in industry manufacture. For unstable precursors such as $SnD_4$, and even $SnH_4$, it is also possible to generate them *in-situ* for the end-of-point use.

**Control of growth dynamics using plasma and atomic hydrogen enhancement techniques:** Both approaches had been proposed to engineer the chemical equilibrium and the surface mobility of radicals to break the existing chemical reaction limit. However, since the implementation is not easily compatible with current industry reactors, they would likely be confined to research. Introducing plasma allows the available Sn radicals to be assembled *in-situ* to radicals with the sp$^3$ structure needed to increase the density of higher mobility radicals (SiH$_3$ and GeH$_3$).[92-95] The hot-filament generation of atomic hydrogen in the reactor could increase the adatom mobility, acting as a catalyst in promoting the correct sp$^3$ bonding for growth, and changing the crystal structure of tetragonal Sn to a sp$^3$ hybridized diamond structure if amorphous growth starts.[96]

**Developing *in-situ* diagnostic tools for understanding the SiGeSn growth mechanisms -** Despite the impressive progress of material development, however, most current growth methodology is still simply a trial-and-error approach that relies on intuition. This sharply contrasts the deep understanding of microscopic growth mechanism for Si and Ge. By imitating how this understanding was developed, customized CVD tools equipped with *in-situ* diagnostics capabilities for SiGeSn growth would benefit the long-term understanding of SiGeSn growth, which is difficult to accomplish in current industry reactors. For example, the use of *plasma emission spectroscopy* for measuring the wavelengths and intensities to identify and quantify radicals, or use of a sampling probe connected to a residual gas analyzer (*mass spectrum analyzer*) to study the radical species ratio and their spatial distributions near the wafer surface. The previous application of both methods to the study of Si and Ge CVD epitaxy has provided a detailed understanding of the surface kinetics and microscopic growth mechanisms that exist today.[97-99]

For MBE, while no foreseeable fundamental scientific challenges limit the growth of broad Si, Ge, and Sn compositions, a practical technical barrier is the identification of suitable substrates to facilitate the growth. There have been significant efforts to use different III-V substrates, or relaxed III-V buffers that provide virtual substrates lattice-matched to SiGeSn with targeted compositions. However, it is non-trivial to prepare a high-quality III-V buffer with well-controlled surface that is readily available for group-IV material nucleation is non-trivial. An interesting idea is to use CVD-grown relaxed GeSn as the virtual substrate for MBE growth.

### 4.2.2 Challenge to improve material quality

Characterization results confirm that line/plane defects, surface states, and point defects play very important roles in limiting the performance of GeSn detectors.[50,88-90] The Table below summarizes the origins of these defects and proposes mitigation approaches. Compared with the multiple clear routes for effectively mitigating the first two defect classes, the critical challenge of understanding and mitigating point defects in GeSn materials may ultimately limit the device performance. For CVD-grown GeSn, Hall and C-V characterization routinely indicate p-type background carrier concentrations of $\sim 10^{17}$, which varies with the Sn composition. If this background originates from vacancies as in epitaxial Ge on Si, it would set a baseline for the point defect density. Approaches such as compensation doping and hydrogen passivation may offer effective routes to mitigation, but they have not yet been implemented since the detector performance is still limited mostly by the structure rather than the intrinsic materials.

Table I. A list of the origin of defects and proposed defect mitigation approaches

| | Line/plane defects | Surface states | Point defects |
|---|---|---|---|
| **Origin** | • Strain relaxation<br>• Thread dislocations from buffer/substrate | • Dangling bonds<br>• Conductive oxides<br>• Residue Sn due to etching | • **Vacancies**<br>• **Anti-site defects**<br>• **What else?** |
| **Defect mitigation approaches** | • Gradient buffer<br>• Superlattice defect filter<br>• Engineering doping profile | • In-situ epitaxial passivation<br>• Doping engineering near surface<br>• Ex-situ GeON passivation | • Surfactant growth<br>• Post growth low temperature RTA |

| | | • Novel low temperature wet etching | • In-situ atomic hydrogen diffusion |
|---|---|---|---|
| | | | • Use thin absorbing region |

## 4.2.3 Understanding Fundamental Material Properties

A list of fundamental material properties (band parameters) that are needed to design the various optoelectronic devices is given below, along with commonly-used characterization techniques for measuring those properties. It is challenging that these basic band parameters have not been systematically measured to form a comprehensive database for the SiGeSn material system. This is largely due to the insufficient quality of available materials, and to complications such as the strain that is intrinsically associated current growths. Only recently has it become clear that Vegard's law is valid for the SiGeSn lattice constant.[100] Also, the exact composition at which GeSn crosses from indirect to direct bandgap remains debatable. In order to accurately measure these band parameters, layer transfer techniques have been used to selectively transfer the top layer to a host substrate to produce a high-quality free-standing layer. This technique holds great promise for allowing these parameters to be measured precisely and systematically. However, with the new development of short-range ordering in SiGeSn, a new of challenge is that the same composition may potentially have distinctly different band structures.

Table II. A summary of band parameters and corresponding characterization techniques [101]

| Band parameters | Characterization techniques and special considerations |
|---|---|
| Lattice constant | Due to controversy regarding whether Vegard's law is valid, theoretical work will guide the inclusion of bowing. The result will be fit with XRD (for the lattice constant) and RBS (for composition). The composition will be cross-checked by APT*, SIMS**, and EDX of XTEM. |
| Bandgap ($\Gamma$, $X$, $L$) and spin-orbit splitting | Temperature dependent PL, spectroscopy ellipsometry, photoreflectance, and electron-reflectance will be used to extract the values of all critical points. |
| | Varshni parameters will be extracted for high signal-to-noise ratio in the range 10 K-100 K, and bandgap parameters will be extrapolated to 0 and 300 K. |
| Electron effective mass ($\Gamma$, $X$, $L$) | Shubnikov-de Haas and magnetophonon measurements of n-type samples at low temperature to ensure long carrier scattering lifetime for sufficient interaction with the magnetic field. Cross-checked with cyclotron resonance. Assuming temperature independence, measure the low energy valleys and use theoretical results for the high energy valleys. |
| Luttinger parameters (heavy-hole, light-hole) and split-off hole mass | Similar to electron measurement. but p-type to simultaneously measure heavy hole and light holes; Cross-checked results with cyclotron resonance; use theoretical result for split-off hole mass, as it has less impact |
| Interband matrix element ($E_P$) and F parameter accounting for remote-bands | The interband matrix element is not independent, and could be calculated using the electron effective mass, band gap, and spin-orbit splitting; Verify by Spin resonance measurement to check self-consistency, and cross-checked with theoretical results; F parameter will be calculated theoretically and verified by spin resonance measurement. |
| Conduction and valence band deformation potentials | Hydro-static measurement, and checked with in-house strain-dependent optical characterization of a set of samples with the same Sn composition. |

## 4.2.4 Multi-disciplinary approach and opportunity to address material challenges

### Opportunities Enabled by Short Range Ordering (SRO)

It has been reported that experimental band structure data do not always agree well with theoretical calculations, particularly when the Sn composition exceeds ~10%. One notable example is that fitted values for the bowing parameter vary significantly across different research groups, which limits the predictive accuracy of material properties. Recent in-depth studies have revealed that this discrepancy originates from local atomic ordering, i.e., short-range order (SRO), within the GeSn lattice. In semiconductor alloys, SRO arises when certain atoms preferentially attract or repel one another, leading to a distribution that deviates from perfect randomness. Since most previous theoretical models assume random atomic distributions, discrepancies with experimental data are inevitable. Importantly, such ordering has been shown to strongly influence material properties, including bandgap energies and band alignment.[102] Figure 7 illustrates that the band structure of a random alloy exhibits a small direct bandgap at the $\Gamma$ point. However, one type of SRO increases this gap. Remarkably, incorporating another type of SRO further enhances the direct gap at the $\Gamma$ point, ultimately driving a transition to an indirect bandgap. This SRO-induced bandgap engineering unlocks a novel type of "heterostructure": lattice-matched and iso-compositional alloys with tailored conduction- and valence-band offsets.

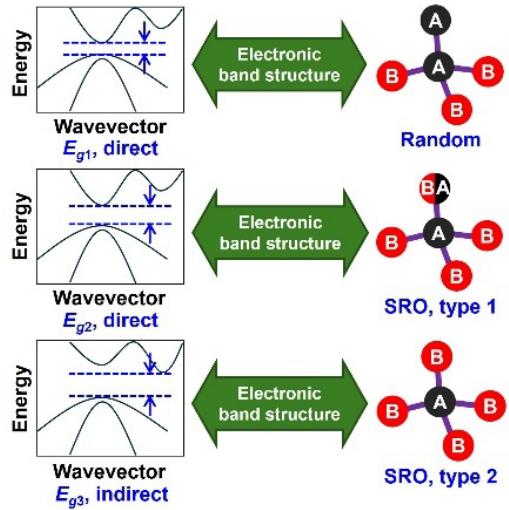

Fig. 7 SRO induced bandgap tailoring

Although theoretical studies have clarified the roles of SRO, experimental characterization remains highly challenging. Difficulties arise from large uncertainties in the atomic coordination numbers, weak signals associated with distinct local configurations, and insufficient spatial resolution at the nanoscale. To address these issues, significant efforts have been devoted to extracting SRO with higher fidelity. For example, in GeSn alloys SRO has been derived from non-ideal atom probe tomography (APT) data by refining $k$-nearest-neighbor (KNN, $k$ = 1, 2, 3…) analysis with physics-informed statistical corrections,[103] enabling reconstruction of the true KNN shells of the diamond cubic lattice.

Recently, the presence of short-range order in SiGeSn has been determined by using advanced energy-filtered four-dimensional scanning transmission electron microscopy and large-scale atomistic models generated by a machine learning neuroevolution potential of first-principles accuracy.[104] Those findings not only confirmed the presence of short-range order, but also directly revealed the actual atomic structure. This sets a foundation for future use of SRO as new degree of freedom to build composition-uniform lattice-matched bandgap-engineered heterostructures.

The goal of synthesizing SiGeSn/GeSn alloys with sufficient structural quality to resolve SRO has motivated the development of two complementary approaches. Spontaneous and stimulated methods have been proposed and implemented to regulate the spatial arrangement of SRO domains, by employing tailored precursors and different thermodynamically-equivalent processing strategies.

**Opportunities Enabled by Selective Area Growth such as Aspect Ratio Trapping Growth**

Despite the great success of (Si)GeSn material development, to realize CMOS monolithically-integrated high-performance (Si)GeSn optoelectronics on Si, several grand technical challenges must still be overcome. These include: i) Material quality: For thin film growth, the large lattice mismatch between (Si)GeSn and the Si substrate results in a high threading dislocation density that limits the ultimate material quality and device performance; ii) Sn incorporation for longwave spectral coverage: Compressive strain due to the lattice mismatch seriously limits the incorporation of Sn. Both the low Sn concentration and compressive strain inhibit bandgap narrowing for LWIR coverage; iii) Less-investigated CMOS integration strategies: SAG may be promising for back-end CMOS integration, which will be highly desirable for CMOS-integrated (Si)GeSn optoelectronics in the long term. However, the prevailing research philosophy of "Material first, Integration second" has precluded its pursuit because the material is considered not "being ready".

There is a great opportunity to utilize the lithography-defined SAG technique of ART growth to break the fundamental barriers to high material quality for relaxed and high-Sn-composition micro-crystal arrays that are easily scaled to large format on 200- or 300-mm wafers. We emphasize, however, that significant work is still needed to develop innovative device architectures that could fully exploit the performance advantages of this growth technique. Although the complexity should not be underestimated, an early investigation may be highly rewarding.

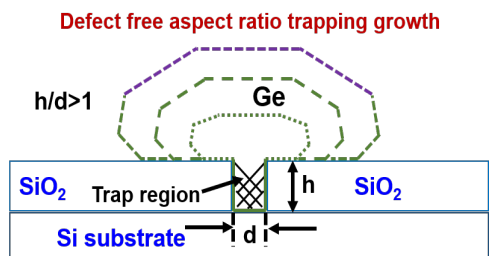

Fig. 8 Schematic Ge ART growth

In order to achieve CMOS integration, high-quality SAG of (Si)GeSn is essential. Considering the similarity of (Si)GeSn growth to Ge growth, and inspired by the huge success of monolithic integration demonstrated by King's Ge-CMOS IR imaging sensor work, our team proposes to accomplish this goal by developing ART growth of (Si)GeSn. ART is a SAG technique that eliminates all defects from the substrate interface in one lithography and epitaxial growth over both the seed window and the epitaxial lateral overgrowth (ELO) regions without increasing the lateral growth rate. This is achieved through a combination of SAG and defect crystallography that forces defects to the oxide sidewall, which results in a perfect top film. In diamond cubic slip systems such as Si, Ge, and (Si)GeSn, misfit dislocations in the (100) growth plane lie along <110> directions while the threading segments rise on (111) planes in the <110> directions. For Ge growth on Si, threading segments on the (111) plane make a 60° angle to the underlying Si (100) substrate. Thus, if the aspect ratio of the holes in the oxide mask is greater than 1.7, crystallography dictates that threading segments will be blocked by the oxide sidewall, resulting in a defect-free top Ge surface on Si (Fig. 8). This process is analogous to bulk Si Czochralski crystal growth, for which the seed crystal is first necked to eliminate defects before the boule is pulled. In comparison, standard SAG (ratio less than 1) combined with ELO is defect-free only in the ELO region, whereas dislocations still propagate through material on top of the window.

The rationales for pursuing ART growth of (Si)GeSn are: i) Potential for high material quality: The trapping mechanism provides a defect-free template for subsequent high quality material growth; ii) Proven accommodation of large lattice mismatch: For example, InP has 8% lattice mismatch with Si, but has been grown successfully using the ART method. Since 8% lattice mismatch between GeSn and Si corresponds to 25% Sn, this provides plenty of room to test the growth conditions for high Sn composition incorporation; iii) Potential for even higher Sn incorporation: A relaxed GeSn layer could serve as the template for growing higher compositions of GeSn; iv) Buffer layer reduction: ART growth could eliminate the need for buffer layers by directly relaxing GeSn in the patterned area. This would not only simplify the growth process, but also save a large amount of precursor; v) Readiness for CMOS integration: While many of the applications for SiGeSn involve integrated photonics, the current research follows a flow of materials, devices, and then integration. Only a small quantity of high-quality GeSn is needed for the two most significant GeSn device applications, IR sensors and lasers. Besides being ideal for small-sized crystal growth, the ART technique also provides an effective route to CMOS integration.

Therefore, ART growth of (Si)GeSn is a paradigm shift that could simultaneously achieve material development and its integration with CMOS. The ART growth for each device application will have its own unique challenges. For example, the detailed device fabrication flow for (Si)GeSn IR sensors grown by ART could follow an approach similar to that used for Ge-CMOS imaging sensors. Since extremely high material quality is required for low dark current, a single threading dislocation could significantly deteriorate the device performance. On the other hand, the use ART-grown (Si)GeSn to build monolithically-integrated lasers could be challenging as the undeveloped technical pathway is not yet clear. An obvious advantage of implementing ART for (Si)GeSn lasers is enhanced optical confinement for much higher modal gain, due to the impossibility of introducing a large contrast of the refractive index using the standard thin film growth approach. However, it is unclear whether all the standard semiconductor laser design rules for carrier confinement, optical confinement, and power scaling will be appliable to ART grown (Si)GeSn lasers, which in particular will be used in conjunction with Si SOI based integrated circuits.

## 4.3 Device challenges and new opportunities

### 4.3.1 Emitter: Interband laser development challenges and opportunities

**Challenge**

SiGeSn materials hold strong promise for enabling the last missing component of the silicon photonics toolbox: an electrically driven CW Si-based laser. While group III–V lasers on Si have already reached a manufacturing level, the realization of Si-based lasers on Si would significantly reduce costs by simplifying on-chip integration and enabling large-scale CMOS-compatible miniaturization. To meet the requirements of an ideal on-chip Si-based light source, two major challenges remain: i) increasing the lasing temperature to room temperature and beyond, and ii) improving the efficiency to enable CW operation.

Two main optical cavity approaches have been explored by the SiGeSn laser research community: micro-cavities and Fabry-Pérot (F-P) cavities. Micro-cavities include microdisks, photonic crystal cavities, and ring resonators. They provide higher quality factors due to improved optical confinement, and a surrounding SiN stressor can enhance the directness of the bandgap.

However, the fabrication is more complex than for a F-P cavity, and it is more difficult to apply well-established laser characterization techniques (e.g., power measurements) that quantify the material gain and provide critical feedback for material and device optimization. F-P cavities exhibit weaker optical confinement, but are simpler to fabricate. More importantly, straightforward characterization of the output from an F-P laser completes the essential feedback loop for material and device development.

A direct comparison between micro-cavity and F-P cavity lasers fabricated from the same SiGeSn material showed only minor differences in the maximum lasing temperature. This indicates that while cavity engineering and thermal management can improve the device efficiency, their impact on the maximum operating temperature remains marginal. One strategy for improving performance has been to apply tensile strain, which can reduce the lasing threshold by up to three orders of magnitude. However, even with this significant improvement, SiGeSn lasers remain restricted to cryogenic operation. An in-depth investigation of the optical transitions identified Auger recombination as the fundamental mechanism that limits the lasing temperature.[105] Specifically, the Auger process becomes resonantly enhanced when the target radiative recombination energy aligns with the energy separation between the heavy-hole (or light-hole) band and the split-off (SO) band. The dominance of non-radiative over radiative recombination channels leads to severe loss of the electrically-injected carriers that quenches the lasing at elevated temperatures.

The most effective strategy for mitigating Auger recombination in SiGeSn lasers is to employ a higher Sn composition in the active region. This reduces the bandgap energy, thereby shifting the radiative recombination energy away from the critical resonance condition that enhances the Auger process. For example, it has been shown that increasing the Sn composition from 15% to 18% raises the maximum lasing temperature from 140 K to 270 K. An alternative approach is to engineer a direct-bandgap SiGeSn alloy through careful selection of the Si and Sn compositions. Tuning of the band structure can shift the intervalence Auger transition to an energy off resonance with the radiative recombination.

Insights from the development of III-V lasers such as InP-based systems highlight the effectiveness of MQW structures in improving device performance. MQWs enable more efficient carrier injection and reduce both heating and free-carrier absorption, leading to lower lasing thresholds and facilitating CW operation. Importantly, the MQW configuration also helps to suppress Auger-related losses, since quantum confinement alters the band structure and reduces the overlap of electronic states involved in Auger transitions. These advantages suggest that implementing MQW designs in SiGeSn-based lasers could provide a pathway not only to higher operating temperatures but also to CW operation, complementing efforts in composition engineering and strain management.

**Opportunities**

The GeSn laser can now be considered a serious competitor to group III-V devices for Si Photonics. The group III–V development trajectory provides a valuable reference, in that strategies like maximizing the optical confinement and modal gain have been beneficial. While these approaches are not directly transferable to GeSn devices due to the intrinsic differences in material and device structure, the group III-V pathway offers a useful framework for both the material development and performance evaluation.

In parallel, advances in the epitaxial growth, particularly exploration of the SAG technique to facilitate the incorporation of higher Sn content without generating excessive strain-induced defects, will open a practical pathway toward more efficient SiGeSn lasers that operate at higher temperatures. An obvious advantage over III-V lasers is the potential for back-end CMOS integration. As a paradigm shift, a co-design strategy for GeSn laser development and CMOS integration should be developed simultaneously. Such an approach would not only accelerate the progress in achieving high-quality, high-Sn active layers, but would also ensure that device architectures and integration schemes are optimized for manufacturability from the outset.

### 4.3.2 Emitter: THz QCL development challenges and opportunities

Terahertz quantum-cascade lasers (QCLs) have emerged as powerful sources of radiation in the 1.2–5.6 THz range.[106] Steady advances have enabled peak output powers exceeding 1 W in pulsed operation and tens of milliwatts in CW operation, with near- diffraction-limited beam quality,[107] single-mode tunability of up to 20%,[108] and broadband frequency-comb operation.[109] Despite these achievements, room-temperature operation without frequency conversion (see Section 1 of this Roadmap) has remained elusive. The highest pulsed operating temperature in GaAs/AlGaAs-based THz QCLs has been ~261 K,[110] while CW operation is limited to 129 K due to excessive power dissipation.[106,111-113] The primary barrier to efficient room-temperature lasing has been a thermally-activated nonradiative scattering process. At elevated temperatures, electrons in the upper lasing subband gain sufficient in-plane kinetic energy to emit polar longitudinal-optical (LO) phonons, causing rapid relaxation to the lower subband.[106] In polar III–V semiconductors, this LO-phonon scattering is highly efficient once energetically allowed, reducing the QCL upper-state lifetime from a few picoseconds at ~10 K to less than 1 ps near room temperature. In contrast, non-polar group-IV semiconductors offer a fundamentally different scattering landscape. In these materials, electron–LO-phonon interactions are mediated by the short-range deformation potential and are therefore significantly weaker. As a result, intersubband (ISB) lifetimes extend to tens of picoseconds and have much weaker dependence on in-plane kinetic energy. This makes group-IV QWs a promising platform for low-threshold, room-temperature THz QCLs.

Early efforts in the 2000s focused on valence-band SiGe-based QCLs.[114,115] However, those devices suffer from strong mixing of the heavy- and light-hole subbands, which complicates both the optical spectra and carrier transport. Furthermore, the large heavy-hole effective mass diminishes ISB gain, suppresses tunneling, and enhances sensitivity to interface roughness scattering. Consequently, although electroluminescence was observed, no p-type SiGe QCLs were ever realized. These limitations underscore the importance of moving from valence-band to conduction-band devices.

The incorporation of Sn into Si and Ge has opened new opportunities. The Ge/SiGeSn material system, composed entirely of group-IV elements, supports n-type ISB transitions in the L-valley and can be engineered in lattice-matched compositions. This combination offers the potential for efficient carrier transport, robust gain, and room-temperature operation. Recently, $Si_{0.20}Ge_{0.75}Sn_{0.05}$/Ge and $Si_{0.40}Ge_{0.50}Sn_{0.10}$/Ge superlattices were successfully grown by MBE. Initial characterization confirms the targeted compositions and lattice-matched heterostructures, and reveals strong PL emission from the superlattices. These results represent an important first step toward realizing group-IV THz QCLs. Further fundamental studies of material and optical properties, such as band offsets, electronic band structure, and ISB transitions, are needed to guide the device design. With continued advances in epitaxial growth, band engineering, and heterostructure optimization, the Ge/SiGeSn platform holds strong promise for the development of compact, low-threshold THz QCLs.

### 4.3.3 Emitter: LED development challenges and opportunities

The single (Si)GeSn material system can cover a broad infrared wavelength range (SWIR, MWIR, and even LWIR). It can be grown on low-cost, large-area Si substrates that are transparent to the emission wavelength. In addition, high quality QWs with high internal quantum efficiency have been grown recently. All of these characteristics indicate that the (Si)GeSn system is a viable candidate for developing high-performance mid-IR light emitting diodes at a cost even lower than LEDs emitting in the visible bands, which are now widely used in display and solid-state lighting. A technical challenge may be the material's high refractive index (~4) that will make it more challenging to extract light efficiently. However, this challenge could be overcome by leveraging existing technical solutions that are well developed in the LED industry. These include roughing the output surface or patterning a metameterial photon management nanostructure (see Section 20 of this Roadmap). And while (Si)GeSn-based mid-IR LEDs will ultimately have a clear cost advantage, development activities are currently far less active compared to efforts developing GeSn lasers and detectors for CMOS-processed imaging. Most current mid-IR LEDs (see Section 7 of this Roadmap) employ III-V and II-VI materials for applications like IR scene projection and low-cost gas sensing. In order to justify the development of low-cost (Si)GeSn LEDs, new applications that utilize millions of devices must be identified. For example, low-altitude drones for delivery, self-driving, humanoid robots, and robotic dogs, may need to adopt IR lighting and display for signaling among themselves and other system components, instead of using visible light that disrupts the humans they serve. Furthermore, (Si)GeSn LEDs will be ideal as mid-IR light sources in low-cost/ultra-compact chemical sensing systems that are monolithically integrated on a single Si-based photonic integrated circuit (PIC) chip (see Section 22 of this Roadmap).

### 4.3.4 Infrared detector challenges and opportunities

**Challenge**

A unique role for GeSn-based detectors is in mid-IR FPAs (see Section 8 of this Roadmap) that are monolithically integrated with the readout integrated circuit (ROIC) on a CMOS-fabricated chip. These can ultimately provide large formats with higher uniformity and lower cost compared to traditional FPAs that require hybridization to integrate the detector array and ROIC. However, the transition to FPAs will require substantial further improvement of the (Si)GeSn dark current (*e.g.*, Fig. 6 above) and other detector characteristics before they are fully-competitive with the more mature mid-IR detector systems discussed in Sections 9 and 11 of the Roadmap. The performance of existing mid-IR GeSn detectors is most often limited by issues such as thin junction/absorption layers and misplaced junctions that are either too close to the surface or buried in the defective region for high dark current. These can be overcome by fabricating more optimal designs using growth techniques such as selective area growth and ART growth, as described in section 4.2.4. We expect future device characteristics dominated by the intrinsic material properties to more realistically represent the actual technology level.

**Opportunities**

The ability to engineer the (Si)GeSn band structure for close energy alignment of the $\Gamma$- and L-point conduction band minima may provide a significant intrinsic advantage over the conventional II-VI and III-V detector technologies. Figure 9 illustrates that if the L-valleys are situated at only slightly lower energy than the $\Gamma$-valley, photocarriers generated by strong absorption at the zone center's direct bandgap will rapidly transfer via phonon relaxation to the 4 L-valleys that have a much higher density of states.[116] This is highly advantageous because the recombination lifetimes for both Auger and SRH processes will be much longer when the indirect bandgap makes it more difficult to conserve both momentum and energy. The longer carrier lifetime will substantially reduce the dark current. Meanwhile for absorption, Fig. 9 (b) shows that the GeSn absorption spectrum near the direct-to-indirect transition has a high value of ~10,000 cm$^{-1}$, and a steep longwave cutoff - twice of that for MCT and T2SLs. Among all the widely-used narrow bandgap materials, only (Si)GeSn provides tuning of both direct and indirect bandgaps in the same spectral range.

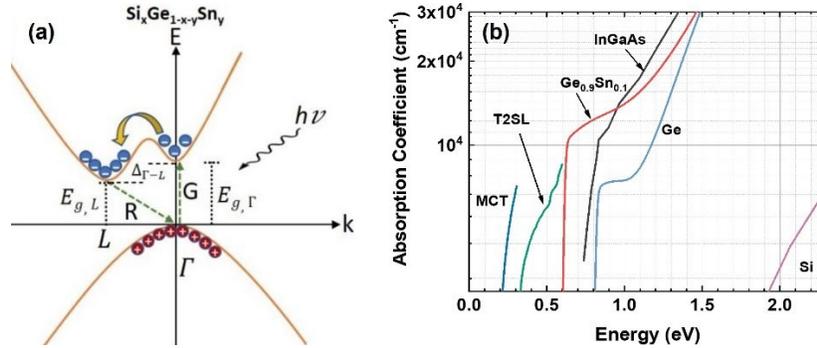

Fig. 9 (a) Schematic of "*Momentum Space Charge Separation*" using SiGeSn band structure as an example; (b) Comparison of absorption spectrum for different dominant infrared materials.

### 4.3.5 GeSn APD challenges and opportunities

Most current GeSn APDs utilize an SACM architecture, to leverage both the long wavelength absorption of the GeSn and the low excess noise factor of the Si multiplication layer. Since the cut-off wavelength is determined by the Sn composition, material quality tends to limit the device performance. To accommodate the relatively-large lattice mismatch between GeSn and Si, a Ge buffer is normally required to maintain high epitaxial quality for the GeSn layer. However, a thick Ge buffer (> 700 nm) dramatically reduces the photocarrier collection efficiency from the front-end GeSn absorber to the Si multiplication layer, leading to poor performance. Therefore, designs with a thinner Ge buffer (~100 nm) are needed, not only for carrier collection but also to achieve punch-through voltage well before the avalanche breakdown, which is ideal for APD operation. It is worth noting that thinning the Ge buffer is non-trivial. The surface strain, which is totally different from that for a thick buffer, greatly affects the subsequent GeSn growth, particularly for Sn incorporation. GeSn APDs with a ~100 nm Ge buffer and using 2% and 5% Sn have shown reasonable avalanche gain at a cut-off of 2.1 μm. However, higher Sn incorporation is necessary to further extend the wavelength cut-off, which poses challenges for optimizing the Ge buffer design and subsequent GeSn epitaxial growth.

Since all the GeSn/Si APDs reported to date have been planar and suffer from a high densify of misfit dislocations at the Ge/Si interface, growing the GeSn/Ge APD by ART could offer the following advantages:[117] i) High absorption, high material quality, and no interface misfit dislocations; ii) Proven CMOS compatibility, ease of scale-up to large array sizes, and unique potential for single-photon counting. When Ge is used for charge multiplication, the small direct bandgap of Ge (0.8 eV) leads to excessive band-to-band tunneling (BBT) that causes extra dark counts.

### 4.3.6 Co-design and integration

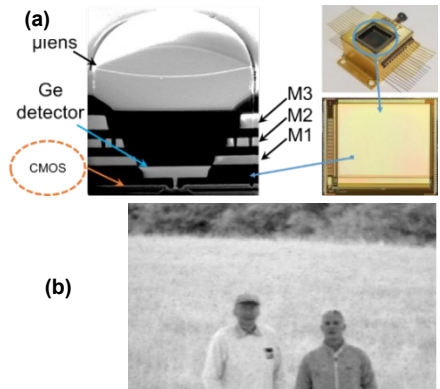

Fig. 10 (a) Ge-on-Si sensor pixel (left) and packaged imager (right); (b) Night imaging performance under moonless conditions.

In microelectronics, the "co-design" concept has emerged as an effective methodology for developing different disciplines and components simultaneously and collaboratively, in order to meet specific performance, efficiency, or functional goals. The envisioned applications of GeSn, such as IR imaging, LiDAR, and integrated photonics, all rely on its intrinsic suitability for monolithic integration. Being in the stage of material development but with a clear application scenario, it is important to place this development in the broader context of integration with Si-based CMOS or PICs. As one example, we highlight the integrated Ge-CMOS imaging sensors based of ART growth, which have enabled high-quality Ge photodiodes to be monolithically integrated and fabricated by a commercial CMOS foundry. Application of the co-design methodology to GeSn imaging sensors would simultaneously address both material quality and integration challenges.

In order to duplicate the great success of Si-CMOS imaging sensors, intensive efforts have been devoted to developing monolithic technologies that integrate IR materials on a CMOS platform for image sensors as well as photonic integrated circuits. Using the Ge/Si ART technique for epitaxy, King and his colleagues successfully demonstrated SWIR image sensors based on the first monolithic integration of single-crystal Ge diode arrays with Si-CMOS ROICs (Fig. 11). These FPAs were fabricated in a CMOS foundry on 200 mm wafers. The process starts with a standard 0.18 μm CMOS front end process and then followed by high-quality ART growth of the Ge to be used for detectors. The ART Ge is planarized by chemical mechanical polishing in order to determine the detector thickness by the well-controlled dielectric deposition process. Surface passivation, ion implantation for junction formation, a 3-metal Al back-end process for inter-connect, and microlens formation are subsequentially conducted to finish the process in the same foundry. At -45°C, a typical Ge diode has dark current of 25 fA per pixel and QE of 44% at λ=1.3 μm. Night imaging under moonless conditions at a resolution of 640 × 480 pixels is shown in Fig. 11 (b).

IMEC's recent production of an III-V laser by a 300 mm CMOS line confirmed the great potential for monolithically integrating lasers with Si-photonics. Potential advantages of SiGeSn lasers grown on Si by ART over III-V counterparts include more straightforward growth without hybridization, mature CMOS compatibility, lower thermal budget than III-V growth, and additional unique processing flexibility such as ion-implantation that can form the 3D doping profile.

## 5. Future developments to address challenges

## 5.1 Materials

Below we describe some approaches to addressing the general material development challenges, or seizing the opportunities discussed in Section 4. 1 and 2.

**Limited growth capability**: New growth infrastructure is expensive, and will require time before it is useable. Although industry CVD tools are designed for 24/7 operation, R&D mode often requires characterization feedback, meaning the growth system is put in idling while waiting. However, this non-intrinsic issue can be mitigated by careful planning between the foundry and customers.

**Extending the material compositions**: Despite the chemical reaction limit, CVD-grown (Si)GeSn compositions depend strongly on the local strain. Compressive strain rejects additional Sn incorporation, whereas tensile strain tends to allow more. It is unclear whether current compositions have reached the chemical reaction limit set by the precursors. Engineering the strain for thin buffers, relaxed buffers, and gradient buffers could be low-hanging fruits as the range of achievable material compositions is expanded. While broadening the precursor toolbox will significantly increase the development cost, applying a different precursor to each specific region, *e.g.*, $GeCl_4$ for the SiGeSn QW barrier, may offer an expanded design space for heterostructures.

**Developing new growth methodologies based on a better understanding of SiGeSn CVD**: A greater in-depth understanding of the microscopic growth kinetics on a level similar to that for Si and Ge growth may be required, perhaps enabled by customized CVD equipment with *in-situ* diagnostic tools. New growth mechanisms such as PECVD and hot-wire CVD could also be investigated. Such studies will not need to target a specific device.

**MBE growth**: MBE can produce short- period superlattices with sharper interfaces than CVD, and thin bulk materials with high Si and Sn compositions that are not easily accessible by current CVD growth. MBE materials may be advantageous in pursuing newly-predicted topological properties, short range ordering, and THz QCLs based on L-valley transitions. MBE growth on CVD (Si)GeSn virtual substrates could simultaneously exploit the strengths of both techniques.

**Attaining and exploiting improved material quality**: It may be useful to explore less conventional approaches that have not been broadly introduced in (Si)GeSn growth. For example, the techniques used to extend the wavelength of InGaAs for detectors and multi-junction solar cells. Once the material quality reaches the hypothesized background-carrier limit imposed by point defects such as vacancies, a variety of techniques such as compensation doping, hydrogen passivation, and RTA could be investigated for further improvement of the material quality.

**Measuring basic band parameters**: Since future device design will depend on the reliability of the input band parameters, it can be hoped that ongoing device development will drive more systematic characterizations of those parameters using high-quality materials. Given that short-range ordering can lead to different band parameters for the same composition, the measurements should ideally focus on the CVD-grown materials used predominantly for current device development.

**Short-range ordering**: Although SRO affects the band structure and therefore the electrical and optical properties of (Si)GeSn, most device studies have not been taken it into account. As the fundamental science of SRO becomes better understood, we expect a greater appreciation of its influence on different aspects of SiGeSn optoelectronics.

**Selective area and aspect ratio trapping growth**: Since many of the applications envisioned for (Si)GeSn involve monolithic integration with Si-CMOS and Si-photonics, a broad approach to selective area growth is needed. To date there have been no systematic investigations of GeSn SAG, despite reports of SAG and ART growths. From a broad material science perspective, we expect GeSn SAG to be especially challenging due to already-known complications such as low solid solubility for Sn segregation, the intrinsic competition of etching and deposition processes, the much different mass transport for high-weight gas molecules in the confined growth region, and the difficulty of controlling the actual growth temperature when patterned substrates with growth templates bring large variations of the emissivity. Although it is unclear how these factors may affect the outcome, control over the growth is likely to be more complicated than for Si, Ge, and SiGe.

## 5.2 Device related

### 5.2.1 Emitters: Interband lasers

The current status of electrically-injected (Si)GeSn-based lasers clearly has headroom for further incremental performance improvements, as quantified by the threshold current density, output power, wallplug efficiency, and maximum operating temperature, with a critical milestone being cw operation at room temperature. In the near term, strain-balanced multiple QW active regions may increase the modal gain and suppress Auger recombination. Selective area growth may relax the buffer for high Sn compositions and reduce the buffer thickness for improved material quality.

### 5.2.2 Emitters: THz QCLs

THz devices based on carefully-controlled superlattice designs should be grown by MBE, although a great deal of material development will be required before actual device structures can be grown. This will include growth of the relevant superlattices, understanding the conduction band offset, demonstrating the intersubband THz transition, and the photoconductive behavior.

### 5.2.3 Emitters: LEDs

The baseline device characteristics of GeSn LEDs should be measured. These include output power, practical spectral coverage, efficiency, and temperature sensitivity. This will provide the basis for establishing the technical pathway to improved material quality and more efficient light extraction.

### 5.2.4 Infrared detectors for imaging

Structures that are limited by the current (Si)GeSn material quality should be grown to establish a baseline for comparison to other mid-IR detectors, and for understanding the technology itself. More complicated structures that incorporate interface defect filtering and transparent windows can be implemented, along with growth approaches that minimize the point defects. Detectors based on selective area growth can be developed, and the approach discussed in Section 4.3.4 for suppressing Auger recombination should be explored.

### 5.2.5 APDs

APDs based on the GeSn/Si SCAM architecture with a thin Ge buffer have demonstrated basic APD characteristics. This architecture may also support even longer cut-off wavelengths if the thin Ge buffer quality does not degrade when absorbers with higher Sn composition are grown, as the Ge buffer could act like a compliant film. From the application perspective, further investigation should determine whether the current device performance can be improved significantly to meet the dark count needs of LiDAR applications for self-driving land vehicles and drones. In the meantime, alternative architectures such as GeSn/Ge, GeSn/SiGeSn, and wafer bonded GeSn/Si may also provide viable routes.

### 5.2.6 Integration on PICs

The development of (Si)GeSn optoelectronics was originally motived by its unique potential for monolithic integration. For many years, the research focused primarily on material and device development. However, with material development now transitioning to selective area growth, and with the broad microelectronics community adopting the "co-design" philosophy, the time has come for small-scale integrations that demonstrate functionality. This could be simple LEDs and photodetectors integrated on a PIC with simple waveguides for sensing applications, or a cryogenic laser integrated with passive waveguides on a PIC. For IR imaging, a SAG-based detector could be monolithically integrated with a CMOS ROIC to demonstrate an FPA or APD array. Or a time-of-flight circuit could be integrated for a LiDAR receiver. Any of these critical milestones could confirm that development has attained the next level, thereby generating more interest in the broader community.

## 6. Concluding Remarks

In conclusion, after decades of groundwork for material development, SiGeSn based optoelectronics have reached a critical point that confirms their potential for substantial impact. The unique electrical and optical properties of this material system, combined with low-cost CMOS-based manufacturing, would significantly lower the barrier to broad exploitation of the mid-infrared spectral band. While tremendous research work is still to be required to mature the technology, pathways to the possible development of early applications have begun to emerge.

### *Acknowledgments*

The authors would like to dedicate this paper to Dr. Gernot S. Pomrenke, who retired from Air Force of Office Scientific Research (AFOSR), for his early vision and decades of persistent support for US SiGeSn research. The authors would also like to thank Dr. Bruce "Chip" Claflin from Air Force Research Laboratory for inspiration and great discussions. The authors acknowledge the funding support from AFOSR (Grant No. FA9550-19-1-0341) and Office of Naval Research (Grant No. N00014-24-1-2651).

### *Disclosures*

The authors declare no conflicts of interest.

# 17. MWIR and LWIR III-V devices grown on silicon-based platforms

ERIC TOURNIÉ,[1,2,*] JEAN-BAPTISTE RODRIGUEZ,[1] AND LAURENT CERUTTI[1]

[1]IES, University of Montpellier, CNRS, F-34000 Montpellier, France
[2]Institut Universitaire de France, F-75005 Paris, France
*eric.tournie@umontpellier.fr

## Overview

The direct growth of MWIR/LWIR optoelectronic devices on Si-based platforms is expected to both make the technology more sustainable and enable new photonic integrated circuits for various applications, in particular portable sensing. Much progress has been made in the growth of these devices over the past decade. While diode lasers and photodetectors are limited by structural defects, interband cascade and quantum cascade lasers are relatively immune to these defects, opening the way to epitaxially integrated systems. However, efficient light transfer from active to passive sections remains a challenge. Work is needed to improve light coupling and to develop new photonic functions, to fully realize the promise of this technology.

## Current status

### 15. Context

Mid-wave infrared (MWIR, 2 – 5 μm) and long-wave infrared (LWIR, 5 – 25 μm) III-V optoelectronic devices such as light emitting diodes (LEDs), diode lasers (DLs), interband cascade lasers (ICLs), all three devices based on interband transitions, quantum cascade lasers (QCLs), based on intersubband transitions, and various photodetectors (PDs) have been under development for several decades [1]. They are now mature enough to be used in a wide range of systems that find application in high resolution spectroscopy for environmental, biomedical, defense/security and process monitoring [2, 3], in countermeasures [4] or in high performance imaging [5]. However, these devices are grown on small, expensive and rare III-V substrates, which increases their cost and raises doubts about the long-term sustainability of these technologies. From another point of view, silicon photonics, a platform originally established to realize near-IR telecommunication and data communication transceivers, is now being extended to longer wavelengths thanks to new designs, or to the integration of other materials such as Ge which is transparent up to 15 μm [6-10]. The direct growth of III-V MWIR/LWIR devices on Si-based platforms is therefore expected to both make III-V technology more sustainable and enable new photonic integrated circuits (PICs) for different applications. In parallel, another strategy has recently emerged which is to develop group-IV optoelectronic devices natively grown on Si substrates which would open a natural route toward MWIR devices epitaxially integrated on Si-based PICs [11,12]. The development of these new materials is the topic of another chapter and will not be discussed any further here.

Given the importance of these issues, the epitaxy of III-V materials and devices on Si (or Ge) substrates has been studied for years [13-15]. However, crystal defects introduced during growth affect the performance and reliability of the devices. Dislocations are linear defects arising from the lattice mismatch between the III-V material and Si and are unavoidable. Antiphase domains (APDs) delineated by antiphase boundaries (APBs), which result from the heteroepitaxy of the polar III-V zinc blende structure on the nonpolar diamond structure of the Si (or Ge) substrate [16], have long been the most problematic defect because APBs form electrically-charged paths within the active layers, creating shortcuts that kill devices [17]. The formation and burying of APDs and APBs has recently been revisited [18,19], and emerging APB free layers have been obtained by molecular beam epitaxy (MBE) of GaSb [20] or GaAs [21-23], as well as by metal organic vapor phase epitaxy (MOVPE) of GaP [24], GaAs [25] or GaSb [26] on Si substrates. This has opened the way to the demonstration of a number of III-V MWIR/LWIR optoelectronic devices grown on Si [27] or Ge [28] substrates. We summarize below the state of the art for LEDs, lasers and photodetectors grown on Si-based platforms.

## 16. MWIR/LWIR light emitting devices grown on Si

The current state of the art for III-V lasers is summarized in Table 1, which shows that different laser technologies have been grown on Si (or Ge) covering a very broad wavelength range spanning the MWIR to the LWIR. The operation regime is indicated as continuous wave (cw) or pulsed operation. In this last case, the output power per facet corresponds to the peak power per facet at the indicated measurement temperature, while the threshold current density and emission wavelength are taken at the same temperature.

**Table 1. Performance of state-of-the-art MWIR/LWIR III-V lasers grown on Si or Ge substrates, organized by increasing emission wavelength.**

| Laser type | Substrate[a] | Emission wavelength (μm) | Threshold current density (kA.cm$^{-2}$) | Output power per facet (mW) | Ref. |
|---|---|---|---|---|---|
| GaSb DL | (001)Si | 2.3 | 0.2 | 15 – 25 (cw, 290 K) | [29] |
| GaSb ICL | (001)Si | 3.3 – 3.5 | 0.3 | 20 – 30 (cw, 290 K) | [30,31] |
| InP QCL | Ge on 6° offcut (001)Si | 4.3 | 2.4 | 10 (pulsed, 170 K) | [32] |
| InP QCL | 4° offcut (001)Si | 4.8 | 4.5 | 1600 (pulsed, 290 K) | [33] |
| InAs QCL | (001)Si | 8.0 | 0.9 | 100 (pulsed, 290 K) | [34] |
| InP QCL | GaP/(001)Si | 8.1 | 1.5 | 1600 (pulsed, 290 K) | [35] |
| InP QCL | 4° offcut (001)Si | 8.3 | 1.3 | 2500 (pulsed, 290 K) 600 (cw, 290 K) | [36] |
| InP QCL | 4° offcut (001)Si | 10.8 | 7.9 | 1500 (pulsed, 290 K) | [37] |
| InAs QCL | 6° offcut (001)Si | 11 | 1.2 | 30 (pulsed, 290 K) | [38] |
| InP QCL | Ge on 6°offcut (001)Si | 11.5 | 4.3 | 3000 (pulsed, 290 K) | [39] |
| InAs QCL | (001)Ge | 14 | 0.85 | 80 (pulsed, 290 K) | [28] |

[a] When not otherwise stated, the substrate orientation is (001)±0.5°

The GaSb DLs, based on GaInAsSb/AlGaAsSb quantum wells (QWs), suffer from high dislocation densities arising from the huge (~12%) lattice mismatch between GaSb and Si.

These defects introduce non-radiative recombination centers that are represented by mid-gap energy levels. In typical type-I QW diode lasers, electrons and holes naturally recombine non radiatively in the QWs and in the barrier/cladding layers (Fig. 1(a)). This increases the threshold current density which is two to three times larger than that of a similar DL grown on GaSb [29]. In addition, it is well known that dislocations in DLs multiply through recombination-enhanced defect reaction, which can lead to progressive device failure [40]. The cw lifetime of QW DLs grown on Si is therefore limited to a few hundred hours.

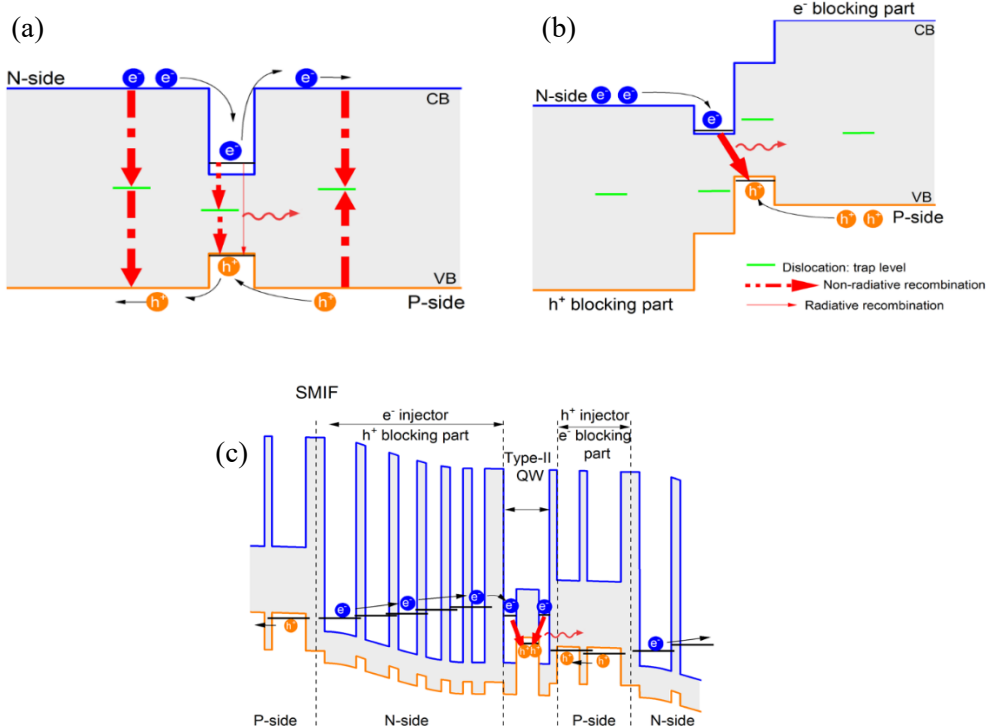

Fig. 1. Representation of the recombination processes in (a) a type-I QW and (b) a type-II QW with electron and hole blocking parts. (c) Band diagram of a type-II QW ICL. From [30] © Optical Society of America.

In contrast, it has been demonstrated that ICLs and QCLs are much less sensitive to dislocations, with performance similar to devices grown on their native III-V substrates [28,30,31,34,35,38], which makes them unique among semiconductor lasers. The origin of this peculiarity may be found in the device band structures. ICLs are based on indirect type-II transitions in Ga(In)Sb/InAs QWs [41], and the radiative recombination path in the QW does not cross the dislocation energy level (Fig. 1(b)). Additionally, the QW is surrounded by carrier injectors that block electrons and holes in the *p*- and *n*-parts, respectively (Fig. 1(c)). Therefore, no electron-hole recombination can occur anywhere else than in the active QWs, which drastically limits non radiative recombination. Preliminary aging studies allowed extrapolation of the device lifetimes to over 30 years [30,31]. On the other hand, QCLs are unipolar lasers that rely on short-lifetime transitions between confined levels in the conduction band and are thus immune to minority carrier recombination [34,35,38]. However, lifetime studies are still lacking.

**Table 2. Performance of state-of-the-art MWIR/LWIR III-V LEDs grown on Si substrates, organized by increasing emission wavelength.**

| Emitting material | Substrate | Emission wavelength (μm) | Output power (μW) @ injection current (mA) at 300 K | Ref. |
|---|---|---|---|---|
| GaInSb/InAs interband cascade | 4° offcut (001)Si | 3.3 | 184 @ 100 | [42] |
| InAs/GaSb superlattice | 6° offcut (001)Si | 4.1 | | [43] |
| InAsSb bulk | 4° offcut (001)Si | 4.5 | 6 @ 190 | [44] |

In contrast to lasers grown on Si, only a few reports on III-V MWIR LEDs grown on Si can be found in the literature (Table 2). However, comparing their performance reveals interesting information. Notably, LEDs based on a bulk absorbing layer have a low output power due to the high defect density resulting from the growth on Si [44]. In contrast, ICLEDs, with an interband cascade emitting zone similar to ICLs active zones, confirm the relative immunity of this design to dislocations [45,46], as was previously observed with ICLs [30]. Additionally, LEDs grown on Si outperformed those grown on their native GaSb substrate at high injection current, thanks to Si's higher thermal conductivity compared to GaSb [43,47]. This effect has yet to be clearly demonstrated with lasers, although there may be some indication in the literature. We show in Fig. 2 L-I-V curves taken from InP (Fig. 2(a)) and InAs (Fig. 2(b)) QCLs grown on their native substrates and on Si. Interestingly, in both cases the threshold current densities are lower on Si than on the native substrates. However, dedicated investigations are needed to confirm this point.

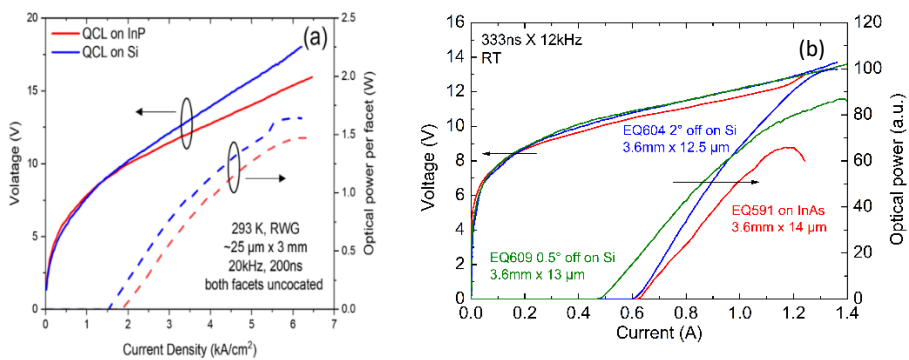

Fig. 2. L-I-V curves taken at room temperature for (a) InP QCLs grown on InP and Si substrates, reprinted from [35] with permission from AIP Publishing, and (b) InAs QCLs grown on InAs, on axis Si and 2°offcut Si (unpublished).

## 17. MWIR/LWIR photodetectors grown on Si

Table 3 shows the state-of-the-art performance of MWIR/LWIR photodetectors grown on Si-based platforms. Dark current densities, peak responsivities and detectivities are given at the same temperature as the cutoff wavelength. Comparing different photodetector technologies is not as straightforward as with lasers or LEDs because the performance criteria depend on the target application and not all research groups perform similar device studies, as shown in Table 3. Also, not all groups perform absolute quantum efficiency measurements, or detectivity evaluations, so the only common figure of merit is the dark current.

**Table 3. Performance of state-of-the-art MWIR/LWIR III-V photodetectors grown on Si substrates, organized by increasing cutoff wavelength. Dark current densities, peak responsivities and detectivities are given at the same temperature as the cutoff wavelength.**

| Photodetector type | Substrate[a] | Cutoff wavelength (μm) @ temperature (K) | $J_{dark}$ (mA.cm$^{-2}$) @ bias voltage (V) | Peak responsivity (A/W) | D* (cm.Hz$^{1/2}$.W$^{-1}$) | Ref |
|---|---|---|---|---|---|---|
| InAsSb bulk | Ge on 6°offcut (001)Si | 4.2 @ 160 | 5 @ -0.2 | | | [48] |
| InAs/InAsSb T2SL[b] | (001) Si | 4.7 @ 150 | 10 @ -0.5 | | | [49] |
| | (001) Si | 5.2 @ 150 | 6 @ -0.5 | | | [50] |
| | 4° off (001) Si | 5.2 @ 160 | 14 @ -0.1 | 0.85 | $3.65 \times 10^{10}$ | [51] |
| InAs/GaSb T2SL | 4° off (001) Si | 5.5 @ 77 | 2300 @ -0.1 | 1.2 | $1.3 \times 10^{9}$ | [52] |
| InAs/GaAs SML-QCD[c] | 4° off (001) Si | 6.2 @ 77 | $2.1 \times 10^{-5}$ @ -0.1 | $5.9 \times 10^{-4}$ | $3 \times 10^{10}$ | [53] |

[a]When not otherwise stated the substrate orientation is (001)±0.5°, [b]T2SL: type-II superlattice, [c]SML-QCD: sub-monolayer quantum cascade detector

The MWIR/LWIR III-V photodetectors are mainly developed for high performance imaging defense applications [5]. When grown on Si, their performance is severely limited by dislocations resulting from the lattice mismatch between the III-V heterostructure and the Si substrate. Dark currents are typically one to three orders of magnitude higher than on the native III-V substrate. This explains why the efforts to grow these devices on Si remain limited, although promising focal plane arrays have been demonstrated [48].

## Challenges and opportunities

Despite the large progress made in the last decade and summarized above, several issues and challenges remain to be addressed at both the discrete device and PIC levels.

Table 1 shows that most QCLs have been grown on large offcut (4-6 °) Si substrates. Although this approach can be used for a laboratory demonstration, it is incompatible with fabrication at a Si foundry that employs on-axis (001) Si. In addition, large miscut substrates tend to generate step bunching and compositional inhomogeneity, both of which are detrimental to device operation [54]. The immunity of QCLs to defects needs further investigation, as some reports claim that they are insensitive to dislocations [28,34,35,38], while others suggest that dislocation densities greater than ~2 x10$^7$ cm$^{-2}$ degrade the device performance [55]. To get a broader view, Fig. 3 provides an overview of the threshold current densities of the various MWIR/LWIR lasers within the 2 – 20 μm wavelength range. It is noticeable that most InP QCLs grown on Si substrates exhibit much higher threshold current densities than their couterparts grown on InP. Only careful growth optimization allowed the demonstration of InP QCLs immune to defects [35] while InAs QCLs appear much less sensitive [28,34,38] (Fig. 3). This probably arises from the fact that InP QCLs rely on ternary GaInAs and AlInAs alloys sensitive to step bunching and prone to composition inhomogeneity whereas InAs QCLs rely on InAs and AlSb compound semiconductors only. More work is needed to clarify this point. In parallel, work is needed to fully understand the origin and limits of ICL immunity.

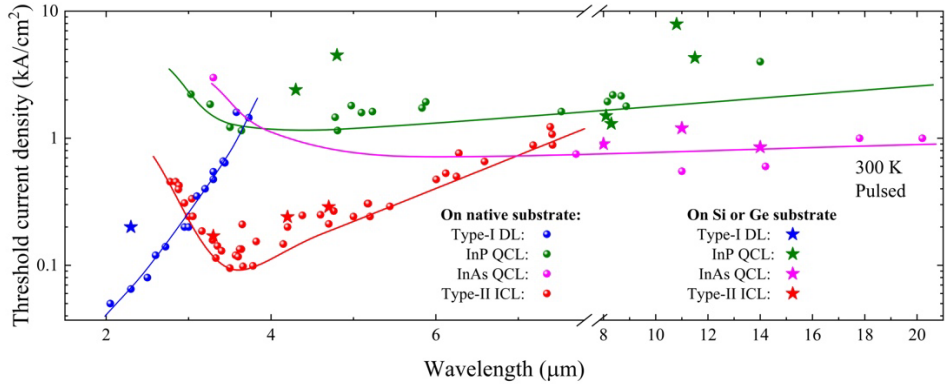

Fig. 3. Evolution with the wavelength of the threshold current densities for type-I QW GaSb DLs (blue), type-II GaSb ICLs (red), InP QCLs (green) and InAs QCLs (magenta). Dots/stars represent devices grown on their native/Si substrates, respectively. All data are for lasers operated in pulsed mode at 300 K. Lines are guides to the eye. Adapted from [56].

It is noteworthy that there is only one report of a QCL grown on Si that operates in continuous wave [36]. This shows that the thermal management of QCLs grown on Si has not been properly addressed yet, although it is a critical issue for QCLs. The lifetime of QCLs grown on foreign substrates, especially Si, remains to be evaluated.

As mentioned above, dislocations arising from the lattice mismatch between the Si substrate and the device heterostructure are a major issue for most devices apart from ICLs and QCLs. Progress has recently been made in the growth of GaSb on Si, and the threading dislocation density has been decreased from the high $10^9$ cm$^{-2}$ to the low $10^7$ cm$^{-2}$ range by inserting AlSb "filtering" layers in the buffer layer [46,57,58]. However, this density is still high and will need to be further reduced before high performance DLs and photodetectors can be fabricated.

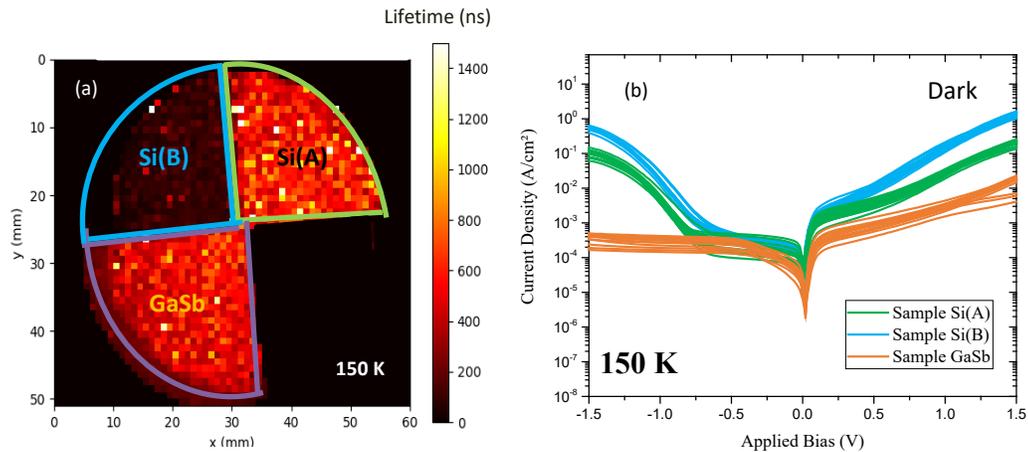

Fig. 4. (a) Maps of the carrier lifetimes and (b) dark current densities measured at 150 K and different bias for InAs/InAsSb nBn structures grown simultaneously on two different GaSb/Si templates and on a GaSb substrate. Adapted from [50].

Fig. 4(a) shows maps of the carrier lifetimes measured for InAs/InAsSb nBn structures grown simultaneously on two different GaSb/Si templates and on a GaSb substrate [50]. Template

Si(A) has one AlSb "filter layer" whereas Si(B) has none. The threading dislocation densities at the surface of the photodetector heterostructures were measured to be 5 x$10^7$ cm$^{-2}$ and 5 x$10^8$ cm$^{-2}$ for the structures grown on Si(A) and Si(B) templates, respectively. Interestingly, the carrier lifetime measured by time-resolved photluminescence at 150 K for the sample grown on Si(A) is similar to that of the sample grown on GaSb, while the lifetime of the carriers in the structure grown on Si(B) is exceedingly low. This suggests that a "threshold-like" behavior may exist regarding the carrier lifetime. Dark current measurements on these devices, designed to operate at -0.5 V, reveal a large dispersion between diodes (Fig. 4(b)). Additionally, leakage currents at large bias are two and three orders of magnitude higher for the photodetectors grown on Si(A) and Si(B) templates than for that grown on a GaSb substrate, respectively (Fig. 4(b)) [50]. Leakage current due to the structural defect as observed with ICLEDs [46] is a plausible source of dark current. Also, spectral responses (not shown) are degraded in the presence of dislocations, and the degradation is more severe at higher dislocation densities [50]. This shows the dramatic impact of dislocations on the properties of high-performance photodetectors. Although these structural defects are unavoidable, it is now clear that strategies must be implemented to decrease their density. This calls for additional work on the growth of GaSb-on-Si templates. This is especially important for diode lasers and photodetectors, but also for ICLEDs, although to a lesser extent, since they have shown variations in device performance when grown on highly dislocated buffer layers [45,46].

Furthermore, to fully benefit from MWIR/LWIR devices grown on Si-based platforms, it will be necessary to switch from discrete optoelectronic devices grown on Si to devices grown on or integrated with PICs. The first challenge is obviously to incorporate epitaxial optoelectronic devices into the PIC without damaging them. The successful operation of III-V lasers grown in recessed areas of Si wafers equipped with passive waveguides has recently been demonstrated in the near IR (1.2 µm) [59,60] and MWIR (2.3 µm) [61]. This should now be extended to more complex PICs, as well as to alternative platforms such as Ge/Si or SiGe/Si to address the LWIR. However, an additional issue arises here. Relaxation of the Ge or SiGe layer grown on Si results in a severely cross hatched surface, which makes it difficult to bury antiphase domains [62].

Moreover, the real bottleneck in the epitaxial integration of lasers on PICs is the ability to efficiently transfer light emitted by the laser into passive waveguides (WGs) [63]. Currently, heterogeneous integration has a clear advantage in this respect, as it allows evanescent light coupling through adiabatic tapers [64]. This strategy is not easily implemented with epitaxial integration, which requires thick cladding layers that prevent mode overlapping between the passive and active WGs. Butt coupling geometries have instead been implemented up to now, which have resulted in the measurement of high insertion losses for all materials systems [59-61].

On a more positive note, the co-integration of MWIR/LWIR optoelectronic devices with Si photonics and other materials such as chalcogenide glasses or 2D materials will open up new possibilities for integrated photonics.

## Future developments to address the challenges

Although these issues will not be easy to resolve, the long-term advantages of processing 300 mm wafers in Si foundries or with microelectronics-grade processing tools will provide unique opportunities for developing complex photonic functions such as integrated combs and spectrometers. It will also make it possible to scale up the production volume of MWIR/LWIR PICs. This perspective provides a strong driving force to address the challenges mentioned above.

As stated in the introduction, dislocations arise from the large difference in lattice parameters between the device or a buffer layer and the Si substrate. Since their formation is inevitable, it is crucial to reduce their influence. It was also mentioned that their density can be reduced, but only to a level that is still too large for high-performance LEDs, DLs and photodetectors. Compared to the work done on near IR InAs/GaAs quantum dot DLs grown on Si, refinement of the dislocation filtering should allow further reduction of the threading dislocation density to the $10^6$ cm$^{-2}$ range [65]. However, this must be accompanied by complementary strategies such as the insertion of misfit dislocation trapping layers to force the misfit segments to form away from the active zone, or the adjustment of growth conditions to minimize the point defect densities and avoid the formation of dislocation loops during device operation [66]. These approaches have effectively improved the lifetimes of near-IR lasers but have yet to be implemented in MWIR/LWIR devices.

Regarding the light coupling from lasers grown on PICs, the main problem arises from the generation of an air gap at the interface between the laser and the passive waveguide facets during the laser processing [59-61,63]. Strategies have been proposed to either fill this gap with an index-matching material [67] or to reduce its width [68]. The former strategy has greatly enhanced the coupling efficiency, whereas for the latter the improvement was only moderate. In fact, a significant improvement would be achieved only by having sub-µm air gaps [63]. Another attractive option, which would completely eliminate the air gap, is to reverse the process flow by first growing and processing the laser before fabricating the passive circuit [67]. This requires that the thermal budget for fabricating the PIC is compatible with the laser. However, all these approaches rely on butt coupling. Further design work is needed to propose new integration schemes and laser cavities to eliminate the coupling bottleneck, and to develop new photonic functions.

## Concluding Remarks

The direct growth of MWIR/LWIR optoelectronic devices on Si-based platforms is being actively pursued, since it will ultimately make III-V technology more sustainable and provide new photonic integrated circuit functionalities for various applications. Fully integrated sensors will enable the implementation of long-awaited on-chip sensor arrays, and the introduction of portable sensors. The growth of III-V devices on Group IV substrates has progressed substantially in the last decade, and the relative insensitivity of ICLs and QCLs to dislocations opens the way to robust photonic integrated circuits. However, MWIR/LWIR PICs are still in their infancy, and sustained research efforts are needed to fully realize their promise.

## Back matter

*Funding*

*Acknowledgments*

As far as our own work is concerned, we thank our coworkers whose names appear in the references for their highly appreciated contribution, and financial support from the France 2030 program (projects EXTRA (ANR-11-EQPX-0016) and HYBAT (ANR-21-ESRE-0026)), the European Union (projects REDFINCH (GA780240) and OPTAPHI (GA860808)) and the French ANR (projects LightUp (ANR-19-CE24-0002) and FILTER (ANR-20-CE92-0045)).

*Disclosures*

The authors declare no conflicts of interest.

# 18. MWIR and LWIR Photonic Integrated Circuits on Silicon

ALEXANDER SPOTT[1,*]

[1]*Mirios, Inc., Santa Barbara, CA 93103, USA*
**spott@mirios.net*

## 1.  Photonic Integration

Photonic integrated circuits (PICs) aim not only to miniaturize systems but also to enable new technologies and improve the performance of existing implementations. This holds additional value in the MWIR and LWIR, where free space photonic elements are expensive and often display suboptimal performance.

Many of the target applications in the MWIR and LWIR are not limited to a single wavelength. Control over multiple wavelengths, sometimes in dramatically different spectral bands, is necessary in practical or ideal technologies. Sensor applications, for example, which very often benefit from the detection of multiple species at once, can employ multiple wavelengths to better isolate target chemicals, or spectroscopically probe a wide spectral range to identify given molecules. Free space optic solutions do not always scale well, particularly if a separate light source is required at each wavelength which prohibitively increases the cost, complexity, and size.

This is both the advantage and challenge of MWIR and LWIR photonic integration. Multiple wavelengths can be addressed feasibly on a single chip, but requires that integration technologies be designed to accommodate broad wavelength ranges. Further complications stem from the diverse selection of applications for MWIR and LWIR light with varying performance requirements.

## 2.  MWIR and LWIR Silicon Photonic Integration

Photonic integration on silicon wafers promises to dramatically reduce the eventual cost and improve the performance of the photonic devices and systems. Fabrication can ultimately be performed with mature silicon foundry processes on wafers up to 24 inches in diameter. Silicon is a robust, mechanically and thermally stable material. Driven by the communications market, silicon photonics has developed into a mainstream technology for near infrared (NIR) and shorter wavelengths.

There are several specific advantages to photonic integration on silicon at longer wavelengths in the MWIR and LWIR. First, the complexity and high growth cost of quantum cascade lasers (QCLs) and interband cascade lasers (ICLs) [both are discussed in separate sub-topic articles of the Roadmap] increase the challenge and expense of developing and commercializing photonic devices alongside cascade lasers, which reduces the inherent value of monolithic integration on a native III-V substrate. The advantage of building PICs on the same substrate as the lasers is further reduced in cases that require multiple wavelengths separated spectrally by more than the laser gain bandwidth. Second, the high optical confinement that can be achieved with silicon and germanium waveguides enables smaller devices with smaller bend radii. This advantage is often quoted in reference to NIR silicon photonics, and becomes increasingly critical at longer wavelengths. Waveguides with lower optical confinement operating in the LWIR, for example, can require bend radii as large as 1-mm, limiting the quantity and size of devices that can be fit in a single die.

MWIR silicon photonic integration initially stemmed from significant advancements of silicon photonics in the NIR telecommunication bands near 1550 nm and 1310 nm. Taking advantage of the high transparency of both silicon and glass at those wavelengths, the field historically used the silicon-on-insulator (SOI) waveguiding platform, composed of a silicon waveguiding "device" layer above a $SiO_2$ lower cladding layer on a silicon substrate. Waveguides are often further encapsulated on the top and sides in $SiO_2$. $Si_3N_4$ waveguides clad with $SiO_2$ are also used to achieve extraordinarily low waveguide propagation losses of 0.5 dB/m in the NIR [1] and for wavelengths below silicon's band gap.

Addressing the longer MWIR and LWIR wavelengths was initially challenging due to the immaturity and cost of optical components needed to characterize waveguides at those wavelengths, including tunable lasers, sensitive photodetectors, optical lenses, and fibers. The first demonstration of silicon waveguiding at wavelengths above 3 μm was reported in 2010 [2]. Numerous waveguides, couplers, resonators, photodetectors, and lasers integrated on a variety of silicon photonic platforms have since been reported.

## 3. Current Status
### 3.1. Platforms

Material absorption is the primary technical challenge facing the transfer of silicon photonics from the NIR to the MWIR and LWIR. Silicon itself is transparent throughout the MWIR, although it begins to pick up phonon-assisted lattice band absorption that reaches 1 dB/cm at 6.7 μm. Optical losses at longer wavelengths are dominated by free carrier absorption, so high purity silicon must be used for implementation in the LWIR. The conventional dielectric glasses, $SiO_2$ and $Si_3N_4$ have strong Reststrahlen absorption bands that extend optical losses to shorter wavelengths. $SiO_2$ reaches 1 dB/cm at 3.6 μm and over 100 dB/cm at 5 μm. While the absorption of $Si_3N_4$ remains below 1 dB/cm until 6.6 μm, its absorption strongly depends on the deposition method, and in many cases is no more transparent than $SiO_2$. Germanium is uniquely suitable for the entire MWIR/LWIR spectrum, maintaining high transparency to 16 μm.

The high optical loss of $SiO_2$ limits the utility of conventional SOI waveguides used in the telecom bands, so numerous alternative platforms have been proposed, demonstrated, and developed over the past 15 years. Si (with refractive index n~3.5) or Ge (n~4.0) have been used as the waveguiding core in most platforms because of their high refractive indices and transparency at longer wavelengths. The choice of lower cladding material has been more problematic, with tradeoffs between fabrication challenges, optical transparency, and thermal properties. Table 1 lists some of the more prominently-reported waveguiding platforms. $Al_2O_3$ [2], $Si_3N_4$ [3,4], $CaF_2$ [5], $Y_2O_3$ [5–7], and porous Si [8] have all been used as alternative lower claddings to $SiO_2$.

Silicon-on-sapphire (SOS) initially stood out as a CMOS-compatible SOI alternative that was historically used in the silicon electronics industry for its high electric insulation and thermal stability. The optical absorption of sapphire stays below 1 dB/cm up to $\lambda \approx 4.3$ μm. SOS wafers have limited commercial availability today. The Si layer is typically grown on a sapphire substrate, which results in dislocation defects that can add to the optical loss, but can alternatively be transferred to the sapphire substrate by wafer bonding. SOS is the only platform discussed here that does not use a Si substrate, which would limit some of the potential commercialization benefits of silicon photonics.

SOI waveguides were initially expected to have excessive propagation loss above ~4 μm. However, minimal optical interaction with the $SiO_2$ cladding can be achieved by engineering large waveguides that maintain reasonable propagation losses within the MWIR. Loss below 1 dB/cm has been achieved at wavelengths up to 4.8 μm [9]. Although SOI is commercially available, less conventional Si layer thicknesses must be used for longer wavelengths. Since SOI is not suitable for LWIR integration, germanium-on-$SiO_2$ (GOI) can alternatively be used, but because the wavelength range is limited by the $SiO_2$ it will not be increased appreciably.

$Si_3N_4$ can be used to create Si-on-nitride (SON) or Ge-on-nitride (GON) [3,4] waveguides, typically by bonding the Si or Ge device layer to a $Si_3N_4$ film deposited on a Si wafer. The $Si_3N_4$ deposition method must be considered carefully to ensure low absorption at longer wavelengths, and hydrogen impurities can cause significant resonant absorption loss in the MWIR.

A pragmatic solution to material loss in the lower cladding is to simply undercut the waveguide so as to provide cladding by air on all sides. This is typically achieved by beginning with SOI or GOI and etching the $SiO_2$ layer below the waveguides with HF. The suspended waveguide is structurally supported on both sides by subwavelength grating structures. Suspended Si or Ge waveguides must be engineered carefully to ensure structural robustness and minimal optical interaction with or leakage through the subwavelength gratings. This platform is highly thermally insulating, which may be preferable for thermal tuning applications but problematic for active devices like QCLs and ICLs that generate heat. In a similar tradeoff, the suspension is ideal for MEMS-based devices but is likely highly sensitive to vibrations.

Germanium-on-silicon (GOS) wafers are manufactured by chemical vapor deposition of crystalline Ge on a Si wafer. Wafers are commercially available from foundries specializing in Ge growth. The wide transparency range of Ge makes this a potentially ideal platform for the full spectral range from 3 μm to over 12 μm, but with a few persisting drawbacks. First, defects due to the significant lattice mismatch at the Ge/Si growth interface cause absorption loss. Films that are too thick can also crack from stress, although this is not always problematic for the 1–3 μm thicknesses necessary for MWIR and LWIR waveguides. Next, the relatively low vertical index contrast between the Ge and Si can cause modal leakage into the Si substrate if not properly designed. Unlike nearly all other MWIR or LWIR waveguide platforms, the Si lower cladding in GOS offers the high thermal conductivity of Si, which is advantageous as a heat sink for lasers.

To address some of these issues, a SiGe alloy can be used to manufacture SiGe-on-Si (SGOS) wafers. By increasing the Si concentration in the Ge layer, better lattice matching reduces the defect density at the growth interface. In many cases, the Ge concentration is linearly graded to be highest at the center of the waveguide core. In that case the refractive index profile should be carefully tailored to minimize substrate leakage losses. The advantages of SGOS over GOS are traded for a reduced optical mode confinement and reduced upper wavelength limit of closer to 8 μm.

Chalcogenide glasses (ChGs) are amorphous semiconductors containing the chalcogen elements sulfur, selenium, and tellurium. ChGs are transparent throughout the MWIR, and into the LWIR to wavelengths as long as 20 μm. They are suitable as waveguide cores or claddings with refractive indices in the range of n ≈ 2–3.5. ChGs are typically deposited by sputtering, thermal evaporation, or chemical vapor deposition. While the wide transparency window of ChGs is attractive, the deposition process is specialized and not CMOS compatible. Nonetheless, significant advances have been made with ChG waveguides on silicon platforms.

The useful wavelength range of any given platform depends on the acceptable waveguide propagation losses (highly application dependent), the material deposition methods (particularly for SiN), accurate material properties, and the waveguide and device geometries. In the MWIR, no single waveguiding platform has fully dominated the literature. With fewer materials available, LWIR demonstrations have primarily focused on the GOS, SGOS, and ChG platforms.

| Platform | Core | Lower Cladding | Cladding 1 dB/cm cutoff wavelength (μm) |
|---|---|---|---|
| Si on Insulator (SOI) | Si | $SiO_2$ | 3.6 μm |
| Suspended Si | Si | Air | - |
| Si on Sapphire (SOS) | Si | $Al_2O_3$ | 4.3 μm |
| Si on $CaF_2$ | Si | $CaF_2$ | 8 μm [10] |
| Si on $Y_2O_3$ | Si | $Y_2O_3$ | 6.3 μm [10] |
| Si on Nitride (SON) | Si | $Si_3N_4$ | 6.7 μm |
| Ge on Si (GOS) | Ge | Si | 6.9 μm |
| Ge on Nitride (GON) | Ge | $Si_3N_4$ | 6.7 μm |
| Ge on Insulator (GOI) | Ge | $SiO_2$ | 3.6 μm |
| SiGe on Si (SGOS) | SiGe | Si | 6.9 μm |
| Suspended Ge | Ge | Air | - |
| ChG on Si | ChG | Si | 6.9 μm |
| ChG on Insulator | ChG | $SiO_2$ | 3.6 μm |

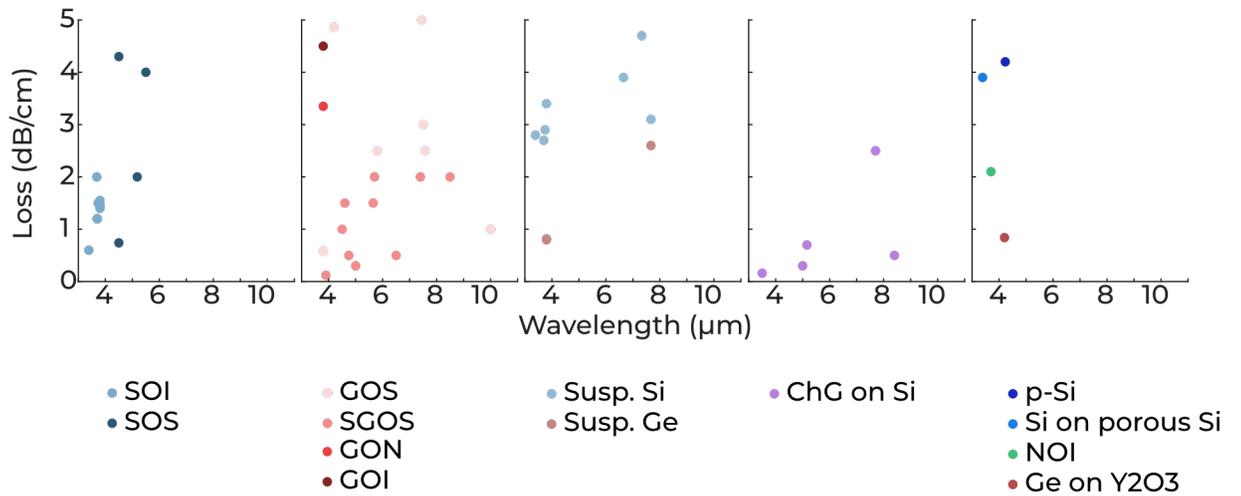

*Figure 24 Waveguide propagation loss as a function of wavelength for various platforms.*

Figure 1 summarizes waveguide propagation loss measurements below 5 dB/cm on silicon platforms at wavelengths above 3 μm. Propagation losses < 1 dB/cm have been reported for SOI, SOS, Suspended Si, Suspended Ge, GOS, SGOS, Ge on $Y_2O_3$ and ChG waveguides. The lowest losses reported to date are 0.12 dB/cm for SGOS waveguides operating at 3.9 μm [11] and 0.16 dB/cm for ChG waveguides at $\lambda$ = 3.4 μm [12]. At longer wavelengths, ~1 dB/cm was reported for GOS waveguides operating at 10.0 μm [13].

### 3.2. Component Devices

Several extensive reviews have overviewed the expansive and growing number of waveguide, device, and application demonstrations [14–20]. Basic building blocks like Y-junctions, multi-mode interferometers (MMIs), evanescent couplers, Mach Zehnder interferometers (MZIs), and slot waveguides are easily translated from NIR silicon photonic developments and are routinely applied to platforms in the MWIR and LWIR. Here we summarize the current development state of several key components.

Grating couplers: Grating couplers are straightforward elements used for vertical coupling and surface emission, and are critical for some laser or detector integration methods. Grating couplers have been reported for SOI [21,22], GOS [23,24], GOI [23], and SOS [25] with coupling efficiencies as low as -4.34 dB [22] and operating wavelengths up to 7 μm [24].

Ring resonators: Ring resonators are useful for wavelength filtering, modulation, and nonlinear comb generation, and can be used as part of a laser external cavity to provide wavelength selection, tuning, and linewidth narrowing. Ring resonators have been demonstrated as an individual component on various platforms: SOI [9], SOS [26], Suspended Silicon [27], ChG-on-Si [28], Ge-on-$Y_2O_3$ [6], GOS [29], SGOS [30–32]. MWIR ring resonator quality factors of over $10^6$ have been measured on SOI (3.5–3.8 µm) [9] and SGOS (3.5–4.6 µm) [30]. In the LWIR, Q values of 113k have been reported at 7.5–9.0 µm [33].

AWGs: AWGs are broadly-implemented integrated optics components that were developed for NIR wavelengths as (de)multiplexers and for optical routing in transceiver applications. In the MWIR and LWIR, AWGs can be used as dispersive elements for spectroscopy, for wavelength beam combining and splitting, or in free space communication applications. AWGs have been reported on SOI [34,35], GOS [36–38], and SGOS [39,40], for wavelengths up to 8 µm [40], with up to 67 inputs [40], 50–300 GHz, crosstalk as low as -32 dB [34], and insertion losses as low as -1.5 dB [38]. Planar concave gratings (PCGs), an alternative (de)multiplexer design, have also been built from GOS waveguides for 5.2 µm [41,42].

Modulators: Modulators will be necessary to address free space communication applications, and can be used to implement high sensitivity spectroscopy techniques like wavelength modulation spectroscopy (WMS) [43] or differential absorption LIDAR. MWIR or LWIR electro-optic (EO) modulators can be fabricated directly from Si or Ge waveguides, taking advantage of the plasma dispersion effect, the thermo-optic effect, the Kerr effect, or the electro-absorption effect [44,45]. EO modulators have been fabricated on SOI [46], GOS [47], and SGOS [48–50], with operating wavelengths as high as 10.7 µm and data rates up to 1.5 GHz [50]. Alternatively, modulators can be fabricated from ferroelectric oxides like lithium niobate (LN) or barium titanate (BTO), or from 2D materials like graphene, black phosphorous (BP) [51], or transition metal ChGs. Modulators based on integrated III-V semiconductors achieve high performance in the NIR but have not yet been demonstrated in the MWIR or LWIR.

### 3.3. Lasers

Integrating a light source at any wavelength onto the silicon photonic system is a significant challenge. As silicon has an indirect band gap, it is intrinsically incapable of making an efficient interband light source so other materials and integration approaches must be considered. Options that have been explored extensively in the broader field of silicon photonics are discussed in [52]. Lasers, broadband light sources, and nonlinear conversion can be used to generate MWIR or LWIR light on the silicon photonic platform.

First and most straightforward, a conventional laser can be constructed separately with minimal or no modifications, and light can be coupled into the silicon photonic chip either directly (butt coupling to an edge facet or vertically coupled to a grating coupler) or indirectly (with free space optics, through an optical fiber, or a photonic wire bond). If the laser is sensitive to optical feedback, an optical isolator may be included. Bonding of the laser to the silicon chip can be referred to as "hybrid integration". Second, unprocessed or partially processed III-V layers can be bonded or transfer-printed to the silicon die where the laser processing is then completed ("heterogeneous integration"). The III-V substrate is removed during this process. Optical coupling can then be achieved with evanescent couplers, on-chip butt-couplers, or grating couplers. Bonding can be achieved by direct bonding or polymer-assisted bonding (e.g., benzocyclobutene (BCB)). Third, the III-V gain material can be epitaxially grown on silicon ("monolithic integration") which is discussed in a separate sub-topic article of the Roadmap.

QCLs and ICLs are the predominant choice of semiconductor lasers for integration with MWIR or LWIR silicon photonic systems, although GaSb-based type-I quantum well lasers can also be considered for wavelengths below ~4 μm. The most significant integration challenge, regardless of method, is the efficiency of coupling light from the laser structure to the silicon photonic waveguides. Butt coupling, vertical coupling, fiber coupling, and lens-based approaches all require careful optical alignments and packaging techniques. Heterogeneous and monolithic integration require carefully-designed integrated components like tapered mode converters or on-chip etched-facet butt-couplers. For QCLs, an additional challenge is dissipating the significant thermal load generated by high drive powers.

Heterogeneous integration approaches are more scalable but require more complex laser designs and fabrication processes. Heterogeneously integrated lasers from separate III-V growths and operating at widely-separated wavelengths can be integrated on the same silicon chip. Lasers can be lithographically aligned to other silicon photonic elements, silicon PICs including lasers can be manufactured at scale on large silicon wafers, and coupling efficiencies do not depend on packaging. Off-chip approaches alternatively allow the integration of conventional, high performance, and optimized laser designs. Laser performance can be individually validated and tested for reliability, and poorly performing lasers can be rejected without discarding the complete PIC. The maturity of conventional laser manufacturing processes also reduces the development barrier-to-entry.

Although fiber and lens coupling are frequently used to characterize silicon PICs, they have not been reported as part of fully packaged solutions in the MWIR or LWIR.

A detriment to nearly all MWIR or LWIR silicon photonic platforms except GOS is the low thermal conductivity of the lower cladding dielectric or glasses. While the thermal conductivity of Si (1.3 W/cm-K) is nearly twice that of InP (0.68 W/cm-K), the lower claddings of platforms like SOI or suspended Si prevent efficient spreading of the heat generated by a laser. This is advantageous for some applications like on-chip heaters, but detrimental for most on-chip light generation. External heat sinks like the AlN mounts used with conventional QCLs, thermal shunts (where heat is transferred to the Si substrate through a metal shunt path), or the integration of lasers below the waveguide layer must be considered for higher heat loads.

Butt coupling has been applied in several demonstrations to create tunable, external cavity MWIR lasers. An integrated external cavity QCL was created by butt coupling a conventional QCL to a Ge-on-SOI (GOSOI) chip with a DBR reflector [53]. Over 50 nm wavelength tuning near 5.1 μm was achieved by thermally tuning the DBR reflectors with on-chip heaters. This tunable laser emitted a maximum peak optical power of 22 mW with side mode suppression ratio (SMSR) > 20 dB. Although a maximum coupling efficiency of 90% was simulated, a gap spacing of < 2.5 μm between the chips was found to be necessary to achieve a coupling efficiency below 3 dB. An ICL was similarly butt coupled to an SOI chip with a microring resonator to demonstrate a tunable laser at ~3.4 μm [54]. A comparable 54 nm tuning was achieved with SMSR of 25 dB. The output power was 0.4 mW.

A hybrid integration approach has also been used to integrate QCLs with GOS waveguides [55]. Two DFB QCLs operating at 7.2 μm, shown in Figure 2(a), were separately fabricated on the native InP substrate and flip-chip solder-bonded to the same silicon chip. Corresponding notches on the surface of the QCLs and the silicon chip were used to self-align the QCL end facet to a Ge waveguide end facet for butt-coupling. 1.5 mm-long lasers had a pulsed threshold of 170 mA and operated CW at 15ºC. The light from both lasers was beam-combined with an on-chip MMI, and 0.7 mW CW power was collected from the chip's output facet.

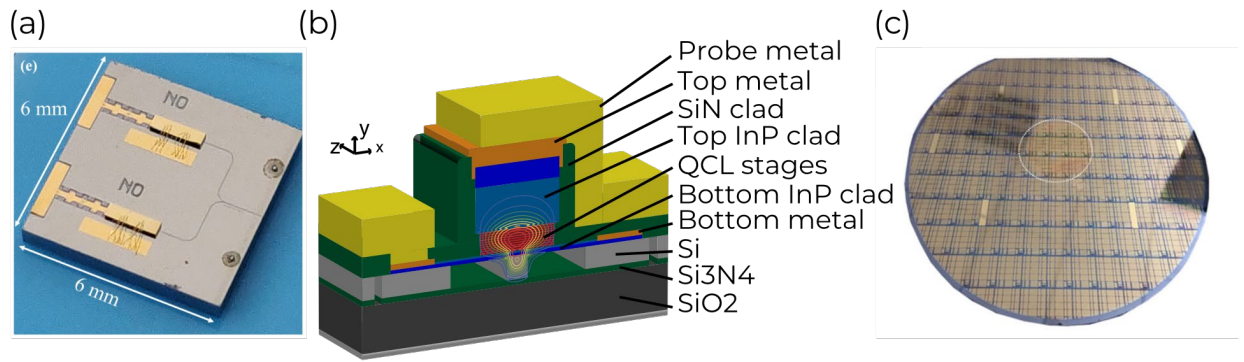

*Figure 25. (a) Two DFB QCLs solder-bonded to a silicon chip. [55]. (b) Cross section of a heterogeneously integrated QCL above a SONOI waveguide [56]. (c) Heterogeneously integrated QCLs manufactured on a 200 mm CMOS/MEMS pilot line. The dotted circle refers to the bonded 2-inch InP wafer. [57]*

Among laser integration approaches, the greatest progress towards fully integrated PICs has employed heterogeneous integration. MWIR QCLs and ICLs have been constructed by direct wafer bonding of III-V layers to the silicon platform [56–60]. Here the laser must be grown with cladding and contact layers designed for integration above silicon or germanium waveguides to ensure that the optical mode overlaps with the QCL active stages and an efficient coupler can be designed. Gratings can be introduced on the silicon photonic waveguide before bonding, to construct distributed feedback (DFB) or distributed Bragg reflector (DBR) lasers. The laser structure is fabricated after bonding the III-V material to the waveguides and removing the III-V substrate. In initial demonstrations, Fabry-Perot [56] and DFB [59] QCLs built on a silicon-on-$Si_3N_4$-on-$SiO_2$ (SONOI) platform (Figure 2(b)) emitted pulsed light near 4.7 μm and operated up to 100ºC. Threshold currents were below 1 kA/cm², 211 mW was emitted from a hybrid III-V/Si facet, and 32 mW was coupled into a silicon waveguide through a tapered III-V mode converter. QCLs for 4.6 μm were also heterogeneously integrated on GOS, emitting 6 mW [60]. The lower performance relative to QCLs on SONOI resulted primarily from the higher refractive index of Ge and the lack of a low-index cladding in the GOS platform, which makes it more difficult to design an efficient coupler. Heterogeneously-integrated QCLs have also been manufactured separately by a different group using a similar process on a 200 mm CMOS/MEMS pilot line, as shown in Figure 2(c) [57].

Continuous-wave (CW) heterogeneously-integrated QCLs have been demonstrated by flip-chip mounting the lasers onto an AlN submount. Figure 3(a) and (b) show the light intensity vs. current density and spectral power of a QCL on Si that emitted 1.2 mW of CW output at wavelengths near 4.8 μm [61].

Heterogeneously integrated QCLs have matured commercially in recent years, and have been implemented in complex PICs. QCLs as part of a spectrometer emitted over 70 mW near 4.6 μm into Si waveguides (further discussed in Section 3.6 below).

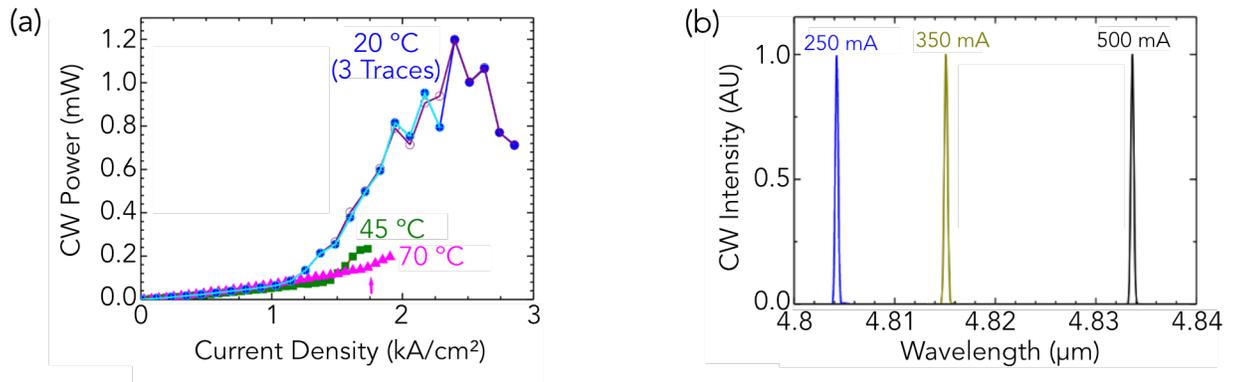

*Figure 26. (a) CW power vs. current density for a heterogeneously-integrated QCL on silicon operating at 20, 45, and 70 ºC. (b) CW intensity vs. wavelength of the heterogeneously-integrated QCL operating at drive currents from 250 to 500 mA [61].*

Similar to the dual-QCL hybrid integration demonstration [55], multispectral MWIR lasers can be constructed by coupling light emit from multiple heterogeneously integrated QCLs. By performing the beam combination on a silicon photonic platform, high efficiency, high resolution beam combiners can be used, and the output facet can be tailored to emit a high-quality single mode beam without impacting the laser performance. The first multispectral MWIR light source on silicon was built by coupling light from 3 DFB QCLs through an AWG to emit from a single waveguide [62]. This multispectral source was constructed on SOI and functioned near 4.7 μm. Fabrication yield limited the number of lasers coupled through the AWG.

By bonding multiple III-V epi dies to the same silicon photonics die, a multispectral source can be constructed to operate across a wavelength range wider than the bandwidth of the gain provided by any individual epi growth. Figures 4(a) and (b) show a microscope image and emission spectral of a multispectral QCL. Two QCL arrays were built by bonding QCL dies from separate III-V growths to one silicon chip. Light from 10 total lasers was combined by on-chip beam combiners for emission from a single facet, covering a wavelength range of ~4.2–4.7 μm [63].

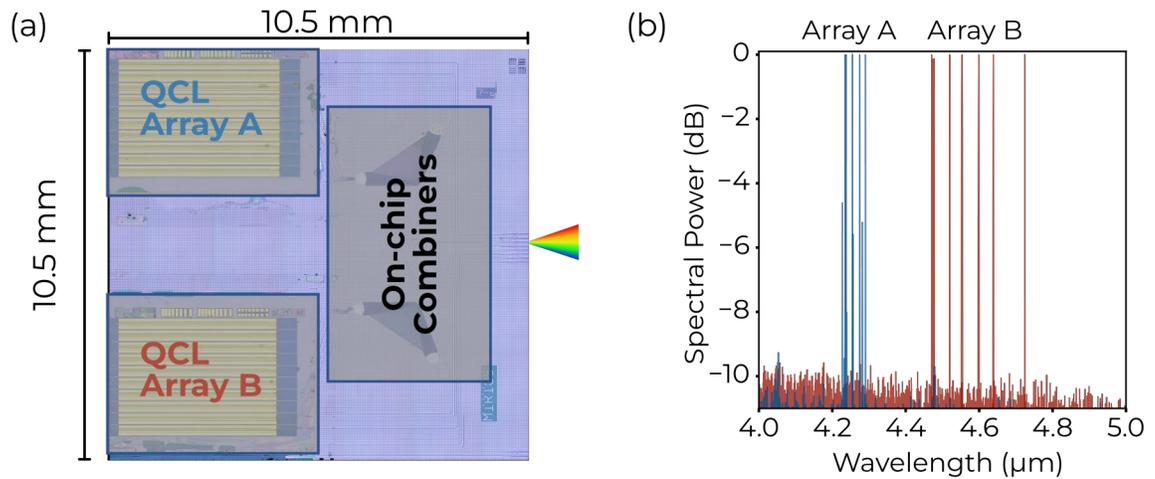

(a) 10.5 mm

QCL Array A

QCL Array B

On-chip Combiners

10.5 mm

(b)

Array A    Array B

*Figure 27 (a) A multispectral MWIR source with two QCL arrays, each integrated from separate III-V wafers. (b) Emission spectra of 10 QCLs collected from the same facet. [63]*

The heterogeneous integration process can be applied similarly to ICLs, although the designs and fabrication processes must be adjusted to accommodate the substantially different refractive indices and fabrication methods of the GaSb-based epilayers. Fabry-Perot ICLs operating at $\lambda \approx 3.6$ µm were built with a similar heterogeneous integration process above SOI waveguides [58]. These lasers emitted 6.6 mW from a hybrid III-V/Si facet, with lower light levels coupled into the silicon waveguide, and operated in pulsed mode up to 50 °C. The primary limitation was sidewall current leakage due to a suboptimal dry etching process.

Heterogeneous integration additionally presents opportunities to improve laser performance with mode engineering. The III-V layers can be bonded to materials with real and imaginary refractive indices that could otherwise not easily be included in a monolithic III-V epilayer stack. As in previous demonstrations, high transparency silicon or germanium waveguides below a QCL or ICL active region can reduce laser internal loss and provide lateral mode confinement. High resolution, low-loss gratings can be placed directly below the laser active region without the need for epitaxial regrowth. By optimizing the thickness, this underlying layer can also be used as a separate confinement layer (SCL) of a weak index guided ICL to provide more stable, higher power, single mode operation [64]. Because the SCL is not electrically pumped, parasitic current injection can be circumvented.

Transfer printing has also been used to heterogeneously integrate QCLs on an SOS wafer [65]. In this case, the QCL structures were first fabricated on the native InP substrate before being transferred to the SOS wafer, after which the SOS waveguides were etched. SU-8 was used to adhesively bond the III-V layers to the Si surface. These lasers had a threshold current density of 5.3 kA/cm$^2$ and emitted 2 mW from a Si waveguide facet.

### 3.4. Photodetectors

Photodetectors can be integrated directly with waveguides, or co-packaged with the silicon PIC. Waveguide-integrated photodetectors can potentially operate with high sensitivities, low dark current, and low power consumption if designed to efficiently maximize the optical interaction with the absorber. Multiple channels on a chip can also be addressed individually, and each photodetector can be optimized for different wavelengths. Waveguide integration approaches that have been explored for MWIR and LWIR silicon photonic waveguides include microbolometers, hybrid or heterogeneous integration of quantum detectors (as with lasers), two-dimensional (2D) materials, defect-mediated absorption, and Schottky diode photoemission. These methods each have intrinsic tradeoffs between sensitivity, bandwidth, and manufacturability.

Shottky diodes and defect-mediated absorption mechanisms can be leveraged with CMOS-compatible Group IV materials without the need to introduce other materials. By metallizing the surface of a GeSi waveguide, a Shottky diode can be formed and used for waveguide-integrated photodetection directly from the GeSi waveguide optical mode [66]. Photodetection in the 5–8 µm wavelength range was demonstrated with this approach with a responsivity of 0.1 mA/W, and for pulses as short as 50 ns indicating 20 MHz operation. PIN SiGe on Si photodetectors, hypothesized to exploit Ge growth defects for photoabsorption, have also been demonstrated, functioning from 5.2 µm to 10 µm with a responsivity of ~1 mA/W and operation to 1.5 GHz [50].

2D materials with layered lattice structures can be engineered with different band gaps for a wide range of wavelengths. The fabrication processes often require the mechanical exfoliation of layers from a bulk material and manually transferring them above waveguides or grating couplers. Black phosphorus (BP) has a narrow band gap of ~0.3 eV in its bulk form, corresponding to a cutoff wavelength of 4.13 µm. MWIR BP-based photodetectors integrated above SOI grating couplers, shown in Figure 5(a), were reported for the 3–4 µm wavelength range with responsivities up to 0.85 A/W [67–69]. The slow light effect in photonic crystal waveguides (PhCWGs) was used to enhance the responsivity of a BP photodetector to achieve 11.31 A/W (Figure 5(b)). Alternatively, the gapless semimetal graphene can exhibit broadband photoresponse from ultraviolet to terahertz. Plasmonically-enhanced graphene photodetectors above Si-on-CaF₂ waveguides showed a responsivity of 8 mA/W under zero bias when measured for 6.3–7.1 μm wavelengths [70].

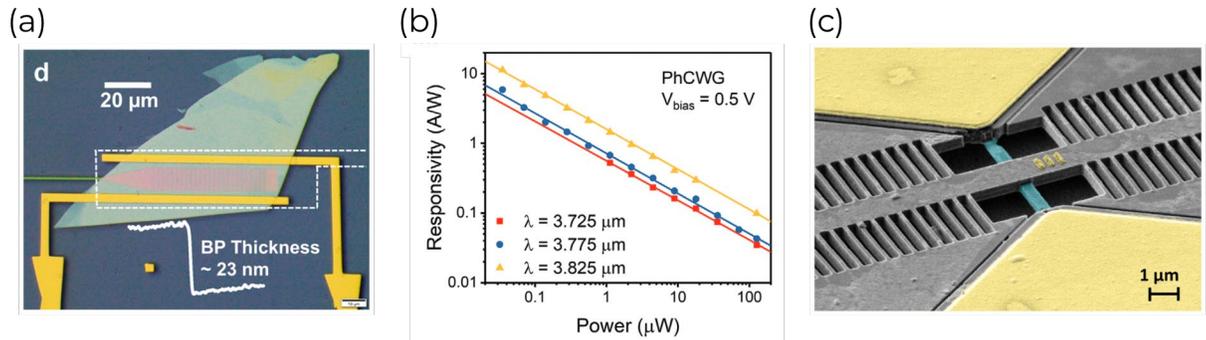

*Figure 28 (a) Photodetector constructed from a BP flake above an SOI waveguide [67]. (b) Responsivity of a BP photodetector on a SOI PhCWG [69]. (c) SEM image of a microbolometer integrated with a suspended silicon waveguide. Yellow regions are gold and blue regions are an a-Si thermometer [71].*

Microbolometers and thermopiles offer broadband photoresponse across the infrared, at the expense of low speed and sensitivity relative to photon detectors. Microbolometers have been constructed by depositing gold plasmonic antennas on the surface of SOI [71], suspended silicon (Figure 5(c)) [71], or a-Si-on-SiO₂ [72] waveguides, which heat when light is absorbed. A nearby thermometer is fabricated from a-Si to measure the change in temperature and deduce the absorbed light intensity. For 3.72–3.88 μm, this approach demonstrated a 24.62% change in resistance per mW of input power. Alternatively, a manufactured MEMS-based thermopile flip-chip thermo-compression bonded above an SOI grating coupler achieved 69 mV/W at 3.72 μm [73].

As with integrated lasers, photodetectors can be built by heterogeneously bonding III-V layers to the silicon photonic platform. This approach has led to high performance integrated detection in the NIR [52] and short-wave infrared [74–76], although few demonstrations have been published thus far for wavelengths above 3 μm. GaSb-based InAsSb nBn photodetectors, GaSb-based type-II "W" absorber InAs/InAsSb/InAs/AlSb photodetectors, InP-based quantum cascade detectors (QCDs), and GaSb-based interband cascade detectors (ICDs) are all promising candidates for integration by fabrication processes similar to those used for heterogeneously-integrated QCLs and ICLs. In one demonstration, an array of InAs₀.₉₁Sb₀.₀₉ photodetectors for 3.7–3.8 µm was heterogeneously integrated above SOI waveguides as part of a spectrometer [35]. The photodetector responsivity was 0.3 A/W at room temperature.

### 3.5. Nonlinear optics

Nonlinear effects can be exploited on-chip to generate broadband infrared light and frequency combs, which may be used as light sources for sensing and spectroscopy [77,78]. The two-photon absorption of silicon, which presents a significant challenge for nonlinear optics, cuts off at 2.2 µm, allowing high performance nonlinear devices in the MWIR and LWIR. Frequency generation has been reported on several platforms, where dispersion-engineered waveguides are pumped with high power sources. Supercontinuum generation spanning 2–5.6 µm was achieved with SOS waveguides pumped at 3.7 µm [79] and 3–8.3 µm was produced by SGOS waveguides pumped at 4.15 µm [80]. MWIR frequency combs have been generated in $Si_3N_4$-on-$SiO_2$ waveguides from 2.5–4.0 µm [81] and by suspended Si across a bandwidth of 2.0–8.8 µm [82]. Optical parametric amplification (OPA) has also generated tunable MIR light from 2.6–3.6 µm in $Si_3N_4$ on $SiO_2$ waveguides [83].

### 3.6. Sensors and PICs

The most straightforward sensor approach is direct absorption spectroscopy (DAS), which can be achieved on a silicon photonic chip by exposing a simple waveguide to a chemical or gas of interest and transmitting light at the characteristic absorption wavelength from a single-wavelength or tunable laser to perform absorption spectroscopy. The sensitivity is fundamentally limited by the waveguide propagation loss and the optical mode interaction, so waveguide geometry is critical. Slot waveguides support optical modes with enhanced electric field amplitudes within a gap region but can have excess sidewall scattering loss. Spiral waveguide patterns or resonators can be used to increase the optical path length and chemical interaction. To justify DAS on-chip, the sensitivity, cost, and total package size must be compared to a simple multipath gas cell with independent laser and photodetector [84].

Numerous waveguide DAS sensors have been reported. As a few examples, sensing of toluene [85] (75 ppm, 6.65 µm, suspended Si), IPA in Acetone [3] (5%, 3.73 µm, GON), methane (0.3 ppm, 3.27 µm, SOI) [86], and $CO_2$ have been performed. Fluid samples, particularly for biomedical sensing applications, can similarly be measured by integrating microfluidics above waveguides. Spectroscopy at ~5.2–10 µm of Bovine Serum Albumin was performed by transferring aqueous analytes through a strip of filter paper to the surface of a GOS waveguide. [87,88]. Absorption spectroscopy of cocaine at 5.71 μm (overlapping the characteristic absorption peak) in saliva samples was achieved by delivering the solution through a polymer microfluidic channel that was bonded above a GOS waveguide. A concentration of 500 μg/mL was measured [89].

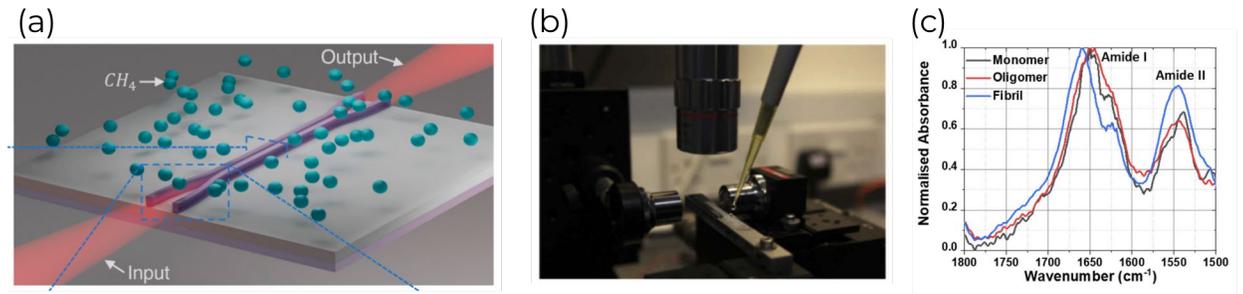

*Figure 29 (a) Direct-absorption methane sensor based on a SOI slot waveguide [86]. (b) Experimental apparatus showing liquid from a pipette introduced onto filter paper above a GOS waveguide [87]. (c) GOS waveguide absorption spectra of aqueous Bovine Serum Albumin (BSA). [87].*

More complex passive spectrometers capable of reconstructing the optical spectrum of a broadband light source have also been developed. While AWGs or PCGs can be used as dispersive elements to separate wavelengths, higher-resolution designs are necessary for practical applications. The footprint and insertion loss scale with resolution and channel count, limiting the achievable performance. Photodetectors can be integrated to individually address each channel [35] (shown in Figure 7(a)), or on-chip switching can be combined with a single off-chip photodetector.

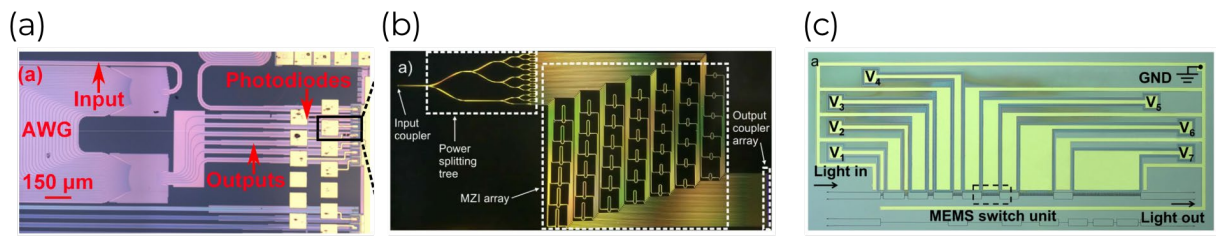

*Figure 30. (a) AWG-based spectrometer with heterogeneously integrated InAs$_{0.91}$Sb$_{0.09}$ p-i-n photodiodes [35]. (b) FT-IR spectrometer based on cascaded asymmetric Mach-Zehnder interferometers [90]. (c) MEMS-based computational spectrometer [91].*

On-chip FT-IR spectrometers can be built by splitting light into a series of Mach-Zehnder interferometers to generate an interferogram from which an input spectrum can be reconstructed [90,92,93]. A resolution of 2.7 nm at 3.75 μm wavelength has been reported with this technique (Figure 7(b)) [90]. Similar to AWGs, the resolution is inversely proportional to the maximum interferometer arm length difference, and the maximum optical bandwidth is proportional to the number of interferometers, so available die space and waveguide loss can put a maximum pragmatic limit on performance. This may be circumvented in some designs by actively tuning the path length difference (e.g., thermally) of one or more interferometers. MWIR computational spectrometers have also been developed using MEMS-actuated waveguide couplers to control waveguide path length differences (Figure 7(c)), an approach that promises a small footprint [91,94]. A resolution of 8 nm was measured for $\lambda$ = 3.65-4.41 $\mu$m.

High sensitivity spectroscopy can be achieved with heterodyne detection, which provides shot-noise-limited gain without optical amplification. To achieve this, a stable, narrow linewidth laser is used as a local oscillator (LO), which is photomixed with a weak source signal to generate an upconverted intermediate-frequency signal that is detected with a fast photodetector. While a high spectral resolution (<1 MHz) can be achieved with heterodyne spectroscopy, the optical bandwidth is limited by the photodetector speed. A wider bandwidth can be achieved by using multiple LOs, but the resulting optics requirements make free space heterodyne systems with more than a few LOs prohibitively complex and expensive. Heterogeneous photonic integration is particularly advantageous for this application, because the number of on-chip lasers can be scaled with no additional cost. Figure 8 (a) shows a prototype on-chip heterodyne spectrometer on Si. An AWG spectrally separates the input light into 8 output waveguides. 8 DFB QCLs in an array then function as on-chip LOs that couple with the AWG outputs through on-chip 2x1 combiners. The photomixed light in each channel is output from a separate waveguide. Figures 8(b) and (c) show the power vs. drive current and spectral power density collected from each LO, respectively [95,96].

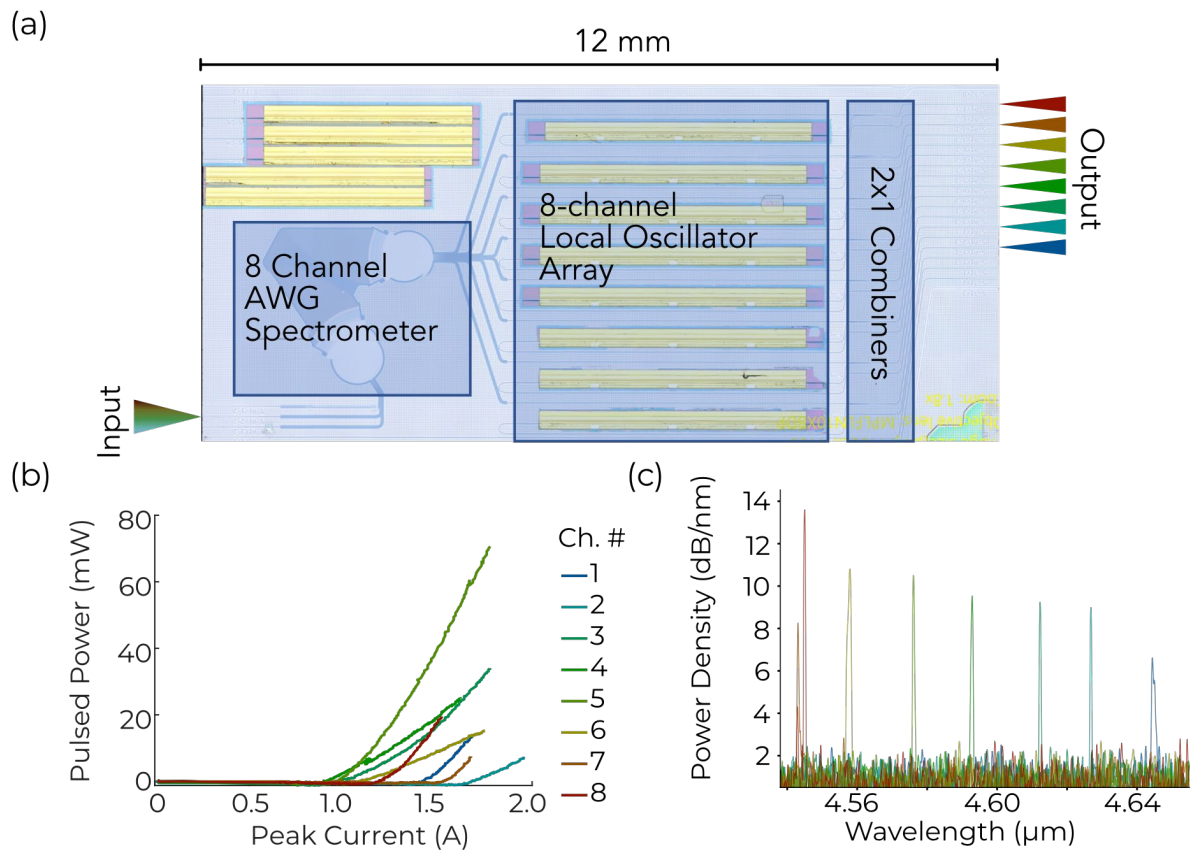

*Figure 31. (a) 8-channel silicon photonic MWIR heterodyne spectrometer. (b) Peak pulsed power vs. peak current of 8 DFB QCLs functioning as LOs in the spectrometer. (c) Spectral power density of the 8 LOs designed for different wavelengths. [96]*

## 4. Challenges, opportunities, and future developments

To date, the primary focus of research on MWIR and LWIR Si-based PICs has been the demonstration of individual passive components. In the MWIR, most of the fundamental building blocks have been demonstrated and advanced to suitable performance levels for many applications. Waveguide propagation losses on several platforms have consistently achieved < 1 dB/cm. Ring resonators display Q values exceeding 1 million and routinely reach at least 100s of thousands, and various couplers, splitters, and multiplexers can be constructed with reasonably low insertion loss. The fundamental limits of device performance may not yet have been reached, and fabrication quality plays a significant role.

Despite 15 years of development for MWIR and LWIR silicon photonics, a dominant waveguiding platform has yet to emerge. Because the application space for these wavelengths is quite broad, no single set of criteria guide the platform choice.

Platforms in component demonstrations are typically chosen either to maximize the performance result, or to leverage previous developments within the given research institution. Although this broad exploration has revealed the benefits and limitations of different platforms, it increases the challenge of future PIC development.

While more direct applications are now being addressed (particularly gas sensors), most of the demonstrated systems are still of relatively low complexity. The largest number of elements reported in a single MWIR PIC remains less than 50 [90,93], and few results include on-chip light sources or on-chip photodetectors. No sensing reports have combined both lasers and detectors integrated on the same chip. This sharply contrasts current NIR silicon PICs that now achieve very large scale integration (VLSI) levels with over 10,000 components [97].

To justify the ongoing pursuit of MWIR silicon photonic integration, higher complexity and higher performance must be demonstrated. Several platforms are mature enough to build multi-specimen chemical sensors, resonant cavity enhanced detection, fully integrated spectrometers, on-chip heterodyne detection, dual-comb spectroscopy, coherent and wavelength beam combination, and tunable external cavity lasers, as a few examples. While preliminary PIC demonstrations have been reported for many of these applications, higher performance is needed and more complete systems should be constructed.

More focus should be placed on the realization and maturation of mutually-integrated photodetectors and lasers. Any immaturity of this development will limit or prohibit scalability in many applications, and negate the economic benefits of photonic integration. Furthermore, the power, thermal, design, and manufacturing requirements of lasers and detectors can be a significant limiting factor and the continued development of PICs without light sources or detection risks reliance on platforms and material choices which are unsuitable or inferior for fully integrated systems. For example, the heterogeneous integration of MWIR QCLs is more difficult to apply to GOS than SOI.

Whereas photodetector approaches like 2D materials, defect-driven, microbolometers, and Schottky diodes hold promise for simplicity and bandwidth, III-V based quantum detectors can potentially attain extremely high sensitivities and speeds but have been explored only minimally. This approach should be further investigated, since applications like optical heterodyne detection will require advanced on-chip photodetectors.

Heterogeneously integrated QCLs now perform well enough for many sensing applications. 10s of mW of pulsed MWIR light has been coupled into silicon photonic waveguides and built alongside other photonic integration components. Still, further development of the on-chip lasers is needed. The power coupled into silicon-based waveguides does not yet match that produced by state-of-the-art conventional QCLs or ICLs. Coupling efficiencies and subsequent wall plug efficiencies should be improved. Integration on GOS and SGOS should also be improved to support longer wavelengths and wider optical transparency.

While most of the laser integration has focused on QCLs, ICLs offer much lower drive powers and voltages. ICLs should be considered for applications at suitable wavelengths below 6 μm, and which do not require the higher optical powers delivered by QCLs. For heterogeneous integration, the challenges of fabricating GaSb-based epilayers on Si will need to be overcome.

Nonlinear devices hold significant promise for broadband applications like dual-comb spectroscopy and frequency generation with parametric oscillation and amplification. These applications will ultimately require the integration of suitable pump sources like mode locked lasers.

It is now feasible to integrate light sources, detectors, and sensing waveguides together on the same chip. To achieve high sensitivities, emphasis must be placed on reducing waveguide propagation loss for waveguide geometries with field confinement in the sensing region, like slot waveguides and resonators, and reducing coupling losses between sensing waveguides, lasers, and detectors. The pragmatically achievable limits of these metrics will determine the ultimate commercial competitiveness of integrated sensors.

5.   Concluding Remarks

MWIR and LWIR silicon photonics has advanced significantly in the last 15 years, progressing from the first waveguide demonstration to complex PICs with integrated lasers. Numerous waveguiding platforms have been matured, and fundamental building blocks are well established. Going forward, we anticipate continued advancements in on-chip lasers, photodetectors, nonlinear devices, and sensor architectures.

Silicon photonics holds dramatic potential for low cost, multi-functional, high sensitivity chemical sensors, non-invasive biological diagnostics, and broadly-applicable infrared spectroscopy. Still, challenges must be overcome before cost effective and commercially-viable technologies are attained. Sensor design, performance, and data analysis techniques will need to be matured and validated in relevant environments. PICs should provide greater functionality by integrating light sources and detectors on a larger scale, and cost-effective packaging must be developed. The ongoing evolution of relevant PIC technologies will ultimately situate MWIR and LWIR silicon photonics as a powerful mainstream technology.

# 19. Mid-Infrared Photonic Integrated Circuits on the Native III-V Substrates

SEUNGYONG JUNG[1,*] AND NISHANT NOOKALA[1]

[1]TransWave Photonics, LLC, 8711 Burnet Road Suite F67, Austin, TX 78757, USA

*sjung@transwavephotonics.com

## Overview

Mid-infrared (mid-IR) photonic integration is an emerging technology that enables fully monolithic photonic integrated circuits (PICs) with unprecedented improvements to system size, power efficiency, and cost for a wide range of applications, including free-space optical communications [1-3], environmental monitoring [4, 5], medical diagnosis [6, 7], and quantum optics [8, 9]. Demonstrating fully monolithic PICs requires the integration of active and passive building blocks such as lasers, detectors, passive waveguides, couplers, modulators, and amplifiers. To this end, an integration architecture should facilitate the seamless integration of light sources with low-loss passive waveguides, efficient coupling between the PIC building blocks, and scalable volume production with high yield and reliability.

Leveraging silicon photonics technology developed for the telecom band appears to be a viable approach for mid-IR photonic integration. However, integration strategies and challenges differ significantly from those in the near-IR regime, where the heterogeneous integration of InP-based diode lasers with Si-based passive waveguides is the de facto standard [10].

For mid-IR electrically pumped semiconductor light sources, InP-based quantum cascade lasers (QCLs) and GaSb-based interband cascade lasers (ICLs) are the primary options. Given the maturity of InP fabrication and the capability of QCLs to cover the entire mid-IR band, most mid-IR active/passive photonic integrations have been demonstrated using QCLs. The first QCL integration was reported using heterogeneous integration, where QCL epitaxial layers were transferred onto Si-on-Nitride-on-Insulator via direct wafer bonding [11] and onto Si-on-Sapphire using adhesive wafer bonding [12]. However, these devices exhibited limited performance due to poor heat dissipation at the bonding interface, highlighting a critical limitation of heterogeneous integration. Unlike near-IR diode lasers, QCLs generate an order of magnitude more dissipated heat, making thermal management a significant challenge.

To address this issue, homogeneous integration platforms based on native III-V substrates are emerging as more suitable solutions for QCL-based PICs [13-15]. This approach enables monolithic active/passive integration without compromising the QCL performance. In addition, $In_{0.53}Ga_{0.47}As$, lattice matched to InP, serves as an excellent passive material, exhibiting propagation loss of 0.5 – 1 dB/cm over 3 – 10 μm range [16-18]. This is lower than any other mid-IR waveguide materials, including Si, Ge, and chalcogenide glasses [19].

Experimental demonstrations of QCL PICs with $In_{0.53}Ga_{0.47}As$ passive waveguides have been conducted at 4.6 μm, where QCL light was coupled to the waveguide via an adiabatic taper[13]. The $In_{0.53}Ga_{0.47}As$ waveguide has also been coupled with QCLs in a butt-coupling configuration using a regrowth and Fe-doped buried-heterostructure technique [14]. Recently, a passivated active region of a QCL was used as a passive waveguide, exhibiting a loss of ~3dB/cm [15].

With this perspective, we review the homogeneous integration of the mid-IR lasers with low-loss passive waveguides, using QCL photonic integration as a primary example. III-V waveguide materials and their potential integration are examined to provide pathways toward monolithic integration. Coupling approaches between active and passive devices are discussed in the context of the integration platforms and architectures. Finally, we present fascinating applications and advancements in mid-IR PICs that are not readily achievable with conventional mid-IR devices, and we discuss integration challenges and future developments.

## Current status

### 1) III-V low loss passive waveguide

A passive waveguide suitable for homogeneous integration must exhibit low optical loss, a high refractive index, and be lattice matched to the target native substrate. For InP-based QCL integration, $In_{0.53}Ga_{0.47}As$, lattice-matched to InP, demonstrates a theoretical loss value of 0.1 – 1 dB/cm at a background doping level of 1x15 cm$^{-3}$ and a refractive index of 3.25 – 3.38 in the mid-IR band, which makes it suitable as a passive waveguide core [13].

Several research groups have characterized the propagation loss of InGaAs waveguides with InP claddings using cut-back [13] or thermo-optic Fabry-Perot oscillation measurements [16]. Typical InGaAs/InP waveguides have core thicknesses in the range of 1 – 2 μm and ridge widths of 5 – 10 μm to support single-mode TM operation. Given the broad spectral range of the mid-IR, optimizing the waveguide dimensions for specific wavelengths is essential. As shown in Fig. 1, the waveguide loss is as low as 0.5 dB/cm near 5 μm, and gradually increases to 1 dB/cm near 9 μm. Further reduction of the loss was obtained from a buried-heterostructure (BH) regime, where an InGaAs core was enclosed by an Fe-doped InP cladding layer. The loss of a BH waveguide device is about 0.7 – 1 dB/cm, which is nearly flat over the 5 – 11 μm range [14].

Recently, $GaAs_{0.51}Sb_{0.49}$, which is also lattice-matched to InP, has been investigated as a passive waveguide material that demonstrates a propagation loss of 2–3 dB/cm over the 5.5–9 μm range [20]. Although GaAsSb exhibits higher optical loss compared to InGaAs, its higher refractive index and enhanced nonlinear properties enable efficient supercontinuum generation, making it a good candidate for nonlinear mid-IR photonics.

An alternative approach eliminates the need for a separate passive waveguide by utilizing the InGaAs/AlInAs QCL active region itself as a waveguide material [21]. To suppress intersubband and free-carrier absorption, the active region has been passivated using ion implantation, yielding a propagation loss of 3.14 dB/cm at 4.8 μm [15] and 1.4 dB/cm at 9.6 μm [21]. This approach simplifies fabrication while enabling monolithic integration of active and passive functionalities within the same material platform.

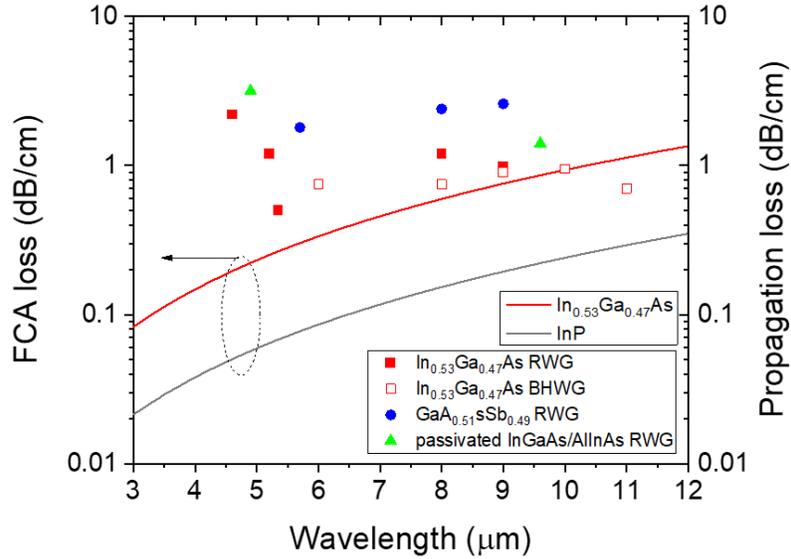

Fig. 1. Calculated free-carrier absorption loss and experimental propagation loss of TM mode from various material systems and waveguide configurations. A doping level of 1e15 cm⁻³ was used for the InGaAs/InP FCA calculations. The experimental data are for InGaAs ridge waveguides (RWGs) from [13, 18, 22], InGaAs BHWGs from [14, 17], GaAsSb RWGs from [20], and passivated InGaAs/AlInAs RWGs from [15, 21].

### 2) Active/passive coupling

Efficient light coupling between the laser and passive waveguide is crucial for determining the performance and practicality of photonic integrated circuits (PICs). The choice of coupling method directly impacts the device design complexity and fabrication feasibility. Depending on the relative positioning of the laser active region and passive waveguide, coupling methods are classified as either vertical or horizontal coupling.

In vertical coupling, the passive InGaAs waveguide layer is grown below the QCL active region, as shown in Fig. 2(a-g), eliminating the need for additional epitaxial growth [13, 23, 24]. Light transfer from the laser gain section to the passive waveguide is achieved via an adiabatic taper, which maximizes power transfer by controlling the mode interaction length near the effective index matching of laser and waveguide modes [13]. This approach has demonstrated coupling efficiencies exceeding 90% in simulations and over 80% in experiments [24].

In horizontal coupling, a selective regrowth technique is employed, where a portion of the QCL active region is etched and replaced with a passive InGaAs layer. The core and cladding layer thicknesses are carefully designed to optimize the coupling efficiency, which depends on the effective index difference and mode overlap between the laser and passive waveguide. The active and passive sections are then enclosed by Fe-doped InP cladding, forming a buried-heterostructure (BH) waveguide. This structure reduces propagation loss, enhances heat dissipation in the active section, and provides dispersion control in the passive section.

An alternative horizontal coupling approach utilizes the QCL active region itself as the waveguide core. To suppress free-carrier absorption, ion implantation is used to deplete electrons within the active region, significantly reducing the optical loss. For a waveguide designed at 9.6 μm, this method reduced the loss from 51.6 dB/cm to 1.3 dB/cm, more than an order of magnitude improvement. Similarly, a 4.8 μm QCL PIC using this technique achieved a propagation loss of 3.14 dB/cm.

Each coupling approach presents unique trade-offs. Vertical coupling simplifies the fabrication by requiring only a single epitaxial growth run, and follows conventional QCL processing steps. However, the integration of an adiabatic taper introduces additional design complexity. Horizontal coupling with selective regrowth offers greater design flexibility but requires intricate fabrication steps, including regrowth for the passive core and overgrowth for the BH structure. These processes are particularly challenging for large-scale PICs with numerous components of varying geometries. While overgrowth can be omitted, this necessitates the use of a taper at the butt-joint interface, as the ridge-waveguide gain section is typically several microns wider than the passive waveguide. The passivated active region approach offers a relatively simpler fabrication process compared to selective regrowth. However, it limits waveguide design flexibility and shows relatively higher propagation loss than other approaches.

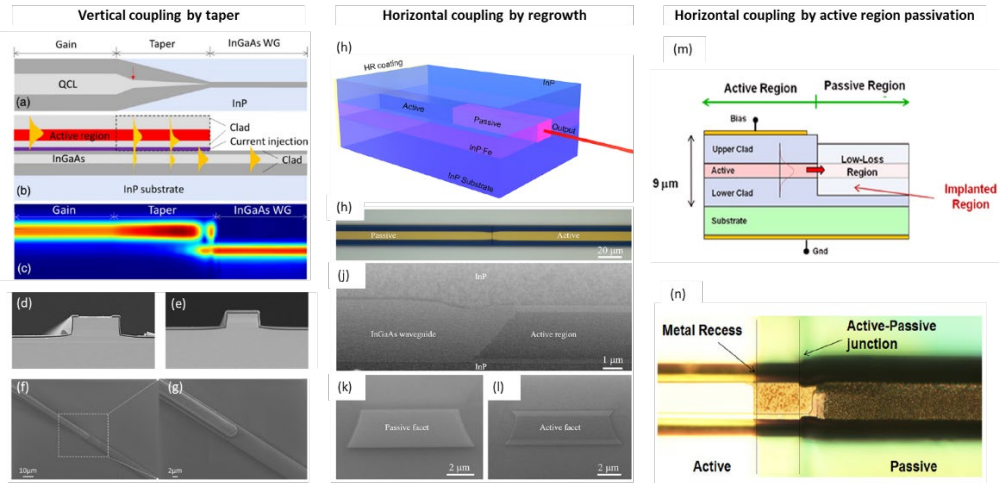

Fig. 2. Active/passive coupling approaches: (a-g) vertical coupling by the adiabatic taper [13], (h-l) horizontal coupling by a selected regrowth technique [14], and (m-n) horizontal coupling by active region passivation [21].

### 3) Mid-IR PIC applications

We present mid-IR PICs developed using the mentioned integration platforms. One example is a single-mode tunable PIC based on the taper coupling approach, as shown in Fig. 3 [24]. This device comprises a QCL gain section, an adiabatic taper, and a passive waveguide. Distributed Bragg reflector (DBR) gratings are implemented atop the waveguide to enable longitudinal mode selection, while a side metal heater adjacent to the grating facilitates wavelength tuning via the thermo-optic effect. The device demonstrated a wavelength tuning range of 7.5 cm$^{-1}$ by applying DC current to the heater, with exceptional power stability, exhibiting only a 10% variation in output power across the tuning range. The integration of a low-loss waveguide allowed for a wide separation (~2 mm) between the gain section and heater, effectively minimizing thermal crosstalk and preserving laser performance.

Another notable demonstration is a monolithic QCL-based wavelength beam combining device, which utilizes horizontal coupling with the passivated active region [15, 25]. This device integrates arrayed waveguide gratings (AWGs) coupled to five QCLs, producing a single-mode emission from a single output port spanning over 60 cm^-1 near the center wavelength of 2040 cm⁻¹ (=4.9 μm). Compared to conventional waveguide beam combining systems that rely on extensive free-space optics, this integrated approach achieves a size reduction exceeding two orders of magnitude, significantly enhancing compactness and scalability.

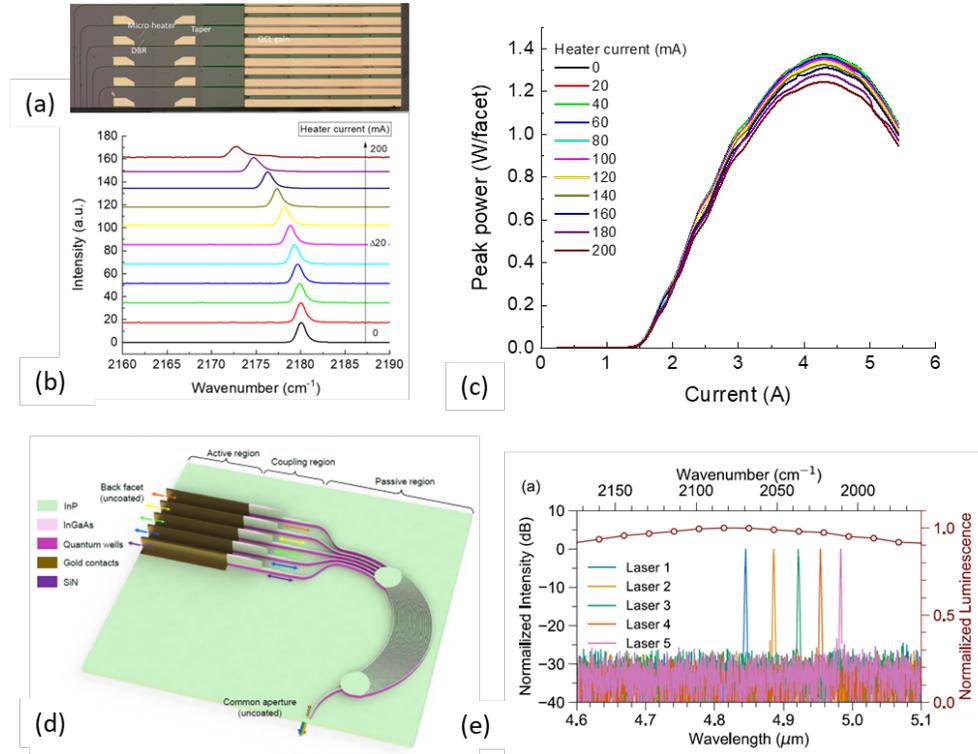

Fig. 3. (a-c) Single mode tunable QCL-PICs consisting of the QCL gain section, taper, DBR, and micro heater [24]. (a) Optical microscope image of the device, (b) lasing spectra measured at different heater currents while the gain section was biased at a fixed current, and (c) output power measured under pulsed current with the corresponding heater currents in (b). (d, e) Monolithic wavelength beam combining QCL-PICs. (d) Schematic of the device consisting of the QCL gain section, taper, and arrayed waveguide grating and (e) output spectra measured from the common aperture under pulsed condition while independently biasing the QCL gain ridges [15, 25].

We also highlight key applications demonstrated on an InP-based passive platform, which holds potential for future monolithic integration. Figures 4(a–d) present an optical phased array (OPA) device for monolithic beam steering at 4.6 μm [26]. OPAs typically require essential building blocks, including beam splitters, phase shifters, and surface gratings. In this work, the device comprises 32 waveguide channels with a channel spacing of 11.5 μm (2.5λ). The waveguide structure consists of a 1.7-μm-thick InGaAs core, a 2.2-μm-thick InP upper cladding layer, and a 4-μm waveguide width. Phase shifters were implemented using metal contacts positioned atop the waveguide, enabling precise modulation control of the phase. A single-mode QCL operating at 4.6 μm was externally coupled to the OPA input. Lateral beam steering was achieved by introducing an incremental phase shift across the waveguide array, demonstrating a total beam steering range of ±11.5°.

Recently, mid-IR supercontinuum generation was demonstrated using GaAsSb/InP waveguides [20], as shown in Fig. 4(e–h). This system achieved nearly one octave of spectral broadening, from 3.5 μm to 6.5 μm, at a pump peak power of 44 W, which is an order of magnitude lower than in previously reported demonstrations. This improvement was attributed to the high Kerr nonlinearity, low group velocity dispersion, and the relatively low propagation loss of GaAsSb waveguides [27], making them a promising platform for nonlinear mid-IR photonics.

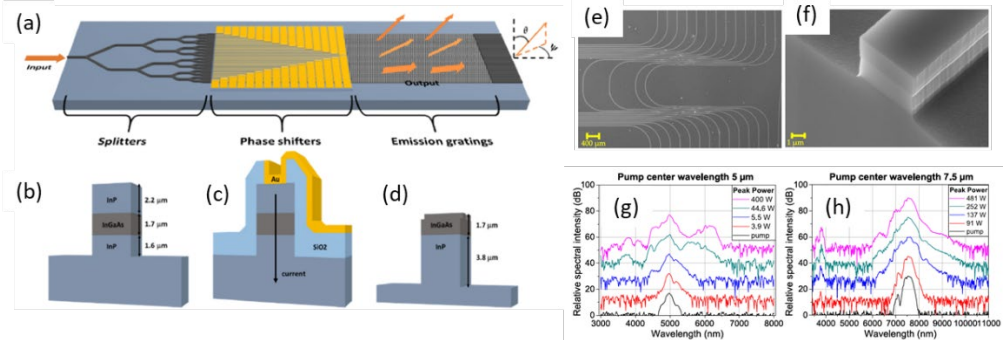

Fig. 4. (a-d) InGaAs/InP based OPA for beam steering [26]. (a) Schematic of the OPA including sections of splitters, phase shifters, and emission gratings, (b-d) cross-sectional schematics of each section. (e-h) GaAsSb/InP based waveguide for supercontinuum generation [20]. (e, f) scanning electron microscope images of the waveguides, and (g, h) experimental results of supercontinuum generation with pumping at 5 μm and 7.5 μm.

## Challenges and opportunities

The field of mid-IR active/passive monolithic photonic integration capable of integrating multiple building blocks with low loss and high efficiency is still in its infancy and faces many of the same challenges as general mid-IR photonics compared to its near-IR counterpart. These challenges include the lack of commercially dominant applications, lower market demand, complex material growth and fabrication processes, and high development costs. Consequently, active/passive integration requires significant investment, limiting research and development activities in this area and resulting in a lack of a comprehensive building block database.

For example, there have been no monolithic integration efforts for photodetectors, semiconductor optical amplifiers (SOAs), high-speed modulators, or Mach-Zehnder interferometers. Additionally, systematic performance databases for passive building blocks, such as multi-mode interferometers, directional couplers, and output couplers, are virtually nonexistent, in stark contrast to the extensive databases available for near-IR silicon photonics.

Furthermore, integrating active and passive subsystems, such as the combination of QCL and OPA discussed in the previous section, is far from a straightforward plug-and-play process. Unlike near-IR photonics, where mature integration strategies exist, mid-IR integration must overcome significant engineering challenges. Achieving seamless compatibility between components demands innovations in material interfaces, fabrication techniques, and system architecture to balance performance trade-offs. As a result, developing robust integration platforms is not only necessary but also a key enabler for the future growth of mid-IR photonic systems.

## Future developments to address challenges

Addressing the challenges requires advancements in the optimization of the existing integration platforms and exploration of new integration platforms or architectures. For example, each of the mentioned QCL-PIC platforms has significant room to improve the waveguide loss, coupling efficiency, fabrication quality, PIC building block functionality, and integration density by exploring various design parameters of the integration platforms. Also, our observation finds that the QCL active regions are likely designed for high power operation, which does not fully explore the capability of the integration approaches and PIC applications. Instead of pursuing high power, the QCL gain can be optimized for low power mid-IR applications [28, 29], which may in turn produce completely different design rules and integration architectures.

Expansion of integration should be explored for GaSb-based platforms to accomplish photonic integration of diode lasers and ICLs [30]. To the best of our knowledge, there has not been a published result of homogeneous active/passive integration of GaSb-based PICs other than demonstration of passive waveguide devices or heterogeneous integrations [31, 32]. Although GaSb-based material systems are relatively less mature than InP-based platforms, the interband nature of these devices can offer better opportunities to translate near-IR PIC technologies, leveraging the low loss of GaSb in the mid-wave band [33] and its potential use for various passive building blocks [34, 35].

In parallel, the development and characterization of fundamental photonic building blocks are essential to establish a comprehensive database akin to that available for the near-IR Si photonics. For QCL-PICs, active photonic elements such as QCDs, SOAs, high-speed modulators, and Mach-Zehnder interferometers should be demonstrated in the compatible integration platform. Systematic efforts to design, fabricate, and evaluate these elements will provide the necessary foundation for future advancements.

Standardization and scalability will further play a pivotal role in fostering widespread adoption. Establishing standardized design frameworks, fabrication processes, and packaging solutions [36] will not only reduce development costs but also enhance reliability and manufacturability. One of the efforts could be the establishment of a mid-IR process design kit (PDK) that is currently not available yet an important step to enable development transition from subsystem to system.

The successful development of mid-IR photonic technologies also depends on identifying and targeting commercially viable applications. Current PIC demonstrations have been limited in scope to enhancing the performance of the light source. Demonstration of sensors for environmental monitoring and medical diagnosis, for example, by incorporating a laser, a detector, and a sensing element with capability of volume production can drive investment, accelerate research and development efforts, and establish robust mid-IR PIC ecosystem.

## Concluding Remarks

Mid-IR monolithic photonic integration, despite being in its early stages, has demonstrated its technical viability and potential to bridge the gap between discrete component integration and fully monolithic platforms. The successful realization of low-loss waveguide platforms and their integration with laser sources marks a critical step toward high-density mid-IR PICs.

These advancements not only validate the feasibility of mid-IR integration but also highlight its transformative potential, unveiling new research directions and underexplored opportunities. As progress continues, the field is poised to expand, driving high-impact research and enabling the development of advanced PICs with both scientific and practical significance. With further innovation in materials, fabrication techniques, and system architectures, mid-IR photonic integration is set to play a crucial role in the future of photonic technologies.

## Back matter

*Funding*

N/A

*Acknowledgments*

N/A

*Disclosures*

The authors declare no conflicts of interest.

# 20. Metamaterials for Mid-Infrared Photonics

ANGELA VASANELLI,[1] BAPTISTE CHOMET,[1] AND CARLO SIRTORI[1,*]

[1]*Laboratoire de Physique de l'École normale supérieure, ENS, Université PSL, CNRS, Sorbonne Université, Université Paris Cité, 75005 Paris, France*
*carlo.sirtori@ens.fr*

## Overview

The tremendous progress of our capacity to fabricate with great accuracy nanometric design has allowed the realization of artificial structures, metamaterials, in which the electromagnetic properties are modified by inscribing in the host material features at a subwavelength scale. In this sense, metamaterials are a new generation of complex optical elements in which the main properties arise from interferential phenomena imposed by the change of subwavelength geometrical features rather than the modification of their chemical compositions. Metamaterials were historically developed in the mm and RF range, and afterwards realized in the optical domain. They allowed the demonstration of fascinating phenomena like cloaking and negative refraction.

The mid-infrared wavelength region, which lies between optics and electronics, can also greatly benefit from the introduction of metamaterials. Moreover, in this range, metamaterials can be realized not only using dielectrics but also by engineering hybrid dielectric/metallic structures. This is because the dielectric functions of metals, commonly used for contacts, induces much less optical loss than in the visible and near-infrared. Metamaterials can therefore be conceived by merging concepts from interference dielectric structures, similarly to photonic crystals, with microwave elements, such antenna, inductors, capacitors, ... Therefore, the metallic regions of mid-infrared photonic devices, embedded in metamaterials, will naturally provide electrical contacts as well as RF functions, in particular antenna.

Figure 1 gives a brief overview of some metamaterial concepts that have been developed in the mid-infrared domain. The main idea is the possibility of designing an artificial structure with engineered optical properties. Such artificial structures are ment to provide optical features beyond those of bulk semiconductors, where they are determined by the material chemical composition and crystal structure, which set the phonon Reststrahlen band (typically in the THz domain) and the energy gap (typically in the near-infrared or optical domain). In a metamaterial, designer optical properties are obtained by combining different dielectric and/or metallic layers with an adequate spatial configuration. Each panel of Figure 1 provides an example of metamaterial employed for mid-infrared photonics. An hyperbolic metamaterial can be obtained by alternating subwavelength layers of undoped and doped semiconductors, giving rise to negative refraction [1] at mid-infrared wavelengths. Subwavelength apertures on the facet of a quantum cascade laser constitute a grating for surface plasmons, allowing collimation of the emitted light [2]. Gratings of metallic antennas deposited on the surface of a semiconductor can be used to generate resonances at a wavelength set by the dimensions of the antenna, while their periodicity controls the coupling to free-space radiation [3]. A combination of lumped elements like inductors and capacitances concentrates the electric and magnetic fields in highly subwalength regions [4].

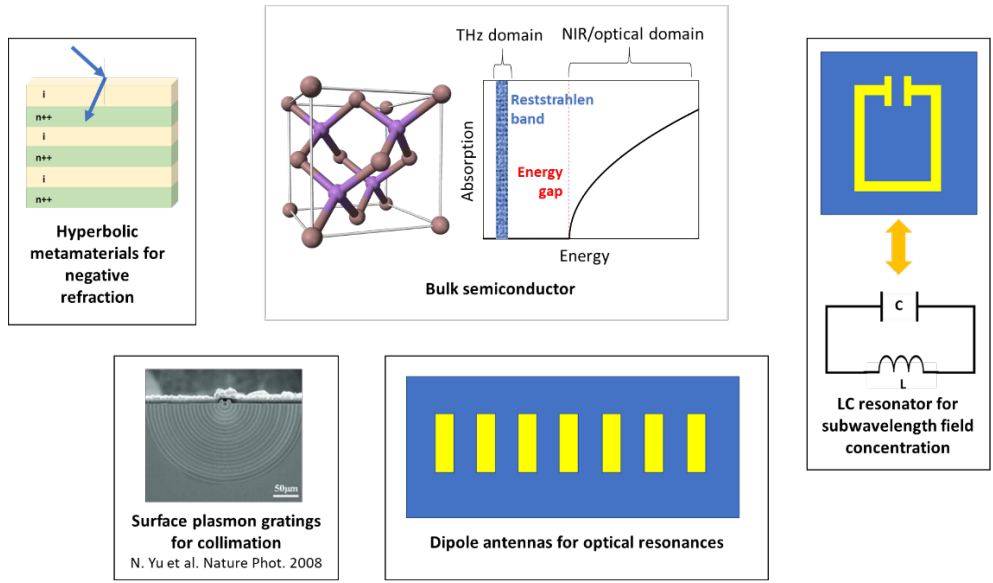

Figure 32. **Examples of metamaterials**. While in a bulk semiconductor (top central panel) optical properties are determined by chemical composition, in a metamaterial they are set by a combination of metallic or dielectric subwavelength structures.

The main role of metamaterials in mid-infrared photonics can be summarized in two main phenomena: a) concentrating mid-infrared radiation into volumes of active material (detectors, modulators, gain region) with highly subwavelength dimensions on the order of nanometers; b) controlling the spectral and spatial properties of the emitted light.

In the following, we summarize the state of the art and provide our vision of future developments in metamaterials for mid-infrared photonics.

## Current status

### 18. Metamaterials for improved detection

The use of metamaterials for mid-infrared detectors is motivated by their ability to focus all the incoming radiation into a deep subwavelength active material volume that forms a microcavity at a given wavelength. As a consequence, light-matter interaction is enhanced, and by properly tailoring the metamaterial geometrical parameters the rate of absorption can match that of the optical insertion. In this case all the energy injected into the microcavity is dissipated by the detector, and the critical coupling condition can be validated by measuring the near-perfect absorptivity. Furthermore, the signal-to-noise ratio can be improved because in metamaterial-based devices the antenna effect extends the detector's photon collection area, making it much larger than the electrical area where dark current is generated. Thus, for the same number of collected photons, there is much lower dark current than when the areas match.

A well-suited metamaterial geometry for detection is based on arrays of metal-semiconductor-metal resonators, where the dielectric part contains the absorbing material for photodetection, *e.g.*, quantum wells [5], [6], [7], type II superlattices [8], [9], or colloidal nanocrystals [10], [11]. Figure 2 presents some examples of metamaterial detectors. Devices are based on arrays of resonators, characterized by a highly subwavelength thickness. For p-polarized radiation, the size of the resonator, *s*, sets the wavelength of the confined radiation, while the period *p* sets the coupling with free space radiation. The coupling can be engineered to achieve almost perfect absorption in the semiconductor layer.

In the case of unipolar detectors, like quantum well infrared photodetectors or quantum cascade detectors that are characterised by excited state lifetimes on the order of ps, metamaterials provide high frequency bandwidth by reducing the parasitic capacitance associated to the size of the electrical area [12], [13].

Metamaterials can also provide existing devices with novel functionalities. As an example, metamaterials can tailor the spectral and polarization response of microbolometers thanks to the integration of laterally structured metal-dielectric–metal resonators onto conventional suspended membranes [14] [15]. The same concept is employed to tailor the optical response of pyroelectric [16] or photothermoelectric detectors [17]. Quantum cascade detectors with dual wavelength operation enhanced by metamaterials have also been reported [18].

In the peculiar case of nanocrystal-based photodetectors, the voltage dependence of the carrier mobility in the structure, combined with inhomogeneous spatial absorption induced by the metamaterial, gives rise to spectral reconfigurability [19].

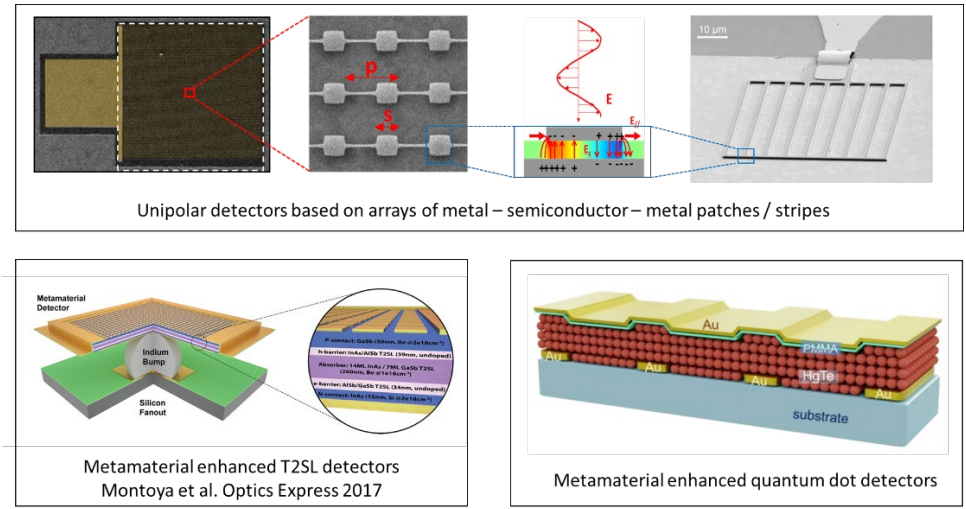

Figure 33. **Metamaterial detectors**. a | Subwavelength metal-semiconductor-metal patches in array configuration exploit antenna to couple the electromagnetic waves into a microcavity mode to realize high-performance mid-infrared detectors at room temperature. b| Despite the thin absorbing material, type-II superlattice infrared detectors for long-wave infrared (8-12 μm) show high responsivity due to the coupling between the Fabry-Perot cavity formed by the semiconductor layer and the resonant nanoantennas on its surface. *[20]* c | A nanocrystal film is coupled to a grating and a top metallic layer to achieve broadband enhancement of short and mid-wave infrared (4-6 μm) absorption *[11]*.

## 19. Metamaterials for controlled emission

Metamaterials can be combined with mid-infrared emitters to provide spectral and shape control of the emitted beam.

An example of spectral control by metamaterials is provided by thermal emitters, where, following the seminal work by Greffet and coauthors [21], metamaterials are used to realize perfect absorbers that intrinsically behave as thermal emitters based on Kirchhoff's law of thermal emission [22], [23], [24], [25], [26].

While first works on this topic involved metallic metasurfaces to control light emission, an alternative approach has pursued the control of incandescence by exploiting so-called hyperbolic metamaterials. [27] These artificial materials are highly anisotropic media with hyperbolic dispersion, characterized by the fact that one of the principal components of their permittivity tensor is opposite in sign to the other two principal components. Hyperbolic metamaterials can be implemented in the mid-infrared by exploiting plasmons in highly-doped semiconductors [1], intersubband transitions in quantum wells [28] or phonon polariton modes [29]. Hyperbolic metamaterials allowed the realization of directional and quasi-monochromatic thermal emitters [30], [31].

Metamaterials play an important role in solving the problem of beam divergence of semiconductor lasers, and in particular quantum cascade lasers [discussed in a separate sub-topic article of this Roadmap], which are the most important semiconductor lasers in the mid-infrared range. Mid-infrared quantum cascade lasers typically employ a dielectric waveguide structure in which the active region is embedded between semiconductor layers of lower refractive index. As the active region is only a few microns thick, quantum cascade lasers suffer large beam divergence induced by diffraction. They thus need collimation in order to be implemented into sensing, free-space communications or ranging systems. Plasmonic collimation [2], first introduced by Capasso's group, solves the divergence problem by exploiting designer surface plasmons. The idea is to pattern a slit aperture and a grating directly on the metal-coated laser facet. The aperture couples part of the laser emission into the surface plasmon propagating along the laser facet. The plasmonic grating scatters the emitted energy into the far field, breaking the diffraction limit set by the emission area. The same concept was subsequently refined by implementing two-dimensional and metasurface collimators [32], [33].

## 20. Metamaterials for non-linear effects

The interest of metamaterials and metasurfaces for non-linear optics stems from the possibility of enhancing the light-matter interaction in small volumes. Consequently, while large samples and complex phase matching techniques are normally employed to observe non-linear effects in natural materials, metamaterials induce a giant non-linear response, especially when combined with quantum engineered semiconductor quantum wells [34]. These provide some of the largest $\chi^2$ factors available in condensed matter systems.

The structures suited for non-linear effects must lack inversion symmetry. For this reason, Lee and coauthors [35] realized a metamaterial based on a metal – semiconductor - metal structure, where the semiconductor region contains doped asymmetric tunnel-coupled quantum wells that display a large non-linear coefficient, while the top surface is a periodic arrangement of L-shaped metallic structures. They demonstrated second harmonic generation starting from a pump at 8 µm. The same concept, but with a different design of the top metasurface and of the tunnel-coupled quantum wells, demonstrated third harmonic generation [36] with a 9 µm pump.

Other strategies involve all-dielectric metasurfaces, based on a high index dielectric guiding layer with a dielectric grating on its upper surface, and where the guiding layer is fabricated on top of an engineered multi-quantum-well system [37]. More recently, inverse design has been employed to optimize the dielectric metasurface [38].

Finally, metasurfaces based on a phase change material induced efficient third harmonic generation, which upconverted mid-infrared radiation at 4.6 µm wavelength to near-infrared at 1.55 µm [39].

## 21. Metamaterials for controlled light-matter interaction

The interaction between light and matter is determined by the Rabi frequency $\Omega_R$, which is proportional to the ratio between the oscillator strength, $f$, of the matter excitation and the volume of the light-matter interaction, $V$: $\Omega_R \propto \sqrt{f/V}$. When the Rabi frequency is lower than the losses, the system is in the weak coupling regime. In this regime, the interaction with the cavity mode can modify the typical time of the light-matter interaction, giving rise to an accelerated spontaneous emission phenomenon known as Purcell effect.

When the Rabi frequency is greater than the losses, the system is in the so-called strong light-matter coupling regime. In this regime the system's states are renormalized by the interaction with the cavity mode, giving rise to new eigenstates, the cavity polaritons.

Finally, when the Rabi energy is comparable to the matter excitation energy, the system enters the ultra-strong light-matter coupling regime, in which anti-resonant and quadratic terms of the light-matter interaction cannot be neglected.

These three regimes have been the object of thorough investigations, not only for fundamental reasons, but also for the novel degrees of freedom they offer for the design of photonic components. Metamaterials are very interesting in this context, as they can provide low-volume modes that bring the system into the strong or ultra-strong coupling regime. Indeed, metallic or dielectric metamaterial modes have been strongly coupled with mid-infrared matter excitations, like intersubband transitions in quantum wells [40], [41], [42], [43], [44], Berreman modes in highly-doped semiconductor layers [45], [46], [47], phonons, or vibrational modes of molecules [48].

Another possible application of metamaterials to the control of the light-matter interaction is generating near field enhancement at molecular fingerprint frequencies. This effect is very interesting for surface spectroscopies like surface-enhanced Raman spectroscopy (SERS) and surface-enhanced infrared absorption (SEIRA) [49]. The general idea is to provide sharp lines, issued from plasmonic nanoantennas or assemblies of gold nanoparticles [50], [51], at specific frequencies that correspond to molecular vibrational modes. The role of the metamaterial is to provide a selective amplification of vibrational signatures of molecules creating hotspots for the electromagnetic field. This allows sensing at lower concentrations than traditional infrared spectroscopy techniques. The frequency and the field enhancement factor are controlled by the geometrical characteristics of the plasmonic material. A SEIRA enhancement factor up to $10^6$ can be achieved by properly designing the metamaterial. Other approaches to improve SEIRA consist in designing metamaterials with a wideband response, for example by using all-dielectric metasurfaces with angle dependent response [52], or designing metamaterials in the over-coupling regime [53].

## Challenges and opportunities

Up to now, metamaterials have been used mostly as *additional elements* to enhance the performance and characteristic of mid-infrared photonic devices. Essentially, they have acted on spectral selectivity, or on funneling mid-infrared radiation into nanometer-scale active volumes. However, metamaterials can also be *active elements* to provide novel functionalities for mid-infrared photonics. In particular, they could provide currently-unavailable building blocks for photonics, like amplitude and phase modulators, reconfigurable filters, or beam shapers. One way to make the metamaterials active is by providing them with metallic contacts that can be biased. The applied voltage acts on the electronic properties of embedded semiconductor nanostructures, which in turn modify the optical properties of the metamaterial.

A different approach for active metamaterials is based on merging them with other technologies like MEMS [54].

As an example of this concept, let us consider the case of amplitude modulators [55], [56], [57]. Figure 3 presents a sketch of a metal – semiconductor – metal metamaterial that contains a set of doped asymmetric tunnel-coupled quantum wells in the semiconductor region. When a voltage is applied to this structure, a linear Stark effect is observed [34]: the energy of the intraband transition from the ground to the first excited state, $E_{12}$, linearly depends on the bias. This effect is due to the structure's asymmetry, and the fact that the electronic distributions of the ground and first excited electronic states have different barycenter. Due to the linear Stark effect, the metamaterial becomes *active*: its absorptivity, and consequently its reflectivity, depend on the bias applied to the structure. Figure 3 presents two typical spectra, calculated at different biases, in the case when the 1→2 electronic transition is strongly coupled with the resonator mode, giving rise to polariton states. At a fixed photon energy, provided by a laser impinging on the active metamaterial structure, the reflectivity changes with the applied bias and modifies the laser intensity. A device based on this concept was recently employed, in combination with a metamaterial mid-infrared detector, to demonstrate free-space communication in one of the transparency windows of the atmosphere, with a bit rate close to 70 Gbit/s. [57] The same kind of active metamaterial, exploiting linear Stark effect and metal-semiconductor-metal antenna-coupled resonators, can also provide electrical control of the beam, diffraction, and polarization and electrical beam steering [58].

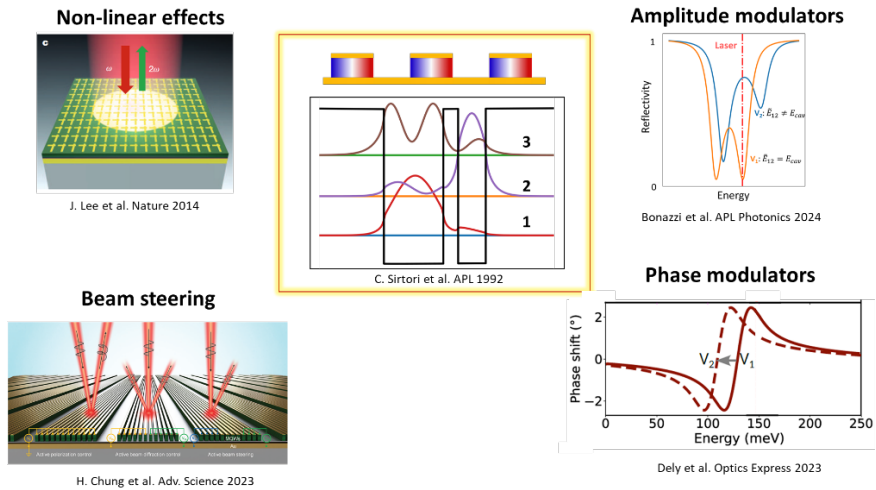

Figure 34 **Active metamaterials.** The ability to tune the material's properties by applying an external bias, together with the strong light-matter interaction provided by the cavity, modifies the optical properties of the metamaterials. Depicted at the center, an electronic potential leads to a linear Stark shift of a mid-infrared intraband transition, which is induced by a bias applied to a metal-semiconductor-metal metamaterial. This concept is used to build amplitude and phase modulators and beam shapers, and to exploit non-linear effects for mid-to-near infrared frequency conversion.

Asymmetric tunnel-coupled quantum wells, combined with metal-dielectric-metal resonators, are also extremely efficient devices for exploiting non-linear effects and particularly for mid-infrared to near infrared frequency conversion. This is a very interesting technique for transferring mid-infrared wavelengths into the realm of near infrared where telecom technologies allow extremely sensitive measurements of phase and amplitude. It may provide an efficient route to extending the frequency range of quantum optics, and open the path to single photon detection for sensing.

For room temperature detectors, 5 orders of magnitude in detectivity separate the 1.5 µm and 4.5 µm technologies. Recent work published by Belkin [33] shows that the 3rd-harmonic generation efficiency can reach $10^{-6}$ in metamaterials. Given these conversion efficiencies, and considering that commercial QCLs can deliver 1 W at 4-5 µm wavelength, non-linear techniques can be considered for high-sensitive detection of mid-infrared signals by exploiting the highly-developed telecom technology.

Another approach to highly sensitive detection consists in exploiting metamaterial unipolar detectors as coherent receivers in a heterodyne set-up [59]. In this configuration, the maximum achievable sensitivity is set by generation – recombination noise originating from dark current. Considering unity heterodyne efficiency, the minimum achievable noise equivalent power is $NEP = E\,\Delta f/\alpha$, where $\Delta f$ is the integration bandwidth, $E$ is the photon energy and $\alpha$ the absorption quantum efficiency. At 9 µm wavelength, the NEP associated with the detection of one photon per second corresponds to 0.02 aW. This is ultimately limited by the absorption quantum efficiency of the detector and by the resolution bandwidth. The net advantage of using a metamaterial is to provide a maximum absorption quantum efficiency, $\alpha$, close to 1.

A final opportunity is the use of mid-infrared metamaterials to enhance emission in the transparency windows of the atmosphere from hot surfaces. This approach may provide the basis for radiative coolers for buildings, on which a large surface of the roof is covered by the metamaterial. The large amount of power received from solar radiation can then be emitted as electromagnetic waves, thus reducing the temperature of the building [60].

**Future developments to address challenges**

In the previous sections we have described metamaterials rising from highly mature technological platforms, well known in the field of optoelectronics. However, recent years have witnessed the appearance of novel materials that display mid-infrared properties. Their integration with conventional III-V and group IV semiconductors gives an excellent opportunity to engineer metamaterials that combine intrinsic material properties with optical phenomena by design.

Photonics based on Silicon and Germanium has had a major impact at telecom wavelengths. The extremely low losses and the possibility of mass production are paradigm changers for photonic integration and hybridization. The mid-infrared cannot stand alone without profiting from this technological breakthrough to realize metamaterials and flat optical components that benefit from the reliable and high-volume fabrication technologies already developed for microelectronic integrated circuits [61]. Intrinsic silicon waveguides face challenges for wavelengths beyond 6 µm, due to the absorption of phonon replicas. However, Germanium, which has transparency up to 15 µm wavelength, is an ideal material for the demonstration of long-wave infrared SiGe photonic structures for metamaterials. Furthermore, Germanium also shows strong third order non-linearities that could be exploited in non-linear frequency conversion. A major development for mid-infrared technologies is the combination of the metamaterial concepts presented in the previous sections, and SiGe photonics that provides extremely high quality factors, up to $10^5$ [62]. Finally, the excellent transparency and the high refractive index of Germanium are also an asset for the development of metalenses [63]. A metalens is a particular type of metasurface engineered to reproduce a phase profile where, as for standard lenses, all rays converge at the same focal point. This is achieved by etching subwavelength elements with different filling factors and arranging them spatially, so that the effective refractive index creates the same phase distribution as a conventional curved lens.

The implementation of electro-optic functions for microwave optics and signal treatment has been improved recently by combining Silicon photonics with materials exhibiting strong RF non-linearities, such as $LiNbO_3$ [64] and GaP [65]. These materials have a promising potential for phase control at mid-infrared wavelengths, particularly if integrated with metamaterials into photonic crystals. Notably, LiNbO$_3$ is very well suited for wavelengths shorter than 4 µm, while GaP is transparent even above 10 µm.

Beyond conventional semiconductor heterostructures, 2D materials such as graphene and black phosphorous have been implemented into metamaterials to strongly enhance the device performance, in particular for mid-infrared detection [66], [67], [68] and non-linear conversion [69]. The ongoing improvement of 2D material growth and processing, together with the possibility of fabricating heterostructures, will allow the realization of novel metamaterial architectures based on planar lumped elements obtained by metal evaporation. These will be more compatible with the physics of two-dimensional systems, but are currently limited to THz devices [70].

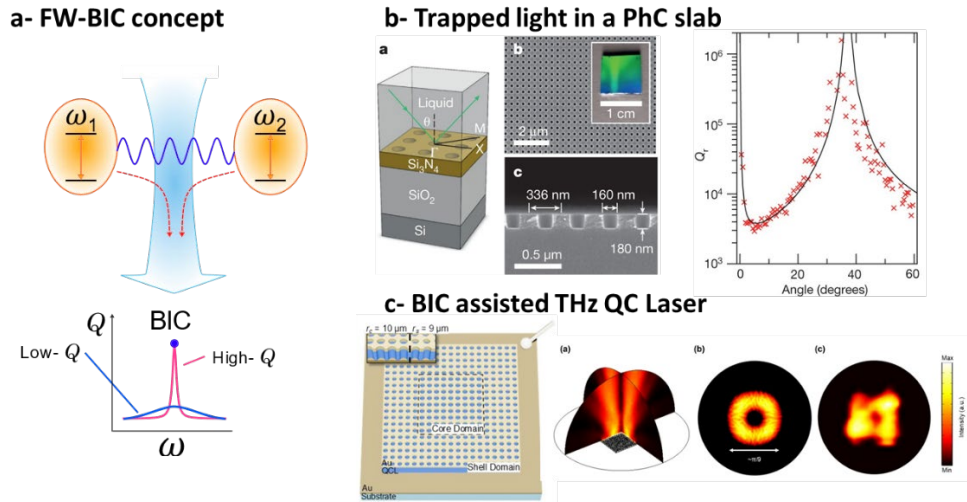

Figure 35 **High quality factor metamaterials.** a- Schematic illustration of the Friedrich-Wintgen bound state in the continuum (FW-BIC). Two near degenerate resonances can be dissipatively coupled if both share a common decay channel. As a result, two supermodes are formed: one is a nondecaying bound state (high Q) while the other one is an increased-decaying state (low Q). b – Realization of an infinite Q-factor BIC supported by a Si$_3$N$_4$ photonic crystal (PhC) slab with a square array of cylindrical holes [71]. c- Electrically pumped core–shell structure BIC photonic laser. The QCL active region is sandwiched between the top and bottom Au layers which act as electrodes. The far-field emission of the laser shows a doughnut shape as the BIC does not radiate [72].

A big challenge for mid-infrared photonics is the achievement of optical modes with high quality factors for the realization of lasers and high-efficiency electroluminescent emitters. The major issue in designing such metamaterial emitters is the reduction of Joule losses induced by metallic contacts and antennae. To address this, the concept of bound states in the continuum (BIC) [73] is particularly promising. These states arise from interactions between trapped electromagnetic modes, which lead to scattering resonances with almost infinite lifetimes [74]. Figure 4a presents a schematic illustration of the emergence of a BIC state from two near degenerate modes sharing a common decay channel. The existence of these high-quality-factor (of order 10$^3$-10$^4$) modes has been demonstrated mostly in dielectric structures, as shown in Figure 4b. However, a few demonstrations in the THz domain already exist in structures containing metals that act as antenna [75], [72] (see also Figure 4c). Similar concepts can be transferred to the mid-infrared by scaling the dimensions and revisiting the coupling between the optical modes that give rise to the interference phenomena that control radiative losses. This will make it possible to reduce the active volume of a metamaterial laser, while strongly reducing the threshold. Moreover, the distributed optical mode of the metamaterial will give rise to modes with controlled spatial profile. Finally, high quality factors will allow also greatly increasing the electroluminescence intensity, thanks to the Purcell effect, resulting in ideal mid-infrared light-emitting devices for spectroscopy.

A further application of high-quality-factor metamaterials can be envisioned in the field of strong light-matter coupling. Indeed, such a structure could answer two major challenges that have existed in the field of polaritonic physics for many years: the realization of mid-infrared lasers based on Bose Einstein condensation [76], [77], [78], and the control of chemical reactions by modifying the ground state through coupling with vacuum fluctuations [79].

## Concluding Remarks

The level of maturity currently reached by metamaterials already allows their implementation in photonic devices for several applications, particularly high sensitivity detectors and more recently low-power-dissipation modulators. We now expect the near future to witness metamaterial light emitters and lasers that will transform the landscape of mid-infrared optoelectronics.

## Back matter

### Funding

This work was supported by ENS-Thales Chair, the ANR project IMPALA and PEPR Electronique.

### Disclosures

The authors declare no conflicts of interest.

# 21. Spectroscopic Chemical Sensing

NATHAN P. LI[1] AND MARK A. ZONDLO[1]

[1]*Princeton University, Department of Civil and Environmental Engineering, Princeton, NJ 08540, USA*

## Current status

Semiconductor photonics have enabled widespread techniques for chemical sensing. Mid-infrared (IR) light (3-14 μm) has frequencies that match the energy levels of the vibrational transitions of most chemicals, superimposed with fine rotational transitions. These transitions allow for highly specific identification and quantification of molecules (Fig. 1) [1-5].

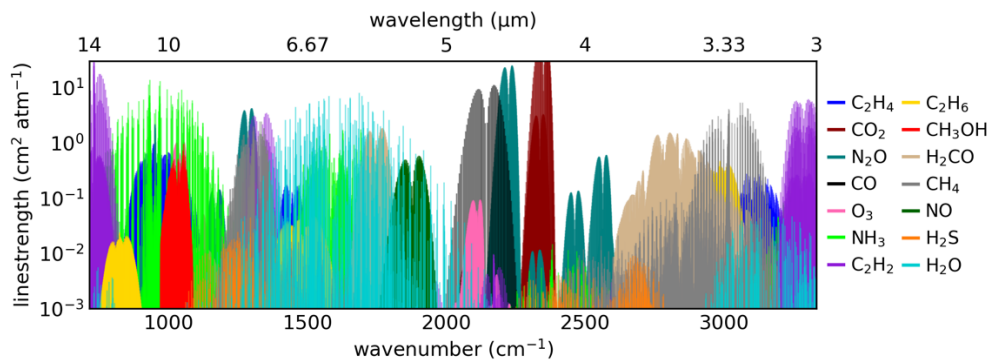

Fig. 1. Examples of mid-IR spectral absorption features (i.e., fingerprints) for chemical species for pure mole fractions at 296 K and 1 atm (data from HITRAN [6] and spectraplot.com).

Every photonic chemical sensing system has three components: 1) a light source, 2) an optical path, and 3) a light detector. The fundamental principle behind photonic chemical sensing is that certain characteristics of light (e.g., intensity, phase) change with the amount of matter with which light interacts in the optical path. These changes can be detected using any of a wide array of spectroscopic techniques [1-5]. It is this fundamental principle—and unique spectral fingerprints (Fig. 1)—that allow chemicals to be sensed using photonics. Eq. (1) is the governing equation for light absorption, the Beer-Lambert-Bouguer law, which states that absorption increases exponentially with the concentration of the target chemical species.

$$I = I_0 e^{-\varepsilon c \ell} \tag{1}$$

Here, $I$ is the amount of light transmitted to the detector, $I_0$ is the amount of light emitted by the source, $\varepsilon$ is the absorption coefficient—dependent on the strength of the targeted absorption feature(s) (Fig. 1), $c$ is the concentration of the target chemical species, and $\ell$ is the optical path length (i.e., the distance that the light travels through the sample assuming a uniform concentration along the path).

Semiconductor materials have been integral in the advancement of chemical sensing and are used in a large fraction of light sources and detectors [7-9]. The invention of the laser—notably quantum cascade lasers (QCLs) and interband cascade lasers (ICLs) [discussed in separate sub-topic articles of this Roadmap]—was a breakthrough for mid-IR semiconductor photonics and chemical sensing. Modern lasers have extremely narrow spectral widths (~$10^{-13}$ m) and can be manufactured at practically any mid-IR wavelength [7, 8], resulting in highly specific light sources that can target any spectral feature (Fig. 1). Lasers have enabled the detection of concentrations at parts-per-trillion (ppt) in many applications [2, 5]. Non-laser (i.e., incoherent broadband) sources have also been important for techniques such as Fourier-transform and non-dispersive IR spectroscopy (FTIR, NDIR) [2-4].

In this sub-topic article, we discuss 1) the current status, 2) challenges and opportunities, and 3) future developments for mid-IR spectroscopic chemical sensing. Each of these three sections contains four subsections: i) light sources, ii) detectors, iii) optical path components, and iv) systems and applications. Our review contains generally applicable material for chemical sensing but focuses in particular on commercially available technologies for trace-gas absorption spectroscopy and environmental applications.

*Light sources*

The wavelength (λ) of a QCL can be tuned by adjusting the layer thicknesses in the engineered quantum-well structure [7]. This is an advantage over the original lasers—invented in the 1960s—which were limited to the natural properties of the semiconductor materials themselves [7, 10]. When QCLs were invented about 30 years ago, they were only available at limited wavelengths and had to be cooled to 77 K for operation [7]. Today, QCLs are widely commercially available, operate in continuous wave (cw) mode at room temperature, and can be made to order at practically any wavelength spanning the mid-IR to THz frequencies [7].

Interband cascade lasers (ICLs) are hybrids between conventional diode lasers and QCLs. ICLs consume less power than QCLs and have been used in numerous low size, weight, power, and cost (SWaP-C) applications [8]. ICLs can also be designed for a myriad of wavelengths, but are not as widely commercially available as QCLs and are typically limited to λ < 7 μm [8]. Table 1 shows the specifications of commercially available "off-the-shelf" (COTS) QCLs and ICLs in 2025. Table 1 does not show state-of-the-art specs, which are discussed in the other sub-topic articles.

**Table 1. Status of COTS QCL and ICL specifications**

|  | QCLs[a] | ICLs[a] |
|---|---|---|
| size | 8×6×4 mm³[b], 32×32×18 mm³[c] | 8×6×4 mm³[b], 32×32×18 mm³[c] |
| weight | 1 to 500 g | 1 to 500 g |
| power (input) | 2 to 10 W | 0.3 to 0.8 W |
| cost[d] | $5k to $12k USD | $5k to $12k USD |
| power (optical) | 40 to 150 mW | 3 to 40 mW |
| wavelength range | 4.0 to 11.0 μm | 2.8 to 6.5 μm |
| max operating temp. | 50 ˚C | 50 ˚C |

[a]distributed feedback (DFB), cw operation, [b]C-mount, [c]HHL package, [d]price per unit

According to archived datasheets and webpages [11], COTS QCLs and ICLs have 1.2 to 10 times stronger output power than ten years ago. Their ranges have been extended by ≈1 μm to reach longer wavelengths. At shorter wavelengths (<4 μm), ICLs have become the preferred COTS option over QCLs due to their manufacturability and low input power. Maximum operating temperatures have increased by 10 ˚C. The number of advertised product options have increased by 2-5×, signaling increased demand and adoption. Otherwise, COTS lasers have not changed appreciably. Notably, SWaP-C has not changed significantly in the last ten years.

QCLs/ICLs have more recently demonstrated operation as frequency combs (FCs) [discussed in a separate sub-topic article], which enable novel techniques for chemical sensing [12, 13]. FCs emit spectra comprising hundreds to millions of equally spaced narrow spectral lines akin to "comb teeth." Not only do the FC "teeth" have narrow width on par with traditionally operated lasers, but they can also span broadband spectral ranges (hundreds of nanometers) that allow for multi-species detection. Since FCs combine the benefits of both QCLs/ICLs and broadband FTIR spectrometers, they have been useful for many diverse applications [12, 13]. A few notable examples are a miniature sub-watt FC for space applications [14], FCs with outdoor measurements for atmospheric science [15] and urban civil security [16], and a high-resolution (1 GHz) ultra-broadband (3-5 µm, 2000-3400 cm$^{-1}$) FC that analyzed kinetics for fast chemical reactions using mid-IR light converted from near-IR lasers [17]. COTS mid-IR FC spectrometers have been available for about ten years [13], but they are still relatively new—all things considered. Mid-IR FCs have not yet seen widespread commercial adoption despite their advantages [13], likely due to their immaturity and added complexities.

For incoherent broadband sources, novel COTS interband cascade light emitting devices (ICLEDs) [discussed in a separate sub-topic article] provide output powers (~mW) that are an order of magnitude higher than conventional mid-IR light-emitting diodes (LEDs) [8, 18], and on par with thermal IR emitters (i.e, incandescent sources) [2]. Recently, ICLEDs have been enhanced with resonant cavities (RC) that narrow their output spectra to ≈50 nm full-width-at-half-maximum (FWHM). RCICLEDs have greater selectivity than ICLEDs while still maintaining high output power [18, 19].

*Detectors*

Photodetectors convert light to electrical signals. HgCdTe (MCT) has been the standard semiconductor material for detectors [discussed in a separate sub-topic article] and has seen widespread use in low SWaP-C applications at wavelengths spanning 2 to 20 µm [9]. Other materials—notably InAsSb—have received attention because they cost less and do not contain mercury and cadmium—heavy metals that are toxic to the environment and human health [20]. The specific detectivity ($D^*$) (i.e., detection limit) for InAsSb is 1-2 orders of magnitude worse than MCT [20], but by substituting a type-II superlattice (T2SL) absorber—similar to the active quantum wells of ICLs—III-V materials, such as InAsSb, can compete with MCT even at longer wavelengths [20, 21]. On the other hand, MCT is still acknowledged to offer the best performance, versatility, and manufacturability [21]. Like COTS lasers, COTS detectors have not seen major changes in the last ten years. The main improvements have been increases in 1) $D^*$ (≈10× increase), 2) the number of products available to customers (≈3×), and 3) the availability of MCT alternatives (e.g., Ga-free T2SL InAsSb) [11].

Resonant cavity IR detectors (RCIDs) in the mid-IR [discussed in a separate sub-topic article] are a recent advancement. By constructing an RC around a thin absorber material, the spectral response of the detector can be narrowed to ≈20 nm FWHM [22]. Traditional detectors are broadband and have very wide spectral responses, spanning 1 to 10 µm [9, 20]. Wavelengths of light irrelevant to the spectral features of interest typically need to be filtered out. Furthermore, traditional filters are discrete components made of transparent crystalline materials with thin-film coatings. RCs enable filtering without discrete components and can be integrated during the semiconductor manufacturing process [23, 24]. Mid-IR RCIDs are not yet commercially available but have similar manufacturing processes to RCICLEDs and could therefore be manufactured easily using current commercial infrastructure.

Graphene and other 2D materials (e.g., black phosphorus) are important semiconductors that are not covered in this Roadmap. 2D materials differ from conventional 3D crystalline materials (e.g., MCT) in that they are only a single atom thick. Since 2D materials are so thin, they can potentially be used in flexible and wearable electronics, as well as low SWaP-C applications. In a few cases, 2D materials have achieved $D^*$ as high as MCT [21]. On the other hand, 2D-material detectors [21, 25]—and light sources [26]—are in their infancy and are far from being COTS products.

*Optical path components (passive optics)*

Although most passive optics are not made of semiconductors, no discussion of photonic chemical sensing systems is complete without optical path components. They play the vital roles of not only determining interactions with the target chemical(s) but also guiding radiation from the light source(s) to the detector(s). Transmissive optics (e.g., lenses and windows) are typically made of transparent crystalline materials, such as ZnSe and sapphire [27]. Windows protect the light source and detector from the environment, while allowing light to pass through and interact with the target chemical. Lenses focus light and increase the amount that reaches the target chemical and the detector. Reflective optics (e.g., mirrors and corner-cube reflectors) are typically made of glass or crystalline materials and coated with gold, silver, or aluminum [28]. They can also be made entirely out of metal and machined with diamond tools to be ultra-smooth and reflective [28, 29]. Reflective optics are used for optical alignment, standoff detection, etc., as well as for cavities and multi-pass cells that enhance the detected signal by increasing the path length and the amount of analyte that the light can interact with (Eq. 1) [1, 2, 5]. Like lasers and detectors, not much has changed for COTS mid-IR passive optics. Arguably, the only advancements have been increases in 1) the selection and availability of anti-reflective (AR) coatings, which reduce reflections of light at undesired wavelengths, and 2) the number of overall product offerings ($\approx 3\times$ increase) [11].

Optical fibers and waveguides can also be components of the optical path in a chemical sensing system. Fibers can be divided into two sub-categories, solid-core and hollow-core. In the mid-IR, solid-core fibers are typically made of chalcogenides, silver-halides, and fluoride metals [30]. Mid-IR solid-core fibers are much more brittle than visible-light and near-IR fibers; they have much higher losses, 1-10 dB/m (i.e., 20-90% loss in one meter) [30, 31]. Hollow-core fibers (HCFs) are essentially thin light pipes that are made of glass or plastic and coated with a reflective material on the inside, such as silver [30]. Engineered microstructures can also be used to confine light within HCFs [2, 32, 33]. COTS mid-IR HCFs have losses <0.1 dB/m ($\approx 2\%$ loss in 1 m) [33, 34], and they have been used in many chemical sensing applications [2, 31, 32, 35], even space exploration [34]. However, these fibers are still relatively immature. In general, mid-IR fiber-optics and waveguides are not widely commercially available and are very limited compared to their shorter-wavelength counterparts, such as the optical fibers used in telecom applications.

Optical fibers are a subset of optical waveguides. Although a full discussion of waveguides is outside the scope of this article, they have been discussed in other recent reviews, such as [31]. In general, mid-IR waveguides have high losses but have important applications in low-volume sensing, liquid-phase sensing, and integrated "on-chip" devices where the waveguides are manufactured in conjunction with light sources and/or detectors [23, 31].

*Systems and applications*

Systems and applications for mid-IR chemical sensing are both large in number and wide in diversity. Applications include—but are not limited to—environmental science; agriculture; medicine/health; industrial processes and safety, oil & gas/petrochemical industries, combustion diagnostics; materials science; aerospace; military/defense/civil security; food quality/adulteration; drug testing, forensics; and molecular and cellular biology [2-5, 7, 8]. Many of these applications overlap.

Mid-IR chemical sensors take advantage of strong absorption features in the mid-IR (Fig. 1) compared to the near-IR. For example, methane ($CH_4$) absorption at 3.27 μm is 50 to 120 times stronger than at 1.65 μm [2, 6], which is the wavelength used by many commercial near-IR sensors [2, 5, 36]. For nitrous oxide ($N_2O$)—the third most important anthropogenic greenhouse gas—absorption in the 4.5 μm region is about 800 times higher than in the strongest near-IR bands around 2.1-2.3 μm [6, 37]. Thus, sensing in the mid-IR allows for two orders of magnitude improvement in sensitivity (i.e., detection limit) than the near-IR—all other things equal. Eq. (1) tells us that the mid-IR also allows for a reduction in size—two orders of magnitude shorter optical path—for the same sensitivity as the near-IR. Therefore, the mid-IR has great potential for high-performance, low-SWaP photonic chemical sensors.

To give an example of how chemical sensors have changed over the past decade, one can compare a state-of-the-art mid-IR sensor used for planetary exploration in 2012 with current commercial offerings (Table 2). The Tunable Laser Spectrometer (TLS), aboard the Curiosity rover, uses an ICL to measure $CH_4$ on the surface of Mars [38]. Aeris Technologies, Inc. (recently acquired by Project Canary) now offers COTS sensors with specifications comparable to the Mars TLS at prices of $25k-$40k. Aeris's sensors are light enough for rotary-wing drones and have been successfully deployed for multiple projects internationally [39, 40]. We acknowledge that space-based instruments with their extreme environments are designed to be far more rigorous than any COTS sensors, but nonetheless Table 2 demonstrates the progress of commercialization of mid-IR sensing technology. Note that Table 2 provides only one example for demonstration purposes. About a hundred companies currently offer mid-IR sensors [41]. COTS sensors have been featured in thousands of academic publications, some of which are discussed in the review papers and books cited in the first paragraph of this subsection. Additional discussion of companies can be found in market research reports such as [41]. It should also be mentioned that mid-IR sensors are not as mature as near-IR sensors. Ref. [41] estimates that >80% of the laser-based gas sensor market still uses near-IR sensors. Ref. [42] implies that there are very few handheld (i.e., low-SWaP) mid-IR sensors compared to near-IR sensors. Reasons for this are discussed in the next section.

**Table 2. Comparison of a 2012 state-of-the-art sensor with a 2025 COTS sensor**

|  | 2012 state-of-the-art | 2025 COTS |
|---|---|---|
| name | Mars TLS | Aeris Strato |
| size (total) | 1000 cm$^3$ (35 × ⌀7 cm)[a] | 4000 cm$^3$ (21 × 21 × 9) |
| size (gas cell) | 405 cm$^3$ | 60 cm$^3$ |
| weight | 3.7 kg[a] | 1.9 kg (incl. 90-min battery) |
| power | 30 W typ. (10 to 50 W) | 15 W |
| cost | N/A (research) | $25k-$45k USD |
| detection limit[b] | 300 ppt (15-min) | 500 ppt (1-sec) |

[a]does not include electronics and pumps shared onboard the Curiosity, [b]1σ std. dev. and integration times for $CH_4$

## Challenges and opportunities

### Light sources

One of the main drawbacks of QCLs is their high power consumption. For example, a generic QCL has an operating voltage of 12 V and current of 0.5 A, equating to a power requirement of 6 W for cw operation. Most of this energy is converted to heat. Additional power is then required for heat dissipation—and the removal of waste heat associated with the inefficiencies of cooling technology—to maintain precise laser temperature. The wavelengths of semiconductor light sources are very sensitive to temperature and require fine temperature control (Fig. 2). The multi-stage thermoelectric coolers (TECs) embedded in most COTS devices struggle to maintain temperature control in outdoor and industrial applications, where temperatures can be below freezing or above 40 °C—especially in the presence of direct sunlight or exhaust from other systems (Fig. 3) [43]. A general rule of thumb is that twice the power of the QCL is needed for heat dissipation (18 W total for a 6-watt QCL) [43]—the actual requirement is system- and environment-dependent and can be modeled using the thermal mass of the heat sink(s) and heat transfer through the three fundamental pathways: conduction, convection, and radiation. Adding other electronics—such as for on-board computing—a generic QCL chemical sensing system would require at a minimum a 100-watt solar panel with dimensions of 0.75 × 0.75 m and a lithium-ion battery weighing about 5 kg—assuming six hours of peak sunlight per day and 24-hours of backup power (≈600 W·h) [37]. The power requirement for heat dissipation can be reduced by using high operating temperature (HOT) lasers in many cases [43]. However, in any case, power consumption and SWaP-C are major challenges for QCL-based chemical sensing systems that use only COTS components.

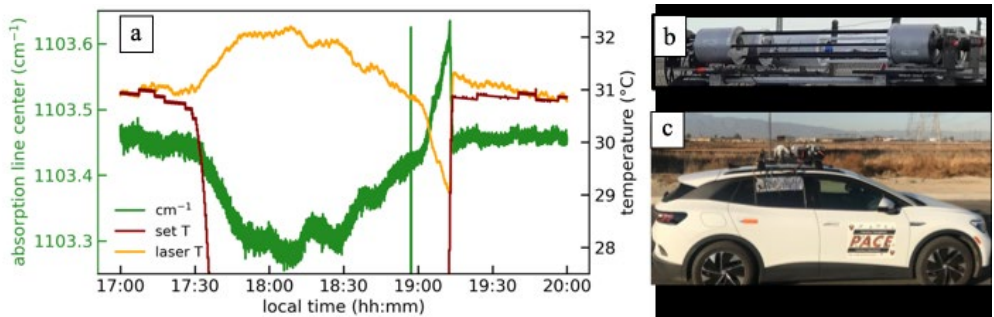

Fig. 2. Example of the precise temperature control required for outdoor chemical sensing applications. (a) Time series of wavenumber and temperature (T) for a QCL targeting absorption lines for ammonia ($NH_3$), centered at 1103.458 cm$^{-1}$. The weather was hot (40 °C) and sunny. At 17:30, the system malfunctioned and could no longer control the laser temperature. Although the QCL overheated by only ≈1 °C, this shifted its wavenumber off the absorption feature. At ≈19:00, the system was fixed and restarted, and it recovered to its typical stability of 0.004 cm$^{-1}$ (1σ). (b, c) Photos of QCL sensors for air quality measurements mounted on top of an electric car. Electronics for power, instrument control, and data processing were stored inside the car.

ICLs consume less power than QCLs but are subject to two challenges: 1) their wavelength range is limited to <7 μm; and 2) their output power is typically an order of magnitude weaker than QCLs (Table 1). Although ICLs use little power, cost is still an issue. QCLs and ICLs have similar prices ($5-12k). Another issue for QCLs/ICLs is that windows on the laser packaging can create optical interference (i.e., fringes) [44]. The quality of mid-IR windows and AR coatings is a problem in general [27, 45].

For incoherent semiconductor sources we primarily discuss ICLEDs, as they are an order of magnitude more powerful than traditional mid-IR LEDs [8, 18]. While ICLEDs have multiple advantages, many challenges must still be overcome before their full potential can be reached in high-precision chemical sensing applications. Firstly, the output power of ICLEDs is more sensitive to temperature than to light absorption for trace-gas concentrations (i.e., ppm, ppb). A 0.01 °C change in device temperature would obscure most chemical signals for environmental applications [35]. Secondly, the energy (i.e., wall plug) efficiency of ICLEDs is low—typically <1%—meaning that <1% of the energy produces light and the rest is converted to heat. Both energy efficiency and heat dissipation are limiting factors that cannot compare with visible-light LEDs, which can be >90% efficient and have replaced incandescent sources for many household and commercial lighting applications [46]. Thirdly, although filters and resonant cavities can narrow the spectrum to 20 nm FWHM, this is still too broad for many chemical sensing applications. In comparison, laser linewidths are ~0.0001 nm. Even at 20 nm FWHM, ICLEDs are still subject to interference from background species such as water vapor [35]. The effects of interfering species can be addressed with reference channels, reference instruments, and routine calibrations, but these measures increase cost and complexity [2, 47-49]. Lastly, ICLEDs cost a few hundred dollars ($USD) per unit and cannot compete with the costs of incandescent micro-bulbs and micro-electro-mechanical system (MEMS) IR emitters [2]. ICLEDs have the potential to be mass-produced using semiconductor manufacturing processes similar to those used for visible-light LEDs that only cost cents (~$0.01). However, to develop new manufacturing processes requires large investments; ICLEDs need further improvements in performance to justify such investments.

While ICLEDs face many challenges for high-precision applications, they may more readily be adopted for qualitative applications, analogous to the numerous qualitative applications that visible-light LEDs currently fulfill. One example of a more qualitative IR chemical sensing application is carbon monoxide and explosive gas leak alarms for home and industrial safety. ICLEDs have three advantages that provide them with opportunities to surpass incandescent sources: 1) ICLEDs have the potential for high-power emission at longer wavelengths (>5 µm) [8]; 2) ICLEDs can be switched on and off at high speeds (~100 kHz) [8, 18], whereas incandescent sources have maximum switching frequencies of only ≈4 Hz and up to ≈50 Hz for state-of-the-art MEMS devices [2, 50]; and 3) ICLEDs can be used in applications with low temperature requirements—for example, industrial applications with flammable gases where the high temperatures (>500 ˚C) of incandescent materials are unsafe.

### Detectors

Arguably, most chemical sensing systems are not limited by detector performance/SWaP. Compared to light sources, COTS detectors are quite prevalent in low SWaP applications [2, 5, 9, 20, 38]. However, cost is still a limitation since COTS detectors are sold for $1-5k USD per unit [11]. Aside from cost, the main limitation of detectors for chemical sensing applications is optical interference (i.e., fringes and unwanted etalons) from detector windows—in addition to laser windows and other components. Unwanted reflections from system components cause interference patterns that can mask target absorption signals [44]. In practice, no system can achieve precision and long-term stability better than ~$10^{-6}$ absorbance due to these interferences [2, 44]—aside from a few cases where it is possible to keep the system in a precisely controlled environment [51]. This has been a long-standing limitation for chemical sensing [44, 51, 52] and continues to be so, preventing systems from achieving precision close to the theoretical noise limits of their semiconductor components (i.e., the lasers and detectors). Any reflective surface is a potential source for optical fringes. The semi-transparent windows for detectors are one such source [44]. Attempts to eliminate fringes have had mixed success. Examples include using off-axis angles, numerical filtering algorithms, subtraction of non-absorbing reference signals, aperture masks, and dithering of optical components [2, 4, 44, 51-54]. On lab benches, fringes can typically be reduced to acceptable levels by changing the alignment geometry, but in many applications—especially those requiring low SWaP-C—space is limited, and fringes cannot be reduced to laboratory levels. Furthermore, in outdoor applications, thermal expansion/contraction results in changes in the optical path that often introduce new fringes that are unpredictable in practice [44, 51].

### Optical path components (passive optics)

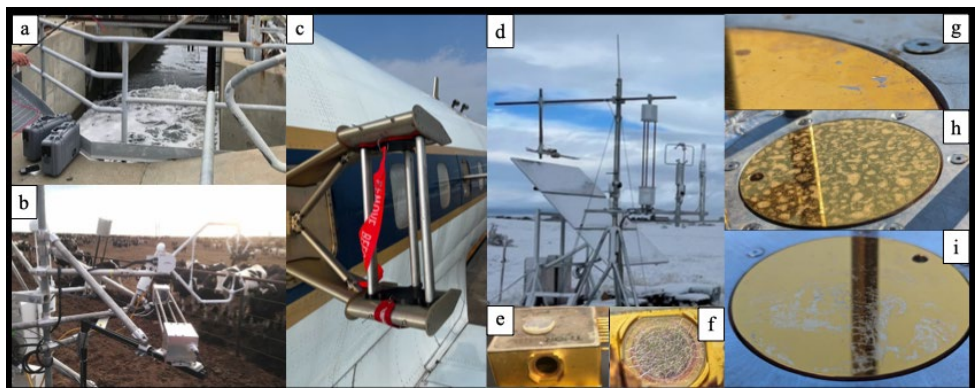

Fig. 3. (a-d) Examples of photonic systems in harsh—but not uncommon—environments for chemical sensing applications: (a) industrial wastewater and resource recovery (high humidity), (b) agriculture – beef feedlot (dust and high temperature), (c) atmospheric science – aircraft exterior (low temperature and high vibrations), (d) snow-covered fields (low temperature and precipitation). (e-i) Pictures of damage to optics after a few days of outdoor use: (e, f) damage to ZnSe windows for QCLs in an agricultural environment, subject to mild corrosion and large swings in temperature and humidity; (g-i) damage to protected-gold mirrors: (g) peeling of the coating, (h) dust, rain, and dirt, (i) scratches, caused by abrasive particles and wind or animals.

With regards to windows themselves, options are limited simply because not many materials are transparent in the mid-IR [27]. Sapphire ($Al_2O_3$) is one of the best materials and can handle harsh environments, such as aerospace applications; however, $Al_2O_3$ is limited to wavelengths <5 μm. ZnSe is typically used for wavelengths >5 μm but is relatively fragile (Fig. 3) [27]. Transmissive optics can have AR coatings applied to reduce unwanted reflections and fringes, but materials for mid-IR coatings are also limited. Furthermore, nearly all COTS mid-IR coatings are fragile when exposed to the environment (Fig. 3) [27, 45]. Robust, military-grade coatings—and commercial (ISO) equivalents—exist [27, 55]. However, military/ISO standards are based on short-term tests and may not apply to outdoor use >6 months [45].

Options for reflective optics (e.g., mirrors) in the mid-IR are limited as well. Glass/crystalline mirrors coated with gold/silver are very fragile and can degrade to <50% reflectivity within just a few days of outdoor use (see Fig. 3 and [43]). Diamond-machined mirrors made entirely of metal have proven to be robust for environmental applications [43], for high-power lasers (e.g., for laser cutting and fusion experiments) [28, 29], and for space telescopes [28, 56]. However, they are also more expensive to machine, and residual grooves from the machining are difficult to remove/polish [28]. For chemical sensing, vendors prefer to offer glass/crystalline mirrors for COTS options [11]. Thus, both cost and robustness are major limitations for COTS mid-IR reflective—and transmissive—optics. These limitations are even more apparent if one compares to the costs and robustness with COTS visible-light passive optics (i.e., household and lab-grade windows, lenses, mirrors, reflectors, etc.).

For mid-IR fiber optics and waveguides, the main challenges are high losses and issues with fiber coupling efficiency and fragility. Mid-IR fiber optics currently cannot compete with their shorter wavelength counterparts in either of these aspects. For example, in [57] and Fig. 3c, a HCF-coupled QCL successfully detected $NH_3$ at concentrations below 500 ppt for atmospheric science, but integration and maintenance on the aircraft was difficult; the fiber broke about a dozen times and had to be replaced, and there were issues with stability in flight. Cost is another limitation for mid-IR fiber optics since these components have not yet benefited from economies of scale.

*Systems and applications*

As alluded to in the previous subsection, mid-IR systems suffer from the lack of availability and selection of high-quality COTS components. Thus, while mid-IR systems have large advantages in sensitivity over near-IR systems, there are several drawbacks to their practical implementations. First, the poor availability of components limits the SWaP-C of mid-IR systems. Power-hungry QCLs require systems with large batteries and bulky thermal management. So far, most—if not all—low-SWaP mid-IR trace-gas sensors, such as the Mars TLS and the Aeris Strato for rotary-wing drones, have employed ICLs. In fact, the Mars TLS employed a custom-built mid-IR ICL at 3.27 μm, but in the near-IR at 2.78 μm, a COTS laser was suitable—the detectors were also COTS products [38]. This design choice was indicative of the technological maturity of near-IR vs. mid-IR light sources at the time, and while mid-IR light sources have improved (Table 1), they still cannot compete with the maturity of near-IR sources used for high-volume telecom applications.

System performance is also limited by the lack of mid-IR components. Mid-IR optics and coatings are relatively fragile (Fig. 3), and they suffer from optical fringes that limit detection to ~$10^{-6}$ absorbance in most applications. Some systems have addressed this by essentially bringing the lab environment outdoors—in other words, isolating the optical path in a precisely temperature- and pressure-controlled gas cell, where the sample gas is pumped through an inlet and often cleaned of any debris and humidity. This method is used by the Mars TLS, Aeris Strato, and many other high-performance mid-IR sensors. Temperature- and pressure-controlled environments and pumps for gas inlets increase the power requirements and size of sensors. Inlets and gas cells can also introduce sampling artifacts and result in slower response times, especially for gases that readily adsorb to surfaces, such as water vapor and ammonia.

A fundamental limitation of mid-IR spectroscopic sensing is that water vapor absorbs strongly and rather ubiquitously in the mid-IR compared to the near-IR. For environmental applications, absorption features of many species of interest have to be carefully vetted against even weak ro-vibrational transitions of water vapor that could underlie the chosen spectral feature, since ambient air typically contains about 1% water vapor by volume (i.e., 10,000 ppm). If one wants to detect trace-gas absorption for concentrations of 1 ppm, water vapor absorption would be ten thousand times stronger, all other things equal. As a result, many strong absorption features in the mid-IR cannot be taken advantage of, due to the presence of water—and other interfering absorbers. Ref. [58] provides an example of selecting absorption lines for ammonia ($NH_3$) at 1103.46 cm$^{-1}$ (9062.4 nm) that are mostly free of lines for other species. Part of the problem with interfering species can be solved with sensing at reduced pressure. Absorption lines become narrower at lower pressures [1, 2] and therefore interfere less with each other. For example, a commercial $NH_3$ sensor from Aerodyne Research Inc. probes one of the strongest lines of the fundamental vibrational band of $NH_3$ at 967.346 cm$^{-1}$ (10337.560 nm) at reduced pressure [59]. However, to do this, the sensor needs a pump; it weighs a total of 50 kg and uses >250 W of power. Furthermore, even with reduced pressure, water is so ubiquitous that in many cases the gas inlet streams still need to be dried with desiccants or cold traps.

Interfering absorbers are an even bigger limitation for other fields such as medicine, molecular biology, and forensics. The Earth's atmosphere is clean compared to these sensing environments. Air is comprised of mostly nitrogen and oxygen, which are transparent in the mid-IR, and gas molecules have relatively simple vibrational modes compared to large complex molecules that exhibit very broad absorption features. For this reason, FTIR spectrometers have been preferred over laser-based sensors outside of trace-gas sensing [3, 4]. Broadband spectral analysis is needed for tackling large molecules, multi-species detection, and complex sampling environments. Probing multiple wavelengths can be done using multiple lasers; however, this increases SWaP-C, especially given the high costs of QCLs/ICLs. External cavities (ECs) can be used to increase the laser wavelength scanning range [7, 8]. Multiple companies offer COTS EC lasers and EC-laser-based sensors, which have been reported in hundreds of publications (see [7] and references therein for examples). However, EC lasers typically span only hundreds of cm$^{-1}$ and are still not as broadband—or technologically mature—as FTIR spectrometers, which span thousands of cm$^{-1}$ [3, 4]. Frequency combs (FCs), as discussed previously, are a promising technology with a broadband spectral span comparable to EC lasers, but with superior spectral and temporal resolution. Current commercial offerings for mid-IR FCs are high in SWaP-C and are mostly limited to laboratory applications [13].

Another fundamental limitation of mid-IR chemical sensing is the exponential relationship between light absorption and concentration, per Eq. (1). This limits the linearity and dynamic range of photonic sensors. Linearity is important because it allows for simpler analysis and calibrations. High dynamic range is necessary for applications where one must detect both very low and very high concentrations. One solution is to probe multiple absorption lines, a weak line and a strong line, such as in [60], although in most cases it is difficult to find suitable adjacent lines. Multiple lasers can be used to probe lines that are far apart spectrally, but this increases SWaP-C. Other spectroscopic techniques, such as dispersion spectroscopy, remain linear at high concentrations and can provide a dynamic range spanning five orders of magnitude [61]. However, such techniques are of no use when the sample is so thick and light absorption is so strong that all the light is absorbed. Frequency combs can tackle this problem since they can simultaneously probe hundreds of absorption lines [17].

**Future developments to address challenges**

*Light sources*

Lower power consumption and better heat dissipation are clear areas of improvement for QCLs. For ICLs, availability at wavelengths >7 μm is the main area for improvement, with higher optical power being a secondary, lower-priority concern. However, output power remains an important factor for long-term operation when exposed to environmental elements such as precipitation, dew/frost, insects, dust, bird droppings, and corrosion, which are common in outdoor (e.g., agricultural, military) applications. For example, if a system is designed in the laboratory to have 100 mW of light reach the detector, then only 1 mW may actually still be available at the detector after long-term operation without maintenance. Therefore, existing mid-IR systems often require routine and costly maintenance for outdoor applications. Higher output powers can also open the door to a large number of remote sensing and standoff detection applications, where the light source is powerful enough that radiation backscattered from the ground or the walls of buildings are sufficient for chemical sensing [62].

One promising development is QCLs/ICLs designed with machine-learning (ML). In 2024, functional QCL designs were made with practically zero human involvement, even resulting in new and unintuitive designs [63]. If these ML-algorithms can be developed further, they could potentially be optimized for energy-efficient designs, usage of non-traditional materials, and more. Even greater potential lies in extending the ML-algorithms to the design of ICLs and other devices that utilize quantum-well structures.

For ICLEDs, they need to outcompete MEMS IR sources to gain more traction. For this, they need to be more energy efficient. ICLEDs ultimately have potential for 1) high output at >5 μm, 2) applications with high pulse rates or flammable-safe requirements, and 3) manufacturability and scalability.

*Detectors*

Infrared detectors will likely see continued growth in the near future. Recent predictions show that even MCT has the potential for at least an order of magnitude improvement in $D^*$ [21]. However, for chemical sensing—especially laser spectroscopy—detectors are not limited by their inherent performance characteristics, but by unwanted optical interference (i.e., fringes) from associated passive components, which also limit lasers and systems in general.

*Optical path components (passive optics)*

The long-standing issue of optical fringes in photonic chemical sensing systems will always be a challenge. The fundamental noise limits of detectors are often orders of magnitude lower than the limits imposed by fringes. One avenue of development is meta-surfaces and meta-materials [discussed in a separate sub-topic article of this Roadmap]. Meta-surfaces have the potential for a paradigm shift as large as the one brought on by the invention of QCLs/ICLs. QCLs/ICLs broke through the limits of naturally occurring materials and allowed for the manipulation of light using artificial structures. Meta-surfaces aim to do the same. Rather than relying on natural material properties for lenses, windows, etc., optics with meta-surfaces use engineered nanostructures to control light (refraction, scattering, etc.) [64-66]. Although meta-surfaces are still in their infancy for mid-IR optics and have their own set of limitations [64, 65], they are a promising area of research.

In the shorter term, one advancement for passive optics could be gold-coated plastic retroreflectors [67, 68]. These do not eliminate optical fringes, but they are at least an order of magnitude cheaper than glass/crystalline and metal retroreflectors. Outdoor experiments have shown performance comparable to COTS reflectors [67, 68]. Another promising and actively researched option is lightweight metal mirrors (MMs) [28, 69]. High-grade COTS flat MMs are currently offered for only a few hundred dollars [11]. Further advancements in diamond-turning processes and additive manufacturing may further lower the costs and increase the availability of different shapes and sizes for COTS MMs.

*Systems and applications*

If chemical sensors are to be deployed in outdoor and/or harsh conditions, which are common in many applications (e.g., environmental, industrial, aerospace, military, etc.), more research on system robustness is needed. Of special concern are transmissive and reflective optics, AR coatings, and light sources, particularly when exposed to sub-freezing temperatures, high temperatures (>40°C), direct sunlight, high humidity, precipitation, dew/frost, salt spray, dust, high winds, and animals (e.g., insects, birds, etc.). While this may seem like a tall order, inspiration can be drawn from detailed testing and robust designs for systems in a multitude of applications such as the automotive and aerospace industries, where many complex and high-precision systems have been successfully built to handle all the harsh conditions that are common in the world. One could also argue that robust testing for mid-IR photonics is absent due to the lack of suitable materials. Consider the commercial methane instrument from LI-COR, the LI-7700, an open-path sensing configuration where near-IR light is passed between two mirrors (i.e., Herriott cell) while exposed to the environment [36]. This sensor uses glass/crystalline mirrors, hard and robust dielectric near-IR AR coatings (i.e., glass and crystalline coatings such as $SiO_2$, $TiO_2$, and $Al_2O_3$), and low-power and high-efficiency lasers [70, 71], which all benefit from the wide selection and availability of materials for the near-IR. The LI-7700 has been successfully deployed over a range of extreme environments [36]. Despite decades of research, no equivalent mid-IR COTS sensor exists. Further research on overcoming the limitations of mid-IR materials—especially for passive optics—is needed in order to unlock the true potential of mid-IR chemical sensors.

In general, the costs for passive optics, light sources and detectors are high. High-precision chemical sensing systems cost thousands to tens of thousands of dollars (USD), which is cost-prohibitive for many applications and for many parts of the world. To reduce costs, photonic systems and components need to be manufactured by more scalable and automated processes. One potential area of advancement is photonic integrated circuits (PICs) [discussed in a separate sub-topic article of this Roadmap], where all three components of the chemical sensing system—light source, optical path, and detector—are combined on a single compact semiconductor chip [23]. PICs are still an immature technology but could see potential investment for low SWaP-C or mass-market applications.

The availability of COTS components for mid-IR sensors will be critical for cost reduction and commercialization. Currently, it is not financially worthwhile for vendors to stock mid-IR QCLs/ICLs and many passive optical components. Most vendors opt to offer customized and tailored solutions [11]. This is partly due to the fact that high-precision trace-gas sensing requires highly specific wavelengths and customized designs for optical paths. A quick scan of the commercial offerings for trace-gas sensors reveals that most sensors are specialized for only one gas—a few at most. Thus, to have a comprehensive, multi-species solution, one would need to procure an entire suite of high-cost sensors. It is for this reason that coherent broadband light sources such as EC lasers and frequency combs have garnered attention for chemical sensing. These sources can probe chemicals with higher spectral and temporal resolution than traditional FTIR spectrometers. Much of the commercial success of FTIR spectrometers can be attributed to the fact that they are practically application agnostic. The same FTIR instrument can be used for vastly different applications. As long as mid-IR chemical sensors need to be highly customized and tailored for each individual solution, it will be challenging for companies to scale up and invest in modular and mass-producible product offerings.

## Concluding Remarks

With all the challenges discussed thus far, it may seem like there are many limitations for commercializing technology for mid-IR chemical sensing. However, one must consider that lasers for spectroscopic sensing have only been commercialized relatively recently, following the inventions of the QCL and ICL in 1994 [7, 8]. In comparison, the "FTIR era" started in 1970 [72]. As highlighted in this review, many promising technologies have emerged in just the last decade. These technologies will only serve to expand the possibilities of mid-IR chemical sensing as they continue to mature.

If costs of laser-based chemical sensors can be reduced to less than $1000 USD, such systems—and components—could become more accessible to markets in places such as Africa, Asia, and Latin America. Cheaper reflectors in the mid-IR could enable applications such as urban monitoring and standoff detection in understudied regions. If the power consumption of QCLs—or the wavelength range and output power of ICLs—can be improved, then chemical sensing systems could become even more widespread, expanding to remote areas such as fence-line monitoring for croplands or environmental measurements in climate-stressed regions such as the tropics and Arctic.

For incoherent sources, we are arguably on the cusp of mass-market photonic chemical sensors. NDIR carbon dioxide ($CO_2$) sensors are available for less than $100 USD each [49]. If costs are reduced further or if demand increases, it is not without question that there could be a $CO_2$ sensor in every home and building to monitor ventilation rates and air quality. At the least, we could see photonic gas sensors become as common as particulate matter (PM) sensors for outdoor air quality. Other gases could be monitored as well, such as carbon monoxide (CO) and methane/natural gas ($CH_4$), to improve safety and overall well-being.

In the farther future, meta-surface mid-IR optics may enable more chemical sensing systems to achieve detection limits better than $10^{-6}$ absorbance. New technologies such as PICs and graphene/2D-material mid-IR devices still need many research advancements and are unlikely to see widespread use anytime soon, but the future is bright. With exponential progress, things like Arduino/microprocessor-compatible mid-IR photonic components, photonics for flexible and wearable electronics, and commercial mid-IR frequency combs fully integrated with waveguides and detectors on-chip may come sooner than we expect.

## Back matter

### Funding

Princeton University; National Science Foundation (NSF) Convergence Accelerator (award ID# 2344395) (https://www.nsf.gov/awardsearch/showAward?AWD_ID=2344395)

### Acknowledgments

The authors thank Vladislav Sevostianov, Yunseo Choi, and Hongming Yi for providing the photos used in Figs. 2 and 3.

### Disclosures

The authors declare no conflicts of interest.

# 22. Harnessing Midwave Infrared On-Chip Technologies for Advanced Chemical Analysis

SOURABH JAIN,[1,2,3] JASON MIDKIFF[4], MAY HLAING[4], KANG-CHIEH FAN[1,2], AND RAY T. CHEN[1,2,4,*]

[1]Department of Electrical and Computer Engineering, University of Texas at Austin, TX 78758, USA
[2]Microelectronics Research Center, The University of Texas at Austin, Austin, TX 78758, USA
[3]Department of Electronics and Communication Engineering, Indian Institute of Information Technology Bhopal, India
[4]Omega Optics, Inc., 8500 Shoal Creek Blvd., Bldg. 4, Suite 200, Austin, Texas 78757, USA
*Corresponding author: chenrt@austin.utexas.edu

## 1. Overview

Optical sensing and spectroscopy in the ultraviolet-visible (UV-VIS) to near-infrared (NIR) range has long been a workhorse for chemical sensing, but these shorter-wavelength techniques face inherent limitations in molecular specificity and sensitivity. Many molecules exhibit only weak overtone or combination bands in the NIR, requiring long path lengths or complex signal processing to detect low concentrations [1, 2]. In contrast, the midwave infrared (MWIR, ~2–20 µm) regions cover the fundamental vibrational absorption bands of chemical bonds – the so-called "fingerprint" spectral region – enabling direct label-free identification of molecular species [3]. MWIR spectroscopy exploits these strong fundamental vibrational transitions, providing much higher selectivity and sensitivity for chemical analysis than NIR methods [4-6]. Moreover, the longwave infrared (LWIR) window ranging from 8–12 µm contains unique low-frequency vibrational modes of large organic molecules and lies in an atmospheric transmission window, making it especially valuable for remote sensing and thermal emission detection [7]. These advantages address many shortcomings of UV-VIS/NIR techniques – MWIR/LWIR spectroscopy offers greater molecular fingerprint specificity, stronger absorption signals, and reduced interference from background matrix absorptions. Despite these advantages, realizing miniaturized MWIR sensing platforms presents significant challenges, including material selection, integration of MWIR-compatible light sources and detectors, and the development of technology-specific engineered solutions aimed at minimizing optical losses and enhancing sensitivity for scalable, chip-integrated sensing platforms. For instance, selecting an optimal material platform requires not only ensuring low-loss optical propagation, but also achieving a sufficient core-cladding refractive index contrast to balance light penetration into the analyte region with fabrication feasibility. This article will highlight recent advancements in overcoming the challenges of on-chip MWIR/LWIR chemical sensing. Techniques like surface-enhanced infrared absorption (SEIRA) [8], Fourier transform spectrometers (FTS) [9], wavelength modulation spectroscopy (WMS) [10], and frequency comb spectroscopy [11] have been adapted to integrated photonics, maximizing the sensitivity in miniaturized sensors. The goal is to develop portable, cost-effective MWIR spectroscopic platforms with high analytical power for applications in industry, environmental monitoring, and biomedicine.

## 2. Current Status of MWIR Plasmonics for Sensing Applications

Plasmonic sensors have long been used for chemical and biosensing, exemplified by surface plasmon resonance (SPR) and surface-enhanced Raman spectroscopy (SERS) [2]. These NIR techniques offer high sensitivity but rely on indirect signals (refractive index shift or Raman scattering) rather than directly probing the molecular fingerprint absorptions. Consequently, NIR plasmonic sensors often lack inherent chemical specificity and may require labels or secondary analyses to identify analytes. As mentioned earlier, MWIR and LWIR photonics can play a crucial role in plasmonics by accessing the vibrational fingerprint regions of the molecules. MWIR plasmonics not only improves specificity by utilizing distinct absorption bands but also allows strong light-matter interactions at wavelengths where many analytes absorb most strongly and with deeper penetration depth. **Figure 1** illustrates the evolution and principles of various plasmonic sensing techniques, including SPR, LSPR, SERS, and SEIRA. Each approach leverages plasmonic field confinement to boost IR molecular signals. SEIRA in particular is one of the most widely used techniques, wherein IR light is coupled to localized surface plasmons (LSPs) on metal nanostructures to greatly amplify molecular absorption in the near-field [12]. Prism-coupled configurations combine these local enhancements with propagating surface plasmons, improving signal intensity for, e.g., biosensing applications [13]. Despite these advantages, several key challenges have hindered the widespread adoption of MWIR and LWIR plasmonic sensors. The following sub-sections discuss how successive breakthroughs build upon one another: improving sensitivity and functionality of IR plasmonic sensors by designing refinements and enhancing performance and integration through new plasmonic materials. We also discuss the various challenges and opportunities to shape the concrete future of on-chip sensing technologies.

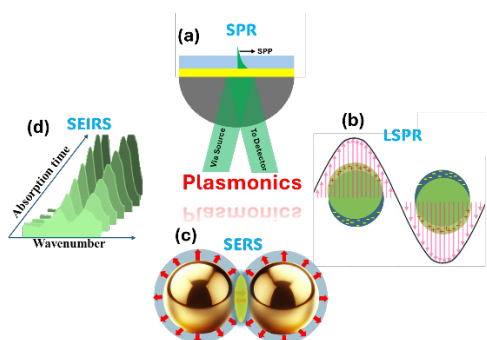

Figure. 36: Schematics of various on-chip IR plasmonic sensing techniques: (a) prism-coupled (Kretschmann configuration) plasmon-enhanced sensing, (b) local surface plasmon resonance (LSPR): collective oscillation of conduction electrons in metal nanoparticles triggered by strong interaction with light. (c) LSPR coupling between adjacent metal nanoparticles creates intense electromagnetic hotspots, amplifying Raman signals in SERS. (d) SEIRA: The absorption signature is enhanced over time as a function of wavenumber. SEIRA leverages plasmonic nanostructures to amplify infrared absorption, enabling sensitive detection of molecular vibrations.

### 2.1 MWIR Plasmonic Materials and On-Chip Integration

Traditional plasmonic materials (noble metals like Au or Ag) exhibit high losses at MWIR frequencies, and are not optimized for longer wavelengths since their plasma frequencies lie in the visible range. This limits the quality factors and confinement achievable in MWIR plasmonic devices. To improve upon this, materials with plasma frequencies in the infrared have been investigated. Heavily doped semiconductors (such as n-doped Si, InAs, InSb or Ge) [14] and polar dielectrics [15] can support plasmonic resonances in the MWIR because their free carrier densities are tuned to yield plasma frequencies in the appropriate range. Sherif et al. developed a MWIR plasmonic gas sensor based on Fano resonance in coupled plasmonic microcavities, addressing the need for high-sensitivity multi-gas detection. Using highly doped silicon (Si) as a plasmonic material, they achieved plasmonic resonance above 3 μm, enabling 6000 nm/RIU sensitivity and a figure of merit (FOM) of 353 at 6.5 μm [16]. In another example, Law et al. demonstrated all-semiconductor plasmonic nanoantennas using doped InAs, which concentrated LWIR light into nanoscale gaps [17]. Semiconductor antennas have detected previously undetectable molecular absorptions, showing their potential to overcome metal limitations and provide tunable, low-loss plasmons in the MWIR. Similarly, doped transparent conducting oxides (like ITO, ZnO) enable tunable IR plasmonics, while transition metal nitrides (like TiN) offer durable, CMOS-compatible options. Lin et al. developed a MWIR SPR fiber optic sensor using a D-shaped silica fiber with a 105 nm ITO coating, achieving 2.7 μm resonance and 1065.70 nm/RIU sensitivity in the 1.33–1.42 RIU range [18]. Doped semiconductors addressed the spectral mismatch issue of metals, graphene tackled the tunability and extreme confinement aspect, and phonon polaritons solved the loss problem to some extent. Together, they form a toolkit of MWIR plasmonic materials from which researchers can choose to meet specific sensing requirements (e.g. a semiconductor nanoantenna for a broad-band, chip-integrated sensor, or a graphene/phonon-polariton structure for ultra-sensitive niche detection). In this direction, Li et al. developed a graphene based hybrid metasurface-based MWIR biosensor using LSPR, plasmon-phonon hybridization, and graphene-assisted tuning, achieving 30 pM sensitivity for protein analysis (6–6.6 μm) [19]. Bareza et al. enhanced MWIR gas sensing by integrating graphene nanoribbons with a polyethylenimine layer, achieving a 390 ppm $CO_2$ detection limit (6.25–7.14 μm) through LSPR tuning and improved molecular interaction [20]. Overall, there have been significant advancements in MWIR/LWIR material development, particularly over the past decade, which have enabled strides in miniaturizing IR chemical analysis. It is expected that the remaining challenges, such as further reducing losses and further enhancing the plasmonic effect, will be addressed through concerted efforts in plasmonic design and materials science. In the next section, we discuss a few important developments in the waveguide engineering to fully utilize the plasmonic effect for a variety of sensing applications.

### 2.2 Plasmon-Enhanced Sensing in the MWIR

Early demonstrations of MWIR plasmonic sensing drew inspiration from NIR plasmonics, adapting techniques like surface plasmons and Raman spectroscopy to the infrared domain.

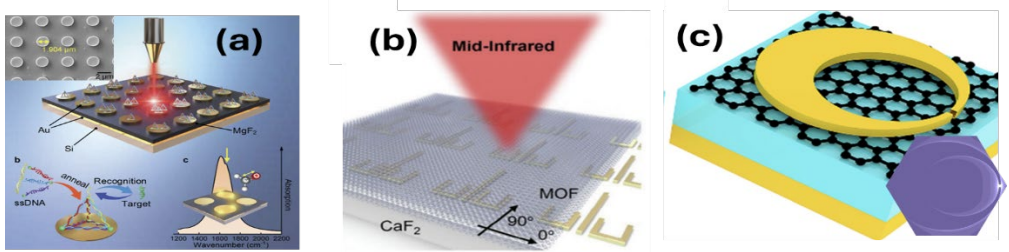

Figure. 37: (a) Depiction of a tetrahedral DNA nanostructure (TDN) as a biosensor on a Au surface with a freestanding probe for target detection. The absorber resonance aligns with the C═O bond, enabling ultrasensitive detection of MIR-155 [8]. (b) Illustration of aMOF-hybrid CMRPN having a Au nanoantenna array on a calcium fluoride substrate, structure [25]. (c) Graphene loaded dynamic plasmon tunable metal-insulator-metal configured crescent resonator for hybrid SPR-SEIRA non-localized field enhancement. The right bottom inset shows the electric field enhancement at optimized parameters [28].

Surface-enhanced infrared absorption is a cornerstone technique, analogous to surface-enhanced Raman but targeting direct IR absorption [21]. In SEIRA, molecules adsorbed on metal nanostructures experience the intensified electromagnetic fields of LSP resonances, leading to greatly magnified IR absorption signals. Recent advancements build on plasmon-enhanced spectroscopy by extending localized field amplification from NIR to the molecular fingerprint region. Improvements like lithographically patterned antennas and flow-cell integrated chips have overcome earlier limitations, expanding SEIRA's applicability. For instance, Osawa et al, showed that even a monolayer of molecules could produce detectable IR absorption when on a roughened gold surface [22]. Early SEIRA substrates used random metal island films for broadband enhancement. Modern sensors use engineered nanoantenna arrays tuned to specific vibrational bands, increasing the sensitivity and selectivity. For example, Ahmadiv and et al. used gold nanoresonators in a toroidal metachip to achieve strong MWIR resonance (5.25 μm) and detect kanamycin sulfate at attomolar levels. In another attempt to enhance the near-field intensity Hui et al., utilized a metamaterial perfect absorber (MPA) with gold (Au) nanoantennas (**Figure 2(a)**), and achieved a 1000-fold field enhancement and an ultrasensitive detection limit of 100 fM ($10^{-13}$ M), which is 5000 times lower than single-stranded DNA probes [8]. Recently, Xie et al. demonstrated coupled multi-resonant plasmonic nanoantennas (CMRPNs) coated with metal-organic frameworks (MOFs, specifically ZIF-67) (**Figure 2(b)**) to trap gas molecules and amplify the near-field enhancement in the molecular fingerprint detection range between 6–10 μm [23]. Researchers have progressively improved SEIRA performance by increasing the field confinement (for example, using gap-enhanced structures or coupled resonators) and by extending the spectral coverage. Hinkov et al. summarized the progress towards SEIRA for liquid sensing in MWIR range in their article [24].

In parallel, researchers explored propagating plasmon modes and metasurfaces for IR sensing. One approach, inspired by the classic Kretschmann SPR configuration, uses prism coupling to excite surface plasmon polaritons (SPPs) on thin metal films at infrared wavelengths [25]. One can also utilize hybrid SPR-SEIRA schemes (**Figure 2(c)**) to leverage the long interaction length of SPPs together with local hot spots, improving the sensitivity for detecting low-concentration analytes on surfaces [26]. Through these iterative advancements, plasmonic sensing in the MWIR has evolved into a versatile toolkit: from broadband, high-enhancement substrates suitable for detecting unknown compounds to narrowband, high-Q metasurfaces optimized for specific analytes. Transitioning from free-space configurations (prisms and external IR microscopes) to planar devices was a critical step that set the stage for full on-chip integration, as discussed next.

### 2.3 Integrated Waveguide Platforms for MWIR and LWIR

A crucial enabler for on-chip sensing is the development of low-loss waveguide platforms that operate across MWIR and into the LWIR. Si photonics is one of the most adapted platforms in the telecom band (~1.3–1.55 μm) band. Si exhibits a low-loss window extending up to 8 μm, but its standard oxide cladding ($SiO_2$) limits its utility beyond ~3.8 μm due to strong absorption [4-6]. However, this window exhibits distinct and robust spectral signatures of numerous crucial chemicals that can be detected effectively using precisely engineered waveguides on this platform. This range encompasses key molecular absorption bands, including O-H, N-H, and C-H stretches, which are characteristic of vital chemical species such as water, ammonia, hydrocarbons, and biomolecules [27]. For instance, an optimized SOI ridge waveguide achieved a high evanescent field fraction (~23% in the analyte) and low loss, enabling sensitive $CH_4$ detection at 3.29 μm [28]. Several material platforms have emerged to span different portions of the MWIR spectrum. To reach beyond 4–5 μm, chalcogenide (ChG) glasses are widely used, as they offer broad IR transparency (up to ~10 μm) and can be deposited on silicon or other substrates [29]. Integrated chalcogenide waveguides have shown low losses and have been used for both gas and liquid sensing in the MWIR [30]. Another approach is to avoid absorbing claddings altogether, such as by using suspended waveguide architectures (e.g. suspended silicon ridges or membranes) [31] or crystalline fluoride substrates [4]. Silicon-on-sapphire [32] and silicon-on-calcium fluoride [4] have extended silicon waveguides into the MWIR and LWIR by using IR-transparent substrates in place of $SiO_2$. Notably, Ma et al. demonstrated silicon waveguides on a $CaF_2$ substrate with low loss across 6.3–7.1 μm [33]. This heterogenous integration, achieved via transfer printing, mitigated substrate absorption and allowed access to the LWIR regime. Hand-in-hand with waveguides, advances in integrated photodetectors have enhanced on-chip MWIR spectroscopy by improving the compactness and performance [34, 35].

Creating lab-on-a-chip IR sensors requires integrating plasmonic structures with on-chip light sources, waveguides, and detectors. However, MWIR sources and detectors like quantum cascade lasers (QCLs) and thermoelectrics are challenging to integrate due to their limited maturity [36]. While longer wavelengths allow easier nanoscale fabrication, the lack of precise, reliable fabrication processes for novel materials remains a significant challenge. On the technology side, strong water absorption in the MWIR can interfere with biosensing in aqueous samples, demanding innovative solutions (e.g. surface coatings or short-path evanescent sensing) to mitigate background interference [37].

In order to have seamless integration capability, dielectric-loaded surface plasmon polariton (DLSPP) based waveguides have shown remarkable performance from telecom wavelengths to MWIR frequencies by confining the light in ultra-small cross-sections over millimeter-scale distances. In this chain, David et al., proposed DLSPP waveguides to achieve efficient plasmonic guiding with strong light confinement on a CMOS-compatible dielectric material [38]. Schwarz et al. realized a bi-functional QCL/detector device integrated with a plasmonic waveguide, creating a self-contained MWIR sensing chip [39]. Such integration is of high importance for monitoring chemical reactions in real-time, since it eliminates the need for external optics and allows in situ measurements. Subsequent research built on this achievement: Ristanic et al. improved the robustness of the chip sensors [40], and Hinkov et al. applied the integrated system to biomolecular diagnostics, demonstrating the practicality of portable MWIR analyzers [41]. Transition statements between these works are evident – each new iteration retained the core idea of on-chip IR plasmonics.

## 3. Current Status on MWIR On-Chip Absorption Spectroscopy

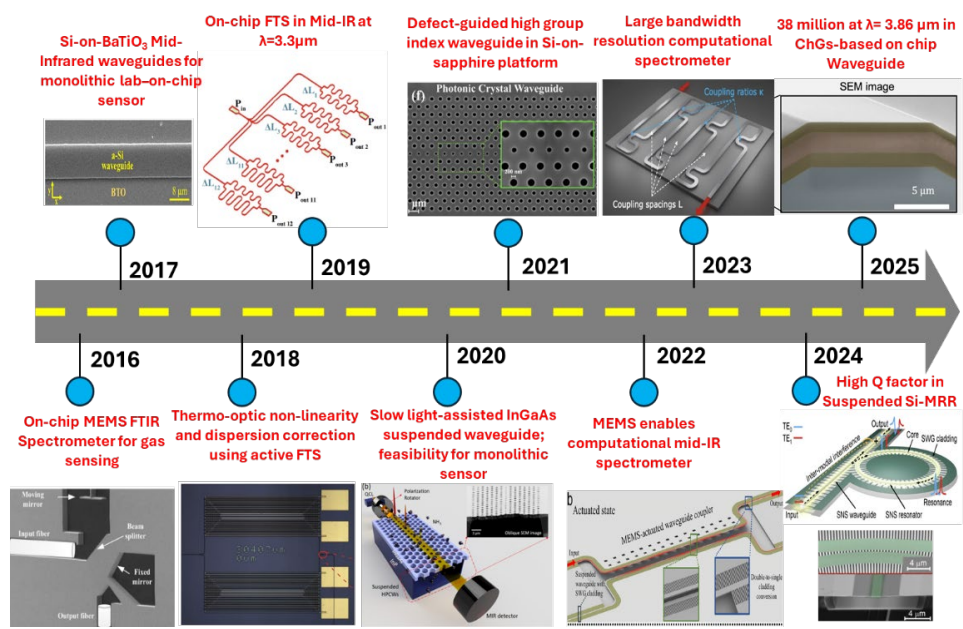

Figure. 3: Timeline of key advancements in on-chip technology for spectroscopy applications over the past decade. The evolution from early gas sensing using on-chip MEMS spectrometers to high-Q resonators, slow-light engineering, monolithic integration, and frequency comb spectroscopy highlights the significant progress in achieving miniaturized, high-resolution, and efficient spectroscopic solutions. [1, 34, 46-53]

Optical spectroscopy techniques have remained versatile tools in the field of chemical sensing for the last several decades [42]. While traditional spectroscopy has been widely employed for chemical analysis in laboratory-based settings and in a wide area, the transition towards compact, on-chip MWIR and LWIR sensing platforms is essential for real-time, miniaturized, and high-sensitivity chemical detection in diverse environments [43]. Recent advances in MWIR spectroscopy have enabled the miniaturization of sensors for highly sensitive, real-time chemical analysis. On-chip techniques like suspended silicon microring resonators with high-Q factors and large bandwidth spectrometers have led to compact, high-resolution, and cost-effective systems, paving the way for portable, scalable spectroscopy platforms. **Figure 3** depicts the past decade of advancements in on-chip technologies for spectroscopy applications, leading to the development of compact, high-resolution, and cost-effective spectroscopic systems [1, 32, 44-51]. Recent breakthroughs include achieving a high-Q factor in suspended silicon microring resonators and large bandwidth resolution spectrometers, demonstrating the potential for highly sensitive, real-time, and scalable spectroscopy on a chip.

### 3.1 On-Chip Direct Absorption Spectroscopy

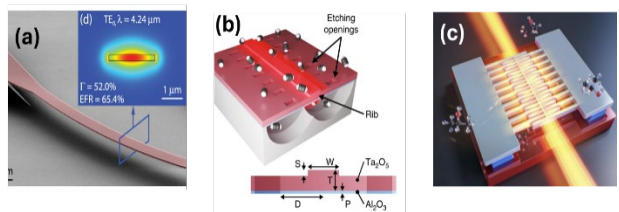

Figure. 4: Advancements in on-chip MWIR Sensing Waveguides. (a) A suspended silicon membrane enables a low-loss optical mode with a high confinement of 52% in air, facilitating enhanced light-analyte interaction. [58] (b) A shallow Ta$_2$O$_5$ optical waveguide operating in TM mode is positioned atop a thin Al$_2$O$_3$ layer, optimizing the light guidance. [59] (c) A silicon-based metamaterial waveguide engineered to achieve exceptionally high confinement within the analyte region, maximizing the interaction efficiency. [61]

The most straightforward spectroscopic sensing method is direct absorption spectroscopy, where a MWIR beam interrogates the sample, and the transmitted intensity is measured as a function of wavelength. On-chip implementations of direct absorption replace the free-space beam path (e.g. a multi-pass cell or fiber optic path) with an integrated waveguide that guides light through or alongside the sample [52]. The sample (gas or liquid) can be introduced either in an open waveguide section (air-cladding or microfluidic channel exposing the core) or in an evanescent sensing scheme, where the analyte surrounds the waveguide and interacts with the evanescent field [52-54]. A variety of integrated absorption sensors have been demonstrated. One approach uses long, folded waveguides to maximize the optical path in a compact area [55] – for example, a serpentine SOI waveguide of 2 cm length was used to detect methane around 3.291 μm, achieving a limit of detection (LoD) of 155 ppm for CH$_4$ in 0.2 s using direct absorption [28]. Another approach to increasing path length is to enhance the light-matter overlap (confinement factor) rather than absolute length [56, 57] (**Figures 4(a) and (b)**). Techniques include using slot waveguides [58] or sub-wavelength lattice waveguides [59] (**Figure 4(c)**) that confine more mode field in the analyte and resonant structures like ring resonators [60] that effectively multiply the interaction length by the finesse of the cavity. Recently Guo et al., reported a suspended nanomembrane silicon (SNS) microring resonator with ultrathin sub-wavelength thickness and obtained an ultrahigh Q-factor ($6 \times 10^5$) and an exceptionally large confinement factor (~80%) in air, enabling over 10× higher gas sensing sensitivity than conventional silicon resonators [50]. The mentioned enhancement in the light-matter overlap is proportionate with the slow-down factor of the optical wave inside the waveguide. Utilizing this approach, a highly miniaturized sensing waveguide can be formed with engineered group index engineering and tapering mechanisms [46, 47, 61-63]. However, direct absorption on the chip often faces limitations from baseline drift and noise, especially when dealing with broadband sources or wide spectral scans [64]. The sensitivity of direct waveguide sensors is limited by their path length and susceptibility to external noise. Researchers are exploring on-chip amplifiers and photothermal micro-calorimetric detection to enhance sensitivity, though these methods are still developing [65]. Current on-chip MWIR absorption sensing achieves detection limits of $10^{-4}$–$10^{-3}$ absorbance, sufficient for applications like methane leak detection and air monitoring. Future efforts aim to reduce noise and improve selectivity using integrated interferometric spectrometers and wavelength modulation techniques.

### 3.2 On-Chip Fourier Transform Spectrometers and Wavelength Modulation Spectroscopy

Fourier transform infrared spectroscopy (FTIR) is the standard for broadband IR chemical analysis. Efforts to miniaturize FTIR for on-chip use focus on replicating the functionality of a Michelson interferometer to generate and analyze interferograms. Common designs use a Mach–Zehnder interferometer (MZI) with adjustable path lengths, achieved through thermo-optic phase shifters or MEMS mirrors, to scan interference patterns [45]. While simple, the resolution of a single MZI is limited by the maximum optical delay; achieving high resolution requires a long delay or large devices. To address this, researchers have developed stationary-wave integrated Fourier transform spectrometers (SWIFTS) and multiplexed interferometers [66, 67]. A notable example is the microring resonator-assisted FT spectrometer (RAFT) by Zheng et al. [68]. This device, which cascades a tunable MZI with a tunable micro-ring resonator on the same chip, achieved an impressive 0.47 nm spectral resolution while maintaining a broad operational band ~90 nm wide. The main trade-off in that approach was a slight reduction in signal-to-noise ratio due to time-multiplexing. To avoid latency issues, another state-of-the-art device is SWIFTS. In an initial demonstration, Coarer and colleagues introduced a compact one-dimensional integrated SWIFTS and achieved a spectral resolution of 4 nm and a bandwidth of 96 nm using a single stationary MZI [69]. However, the bandwidth of on-chip FTIR spectrometers is limited by the optical path delay (OPD) and detector pitch. Some designs use multiple detectors or fixed OPDs, avoiding physical path tuning. In contrast, spatial heterodyne FTS (SHFTS) achieves high resolution using arrays of stationary MZIs with increasing OPD. For instance, Heidari et al., demonstrated an on-chip FTS on a silicon-on-sapphire platform with 12 MZIs of progressively increasing OPD [32]. The challenge with such an OPD scheme is that each interferometer's output needs to be read out (either via multiple on-chip detectors or by multiplexing signals), adding complexity.

FTIR captures broadband spectral fingerprints but also includes noise, while WMS targets weak absorption lines to improve the signal-to-noise ratio. WMS modulates a laser wavelength near an absorption line and detects intensity changes at harmonic frequencies (typically the second harmonic, 2f), reducing low-frequency noise. Traditionally, WMS uses free-space or fiber-coupled lasers and detectors [70]. The recent trend is to implement WMS in an integrated fashion for MWIR gas sensors, leveraging on-chip lasers (or laser feeders), waveguides, and detectors to create a compact sensor with parts-per-billion (ppb) level sensitivity. Recent advancements have enabled the implementation of WMS on chip-scale platforms. For instance, Pi et al. investigated WMS in waveguide sensors and reported significant sensitivity enhancements over direct absorption spectroscopy [10]. Similarly, Zhao et al. developed on-chip SOI waveguide methane sensors operating at 3.291 µm, utilizing both DAS and WMS techniques. They achieved a limit of detection (LoD) of 155 ppm with direct absorption spectroscopy (DAS) and an improved LoD of 78 ppm using WMS, highlighting the superior sensitivity of WMS in integrated waveguide sensors [28].

### 3.3 Frequency comb technique

Frequency comb spectroscopy enables precise molecular absorption measurements by generating evenly spaced spectral lines [71, 72], while dual-comb spectroscopy (DCS) offers superior resolution and acquisition speed by eliminating the need for scanning mechanisms. Recent advancements in MWIR DCS have focused on QCL and interband cascade laser (ICL) frequency combs [73, 74], subharmonic optical parametric oscillators [75], and broadband feed-forward stabilization [76], significantly enhancing spectral coverage and coherence. A major breakthrough was introduced by Liu et al., who demonstrated MWIR cross-comb spectroscopy, overcoming challenges like low detector sensitivity and strong excitation background by upconverting MWIR signals to the NIR using sum-frequency generation, achieving high signal to noise ratio, dynamic range, and tunability [11]. Furthermore, advancements in integrated lithium niobate photonics [77, 78] and efficient comb sources [79] have paved the way for compact and field-deployable MWIR spectrometers. Recently, Hoghooghi et al. developed a MWIR dual-frequency-comb absorption spectrometer for high-speed molecular diagnostics, addressing limitations in detecting fast chemical reactions. Using mode-locked frequency combs with GHz repetition rates, the system spans 3–5 µm, enabling microsecond-scale, high-resolution tracking of reaction dynamics [80]. At the chip scale, MWIR frequency comb spectroscopy (FCS) is still emerging but is expected to advance rapidly in the next five years. AI-assisted data processing, enhanced nonlinearity for broadband detection, and fully integrated photonic architectures will drive progress for high-sensitivity trace gas analysis, environmental monitoring, and real-time biomedical diagnostics.

## 4. Challenges, Opportunities, and Future Directions

Despite significant advancements, on-chip plasmonic and spectroscopy sensing still faces notable challenges that hinder its widespread adoption. As witnessed, plasmonic materials like gold and silver, which are commonly used in visible and near-IR applications, do not exhibit a strong plasmonic effect at MWIR and LWIR wavelengths, limiting their efficiency. This opens opportunities to explore alternative materials that demonstrate strong plasmonic behavior at longer wavelengths while being easier to fabricate with high precision, repeatability, and mass manufacturing capability. These materials should maintain performance even under uncontrolled conditions like temperature fluctuations, mechanical vibrations, and environmental variations. Integration complexity is another significant barrier in the plasmonics domain. Current fabrication methods often rely on custom processes that are not compatible with commercial foundries, making large-scale, cost-effective production difficult. Additionally, strong water absorption in the MWIR range complicates biosensing in aqueous environments, necessitating innovative solutions like advanced coatings or evanescent sensing techniques to reduce background interference. In non-plasmonic absorption techniques like DAS, FTIR, WMS, and frequency comb spectroscopy, the primary challenge lies in maximizing light-analyte interaction on-chip while maintaining low optical losses and achieving an exceptionally high signal-to-noise ratio. Although significant progress has been made over the past decade with innovative approaches like suspended waveguides and dispersion-engineered metamaterials to slow down light, achieving ppb sensitivity—particularly for trace gases and volatile organic compounds (VOCs) with weak absorptive signatures, even in their fundamental vibrational-rotational wavelengths—remains challenging. Additionally, the reliable recovery of weak signals buried in noise has been a longstanding issue, necessitating more advanced signal processing techniques. Addressing these challenges could pave the way for lab-on-a-chip diagnostics and lightweight, drone-embedded environmental monitoring systems capable of both in situ and remote sensing, bringing us closer to real-time, highly sensitive, and portable analytical solutions.

The future directions aim to address these challenges through advanced material platforms like SiGe waveguides, III–V semiconductors, and two-dimensional materials such as black phosphorus for improved MWIR performance. Developing monolithically integrated MWIR lasers using materials like GaSb, and GeSn could facilitate better compatibility with silicon-based platforms. To achieve scalability and manufacturability, there is a need for foundry-compatible fabrication techniques and advanced packaging methods. Computational approaches like machine learning can further enhance the sensitivity and selectivity of spectral data analysis, while compressive sensing techniques can optimize spectral reconstruction with fewer measurements. The integration of photonics, electronics, and microfluidics on a single platform can create comprehensive, lab-on-chip solutions for applications in healthcare, environmental monitoring, and security. Overcoming these challenges can lead to compact, real-time, and highly sensitive diagnostic tools, expanding the scope and impact of on-chip plasmonic and spectroscopy sensing.

## 5. Concluding Remarks

In summary, on-chip MWIR/LWIR chemical sensing has rapidly evolved from a conceptual possibility to a flourishing research field with tangible prototypes. By leveraging advanced integrated biosensing and spectroscopy techniques, researchers have surmounted many initial barriers and demonstrated that chip-scale devices can rival traditional sensors in performance. Continued interdisciplinary efforts in photonic engineering, materials science, and signal processing are driving this progress and are expected to resolve the remaining challenges of integration and sensitivity up to the ppb level. The coming years will likely witness the first commercial MWIR lab-on-chip analyzers, bringing the power of infrared chemical detection out of specialized labs and into broad use.

# 23. Midwave and Longwave Infrared Free-Space Optical Communication Systems

**Frédéric Grillot[1,2,†,*] AND SARA ZAMINGA[2,†]**

[1]Centre d'Optique Photonique et Lasers, Université Laval, G1V 0A6 Québec City, Canada
[2]LTCI Télécom Paris, Institut Polytechnique de Paris, Palaiseau, France
[†]These authors contributed equally to this work.
[*]frederic.grillot@phy.ulaval.ca

## A. Introduction

Free-space optical (FSO) communication enables high-capacity wireless data transmission via optical carriers propagating through the atmosphere, serving as a critical alternative to guided fiber links for terrestrial, airborne, and satellite applications [1]. Compared to radio-frequency (RF) systems, optical carriers offer significantly higher bandwidth, lower beam divergence, and enhanced spatial confinement, which collectively improve energy efficiency and reduce the probability of interception—key factors in both military and high-density civilian networks. Actual systems operating at near-infrared (NIR) wavelengths—particularly around 1.55 µm—have achieved terabit-per-second data rates using coherent modulation, amplification and polarization multiplexing, but remain highly susceptible to fog-induced Mie scattering, turbulence-induced scintillation, and atmospheric absorption [2]. One of the main challenges facing FSO systems is their ability to withstand very hazy conditions, which result from the accumulation of water droplets in the atmosphere. This phenomenon leads to considerable scattering and attenuation, which particularly affects any FSO system. Not only absorption but also signal distortion decrease at longer wavelengths. Indeed, turbulence on the propagation path is known to considerably deteriorate the optical signal, causing, for example, beam broadening, beam wandering, scintillation or loss of spatial coherence. In this case, scintillation will be the predominant phenomenon, corresponding to fluctuations in the intensity of the propagating beam. In contrast, the mid-wave (MWIR, 3–5 µm) and long-wave infrared (LWIR, 8–14 µm) bands reside within atmospheric transmission windows and experience weaker turbulence that scales as $\lambda^{-7/6}$, while also offering reduced beam divergence and enhanced spectral discretion. These spectral regimes are especially promising for covert, resilient communication in degraded environments.

To unlock the full potential of FSO systems in the MWIR and LWIR regimes, careful consideration must be given to the choice of light source—particularly the type of laser employed. Among the various selection criteria, the operating wavelength is paramount, as it determines both the compatibility with atmospheric transmission windows and the range of available laser technologies. For the LWIR and beyond, quantum cascade lasers (QCLs) are the only mature option. In the MWIR range, both QCLs and interband cascade lasers (ICLs) can be considered (both are reviewed in separate sub-topic articles of this Roadmap), and the choice depends on other criteria such as energy consumption, optical output power, electro-optical bandwidth, and spectral purity. QCLs generally consume more electrical power but provide higher optical output, while ICLs are more energy-efficient and better suited for compact systems, as long as the optical power requirement is moderate. Although both technologies have similar wall plug efficiencies, QCLs require active thermal management, adding complexity and energy cost. Another important aspect is the electro-optical bandwidth, especially for applications needing fast direct modulation. Thanks to their unipolar design, QCLs offer much higher intrinsic bandwidths (hundreds of GHz) compared to ICLs, which are limited by their bipolar nature. Therefore, QCLs are preferred when very high-speed modulation is required. Overall, QCLs are the main option for LWIR applications, for high-power needs, and when very fast modulation is critical, while ICLs are better suited for MWIR applications where low energy consumption is more important than achieving very high output power or ultra-fast modulation. This roadmap examines the state of MWIR and LWIR photonic platforms for high-speed coherent and chaos-based transmission, encrypted communication, LiDAR ranging, and entropy-rich random number generation—functionalities where NIR and RF systems currently face some intrinsic limitations. Expanding free-space photonics into these bands requires dedicated advances in laser dynamics, detector responsivity, and modulation bandwidth tailored to MWIR and LWIR propagation physics.

## B. State of the art

### 1. High-speed data throughput

Recent advances in MWIR and LWIR free-space optical communications have enabled multi-gigabit-per-second data transmission, powered by progress in ICLs and unipolar QCLs. In the MWIR range, ICLs have demonstrated energy-efficient, directly modulated links with excellent performance. A 4.2 μm ICL on native substrate paired with an interband cascade infrared photodetector (ICIP) achieved up to 16 Gbit/s using PAM-4 and 14 Gbit/s with OOK over 2 m, maintaining bit error rates (BER) below 4% with digital signal processing (DSP) [3]. Epitaxial ICLs on silicon have also shown promise, with OOK transmission at 8 Gbit/s and PAM-4 at 10 Gbit/s achieving BERs as low as 0.0022% and 0.11%, respectively [4,5]. In all cases, feed-forward equalization played a key role in mitigating inter-symbol interference. In the LWIR domain, QCLs—traditionally limited by modulation bandwidth—have undergone significant breakthroughs. Externally modulated QCLs at 9 μm have enabled 30 Gbit/s PAM-4 transmission [6], while directly modulated distributed-feedback (DFB) QCLs have supported advanced formats like PAM-6, PAM-8, and DMT up to 11 Gbit/s [7]. Another demonstration achieved a record 65 Gbits/s transmission using a directly modulated DFB QCL [8], illustrating the platform's potential for ultra-high-speed LWIR links.

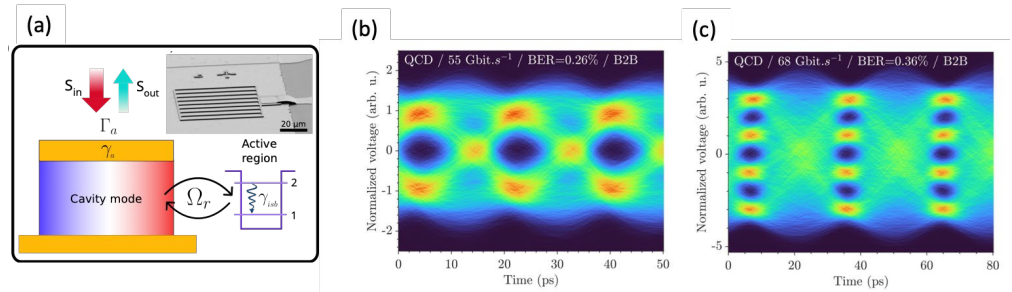

**Fig. 1:** Metamaterial unipolar quantum devices. (a) Coupled mode theory representation of the system. The incoming field couples to the cavity electromagnetic mode, which is coupled to the active region polarization mode. At upper right is a scanning electron microscope image of the device made with an array of stripes. Eye diagrams from a data transmission experiment illustrating free-space propagation for distinct modulation formats: (b) 55 Gbit/s NRZ

transcription after DFE equalization, and (c) 68 Gbit/s PAM-4 transmission after DFE equalization. Reproduced with permission from [9].

Very recently, it was shown that the use of metamaterials enables a significant enhancement of the functionality of unipolar devices (Fig. 1). Owing to their efficient electromagnetic energy confinement and reduction of the electrical area, they promote high-frequency detectors with better detectivity and modulators with low power consumption. These devices demonstrated a single channel data transmission link within the 8–14 µm atmospheric transparency window, with a capacity close to 70 Gbit/s using PAM-4 modulation [9]. Though these demonstrations rely on short-range links and extensive DSP, they validate the feasibility of high-capacity thermal-infrared communication. While MWIR ICLs offer energy efficiency and integration for compact systems, LWIR QCLs deliver unmatched bit rates and atmospheric resilience. Together, these technologies represent a critical step toward scalable, high-speed free-space communication across the infrared spectrum.

## 2. Secure communications

While FSO communication offers high bandwidth and deployment flexibility, ensuring physical-layer security remains essential—especially in dynamic or contested environments. Indeed, developing mission capability of secure free-space MWIR and LWIR communications that optimize data transfer rates and BER while achieving physical-layer security such that eavesdroppers cannot decipher intercepted messages is of paramount importance. Traditional encryption and key exchange protocols are often impractical at the physical layer, and quantum communication systems, while promising, remain limited to near-infrared wavelengths due to the lack of compatible hardware in the MWIR and LWIR regions. Photonic chaos presents a compelling alternative, offering inherent unpredictability and high-dimensional complexity ideal for secure encoding [10]. In this approach, information is embedded in a broadband chaotic carrier, rendering it unintelligible without precise chaos synchronization between matched transmitter and receiver hardware. The security key of the system relies on chaos synchronization (or anti-synchronization) between a transmitter and a receiver. This technique is well established in the field of fiber optics [11], while MWIR and LWIR implementations further enhance covertness and resilience due to lower detectability and stronger atmospheric stability. Chaos-based encryption has achieved coherent private communication at 100 Gb/s over 800 km in fiber systems [12]. In free space, challenges such as turbulence and beam distortion can hinder synchronization, but recent work has shown that programmable optical processors can recover degraded chaotic signals in real time, restoring secure transmission even under strong turbulence [13]. Extending this paradigm to the MWIR, a secure indoor link at 5.6 µm using two DFB QCLs was first demonstrated in 2021 [14]. More recently, LWIR secure communication at 9.3 µm was achieved over 31 m at room temperature, maintaining BERs below 4% and enabling reliable decoding with forward error correction (fig. 2) [15]. While current data rates remain in the Mbit/s range, advances in QCL dynamics and chaos bandwidth are expected to push speeds toward 100–200 Mbit/s [16]. With their combined robustness, low intercept probability, and spectral discretion, MWIR and LWIR chaos-based links offer a powerful route to secure communication in harsh or visibility-limited environments where conventional and RF-based systems fall short.

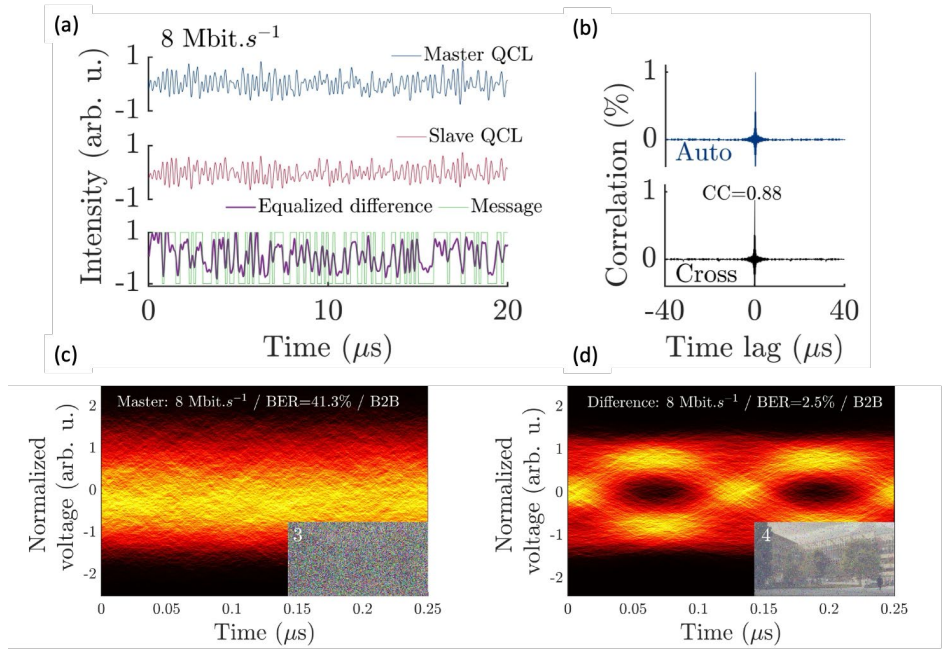

**Fig. 2.** (a) Intensity time traces of the master QCL (blue), the slave QCL (red) and the equalized difference (purple) with the 8 Mbits/s initial message (green) as a reference. (b) Auto-correlation of the master signal (blue) and cross-correlation between the master signal and slave signal (black), illustrating the high degree of synchrony. (c) Eye diagram and image recovery (inset) for an illegitimate receiver. (d) Eye diagram and image recovery (inset) for a legitimate receiver. Reproduced with permission from [15].

## 3. Remote sensing (LiDAR)

Light Detection and Ranging (LiDAR) enables precise distance measurements through time-of-flight or frequency-based analysis of laser light and is widely used in navigation, mapping, automation, and monitoring. While traditional systems rely on NIR wavelengths, their performance is limited in real-world conditions by scattering, turbulence, and absorption. This has motivated the shift to MWIR and LWIR, which offer better atmospheric transmission and reduced signal degradation. A key innovation in this transition is chaos LiDAR, which uses broadband chaotic waveforms for correlation-based ranging. Originally proposed by Fan-Yi Lin at 1.55 µm [17], chaotic LiDAR combines implementation simplicity with high precision. Recent MWIR demonstrations using ICLs achieved sub-centimeter accuracy over 3 m in the 3–5 µm window, showing strong potential for compact, low-noise ranging systems [18]. In parallel, frequency-modulated continuous-wave (FMCW) LiDAR using QCLs in the LWIR has achieved outdoor operation up to 54 m with precision under 2% of the absolute range [19]. These systems, however, still rely on complex frequency-sweeping electronics and remain vulnerable to external interference. To address this, the first chaos LiDAR in the LWIR has been demonstrated using a 9.34 µm DFB QCL under external optical feedback (fig. 3a and 3b). This system achieves sub-centimeter precision over 10 m with no need for modulation or frequency tuning [20]. Its simplified architecture enhances stealth, environmental resilience, and system compactness, which position LiDAR as a strong candidate for medium- and long-range sensing in both defense and commercial applications.

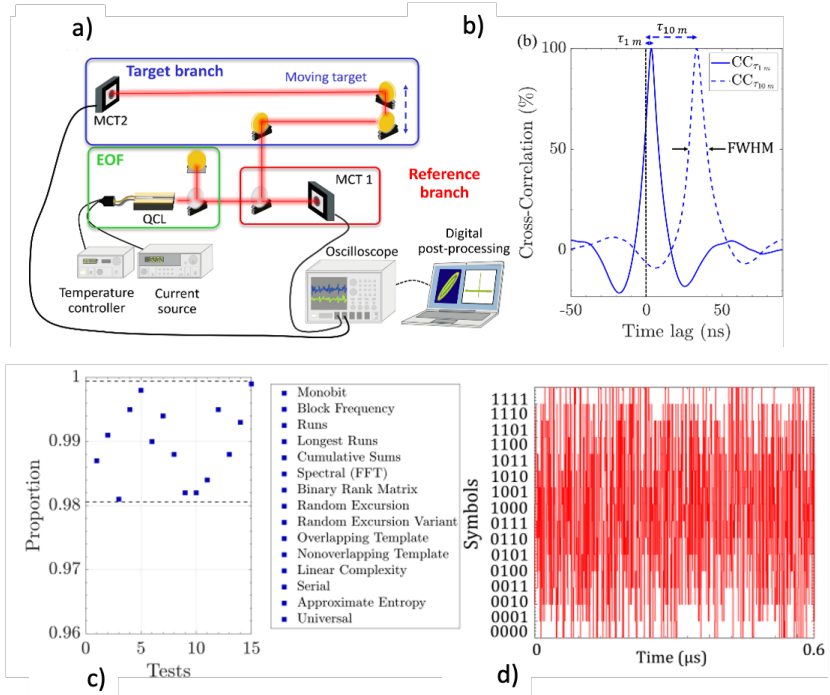

**Fig. 3.** (a) Experimental setup of LWIR chaotic LiDAR, with a QCL in external optical feedback configuration for chaos generation as the light source. (b) Cross-correlation function at target distances of 1 m (solid curve) and 10 m (dashed curve), which correspond to time lags of $\tau_1$=3.416 ns and $\tau_2$=33.424 ns, respectively. Reproduced with permission from [20]. (c) Proportion of sequences that pass the randomness verification for each of the 15 tests proposed by NIST. The dashed black lines delimit the confidence interval. (d) Random sequence after PAM-16 application. Reproduced with permission from [26].

## 4. Random number generation

Random number generators (RNGs) are essential for securing digital systems, enabling encryption, authentication, and stochastic modeling. While early physical RNGs based on thermal or quantum noise struggled with low signal levels and slow bitrates, the introduction of chaotic semiconductor lasers revolutionized the field [21,22], pushing generation speeds into the terabit-per-second regime through advanced digitization techniques, cascaded architectures, and parallelization. Fiber-based and silicon-integrated approaches have further expanded capabilities [23,24], but most efforts remain limited to the near-infrared. Recently, attention has shifted to the MWIR and LWIR, where enhanced atmospheric transmission and natural stealth make photonic chaos particularly attractive. In this context, Spitz et al. demonstrated RNG at 3 Gbit/s using a chaotic ICL at 4.1 μm [25]. Building upon this, high-speed RNG has recently been achieved in the LWIR using QCLs (fig. 3(c)). In particular, a system based on a 9.14 μm DFB QCL under external optical feedback, operating at room temperature, enabled the generation of 12.5 Gbit/s random bitstreams [26]. To do so, low-frequency thermal effects (<200 kHz) are filtered using a digital high-pass Butterworth filter, and wavelet denoising is applied to reduce white Gaussian noise.

Digitization involves representing each sample with 8 bits, from which the 4 least significant bits (LSBs) are extracted to enhance statistical randomness by removing residual correlations. PAM-16 is eventually applied to improve transmission efficiency (fig. 3d). These results highlight the ability of photonic chaos sources in the MWIR and LWIR to support high-speed, stealth-compatible RNG—an emerging technology well aligned with the needs of secure, long-range free-space communication.

## C. Future directions and outlook

One of the primary challenges in FSO communication lies in the limited transmission bandwidth. While detector capabilities have historically defined the upper limit of mid-infrared communication speeds, the performance bottleneck is now increasingly shifting toward lasers and modulators. Enhancing these components to match the bandwidth of state-of-the-art detectors could pave the way for 100 Gbit/s transmission rates—even without relying on complex signal processing—a milestone that would mark a significant technological breakthrough.

Encouraging progress has already been made: amplitude modulators with bandwidths in the 10 GHz range have recently been demonstrated [27], yet their operation across the full mid-infrared spectrum remains limited [28]. By contrast, phase modulation—essential for advanced coherent schemes—remains in its early stages, with only narrow bandwidths demonstrated to date [29]. Unlocking the full potential of coherent detection, which relies on both amplitude and phase encoding, will be essential for pushing toward higher data rates and improved robustness.

To overcome these limitations, several promising technologies are emerging. Lithium niobate-based modulators, for example, have already demonstrated bandwidths beyond 10 GHz and operation up to 6 µm [30]. However, the key challenge lies in integrating lithium niobate onto platforms compatible with wavelengths beyond 3 µm. Parallel progress in quantum-confined Stark-effect modulators, optimized for LWIR, is pushing bandwidths past 100 GHz, enabling compact, integrated systems for RF-to-optical conversion.

On the detection side, innovations are equally dynamic. Patch-array quantum well infrared photodetectors (QWIPs) have demonstrated frequency responses up to 220 GHz via mid-infrared photomixing, particularly when paired with advanced antenna designs [31]. Despite recent progress in terms of bandwidth enhancement, MWIR and LWIR detectors remain a key bottleneck—particularly for emerging applications such as chaos-based LiDAR and random number generation. Improving responsivity and detectivity is essential, as these parameters fundamentally determine system-level performance. Crucially, their values must remain stable across the GHz range, without degradation relative to those observed at lower frequencies (kHz–MHz). This frequency-dependent behavior, often insufficiently characterized, will be rigorously addressed in an upcoming publication.

However, another major obstacle is the constrained operating environment: nearly all QCL-based FSO experiments have been limited to short distances and indoor setups. Only a handful of early outdoor trials have been conducted, and these achieved only modest Mbit/s data rates [32,33]. Given that recent lab-scale experiments are approaching 100 Gbit/s, the need to translate these advances into real-world outdoor scenarios is both timely and urgent. Indeed, field validation is essential in proving the robustness and utility of MWIR and LWIR wavelengths under atmospheric conditions that would debilitate traditional NIR systems. Controlled outdoor experiments will be crucial for benchmarking the performance under real-world turbulence—where reduced scintillation, lower beam divergence, and high spatial discretion offer decisive advantages. Extending transmission distances to hundreds or even thousands of meters will also necessitate beam shaping techniques. Here, adaptive optics—well-established in other spectral domains—can be repurposed to correct for wavefront distortions in the MWIR and LWIR.

Looking ahead, the evolution of FSO systems will increasingly involve a convergence of MWIR and LWIR technologies with coherent communication strategies, photonic-assisted THz generation, and integrated unipolar optoelectronics [30]. Coherent systems, especially those using quadrature amplitude modulation (QAM), offer phase-sensitive encoding that enhances spectral efficiency and noise robustness. In the MWIR and LWIR domains, QCLs and quantum cascade detectors (QCDs) offer a promising hardware platform due to their shared intersubband transitions, passive operation, and compatibility with room-temperature conditions, which simplifies system architecture and reduces energy demands.

At the same time, momentum is building around photonic-assisted THz communication, enabled by heterodyne mixing of two narrow-linewidth QCLs. When combined with external modulation, this setup enables RF carrier generation in the 0.1–0.5 THz range, overlapping with key atmospheric transmission windows. Recent demonstrations have achieved data rates in the hundreds of Gbit/s over short distances.

Finally, QCLs and ICLs are poised to play a central role in the development of FSO quantum communication systems. By encoding information into the field quadratures using coherent or squeezed states of light, these technologies can support fundamentally secure, high-capacity links—pushing the frontiers of both classical and quantum communication.

### D. Conclusions

The development of MWIR and LWIR FSO communication systems represents a significant leap forward in the quest for high-performance, secure, and resilient photonic technologies. Operating in these atmospheric transparency windows not only improves robustness to turbulence, scattering, and absorption, but also enhances covertness and energy efficiency, positioning MWIR and LWIR as ideal candidates for next-generation communication and sensing platforms. Recent advances in quantum cascade lasers, modulators, and detectors have enabled high-speed data transmission, chaos-based encryption, physical random number generation, and precision LiDAR—all within compact and integrable architectures. Furthermore, the introduction of coherent detection schemes and heterodyne photonic-assisted THz generation has expanded the horizon of what MWIR systems can achieve. Yet, challenges remain in scaling data rates, optimizing modulation bandwidths, and ensuring long-range performance under real atmospheric conditions. Continued research is needed to bridge the gap between laboratory demonstrations and robust, deployable systems. In this context, MWIR and LWIR FSO technology is no longer a distant prospect—it is an emerging reality with transformative implications for secure communications, environmental monitoring, and defense applications.

*Acknowledgments*

The Authors acknowledge Pierre Didier, Olivier Spitz, Thomas Poletti and Huynah Kim. We also acknowledge the financial support of mirSense, the Institut Mines-Télécom, the Ecole Normale Supérieure, the University of Montpellier, the French National Agency (ANR), the Direction Général de l'Armement (DGA), and the Air Force Office for Scientific Research (AFSOR).

# 24. Mid-infrared integrated photonic LiDAR

**PO-YU HSIAO,[1†] PATRICK T. CAMP,[1,†] JASON MIDKIFF[2] AND RAY T. CHEN[1,2,\*]**

[1] *Chandra Department of Electrical and Computer Engineering, The University of Texas at Austin, Austin, TX 78712, USA*
[2]*Omega Optics, Inc., Austin, TX 7875, USA*
[†]*These authors contributed equally to this work.*
[\*]*chenrt@austin.utexas.edu*

## Introduction

LiDAR (Light Detection and Ranging), which is analogous to RADAR (Radio Wave Detection and Ranging), is a remote sensing application that utilizes laser emission and detection technologies to measure the distance of objects. The key difference between the two lies within the operational wavelength. RADAR is typically performed using the microwave spectral band, whereas LiDAR is typically performed using the infrared spectrum. Since the operating wavelength of LiDAR is much shorter than that of RADAR, LiDAR can provide high-resolution images and accurate spatial data [1]. LiDAR has been widely applied across various fields such as agriculture [2] and autonomous driving [3], due to its precise and real-time detection.

LiDAR systems are broadly classified by the specific process used to perform distance ranging. The two dominant approaches are Time-of-Flight (ToF) LiDAR and Frequency Modulated Continuous Wave (FMCW) LiDAR. ToF LiDAR determines distance by measuring the time delay between emission from a laser and reflection from a target. This approach to performing the time measurement can be further divided into direct ToF, which utilizes a pulsed laser source, and indirect ToF, which utilizes an amplitude-modulated or frequency-modulated laser source. An additional advantageous application of LiDAR allows for long distance and remote material detection, which utilizes differential detected power monitoring, aptly named differential absorption LiDAR (DIAL). The distinctions between these techniques have significant implications towards applicative use and complexity, and the operational differences will be discussed in the following sections.

## Direct ToF (Pulsed)

Direct ToF LiDAR transmitter systems emit a pulsed laser signal towards a target and the receiver system detects the backward reflected signal. Afterwards, the distance can be directly calculated by the using the measured time interval between two pulses via the following expression (1):

$$Distance = c \times \frac{\Delta T}{2} \qquad (1)$$

where $\Delta T$ is the time interval and $c$ is the speed of light. The main application for this approach is direct target detection. By taking multiple measurements on a moving target, the velocity of the target's movement can be determined as well. Additionally, by using multiple measurements over the surface of a targeted area, a topological mapping of the targeted area can be acquired. Generally, high-powered pulsed lasers are used, which can provide long range distance measurements. Furthermore, the low complexity of acquiring the ToF, when compared to the method described below, allows the use of a simple setup. Potential drawbacks are the power required to emit the pulsed signal, especially for high powered pulses, the necessity of high-speed time-to-digital converters (TDCs) or analog-to-digital converters (ADCs), which influence the precision and overall resolution, and potential accuracy limitations due to environmental factors such as temperature variations and signal interference.

## Indirect ToF - AMCW

Instead of emitting a pulse, a ToF LiDAR system emits an amplitude-modulated continuous wave (AMCW). By measuring the phase shift between the emitted and reflected continuous waves, the distance can be derived from the following equation (2):

$$Distance = c \times \frac{\Delta\varphi}{4\pi f_M} \tag{2}$$

where $\Delta\varphi$ is the phase shift (in radians) and $f_M$ is the modulated frequency (in Hz) of the emitted wave. This method is also known as phase shifted measurement and indirect detection. Typically, AMCW LiDAR has been utilized in applications that require short-to-medium ranged distance measurements. Key benefits are superb precision and fast measurement capability. However, a major drawback is more limited range when compared to pulsed ToF, which can limit the accuracy of long-distance measurements.

## Indirect ToF - FMCW

Frequency modulated continuous wave (FMCW) LiDAR is another approach that also utilizes the indirect ToF method. Unlike AMCW LiDAR, FMCW LiDAR emits a linearly-frequency-modulated (chirped) laser signal. The distance to the target is then calculated from the frequency difference, called the "beat frequency", between the transmitted and received laser signals. This relation can be expressed as:

$$Distance = c \times \frac{f_B \cdot T}{2B} \tag{3}$$

where $f_B$ is the beat frequency (in Hz), $B$ is the chirp bandwidth (in Hz), and $T$ is the chirp period. The most important benefit of this approach is the capability of measuring the range and velocity of the target simultaneously, since the beat frequency will vary due to the Doppler effect [4]. Due to the dual distance-velocity measurement capability of FMCW LiDAR systems, they are generally used only when both parameters are necessary. Despite the major benefits of this method, its complexity can increase the cost when compared to the other two, and it suffers from low signal-to-noise ratio (SNR) when applied to high-speed targets.

Fig. 1 provides a visual breakdown of the operating principles for pulsed, AMCW, and FMCW LiDAR.

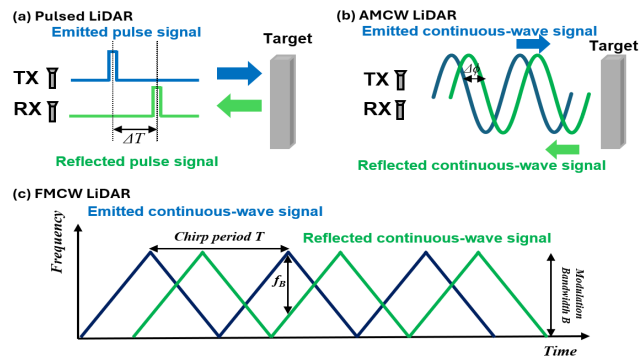

*Fig. 1. Schematics showing the working principles of (a) Direct ToF (Pulsed), (b) Indirect ToF (AMCW), and (c) FMCW LiDAR.*

## Differential Absorption LiDAR (DIAL)

Differential Absorption LiDAR (DIAL) is a remote sensing technique specifically designed for the quantitative detection of trace gas species by leveraging their unique optical absorption features. Unlike conventional LiDAR systems that determine range or velocity based on reflected signal timing or frequency shifts, DIAL systems are spectroscopically tuned to interrogate specific molecular transitions.

As depicted in Fig. 2, DIAL operates by transmitting two co-aligned laser beams toward a target or atmospheric column, where the "online" wavelength coincides with a strong absorption feature of the targeted analyte and the "offline" wavelength lies outside the absorption band.

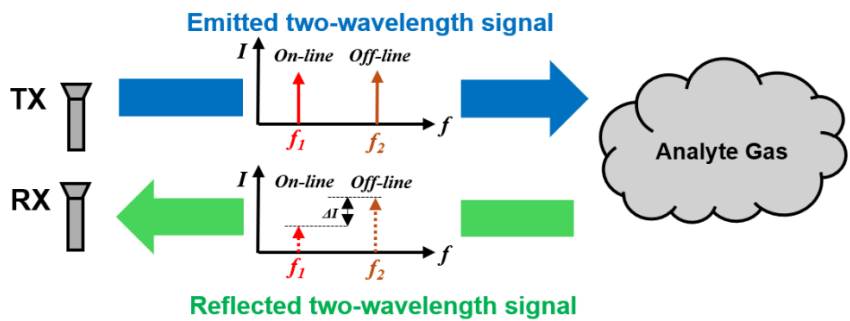

Fig. 2. Schematic showing the working principle of Differential Absorption LiDAR (DIAL).

By comparing the relative backscattered signal intensities for each wavelength, the system can infer the path-integrated concentration of the target gas. This differential measurement approach makes DIAL highly resistant to baseline fluctuations and system drifts, offering exceptional specificity and sensitivity in gas detection.

The mid-wave infrared (MWIR) spectral region (2–10 µm) is particularly advantageous for DIAL because many gases exhibit their fundamental vibrational absorption bands here. In contrast, near-infrared (NIR) DIAL systems rely on weaker overtone or combination bands. The significantly stronger absorption cross-sections of MWIR DIAL enable lower minimum detection limits, shorter path lengths, and enhanced selectivity.

Key analytes with high-value absorption features in the MWIR are shown in Table 1, where the analyte is provided in the first row and the peak absorption wavelength is provided in the second row:

**Table 1. Analytes of Interest for MWIR DIAL[a]**

| Carbon monoxide (CO) | Methane ($CH_4$) | Nitric oxide (NO) | Ammonium nitrate ($NH_4NO_3$) | Benzene ($C_6H_6$) | Nitrous oxide ($N_2O$) | Formaldehyde ($CH_2O$) |
|---|---|---|---|---|---|---|
| ~4.6µm | ~3.3µm | ~5.5µm | ~3.2µm | ~3.3µm | ~4.5µm | ~3.6µm |

[a] From NIST Standard Reference Database 35

These spectral targets are critical in applications such as environmental air quality monitoring [5], methane leak detection [6], standoff detection of explosives [7], and industrial emissions compliance [8].

Modern implementations have leveraged quantum cascade lasers (QCLs) for their narrow linewidth, high brightness, and wavelength tunability in the MWIR. Integrated photonic versions of DIAL using InP-based dual-laser transmitters [9, 5] are actively under development, targeting compact, high-speed solutions for on-field deployment.

## Current Status of MWIR LiDAR

Conventional LiDAR systems combine a mix of bulk optical components, such as mechanical beam-steering and lens systems for receiving the returned light, as well as on-chip integrated components, integrated emitters and detectors. However, these bulky components reduce the robustness and scalability. On the other hand, with the advance of photonic integrated circuit (PIC) technology (which is discussed in a separate sub-topic article of the Roadmap), integrated LiDAR has become crucial for enabling compact, cost-effective, and scalable sensing solutions across various industries.

Currently, near-Infrared (NIR) is the dominant spectral range for integrated LiDAR systems. This may be attributed to its technological maturity and ease of fabrication, due to cross-platform designs between conventional semiconductor fabrication techniques and NIR PICs. Additionally, the materials typically used to fabricate the optical components within a LiDAR system are highly cost effective, due to the widespread availability of NIR capable materials and the numerous PIC-capable CMOS and GaAs foundries when compared to the MWIR.

Due to its ready availability, multiple applications for integrated NIR LiDAR have been proposed and implemented commercially, including use within the automotive industry for autonomous driving. As an example, there currently exists a company that specializes in solid-state beam steering NIR LiDAR using its liquid crystal metasurface (LCM) technology that enables scalable, compact, and high-performance LiDAR solutions (LM10) for applications in autonomous vehicles, industrial automation, and consumer electronics [10]. A separate company in the commercialized NIR integrated LiDAR industry, with commercial specialty in silicon photonic optical phased array (OPA) FMCW LiDAR, currently offers solid-state beam-steering solutions for applicative use cases that include autonomous vehicles and robotics [11].

Despite the much greater maturity of LiDAR systems operating in the NIR, an MWIR LiDAR technology would have a lower level of solar background noise within multiple highly-transparent atmospheric windows [12, 13]. These inherent advantages may potentially make a fully-integrated MWIR LiDAR system more suitable than NIR LiDAR, especially in adverse weather conditions, for usage within the automotive industry for autonomous vehicles, within the datacom industry for low-loss point-to-point and point-to-net data transmission, and within the defense industry for long-range target distance acquisition and identification.

Because of these motivations, there have been numerous recent advances toward the development of integrated MWIR LiDAR systems. These include:

(1) MWIR-capable integrated non-mechanical 2D beam steering, by using optical phased arrays (OPAs) on an InGaAs/InP platform, which feature an array of non-redundantly placed emitters [14, 15].

(2) The incorporation of nonlinear optical processes, nominally photonic up-conversion through sum frequency generation (SFG), permitting the use of already-existing NIR detectors to develop an MWIR LiDAR system [16, 17, 18].

(3) The use of superconducting nanowire single photon detectors within MWIR LiDAR, which could be implemented in a more compact integrated LiDAR system [18, 19].

(4) The implementation of FMCW-based integrated MWIR LiDAR, which provides dual measurement of the target distance and velocity, and also keeps the progress of MWIR LiDAR in a relatively comparable trajectory to NIR LiDAR systems [20, 21].

## Current Status of MWIR DIAL

Recent research has demonstrated strong progress in realizing MWIR DIAL systems using QCLs, PICs, and advanced detection schemes tailored for compact, highly selective gas sensing.

A key development was the demonstration of integrated-path DIAL (IP-DIAL) using InP-based dual-laser transmitters, designed to operate near 3.2 μm for benzene detection [6]. The results of this research showed that such systems could achieve sub-ppm·m sensitivity using compact, modulated sources integrated with beam-scanning elements, positioning them for deployment in field environments for volatile organic compound (VOC) monitoring. The system used a dual-wavelength QCL emitter on an InP platform, emphasizing potential scalability and miniaturization using telecom-inspired fabrication strategies. In a separate study, a high-repetition-rate MWIR DIAL system operating at 500 Hz demonstrated precise, path-resolved detection of $NO_2$ with excellent signal-to-noise performance under laboratory and outdoor conditions [5]. These results validated the use of QCLs in combination with reflective telescope-style receivers and continuous wave operation modes.

Additional studies were performed to take advantage of open-path designs, where a QCL-based multi-gas DIAL platform capable of simultaneously detecting $CH_4$, $N_2O$, and water vapor ($H_2O$) across atmospheric columns was reported [8, 22]. Using pulsed QCLs and high-resolution digitizers, this design avoided mechanical scanning altogether, proving the viability of rapid, multiplexed atmospheric monitoring in industrial and environmental applications.

While most early-stage MWIR DIAL platforms were bulk-based, integration efforts have accelerated. For instance, one study demonstrated the simulation-driven design of modulated IP-DIAL systems for benzene, showing that optimized emitter modulation schemes and receiver optics could reduce minimum detection thresholds by over 40% in compact geometries [23]. At the same time, dual-laser photonic integration on InP and GaSb substrates has progressed toward designs with sub-GHz linewidth, tunability >30 cm$^{-1}$, and output powers ≥300 mW—key specifications for real-world DIAL operation [9]. Aside from investigating integrated emitter systems, DIAL systems incorporating upconversion-based MWIR photodetectors have been investigated for usage with DIAL. One such study demonstrated both IP-DIAL and DIAL systems with photodetector Noise Equivalent Power (NEP) of $240 \frac{fW}{\sqrt{Hz}}$, which displayed the potential for using this detector setup without the necessity of cryogenic cooling [24]

Collectively, these results emphasize that MWIR DIAL is no longer limited to research labs. The development of compact QCLs, integrated photonic transmitters, and high-performance detectors is pushing MWIR DIAL toward full system-level integration, enabling its use in industrial emissions sensing, chemical threat detection, and precision environmental monitoring.

## Challenges and opportunities

Despite the strong motivations and promising demonstrations highlighted previously, fully integrated MWIR LiDAR systems, particularly those tailored for DIAL, face technical and economic challenges. The complexity of integrating high-performance QCLs, precise beam steering elements, and ultra-sensitive MWIR detectors on a compact PIC platform remains a major barrier. As illustrated in Fig 3, integrated MWIR LiDAR systems comprise: an emitter (object a), typically a QCL or interband cascade laser (ICL); a beam-steering module (object b), such as an optical phased array (OPA), liquid crystal (LC), or MEMS-based device; and a detection component (object c). The emitting and beam steering components are typically interconnected via low-loss waveguides (red line), and their integration presents both engineering precision and material compatibility challenges that increase the development cost and complexity.

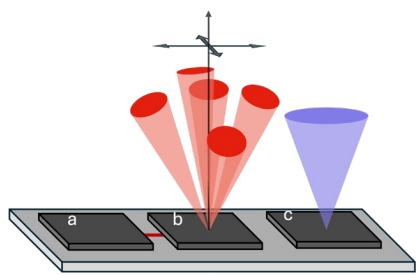

*Fig. 3. Conceptual schematic highlighting key components within an integrated 2D steerable LiDAR system: (a) the emitting component, (b) the beam steering element, and (c) the receiving component.*

One of the primary hurdles lies in the integration of MWIR emitters, primarily due to their power inefficiency, thermal demands, and material limitations. QCLs and ICLs, widely used in MWIR applications, require high input power to achieve the necessary optical output, typically 500 mW to 2 W per channel, for sufficient backscatter detection, with wall-plug efficiencies averaging 5–18% [26, 8]. This substantial power draw results in significant heat generation that will necessitate efficient thermal management to prevent thermally-induced wavelength drift or damage. Current cooling methods, such as cryogenic cooling or Peltier devices, add to the system bulk and cost, posing further challenges for integrated, field-deployable systems. While Si-based PICs efficiently manage thermal loads for NIR applications, their bandgaps corresponding to shorter wavelengths than the MWIR require the use of materials like InP [27], GaSb [28], and GeSn [29] for MWIR emission.

In addition, effective beam steering at MWIR wavelengths will introduce another set of critical challenges, particularly in relation to material losses, modulation techniques, and device scalability. Non-mechanical beam steering approaches, such as OPAs or LC metasurfaces, require materials with high electro-optic coefficients and low propagation losses in the 3–5 μm range, e.g., $LiNbO_3$ and InP. However, when scaled for integration, these materials experience degraded electro-optic performance and increased insertion losses. Furthermore, high-speed operation, ideally exceeding 10 kHz modulation frequency at voltage swings below 20 V, is required for real-time DIAL scanning applications, a specification that remains unachievable with current MWIR modulator technologies. Fabricating such high-performance MWIR modulators remains significantly more complex and cost-intensive than their NIR counterparts, further hampering the scalability of integrated beam steering solutions in this spectral region.

Detector integration in MWIR DIAL systems is equally challenging due to stringent requirements on sensitivity, thermal stability, and fabrication compatibility. DIAL systems rely on detecting small differences in signal strength between "on-line" and "off-line" wavelengths, which can be 1–3 orders of magnitude weaker than typical LiDAR returns. As such, detectors must achieve minimum detectable power (MDP) below 1 nW, with noise-equivalent power (NEP) in the range of $10^{-15}$ W/Hz$^{1/2}$ to resolve gas concentrations at parts-per-billion levels. Conventional MWIR detectors, including HgCdTe (MCT), type-II superlattice (T2SL), and InSb devices, can reach these sensitivities but require cryogenic cooling (typically <77 K) to maintain low noise and high quantum efficiency. This cooling necessity adds size, weight, and power constraints that are impractical for mobile or UAV-based systems. Upconversion detection schemes, which convert MWIR photons to NIR for detection by Si-based avalanche photodiodes (APDs), have shown promise, offering room-temperature operation with high dynamic range [24]. However, while proven in laboratory settings, upconversion detectors are still in early stages of integration with QCL-based DIAL systems.

Moreover, DIAL systems impose unique spectral and power constraints beyond those of conventional LiDAR. Accurate gas detection requires mode-hop-free QCL operation with dual-wavelength emission and precise alignment to narrow gas absorption features (typically <1

cm$^{-1}$). For example, CH$_4$ (3.3 μm), CO (4.6 μm), and NH$_4$NO$_3$ (3.2 μm) have highly specific absorption lines, necessitating high spectral purity and thermal stabilization in the QCL designs. While dual-emitter QCL architectures are being developed, maintaining stable wavelength separation and low linewidth (<1 MHz) under integrated conditions remains a significant challenge, especially when combined with the need for >500 mW optical output per channel [26].

Material systems and waveguide designs are also under active development to reduce propagation losses and enable broader MWIR coverage. GaSb-based PICs extend operation into the 5–5.5 μm range, but are limited by poor CMOS process compatibility and restricted foundry availability. In contrast, hybrid InP-on-Si integration offers better scalability but suffers from attenuated transmission efficiency beyond 3.5 μm. Recent innovations in chalcogenide glass waveguides and dielectric metasurface coupling provide new avenues for minimizing insertion loss and enhancing the mode confinement in MWIR PICs [25,22]. This offers optimism for compact, high-performance DIAL integration.

Despite these formidable challenges, the value of MWIR DIAL systems for selective remote gas detection creates a compelling opportunity for early commercialization and leadership in integrated MWIR photonics. Applications such as methane leak detection in energy pipelines, airborne monitoring of combustion gases in wildfire plumes, and standoff explosives identification (e.g., ammonium nitrate) highlight the practical demand for this technology. If systematically addressed, the integration of high-efficiency emitters, low-loss modulators, and room-temperature detectors can establish MWIR DIAL as the first widely deployable class of integrated MWIR LiDAR, enabling unmatched selectivity, trace gas sensitivity, and environmental resilience beyond the capabilities of NIR systems.

## Future developments to address challenges

The path toward realizing fully integrated MWIR LiDAR systems, particularly DIAL, hinges on overcoming a well-defined set of challenges in emitter efficiency, detector sensitivity, beam steering scalability, and material compatibility. These technical barriers, while significant, are being actively addressed through advancements in materials science, photonic device engineering, and scalable fabrication. A collaborative, cross-disciplinary approach will be essential to transition MWIR DIAL from laboratory prototypes to commercially viable, compact sensing platforms within the next three to five years.

A primary area of development is the creation of more efficient and thermally manageable MWIR laser sources. Current QCL and ICL technologies (which are discussed in separate articles of the Roadmap), while capable of high-power MWIR emission, are constrained by wall-plug efficiencies typically below 18%, necessitating complex thermal management solutions that hinder integration. Future integrated MWIR DIAL systems will depend on QCL architectures with enhanced wall-plug efficiencies, targeting values above 25%, while maintaining stable output powers exceeding 1 W per emitter. Advances in InP- and GaSb-based material systems are expected to play a key role in this improvement, enabling lower thermal resistance and better heat dissipation [26-28]. Additionally, mode-hop-free dual-wavelength QCL and ICL systems with narrow linewidths (<1 MHz) and spectral tuning over 15–50 cm$^{-1}$ will be required to accurately target narrow absorption bands of critical analytes such as methane, carbon monoxide, nitric oxide, and ammonium nitrate.

Simultaneously, future developments in MWIR photonic integration will focus on creating ultra-low-loss waveguides and packaging techniques that allow for compact, efficient routing of MWIR signals. Material platforms such as InP-on-Si and GaSb bonded to chalcogenide glass are being refined to support waveguide losses below 1.5 dB/cm [22]. These platforms offer the potential for dual-wavelength coupling, high confinement, and compatibility with on-chip thermal tuning elements. It is anticipated that optimized etching, sidewall smoothing, and advanced lithography techniques will further improve propagation efficiency and enable tight integration of dual-wavelength DIAL systems.

Detector-side innovations are expected to alleviate current limitations in sensitivity and operating temperature. While traditional MCT and T2SL detectors provide sufficient NEP levels for DIAL applications, their reliance on cryogenic cooling restricts portability. One of the most promising directions involves photonic upconversion detectors, which shift MWIR signals into the near-infrared (NIR) range to allow detection by low-cost, room-temperature Si-based avalanche photodiodes. Initial demonstrations of such upconversion systems have achieved NEP values below $1\times10^{-15}$ W/Hz$^{1/2}$, which are suitable for detecting path-integrated gas concentrations at the ppb·m level [16-18, 24]. Concurrent research into superlattice-enhanced photodiodes and superconducting nanowire single-photon detectors (SNSPDs) may further extend the sensitivity and dynamic range of integrated MWIR DIAL systems, although these remain less mature for practical deployment. Resonant cavity infrared detectors (RCIDs, discussed in a separate article of the Roadmap) that enhance the sensitivity to a narrow-band laser signal may also play a role.

Beam steering technologies are poised for substantial evolution, moving away from bulky mechanical systems towards fully integrated, non-mechanical solutions that are compatible with MWIR wavelengths. Metasurface-based reflectarrays and electro-optic phased arrays (eOPAs) operating in the 3–5 μm range are under development, aiming to provide field-of-view (FOV) coverage exceeding 30°, steering speed greater than 1 kHz, and angular resolution finer than 0.5°. These developments will enable DIAL systems to dynamically interrogate targets over wide spatial domains, which is essential for applications such as topographic gas mapping and standoff detection. Further material innovations, such as the use of chalcogenide or telluride-based electro-optic films, are expected to improve modulation speed and reduce power consumption, facilitating real-time beam steering in portable systems.

At the system integration level, compact MWIR DIAL platforms are anticipated to combine high-power dual-section QCLs and low-loss PICs with on-chip filtering, metasurface-based beam scanners, and upconversion detector arrays in packages with volumes under 2 liters. These systems will leverage FPGA-driven control electronics for real-time signal processing, enabling effective path lengths over 100 meters and detection thresholds below 50 ppb·m for target gases such as $CH_4$, CO, NO, $H_4NO_3$, $C_6H_6$, $N_2O$, and $CH_2O$. Such performance benchmarks would not only rival but exceed those of current NIR-based DIAL systems, while opening new detection capabilities for MWIR-specific analytes critical to environmental monitoring, industrial safety, and defense applications.

In conclusion, MWIR DIAL currently represents the most attainable and application-ready frontier for integrated MWIR LiDAR technology. Its intrinsic alignment with niche but high-value use cases, combined with rapidly advancing component maturity, positions it to be the first integrated MIR LiDAR system to reach widespread deployment. Continued investment in materials, device architecture, and scalable manufacturing will ensure that MWIR DIAL can deliver on its promise of unmatched selectivity, sensitivity, and operational versatility in real-world environments.

## Concluding Remarks

In summary, LiDAR is a remote sensing technology that utilizes lasers and photodetectors to measure the distance to a target by determining the time-of-flight (ToF) of an emitted signal. The core of ToF LiDAR operation involves either direct detection, typically facilitated by a pulsed laser source, or indirect detection, which is accomplished through an amplitude-modulated (AMCW) or frequency-modulated (FMCW) continuous-wave technique. Direct ToF LiDAR is favored for its straightforward design but suffers from low SNR at longer distances as well as the requirement of high-speed electronics. AMCW ToF LiDAR, which has an established commercial presence, is often limited in long-range sensing. FMCW ToF LiDAR stands out by enabling simultaneous measurement of both distance and velocity, yet it is limited by coherence length, system stability, and implementation complexity.

MWIR DIAL offers unique advantages in gas sensing and remote chemical detection, enabled by the strong fundamental vibrational bands of analytes such as methane, carbon monoxide, nitric oxide, benzene, and ammonium nitrate. The integration of MWIR DIAL—using dual-laser quantum cascade emitters, low-loss photonic platforms, and room-temperature upconversion detectors—has shown promising experimental demonstrations and is on a trajectory for deployable systems within the next 3 to 5 years.

The achievement of integrated MWIR DIAL could revolutionize multiple industries. Autonomous vehicles would benefit from the MWIR's resilience to fog, smoke, and rain, enabling robust navigation in degraded conditions. Energy sector and pipeline monitoring could leverage high-specificity open-path detection of trace leaks such as methane or hydrogen sulfide. Defense and security applications would gain from standoff detection of hazardous chemicals, explosive precursors, and trace atmospheric pollutants. Even aerospace and satellite-based Earth observation platforms could incorporate MWIR DIAL for long-range monitoring of volatile gases and climate indicators.

As research continues and barriers are methodically overcome, the successful development of fully-integrated MWIR LiDAR systems—and in particular MWIR DIAL—will mark a major milestone in the evolution of photonics, spectroscopy, and remote sensing. Beyond the lab, it will transform how we perceive, map, and respond to our environments, offering next-generation sensing capabilities across infrastructure, defense, energy, and climate.

## Funding

Content in the funding section will be generated entirely from details submitted to Prism. Authors may add placeholder text in the manuscript to assess the length, but any text added to this section in the manuscript will be replaced during production and will display official funder names along with any grant numbers provided. If additional details about a funder are required, they may be added to the Acknowledgments, even if this duplicates information in the funding section. See the example below in Acknowledgments.

## Acknowledgments

Acknowledgments should be included at the end of the document. Additional information crediting individuals who contributed to the work being reported, clarifying who received funding from a particular source, or other information that does not fit the criteria for the funding block may also be included; for example, "K. Flockhart thanks the National Science Foundation for help identifying collaborators for this work."

## Disclosures

The authors declare no conflicts of interest.

# 25. Mid-Infrared Thermophotovoltaics

## GÉRARD DALIGOU,[1] SEAN MOLESKY,[1] AND OUSSAMA MOUTANABBIR[1, *]

[1]*Department of Engineering Physics, École Polytechnique de Montréal, C.P. 6079, Succ. Centre-Ville, Montréal, Québec, Canada H3C 3A7*
*oussama.moutanabbir@polymtl.ca*

## Overview

The overlap of blackbody emission from 500-2000 K heated objects with the midwave-infrared (MWIR) spectral band has sparked a surge of interest in developing thermophotovoltaic (TPV) cells for specific applications such as harvesting waste heat in industrial, automotive, and aerospace settings, remote power generation, compact power supplies, and self-powered sensors. Addressing these emerging applications requires the incorporation of narrow bandgap semiconductors to ensure optimal overlap between the TPV cell's absorption range and the blackbody emission spectrum. Figure 1(a) illustrates the overlap of photoresponsivity for an InAs cell with peak at 3.35 μm with blackbody emission spectra at relevant source temperatures of 865 K and 1500 K. The 1500 K spectrum reflects the approximate emission characteristics of the silicon carbide (SiC) thermal source typically used to test TPV cells. Figure 1 (b) displays the cell's room-temperature I-V characteristics measured both in the dark and under illumination. For these measurements, the SiC thermal source was placed at a distance ($d_{sd}$) about 3 cm above the top of the device. The illuminated curve demonstrates increased photocurrent with a short-circuit current of 1.68 mA. However, the measured open circuit voltage of approximately 13.9 mV was limited by significant defect-related recombination losses and the high dark current typical of a narrow-bandgap semiconductor. These measurements underscore both the potential and the challenges associated with MWIR TPVs. In fact, the need for optimized narrow bandgap semiconductors in this range of the electromagnetic spectrum raises specific questions pertaining to materials development, radiation-matter interactions, carrier transport, and device processing [1]. Herein, we discuss these aspects, review the current state-of-the-art, and outline potential future directions to advance the design and implementation of efficient MWIR TPV cells.

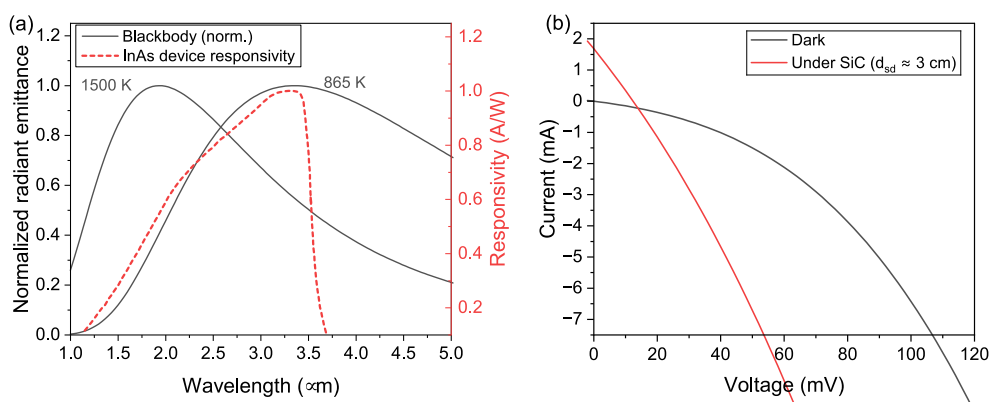

**Figure 1. Optoelectronic characterization of an InAs TPV cell. (a) Spectral photoresponsivity (dashed red), overlaid with normalized blackbody emission spectra at 865 K and 1500 K (black curves). (b) I-V curves for the cell measured at room temperature in the dark (black) and under radiation from a SiC thermal source at approximately 1500 K (red).**

Historically, TPV device design has been heavily influenced by solar photovoltaic (PV) technology. Early TPV devices, such as those proposed by Wedlock in 1963 [2] and Swanson in 1978 [3], were straightforward adaptations of PV structures that were optimized for thermal emission rather than solar radiation. The key adaptations relied on the choice of semiconductor materials suitable for the efficient absorption of infrared radiation, and on finding a way to exploit low energy photons that are not absorbed and lost in a conventional PV device. Despite recent strides, the development of TPV devices for practical applications is still confronted with multifaceted challenges that demand interdisciplinary approaches. Figure 2 illustrates the main building blocks of a TPV system: (1) A thermal emitter that converts heat from a hot source (combustion, nuclear, concentrated solar cells, ambient heat waste, etc.) into infrared radiation; (2) A TPV cell that converts the infrared radiation into electrical power; (3) A filter or integrated spectral control to recapture or restrict the emission of photons outside the cell's absorption band; and (4) An electrical convertor that collects generated charge carriers and routes them to an external circuit. A fundamental figure of merit for TPV systems is the power conversion efficiency (PCE), typically defined as the ratio of electrical output power to the net radiative power absorbed by the cell under illumination from the emitter. This metric, sometimes referred to as the pairwise efficiency or radiative heat conversion efficiency, captures how effectively the cell converts absorbed photons into electrical energy, accounting for internal factors such as carrier extraction, recombination losses, and photon absorption [4, 5]. While this definition does not include broader system-level losses, such as photo-inactive areas on the cell or convection losses at the emitter, it provides a useful means of decoupling material and device-level limitations from system integration losses [5]. In the following, we discuss recent progress in addressing the challenges facing the optimization of each of these main building blocks and outline evolving approaches to achieving higher efficiency TPV cells with focus on operation in the MWIR range.

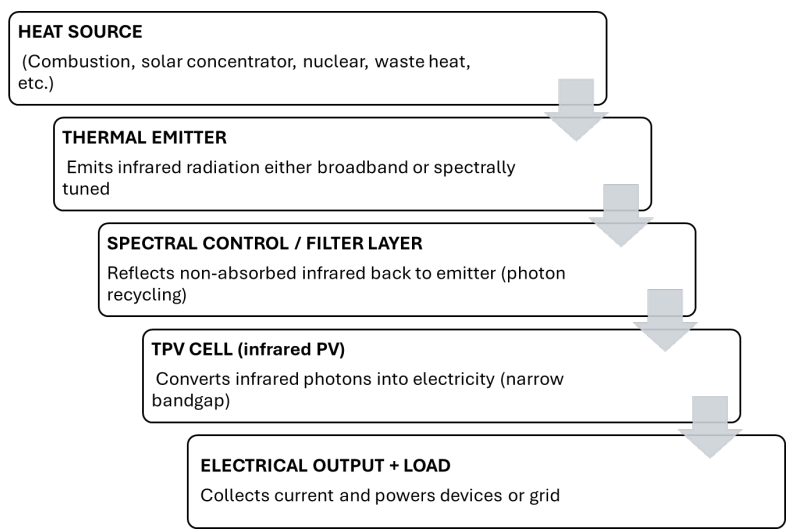

**Figure 2**: Illustration of the main building blocks of a TPV module.

**Current status, Challenges, and Opportunities**

**1. Basic physics of thermophotovoltaic systems**

The behavior of a p-n diode under external illumination is often analyzed using the idealized Shockley-Queisser model of perfect internal quantum efficiency and perfect electrical transduction (infinite carrier mobilities) [6]. Within this framework, two central mechanisms restrict the achievable power extraction in a homojunction p-n diode: the breadth of the blackbody distribution [7], and the low power density available in typical terrestrial heat sources and solar radiation [8, 9]. The spectral extent of the thermal distribution limits the efficiency via a simple trade-off. If the position of the photovoltaic bandgap is set too high, a large portion of the spectrum will not have sufficient energy to create charge carrier excitations [10]. If the position of the bandgap is set too low, however, the majority of excited charge carriers will have energy in excess of the bandgap. Unless these hot carriers can be extracted from the cell before they thermalize with the semiconductor material, this energy will be lost as irrecoverable heat. The limitation set by the accessible spectral power density of the source is, in contrast, unidirectional. Arguing from the principle of detailed balance [11, 12], the amount of power that can be extracted from a system is generally related to the deviation of its population distribution from the equilibrium distribution that maximizes entropy. In the context of photovoltaic energy conversion, if the density of excited charge carriers does not substantially exceed the ambient thermal population, then a large percentage of the absorbed power must remain trapped within the photovoltaic material [13], which manifests as a reduction of the operational voltage and current at the power-point.

The motivation of the filter or spectral control element of a MWIR TPV (Fig.2) can be understood as a direct means of circumventing this first restriction. Through material resonances or the design wavelength scale of the geometric features, both filter and radiative thermal emission characteristics can be shaped to provide remarkable spectral selectivity [14, 15, 16, 17, 18, 19, 20, 21]. By creating a narrow peak just above the bandgap energy, or more simply suppressing all radiative emission below a specified cut-off, the thermalization losses that would otherwise undercut the efficiency of a single junction cell can be substantially mitigated [22].

The second loss mechanism can be addressed by bringing the cell into the near-field of the emitter. When applied in the far-field, spectral shaping can only reduce the total radiation incident on the cell compared to an idealized blackbody [23], which reduces the overall power input to the cell and hence its thermodynamic efficiency. Accordingly, to obtain useful system efficiency and power density metrics, emitter temperatures exceeding 1200 K must typically be supposed [5, 15]. However, if the spatial separation between the emitter and the cell is only a small fraction of the bandgap wavelength, this barrier is lifted: Although a dipole radiates only a finite amount of power to the far-field, its electric field supports an unbounded energy density. As a result, when a lossy material is introduced into the near-field of a stochastic current fluctuation, the amount of power that may be transferred via the electromagnetic field can greatly surpass expectations based on Planck's law [24, 25, 26, 27]. This mechanism underpins phenomena such as Förster resonance energy transfer [28] and fluorescence quenching in the presence of metallic films or nanoparticles [29]. For example, at the gap phonon-polariton resonance of 10.55 μm, stochastic electrodynamics predicts that the radiative heat transfer between two SiC half-spaces separated by a 10 nm vacuum gap is roughly four orders of magnitude higher than the far-field blackbody limit [30].

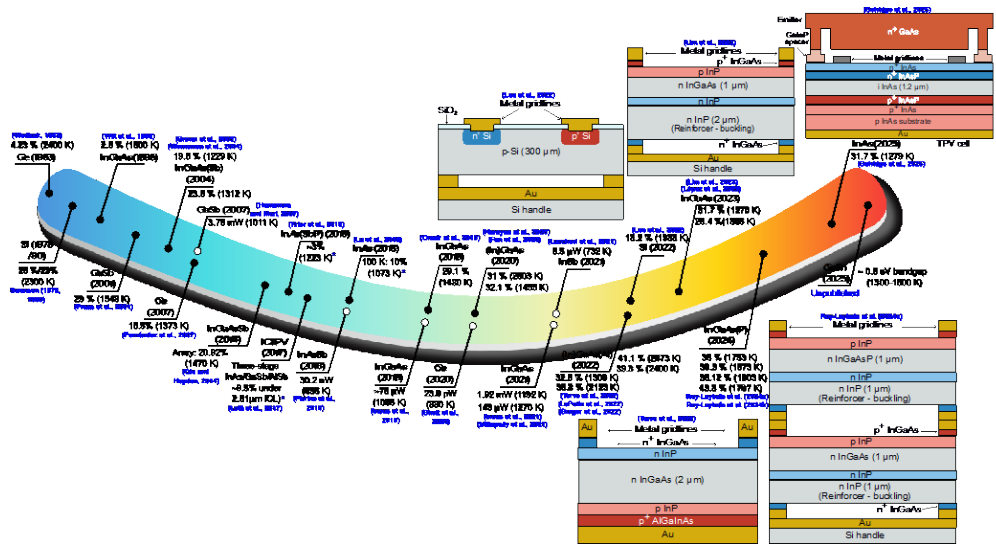

Figure 3. Timeline of TPV device development, summarizing key achievements since the introduction of the first Ge-based TPV device in 1963 [2]. Filled circles represent far-field TPV devices, open circles denote near-field TPV devices, and asterisks (*) highlight devices using a PCE definition similar to that of solar cells, i.e., the ratio of electrical output power to the power incident on the TPV cell surface under emitter illumination. GeSn-based TPV devices with an active area diameter of about 1 mm were fabricated in 2025, although the estimation of their PCE is still under investigation. Citations follow author-year format for clarity (a/b indicate multiple works by the same author and year). Device schematics are adapted from references [31] (Adapted with permission from [31]. Copyright © 2022 Elsevier Inc.), [16] (Adapted with permission from [16]. Copyright © 2022, American Chemical Society), [18] (Adapted with permission from [18]. Copyright © 2023, American Chemical Society), [19] (Adapted with permission from [19]. Copyright © 2024, American Chemical Society), and [32] (Adapted with permission from [32]. Copyright © 2024 The Author(s). Advanced Materials published by Wiley-VCH GmbH). Additional references used only in this figure: [33, 34, 35, 36, 37, 38, 39, 40, 41]

## 2. Practical considerations and challenges

### Cell modelling

Over the years, several materials and device designs have been developed and tested to engineer MWIR TPV cells. Figure 3 summarizes the evolution of TPV devices since 1963, highlighting both far-field and near-field architectures. Regardless of the operating regime, far-field or near-field, the accuracy of modeling a TPV system depends critically on the electronic and optical properties of the PV cell. While the Shockley-Queisser model is useful for estimating the efficiency limit, it does not accurately describe real devices. A more complete formalism solves the Poisson and drift-diffusion equations to allow a coupled analysis of the device's optical and electrical properties. The Poisson equation relates the spatial distribution of charge to the resulting electrostatic potential, while the drift-diffusion equations describe variations of the quasi-Fermi levels, and thus current densities, under external illumination and bias. While this formalism is generally applicable to PV devices regardless of the bandgap, it becomes particularly important for accurately capturing the carrier dynamics in narrow bandgap semiconductors. In devices based on wide bandgap materials, such as silicon, intrinsic recombination mechanisms are often negligible and have limited impact on the conversion efficiency [42]. Conversely, in narrow bandgap materials these loss mechanisms can become dominant, depending on material quality, and must be carefully accounted for.

Moreover, a precise description of the radiative processes requires a thorough knowledge of the complex refractive indices and absorption coefficients across the relevant spectral range. In practice, the material absorption spectra must be carefully characterized, especially when dealing with narrow bandgap semiconductors in which parasitic absorption mechanisms can significantly degrade the device performance. One such mechanism is free carrier absorption, which becomes particularly relevant in doped semiconductors and heavily photoexcited materials [43]. Free carrier absorption reduces the number of photons that contribute to photocurrent generation, since part of the absorbed incident radiation does not produce any electron-hole pairs [44]. This non-productive absorption also leads to undesired heating of the cell, which can potentially increase the recombination rates and further reduce efficiency [45]. It is often modeled using the Drude-Lorentz formalism or a generalized parameterization, which may require reliable experimental fitting to improve the accuracy [44, 46]. However, this formalism has been shown to overestimate free carrier absorption both above and below the bandgap, especially for highly doped InAs [47]. Thus, for realistic device modeling, accurate experimental characterization of the optical constants is indispensable. When such data are unavailable or incomplete, the resulting uncertainties can propagate through modeling of the thermal emission and generation rates, ultimately impacting the performance predictions for both far-field and near-field TPV (NF-TPV) devices.

## Spectral selectivity

Practically, a key challenge in designing an MWIR TPV systems is to realize a thermal infrared emitter that suppresses low energy photons [21]. Selective emitters, which integrate the thermal emitter with optical filters into a single component, have been developed using rare-earth oxides [48], photonic crystals [49, 50, 51], metamaterials [52, 53, 54], or bulk materials combined with anti-reflective coatings [21]. While this approach can effectively suppress out-of-band photons, it also reduces the intensity of radiation with energy above the bandgap [50]. Moreover, as discussed above, these emitters must operate at temperatures exceeding 1000 K. At such extreme temperatures, many materials undergo irreversible structural changes that hinder their selective emissivity. For instance, the metal layers used in emitter fabrication are prone to oxidation at high temperatures, and some materials may even evaporate [5]. One approach proposed to mitigate this issue involves depositing a protective coating, such as $HfO_2$, on the emitter surface to maintain stable optical performance at elevated temperatures [49, 55]. Another solution is to use a refractory materials like TiN in the emitter fabrication to ensure thermal stability [56]. While overcoming the oxidation of metal-containing selective emitters remains challenging, promising solutions are under active investigation. These include emitters based on perovskite and rock-salt heterostructures [57], all-oxide selective emitters [51], cuboid metamaterials [54], Tamm plason polaritons structures with distributed Bragg reflector front surface [58], and dual-coherence enhanced absorption [59], all of which exhibit stability at temperatures around 1273 K.

Spectral control can also be achieved by using optical filters as separate components. These filters are positioned either between the emitter and the photovoltaic cell or directly on top of the TPV device [14]. This method, commonly referred to as reflective spectral control, aims to prevent low-energy photons from reaching the TPV cell by reflecting them back toward the emitter, which, like the selective emitter approach, reduces the number of high-energy photons reaching the TPV device and thereby decreases the output power density. While this reduction is less significant than with selective emitters, optical filters face similar physical limitations. Namely, they exhibit performance degradation over time, especially under high-temperature operation where severe deterioration can occur [14]. Unlike the previously-discussed strategies, a third method, referred to as transmissive spectral control [5, 31, 60], does not block any photons from entering the TPV cell. Instead, all of the emitted infrared radiation is directed toward the cell, allowing high-energy photons to be absorbed for the generation of electron-hole pairs (EHPs) as it is essential for power production. Meanwhile, low-energy photons that are not absorbed pass through the device and are reflected to the emitter using a back-surface reflector (BSR) or highly reflective rear mirror. For this approach to be effective, parasitic absorption, particularly due to free carriers within the TPV cell, must be minimized [17].

Transmissive spectral control has emerged as the leading strategy in recent years, especially for TPV devices based on III-V semiconductors [19, 20]. However, it faces its own set of challenges. Apart from free carrier absorption within the device, a key concern is minimizing optical losses at the interface between the substrate and the reflective mirror. For instance, conventional reflectors such as gold (Au) introduce a loss of approximately 5% at the substrate-Au interface with each reflection cycle [15]. Several mitigation strategies have been explored, with the most widely adopted being to introduce an air gap between the TPV device and the metal reflector to form an optical cavity. This configuration enables near-lossless Fresnel reflection at the TPV-air interface to effectively reflect out-of-band radiation [15, 20]. The air cavity is typically created by bonding a gold support grid, previously deposited and patterned on the backside of the TPV material, onto an Au/Si handle using thermocompression bonding [15]. While this technique offers excellent spectral control, the resulting air-bridge structure can suffer from mechanical instabilities such as buckling and cracking. Buckling occurs due to a combination of compressive distortion in the Au support grid and thermally-induced pressure buildup in the air cavity during bonding, which pushes the TPV film away from the substrate [61]. In such cases, the deformed TPV membrane forms an optical cavity of variable length to the bottom Au mirror, leading to the emergence of multiple cavity modes in the reflection spectrum and resulting in parasitic reflection losses at sub-bandgap energies [18]. The severity of buckling can be mitigated by adjusting the spacing of the support grid lines and modifying the thickness of the TPV structure. For example, Lim et al. realized flat, single-mode cavities by mechanically reinforcing the TPV membrane with a thick transparent n-InP buffer layer [18]. This buffer provided sufficient mechanical stiffness while maintaining enough electrical conductivity to avoid a significant increase in the series resistance [18]. Using a similar approach, Roy-Layinde et al. demonstrated 0.9 eV $In_{0.69}Ga_{0.31}As_{0.67}P_{0.33}$ TPV square devices with a maximum power density of $\approx 0.9$ W/cm$^2$ and PCE of 42% under illumination from an 1805 K gray body radiation [20]. However, a trade-off remains between enhancing the mechanical stability and minimizing free-carrier absorption and recombination losses.

Enhanced spectral utilization can also be achieved by integrating multiple absorbers within the TPV device. Tandem TPV cells, comprising multiple junctions, can potentially surpass the spectral efficiency of a single-junction cell. This improvement stems from a reduction in hot-carrier thermalization losses and lower resistive losses, as the device operates at reduced current densities [19, 22]. For example, LaPotin et al. reported the fabrication and measurement of two-junction $Al_{0.15}Ga_{0.55}In_{0.3}As$ (1.2 eV)/$In_{0.3}Ga_{0.7}As$ (1.0 eV) TPV devices that reached a maximum efficiency of 39% and an electrical power density of $\approx 1.8$ W/cm$^2$ under an emitter temperature of 2400 K [22]. Similar to their PV counterparts, standard tandem TPV cells typically employ a heavily doped semiconductor tunnel junction to connect the top and bottom sub-cells in series. However, this approach can lead to considerable parasitic free carrier absorption at sub-bandgap energies, which negatively impacts the reflectance and overall device efficiency [45]. Additionally, the complex fabrication of these devices poses cost and scalability challenges, particularly for III-V materials. To address these limitations, Roy-Layinde et al. proposed an alternative air-bridge tandem architecture based on multiple air cavities, which enables the integration of materials with different bandgaps without relying on tunnel junctions [19]. This design uses mechanical bonding to stack electrically independent sub-cells, with patterned metal grid lines replacing the tunnel junctions to minimize free carrier absorption of the sub-bandgap photons. In this configuration, each cell is separated by an air-bridge that significantly enhances the reflectance of low-energy photons compared to conventional back reflector designs. Moreover, the use of three distinct metal layers enables flexible interconnection schemes. This design supports two-, three- or four-terminal operation by contacting the appropriate electrodes. It also eliminates the need for epitaxial growth between the sub-cells, offering more freedom in material selection. Using this strategy, $In_{0.69}Ga_{0.31}As_{0.67}P_{0.33}$ (0.9 eV)/$In_{0.53}Ga_{0.47}As$ (0.74 eV) hetero-tandem TPV cells have been demonstrated with a maximum power conversion efficiency of 39.3% under a multi-terminal operation [19]. Nonetheless, this architecture presents several challenges. A key trade-off remains between improving the mechanical stability and minimizing free-carrier absorption and recombination losses. In addition, the three metal grid layers must be precisely aligned. Misalignment, especially at the junction between sub-cells, can cause shadowing of the bottom cell, increase the series resistance, and reduce the quantum efficiency [19].

In addition to tandem architectures, interband cascade photovoltaic (ICPV) devices have been shown to achieve enhanced spectral utilization through the stacking of multiple absorber stages without the need for conventional tunnel junctions. Originally inspired by developments in mid-infrared interband cascade lasers (ICLs) [discussed in a separate sub-topic article of this Roadmap], ICPV devices connect multiple absorber stages via interband tunnel junctions, leveraging the type-II broken-gap alignment of the InAs/GaSb/AlSb material system. This architecture alleviates several limiting factors of bulk narrow bandgap TPV devices, including high saturation dark current density, relatively low absorption coefficient, short diffusion length, and material quality issues [62, 63, 64]. The cascade structure also enables high carrier collection efficiency even in materials with weak absorption and a short diffusion length, while simultaneously reducing the series resistance by trading photocurrent for operating voltage without sacrificing power conversion efficiency [65]. Moreover, the absorber bandgaps in the different stages can be tailored to achieve spectral filtering and efficient photocurrent generation from a broad infrared spectrum. Under blackbody radiation [66, 67], ICPV devices based on InAs/GaSb type-II superlattices (T2SLs) and operating at 80 K exhibited high open-circuit voltages ($\sim 1.1$ V $> E_g/e$), demonstrating that the cascade stages contributed additively to the voltage. However, the performance at 300 K degraded significantly ($V_{oc} \sim 5.7$ mV, power density < 0.003 mW/cm²), due to high saturation current densities and short carrier lifetimes driven by non-radiative recombination. These limitations are exacerbated by the need to match current across all the stages, which is particularly challenging under broadband illumination [65]. To mitigate this, mid-infrared ICLs have been employed as narrowband, high-power sources [68, 69, 70, 71, 72, 73, 74, 75]. For instance, Lotfi et al. demonstrated that with illumination near 4.3 μm, ICPV devices, based on similar 7-stage wafer reported in [67], operating at 300 and 340 K benefited from voltage scaling to attain $V_{oc}$ values exceeding the bandgap of a single stage [72]. Yet, these devices still suffer from high dark current densities, conversion inefficiencies, and photocurrent mismatch due to uniform absorber thicknesses. Subsequent designs addressed these issues by varying the absorber thickness across the stages to improve current matching [68, 69, 70, 71, 73, 74, 75]. This resulted in higher short-circuit current densities and improved efficiencies of up to 3.6% in 5-stage devices, especially when a thinner absorber enabled efficient carrier collection [68]. However, attempts to scale up to 16- and 23-stage devices encountered performance degradation linked to increased current mismatch and shorter carrier lifetimes, both of which were attributed to material quality limitations and fabrication defects such as sidewall leakage and insufficient passivation [65, 74].

## Near-field

The transition to near field TPV systems, while promising for ultra-compact and high-performance energy conversion, creates additional issues. Pointedly, the need to maintain a high-fidelity, nanometer-scale gap between the emitter and the NF-TPV cell is technically daunting [76, 77, 78, 79, 80, 81, 82, 83, 84, 85]. Although piezoelectric micro-positioners have demonstrated control of the emitter-cell separation with sub-nanometer resolution and near-field power densities [76, 80, 81, 82], slight angular misalignment can cause the gap to vary substantially across the device area, limiting the effective emitter size to about a 75 µm radius [81]. As a result, despite their effectiveness in enhancing performance at small scales, these configurations are unsuitable for applications requiring high total converted power. Alternative approaches involve the integration of nanometer-scale spacers directly between the emitter and the cell [79,83,84], the use of nano-electromechanical system designs [85], or the fabrication of thin supporting columns through highly selective etching processes [86, 87, 88, 89]. However, all of these strategies introduce parasitic conductive heat transfer that must be strictly minimized to preserve overall system efficiency [84, 90, 91, 92, 93, 94]. Deeply subwavelength gaps also make it more difficult to shape the spectral distribution of the radiative power transfer [95, 96, 97]. Even assuming an emitter operating temperature below 1000 K, a significant portion of the thermal spectrum lies at photon energies well below the semiconductor bandgap. Because the tunneling barrier strength decreases exponentially with decreasing photon energy [98], low-energy sub-bandgap photons can contribute disproportionately to energy transfer, resulting in the creation of heat loads that largely offset the benefits of increased carrier generation.

To enable large-area NF-TPV operation with high power density under relatively low emitter temperature, Selvidge et al. introduced an innovative epitaxial co-fabrication approach [32]. Their device featured a 0.28 cm² self-aligned emitter-cell pair separated by a nominal nanoscale vacuum gap of approximately 150 nm. The emitter structure consisted of an n$^+$ GaAs substrate with an epitaxially grown GaInP spacer layer. Selective etching of the GaInP layer defined four supporting posts, which simultaneously determined the vertical separation between the emitter and the TPV cell. The cell, based on a p-i-n InAs/InAsP structure, delivered approximately 1.22 mW of power at an emitter temperature of 733 K. This corresponded to a 25-fold improvement compared to the same cell operated under far-field conditions with a 10-15 µm gap. However, the elevated radiative flux in near-field operation was shown to degrade the performance due to cell heating and increased series resistance. In such cases, robust thermal management and careful device design are required to mitigate resistive losses and parasitic absorption.

Beyond these practical fabrication and thermal challenges, NF-TPV systems also present deeper conceptual and modeling difficulties. Computational studies have consistently shown that near-field devices are highly interdependent, in the sense that an accurate picture of the overall system cannot be formed by considering characteristics of the isolated subsystems. Physically, the features of the emitter cannot be separated from those of the cell because the evanescent transfer phenomenon lying at the heart of the NF-TPV value proposition is intrinsically non-local. It is a phenomenon of sets of objects, not single objects [99, 100, 101]. Conceptually, the concomitant photonic, electronic, and heat transfer physics cannot be fully separated because they all reside downstream from the same choices of material and geometric structuring control parameters [85, 102, 103].

## Carrier transport

Photon management, whether in a far-field or near-field configuration, is only one component of the TPV optimization. To fully leverage the benefits of enhanced photon harvesting, it is also essential to optimize the internal electrical performance of the device through effective charge carrier management. Once electron-hole pairs are generated in a TPV device, they must be separated and collected efficiently to contribute to the electrical current. Strategies for enhancing the carrier collection involve tailoring the device structure to reflect the material quality and reduce the losses from recombination and resistive effects [5, 104]. Ohmic losses are of particular concern, as they increase quadratically with current density and significantly impact the fill factor [5]. Reducing the series resistance may require further development of transparent lateral conduction layers, low-resistance interfacial contacts, or optimized metal grids [5]. For instance, Tervo et al. demonstrated that thick electroplated back reflectors and front gridlines (4 and >6 μm, respectively) helped to lower the series resistance of InGaAs-based TPV cells to approximately 6.5 mΩ/cm² [31]. High internal quantum efficiency (IQE) also depends on minimizing the non-radiative bulk and surface recombination, which requires high material quality with minimal defects and impurities. Achieving such quality is often challenging and costly, particularly for complex or metastable materials. Another important consideration is thermal management, as elevated cell temperatures degrade performance by increasing the dark current, lowering the open-circuit voltage and shortening carrier lifetimes [5, 105]. For example, Wernsman et al. reported a 1.4 V drop in $V_{oc}$ and 3.6% lower absolute efficiency when a 0.6 eV InGaAs monolithic module was heated from 297 K to 337 K under constant illumination [14]. The challenge is even greater for materials such as GeSn, which can be sensitive to thermal annealing [106, 107, 108]. Maintaining a stable operating temperature, ideally near room temperature, is therefore crucial [15, 60].

## Performance assessment

The absence of a standardized approach for estimating power conversion efficiency also poses a significant challenge, since it hinders meaningful comparisons and slows the advancement of TPV technology. In TPV devices, PCE is often influenced by multiple factors that include material quality, thermal management, and the geometric configuration of the emitter-cell system. However, there is currently no universally-accepted method for accurately quantifying these variables across different experimental setups [20, 31, 109]. Two main approaches are broadly used to estimate the PCE of TPV devices [4, 109, 110]. The first relies on measuring the radiative properties of the cell and emitter, specifically the emitter's spectral emissivity and the cell's reflectance, to evaluate the net radiative power $P_{abs}$ absorbed by the cell. In this method, $P_{abs}$ is inferred from the difference between the incident power from the emitter and the portion reflected by the device. For this approach to be accurate, the optical properties of the emitter-cell pair must be evaluated under realistic operating conditions, while accounting for geometric effects such as the view factor, i.e., the fraction of emitter radiation that reaches the cell [4]. Notably, this method does not directly measure efficiency. Instead, it combines the measured electrical output corresponding to a hot emitter with a separate reflectance measurement to estimate the absorbed power. While relatively straightforward, the approach has significant limitations. It assumes a two-way radiative exchange between the cell and thermal emitter, and uses the effective emissivity $\varepsilon_{eff}$ of the pair, as presented by Mahorter et al. [4]. However, as Tervo et al. noted, this definition holds only when the two surfaces are infinite and parallel, or specular concentric cylinders or spheres [31]. Moreover, this method is highly sensitive to the reflectance $R$. Even a small uncertainty in $R$ can lead to substantial error in estimating $P_{abs}$. This issue is especially relevant when $R$ is measured externally, without properly accounting for the angle of incidence and temperature dependence [110]. To address these challenges, Mahorter et al. recommended using the angle-of-incidence-weighted total (specular and diffuse) reflectivity of the TPV device [4], while Roy-Layinde et al. proposed in-operando measurement of the emitter's spectral emittance, controlling the cell temperature to minimize bandgap narrowing, and verifying the consistency of radiative properties before and after testing [20]. Still this method remains relevant for an apparent view factor below unity, where the cell area is larger than the emitter area or the aperture defining the emitter beam size [109].

The second method for estimating $P_{abs}$ relies on calorimetry, and quantifies the absorbed energy by measuring how much heat flows through the device [3, 22, 31, 109]. The principle is that all the absorbed incident energy is either dissipated as heat or converted to electrical power. A typical implementation uses a metal post with regularly-spaced thermistors, mounted on a temperature-controlled heatsink [31]. The parasitic heat to be quantified is the sum of the heat flowing through the metal post and the heat conducted through the electrical probes, with the latter estimated by placing additional thermocouples on the probes. This method is particularly suitable when the emitter–cell view factor approaches unity (i.e., small separation and comparable areas). In such cases, it can capture losses due to series resistance at high photocurrent and account for variations in the radiative properties [109]. However, practical implementation is rare, as maintaining a high view factor requires careful thermal and mechanical engineering to avoid losses due to overheating, convection, or material deposition from the emitter [109]. In systems with a low view factor, high photocurrent cannot be achieved and series resistance losses are underrepresented [22, 31]. Moreover, the use of baffles, defining apertures, and protecting electronic components limits the solid angle of the emitter, reducing the angular coverage of radiative exchange between emitter and cell [20, 31, 109].

## Future developments

MWIR TPV devices present an exciting opportunity for high-efficiency radiation harvesting and energy conversion, particularly in environments where conventional technologies fall short. However, unlocking their full potential requires overcoming a series of interrelated challenges. In fact, optimization efforts are needed at multiple levels to advance MWIR TPV performance. A primary obstacle lies in achieving precise spectral matching between the thermal emitter and photovoltaic cell. Efficient MWIR TPV operation demands that the emitters and absorbers be optimized for harvesting low-energy photons (<0.5 eV), necessitating the development of tailored narrow band emitters alongside PV cells with tunable ultra-low bandgaps. In this regard, material development remains crucial. While semiconductors such as InSb, HgCdTe, and Pb-based semiconductors offer low bandgaps, issues of toxicity, stability, cost, and scalability persist. Emerging materials, such as GeSn alloys, which are compatible with silicon processing, offer a high-potential route but still face hurdles in lattice matching, carrier mobility, and stability. Carrier recombination losses present another critical challenge, especially as lower photon energies heighten the sensitivity to non-radiative decay. This demands advances in high-quality epitaxial growth and surface passivation to ensure long minority carrier lifetimes. Additionally, thermal management and optical concentration strategies will be vital to maximize the photon flux from modest-temperature emitters. Solutions include the development of high-emissivity selective surfaces, optical concentrators, and advanced thermal insulation to minimize parasitic losses. To further enhance efficiency, back reflection or photon recycling architectures must be integrated to recover sub-bandgap photons, employing innovations like cold-side spectral filters and highly reflective mirrors.

In parallel with material and photonic optimization, advanced far-field architectures such as tandem and ICPV devices present a promising direction to improved spectral utilization and voltage scalability. Future progress will likely hinge on the development of novel superlattice materials such as InAs/InAsSb, which exhibit longer carrier lifetimes and the potential for improved device performance; however, their performance and stability under high-temperature operation remain largely uncharacterized and require further investigation [65, 111]. The integration of absorbers within resonant cavities or the implementation of sophisticated light-trapping structures can further enhance broadband absorption and suppress recombination losses, pushing device efficiencies higher [65, 112, 113]. Expanding cascade designs to incorporate diverse absorber types, including type I and II quantum wells, promises broader spectral coverage, although this will require careful engineering to overcome current-matching constraints [65, 114, 115]. Ultimately, sustained progress in high-quality epitaxial growth and fabrication processes will be necessary to fully unlock the potential of narrow-bandgap semiconductors for TPV applications.

**In ref. [116], it was shown that no amount of nano-structure engineering can enhance the near-field radiative heat transfer much beyond what is observed between half-spaces supporting a hybridized gap polariton mode. Taken at face value, this result suggests that multilayer configurations represent an optimal design strategy for NF-TPV systems, and consequently, that more complicated alternatives may offer limited benefits. While there is some wisdom in this conclusion, we should remember that the magnitude of the radiative heat transfer is only one of many factors that determine the viability of NF-TPV architectures [81, 117, 118].**

Although some recent works have incorporated more realistic electronic descriptions [119, 120, 121, 122, 123], current efforts remain far from capturing the totality of the underlying physics and inherent fluctuations in device processing, as sizeable gaps persist between the measured and simulated efficiencies. For example, ref. [124] reports record power densities of ~5,000 W/m² at an efficiency of 6.8%, whereas idealized models predict up to 53% conversion efficiency and 150,000 W/m² output at 1300 K [123]. For example, beyond simply supplementing basic drift-diffusion with frequency-dependent parameters [125, 126, 127], few concrete results exist regarding how the electronic and thermal transport properties of polaritons could influence NF-TPV performance [128, 129]. More broadly, to the best of our knowledge, a complete co-optimization of electronic, photonic, and thermal properties in a multilayer NF-TPV configuration with realistic electrical contacts has yet to be achieved. Existing optimization strategies typically switch sequentially between domain-specific solvers for predefined material stacks, rather than simultaneously addressing the full multi-physics of the system. The reason for this persistent gap, despite the long history of interest [130], is obvious but nevertheless worth stating. Namely that concurrently optimizing the multi-physics involved in a generalized NF-TPV system would require a unified computational framework for stochastic electromagnetic, charge transport, and heat transport that is capable of performing simulations at least as well (and likely better) than the current state-of-the-art for each of these domains individually. While this assertion may appear discouraging, we take it as one of the greatest reasons for optimism regarding the long-term prospects of NF-TPV technologies. The past decade has seen tremendous advances in photonic simulation and inverse design tools [131, 132, 133]. For the design of nanophotonic components such as resonant cavities [134], grating couplers [135], and lenses [136], these techniques have led to the discovery of novel architectures that greatly outperform traditional design templates. Given the intricate interplay of design strategies needed to optimize NF-TPV systems, there is every reason to believe that similar improvements will be realized once these increasingly-powerful numerical techniques are finally brought to bear [137]. From the electromagnetic perspective alone, the question of how to approach the fundamental limits for radiative heat transfer in the absence of materials directly supporting surface polaritons is highly compelling, both in its own right and for its deep connection to questions concerning information transfer and multiple-input/multiple-output systems [138, 139, 140]. When further coupled with the creation of electronic excitations and charge transport phenomena, there are also immediate ties to detection, imaging, and parameter estimation problems [141, 142, 143]. Moreover, recent work suggests that multi-physics formulations that couple electromagnetics to auxiliary thermal physics can act as effective surrogates for discovering connected and easily fabricable device layouts [144]. With this understanding in mind, the complexity that has forestalled the development of commercially viable NF-TPV devices could well become an engine driving its progress. The problem of efficient energy extraction essentially concerns all of the engineering.

Finally, system-level considerations, including packaging, vacuum maintenance, mechanical robustness, and thermal cycling resistance, must also be considered for practical deployment. Scalability and cost reduction will ultimately determine the viability of MWIR TPV systems for widespread applications. Maintaining stable nanogaps, managing nanoscale thermal effects, and advancing the understanding of near-field radiative transfer are major challenges that must be addressed in NF-TPV systems. Regardless of the design, the long-term durability of MWIR TPV devices under high thermal and oxidative stresses must be ensured to achieve practical operational lifetimes and power conversion efficiencies. Success in these areas would position MWIR TPV systems as compelling competitors to batteries and traditional thermoelectric converters, and open new frontiers in portable power generation, industrial waste heat recovery, and beyond.

## Concluding Remarks

With its standing at the intersection of several rapidly evolving disciplines: photonics, materials science, and thermal engineering, MWIR TPV offers a unique opportunity for efficient heat-to-electricity conversion in scenarios inaccessible to conventional energy technologies. Here, we have highlighted the multifaceted challenges that hinder the realization of high-efficiency TPV systems, including spectral mismatch, parasitic absorption, non-radiative recombination, fabrication constraints, and the inherent difficulty of achieving stable nanoscale emitter-cell gaps for near-field devices. Despite these hurdles, recent breakthroughs in materials development and device architectures, such as tandem and cascade configurations, air-bridge designs, and resonant photonic structures, have demonstrated clear pathways toward enhanced spectral utilization and higher power conversion efficiency. The emergence of novel low-bandgap materials like GeSn alloys and type-II superlattices promises broader spectral coverage, while innovations in optical design, such as transmissive spectral control and near-field enhancement, continue to push performance boundaries. Still, TPV development is far from mature. A fully integrated approach that encompasses material growth, photon management, carrier dynamics, and thermal regulation is critical to unlocking the true potential of both far-field and near-field systems. Furthermore, robust modeling frameworks that simultaneously capture electromagnetic, electronic, and thermal transport physics are urgently needed to bridge the gap between simulation and experiment. Ultimately, the practical deployment of MWIR TPV systems will require addressing system-level considerations such as packaging, mechanical robustness, cost, and scalability. Nonetheless, the compelling prospects of compact, high-efficiency, and non-mechanical power sources capable of harvesting ubiquitous thermal energy make TPV a fertile ground for innovation. As computational tools, material platforms, and device concepts continue to evolve, MWIR TPV is poised to emerge as a disruptive technology for next-generation energy conversion.

*Funding*

NSERC Canada (Discovery and Alliance Grants), Canada Research Chairs, Canada Foundation for Innovation, Mitacs, PRIMA Québec, Defence Canada (Innovation for Defence Excellence and Security, IDEaS), the European Union's Horizon Europe research and innovation program under Grant Agreement No. 101070700 (MIRAQLS), the US Army Research Office Grant No. W911NF-22-1-0277, and the Air Force Office of Scientific and Research Grant No. FA9550-23-1-0763.

*Disclosures*

The authors declare no conflicts of interest.